\documentclass[11pt]{ar-1col}
\usepackage{url}
\usepackage[numbers, sort&compress]{natbib}
\usepackage{amsmath}
\usepackage{amssymb}
\usepackage[letterpaper, margin=1.33in]{geometry}
\usepackage{hyperref}

\renewcommand\[{\begin{equation}}
\renewcommand\]{\end{equation}} 

\newcommand{\sgn}{\mathcal{\text{sgn}}}

\jname{Annu. Rev. Condens. Matter Phys.}
\jvol{AA}
\jyear{202Y}
\doi{10.1146/((article doi))}

\begin{document}

\markboth{Shaffer and Levchenko}{Theories of SDE}

\title{Theories of Superconducting Diode Effects}


\author{Daniel Shaffer,$^1$ Alex Levchenko,$^1$
\affil{$^1$Department of Physics, University of Wisconsin-Madison,\\
Madison, Wisconsin 53706, USA}
}

\begin{abstract}
Superconducting diode effects (SDE), both in bulk superconductors and in Josephson junctions, have garnered a lot of attention due to potential applications in classical and quantum computing, as well as superconducting sensors. Here we review various mechanisms that have been theoretically proposed for their realization. We first provide a brief historical overview and discuss the basic but subtle phenomenological Ginzburg-Landau theory of SDE, emphasizing the need to the simultaneous breaking of time-reversal and inversion symmetries. We then proceed to more microscopic treatments, focusing especially on implementations in noncentrosymmetric materials described by the Rashba-Zeeman model. Finally, we review proposals based on other condensed matter systems such as altermagnets, valley polarized and topological materials, and systems out of equilibrium.
\end{abstract}

\begin{keywords}
Superconducting diode effect, Josephson diode effect, nonreciprocal supercurrent, finite-momentum pairing, noncentrosymmetric materials, anomalous Josephson effect, Ginzburg-Landau theory, time-reversal symmetry breaking, inversion symmetry breaking, spin-orbit coupling.
\end{keywords}
\maketitle


\section{Introduction}

The superconducting diode effect (SDE) and the closely related Josephson diode effect (JDE) refer to the nonreciprocity of the critical supercurrents \(J_{c+}\) and \(J_{c-}\) flowing in opposite directions in bulk superconductors (SCs) and Josephson junctions (JJs), respectively: namely, SDE and JDE are defined by \(J_{c+}\neq -J_{c-}\).\footnote{Here we choose opposite currents to have opposite sign; some authors use a different convention.}
For purposes of this review, we will reserve the term SDE for the effect in bulk materials, though in the literature it is often applied to JDE as well. Both effects require time-reversal and inversion symmetries (TRS and IS), as well as any other symmetry exchanging \(J_{c+}\) and \(J_{c-}\), to be broken.\footnote{The normal state diode only requires inversion symmetry breaking because TRS is broken by the ensuing dissipative current. An additional TRS breaking is thus necessary for SDE and JDE due to the absence of dissipation in a supercurrent.} One reason that SDE in particular is quite rare in nature is that those are precisely the symmetries that guarantee the pairing in stability in BCS-type superconductors. On the other hand, both TRS and IS breaking (TRSB and ISB) can be achieved by a plurality of different means. This has led to an ever-increasing list of different theoretical proposals for realizing SDE and JDE in recent years, which has not been fully covered by several reviews that already exist on the subject \cite{Nadeem23, MollGeshkenbein23, NagaosaYanase24, Ma25}. Though this list keeps expanding, we attempt to capture the main entries in this review.

The main dichotomy in the list is the distinction between intrinsic and extrinsic effects, though the definitions of these terms are not universally agreed upon in the literature, especially for the case of JDE. In particular, either TRS or IS can be broken intrinsically by the material or extrinsically by an applied field or current, for a total of four possible categories. The term intrinsic SDE was first introduced in the context of noncentrosymmetric SCs (NCSs) that exhibit intrinsic IS breaking \cite{DaidoYanase22}, with TRS broken extrinsically by an applied magnetic field. We will generally stick to that terminology. By this definition, the intrinsic SDE was first noted by Levitov et al. in 1985 \cite{LevitovNazarovEliashberg85} and reported experimentally by Ando et al. in 2020 \cite{AndoYanase20}. Since then the so-called field-free realizations of intrinsic SDE had also been proposed \cite{ScammellScheurer22, ChazonoYanase23, BanerjeeScheurer24, BanerjeeScheurer25, ChenSchrade25, Yoon25, ChenScheurerSchrade25, DaidoYanaseLaw25,  SamantaGhosh25, BhowmikGhosh25} and observed \cite{LinScheurerLi22, NaritaYanase22, Qi25, NagataYanase25}.
The case of SCs with intrinsic (usually spontaneous) TRSB such as chiral topological SCs (TSCs) with extrinsic ISB due to boundary conditions have been considered less often \cite{ZinklSigrist22}.
Historically, on the other hand, completely extrinsic SDE had been the earliest category to be observed \cite{EdwardsNewhouse62, SwartzHart65} but had not attracted as much attention, as we discuss in Sec. \ref{GLSec}.

In JJs -- since IS is almost always extrinsically broken by imperfect device geometry and magnetic fields are commonly applied in device operation -- extrinsic JDE was technically observed shortly after the Josephson effect itself and potentially considered an undesirable quality \cite{DeWaele67, Goldman67}. Intrinsic JDE, furthermore, can be further subdivided by whether TRSB and ISB occur in the SC regions or the tunneling barrier region, and moreover the two SC regions may be distinct (which in itself results in ISB).
The term JDE has also been applied to characterize the nonreciprocity of the retrapping current \(J_{r\pm}\) that can occur in presence of dissipative currents and does not require TRSB (see Sec. \ref{NonEqSec}). As the critical currents remain reciprocal in the absence of TRSB, the nonreciprocity of retrapping currents may be considered to be distinct from JDE, and can instead be referred to as pseudo-JDE following Ref. \cite{WangWangWu25}.
We will mostly restrict \emph{the} intrinsic JDE to refer to the case of ISB (and possibly TRSB) in the SC regions.

SDE in bulk materials is closely linked to nonuniform SC states, i.e., SCs with Cooper pairs that carry nonzero total momentum \(\mathbf{q}\). Nonuniform SCs include helical SC states in NCSs, still the most widely-studied system in the context of SDE, as well as the Fulde-Ferrel-Larkin-Ovchinikov (FFLO) state and pair density waves (PDW) \cite{FF, LO, Agterberg20}. In particular, either TRS or IS imply \(\mathbf{q}=0\), and so it is typical for SCs with SDE to be nonuniform. In that sense, SDE plays a similar role to nonuniform SCs as the quantum spin/anomalous Hall effect (QSHE/QAHE) does to topological/Chern insulators (TI/CI). However, it should be emphasized that, in principle, nonuniform SCs need not exhibit SDE, especially when they are stabilized by nesting conditions under which the system is (approximately) invariant with respect to an inversion-like symmetry that takes \(\mathbf{q}+\mathbf{q}_0\) to \(-\mathbf{q}+\mathbf{q}_0\). Nor do SCs with SDE necessarily must be nonuniform, for example in 2D systems with three-fold rotation symmetry. However, even if the ground state of the SC is uniform, excited supercurrent-carrying states, which generally carry momentum, are not: indeed, the critical currents \(J_{c\pm}\) are computed as maxima and minima of the supercurrent as a function of the Cooper pair momentum, \(J(\mathbf{q})\).

JDE is similarly related to the anomalous Josephson diode effect (AJE), first predicted by Geshkenbein and Larkin in 1986 in the context of junctions between singlet and triplet SCs \cite{GeshkenbeinLarkin86}. In JJs, the role of the Cooper pair momentum \(\mathbf{q}\) is played instead by the SC phase bias \(\varphi\) across the junction, with the current being determined as a function of the phase bias in the current-phase relationship (CPR) \(J(\varphi)\). AJE is defined as \(J(0)\neq 0\), i.e., a nonzero supercurrent flowing at zero phase bias, which is equivalent to the statement that \(J(\varphi_0)=0\) for \(\varphi_0\neq 0\), i.e., that the zero current state of the JJ requires a nonzero phase bias.  AJE has been observed in induction measurement experiments in 2016, as reported in Ref. \cite{SickingerGoldobin12, SzombatiKouwenhoven16}, and has been extensively studied in the past \cite{Buzdin05,Shukrinov22}.
JDE, on the other hand, is again defined in terms of the maxima and minima of \(J(\varphi)\), and thus is not directly linked to AJE: logically, the two can occur independently. However, since AJE also requires both TRSB and ISB, the two effects typically occur together.

From a practical standpoint, and as suggested in their name, both SDE and JDE have attracted a lot of interest due to their potential applications in superconducting circuits for either classical or quantum computation.
From that perspective, the distinction between intrinsic and extrinsic effects, or even between SDE and JDE, is not as important. So far, however, few practical applications of SC diodes have been realized, one of the main limitation being the diode efficiency. It is now standard to define the SC diode efficiency coefficient \(\eta=(J_{c+}+J_{c-})/(J_{c+}-J_{c-})\) as a figure of merit of the effect: \(\eta=0\) in the absence of the effect and \(\eta=\pm 1\) if either \(J_{c+}=0\) or \(J_{c-}=0\), in which case the diode effect is said to be perfect. For proper diode operations the perfect diode effect is highly desirable, but so far has not been conclusively reported (outside of multiterminal and nonequilibrium devices). Generally, JDE exhibit higher \(|\eta|\), in the \(30-50\%\) range, while most reported SDE efficiencies lie in the \(5-25\%\) range. However, SDE offers an advantage in terms of the magnitude of the critical currents, on the order of mA, compared to \(\mu\)A critical currents in most JJs.

From the standpoint of basic condensed matter physics, the intrinsic bulk SDE is of greater fundamental interest, as it directly reflects the microscopic properties of the material system. This will therefore be our main focus below, particularly in the context of noncentrosymmetric superconductors discussed in Sec.~\ref{NCSSec}. Nevertheless, even from this perspective, there are many useful parallels with extrinsic effects and with the JDE-especially in SQUIDs, as illustrated by the basic phenomenology discussed in Sec.~\ref{GLSec}. Moreover, Josephson-junction experiments can themselves serve as sensitive detection devices for identifying properties of their constituent materials. Studying the various possible mechanisms for realizing both SDE and JDE, which we review in Sec.~\ref{OtherSec}, is valuable not only for potential applications but also for developing new probes of condensed matter systems, in particular unconventional superconductors.

\section{History and Phenomenology}\label{GLSec}

Although the intrinsic SDE was first observed in 2020 \cite{AndoYanase20} and had been theoretically predicted much earlier -- by Levitov, Nazarov, and Eliashberg in 1985 \cite{LevitovNazarovEliashberg85}, and independently by Edelstein in 1996 \cite{Edelstein96} -- we must go back even further in time for the discovery of both the extrinsic SDE and JDE. This takes us at least to 1962, when Edwards and Newhouse, working at General Electric, measured the critical current in thin Sn films without any externally applied field. They used normal-metal wires running parallel to the current on both sides of the film, thereby producing “pinch” currents \cite{EdwardsNewhouse62}. From their data, a notably large asymmetry parameter, $\eta\sim92\%$ , can be estimated. Soon after, Swartz and Hart, also at GE, reported similar asymmetric critical currents in Pb-alloy thin superconducting films in 1965 \cite{SwartzHart65} and 1967 \cite{SwartzHart67}, using a setup similar to that of Ref.~\cite{EdwardsNewhouse62}. They also observed the effect without pinch currents, in the Saint-James–de Gennes superconducting state nucleated at the film surface \cite{SaintJamesDeGennes63, Tinkham04}, as well as in samples with triangular cross-sections.

The discovery of the extrinsic SDE was largely overshadowed at the time by other seminal breakthroughs of 1962, including the prediction of the Josephson effect \cite{josephson_possible_1962} and the observation of the Little-Parks effect \cite{LittleParks62}. It did not help that the theoretical framework needed to properly explain the experiment had not yet been fully developed. The most likely mechanism behind the SDE observed in Refs.~\cite{EdwardsNewhouse62,SwartzHart65,SwartzHart67}, as later suggested in Ref. \cite{VodolazovPeeters05}, is the vortex diode effect (VDE), arising from either an asymmetric Bean-Livingston barrier or an asymmetric pinning potential for Abrikosov vortices in type-II superconductors (we discuss the theory of VDE in Sec.~\ref{VDESec}). The Bean–Livingston theory and the theory of vortex flow were published only in 1964 \cite{BeanLivingston64} and 1965 \cite{BardeenStephen65, Kim65}, respectively. Even the validity of Abrikosov's vortex theory -- originally published in the Soviet Union in 1957 -- was not fully recognized until after the experimental observation of superconducting vortices in 1967 \cite{Cribier67, Essmann67, Abrikosov04}. It was also not firmly established at the time whether the thin Sn films used by Edwards and Newhouse were indeed type-II superconductors \cite{SwartzHart67, Miller68}. Consequently, although the VDE has been rediscovered several times since, the original discovery remained largely forgotten until recently.

Coincidentally, the extrinsic JDE was also discovered around the same time, in 1967, in SQUIDs by De Waele et al. \cite{DeWaele67, DeWaele69} and in wide asymmetric Josephson junctions (JJs) by Goldman and Kreisman \cite{GoldmanKreisman67}, shortly after the invention of the SQUID in 1964.\footnote{Foreshadowing the intrinsic effects, the FFLO state had also been predicted in 1964 \cite{FF,LO}.} In both cases, extrinsic orbital effects -- namely, inductance in SQUIDs and Meissner screening currents in wide JJs -- play a crucial role in producing the diode effect. The JDE in a SQUID, in particular, provides an instructive example showing that time-reversal and inversion symmetry breaking (TRSB and ISB) alone are not sufficient to guarantee a diode effect -- a lesson that generalizes to both intrinsic JDE and SDE.

We therefore begin our theoretical overview with the phenomenological Ginzburg–Landau (GL) theory of the JDE in single-channel and multichannel JJs, including SQUIDs, in Sec.~\ref{GLJDESec}. We then turn to the GL theory of the bulk SDE in Sec.~\ref{GLSDESec}, which, as we will see, shares several important similarities.

\subsection{GL theory of AJE and JDE}\label{GLJDESec}

The GL theory for single channel JJs can be formulated in terms of the complex SC order parameters \(\Psi_L=|\Psi_L|e^{i\phi_L}\) and \(\Psi_R=|\Psi_R|e^{i\phi_R}\) (\(L/R\) for the left/right terminals of the junction). The GL free energy of the JJ can then in general be written as
\begin{align}
    \mathcal{F}[\Psi_L,\Psi_R]&=\alpha_L|\Psi_L|^2+\alpha_R|\Psi_R|^2+\left(\alpha_{LR}\Psi_L^*\Psi_R+\text{c.c.}\right)+\sum_{ABCD=L,R}\beta_{AB}^{CD}\Psi_A^*\Psi_B^*\Psi_C\Psi_D+\dots\nonumber\\
    &=\alpha_L|\Psi_L|^2+\alpha_R|\Psi_R|^2 +2\Re[\alpha_{LR}]|\Psi_L\Psi_R|\cos\varphi+2\Im[\alpha_{LR}]|\Psi_L\Psi_R|\sin\varphi+\dots
\end{align}
where \(\varphi=\phi_R-\phi_L\). TRS acts as \(\Psi_A\rightarrow\Psi_A^*\) while IS acts as \(\Psi_L\leftrightarrow\Psi_R\).
The current is computed as \(J=\delta\mathcal{F}/\delta A=-2e\partial_\varphi\mathcal{F}\) where \(e\) is the electron charge (we set \(c=\hbar=1\); fist equality is by definition, second equality follows from minimal coupling since the gauge-invariant quantity is \(\varphi - 2e \int\mathbf{A}\cdot d\mathbf{l}\)). To leading order in the order parameters, the CPR is thus given by
\[J(\varphi)=4e\Re[\alpha_{LR}]\sin\varphi-4e\Im[\alpha_{LR}]\cos\varphi\]
Note that either TRS or IS require \(\Im[\alpha_{LR}]=0\), and more generally in \(J(\varphi)=-J(-\varphi)\). \(\Im[\alpha_{LR}]\neq0\) is thus the simplest term one can introduce in the GL theory to break both of these symmetries. In particular, this leads to AJE:
\[J(\varphi)=4e|\alpha_{LR}|\sin(\varphi-\varphi_0)\]
where
\[\varphi_0=\arctan(\Im[\alpha_{LR}]/\Re[\alpha_{LR}])\]
is the anomalous phase.
Importantly, however, there is, at this order, \emph{no JDE}: the critical currents are just the maximum and the minimum of the CPR, so \(J_{c\pm}=\pm4e|\alpha_{LR}|\).

More generally, the CPR can be expanded in harmonics \cite{GolubovKupriyanov04}:
\[J(\varphi)=\sum_{n}(J_n\sin n\varphi+I_n\cos n\varphi)\,,\label{CPR}\]
where the presence of \(I_n\) terms require broken symmetries. 
The possibility of a nonvanishing first harmonic \(I_n\)	
  in the absence of both TRS and IS was first noted by Geshkenbein and Larkin \cite{GeshkenbeinLarkin86}, who thereby predicted the anomalous Josephson effect (AJE).
Higher-order harmonics arise from terms of higher order in the superconducting order parameters -- for example, \(\beta_{LLRR}|\Psi_L|^2|\Psi_R|^2\cos2\varphi\) -- and are usually subleading, particularly in short junctions.
In SNS junctions, such contributions correspond to multiple Andreev reflection (MAR) processes, as discussed in Ref.~\cite{ZazunovEgger24}, which also emphasized the necessity of higher harmonics for realizing the JDE.
As further pointed out in Ref.~\cite{FominovMikhailov22}, the presence of multiple harmonics is indeed a crucial ingredient for achieving the JDE in these systems.
The importance of the second harmonic has also been recognized in planar JJs based on noncentrosymmetric heterostructures \cite{JeonParkin22} and in twisted cuprates \cite{ZhaoKim23, Zhu23, Ghosh24, VolkovFranz24}.

\begin{figure}[t]
\includegraphics[width=0.99\textwidth]{JDEfig.pdf}
\caption{JJ energies (top, blue) and CPRs (bottom, orange): (a) with a single harmonic \(J_1\), without either AJE or JDE; (b) with a an additional first harmonic \(I_1\) with ISB and TRSB resulting in a single harmonic CPR shifted by \(\varphi_0\neq0\) (i.e., AJE), but no JDE; and (c) with an additional second harmonic \(J_2\) resulting in JDE.}
\label{JDEfig}
\end{figure}

A minimal realization of JDE thus requires \(J_1, I_1\) and \(J_2\) to be nonzero, as shown in Fig. \ref{JDEfig}. This case has been considered in Ref. \cite{Goldobin11} in the context of realizing \(\varphi\)-JJs (i.e., JJs with AJE), who noted the asymmetry of the CPR and studied the critical currents, but did not identify the JDE at the time.
An alternative scenario is possible even if the JJ does not explicitly break either TRS or IS, such that free energy \(\mathcal{F}\) is still symmetric under \(\varphi\rightarrow-\varphi\), but instead breaks them spontaneously due to the ground state of the JJ lying at \(\varphi\neq0,\pi\). Such a scenario for a spontaneous AJE is realized for example with \(J(\varphi)=J_1\sin\varphi+J_2\sin2\varphi\) for sufficiently negative \(J_2\), as studied in Ref. \cite{GoldobinBuzdin07}.
A similar spontaneous bulk SDE can also occur due to spontaneous TRSB/ISB, for example in an FFLO state (see Sec. \ref{FFLOSec}). In both cases, due to an energy barrier between two symmetry-breaking ground states a dissipative current may start flowing at a lower critical current corresponding to the local minimum nearest the initial ground state, resulting in a diode effect. Alternatively, at lower damping the system may instead exhibit a memory effect.
In all cases, we can make the general statement about the diode efficiency \(\eta=(J_{c+}+J_{c-})/(J_{c+}-J_{c-})\): assuming the JJ is stable, i.e., \(\mathcal{F}(\varphi)\) is bounded from below, \(|\eta|<1\). This follows from the fact that \(J(\varphi)=0\) and \(\partial_\varphi J(\varphi)>0\) in the ground state, so that \(J_{c-}\leq 0 \leq J_{c+}\), from which the bounds follow. As we will see in Sec. \ref{NonEqSec}, this bound does not necessarily hold in open systems driven out of equilibrium.

\subsubsection{SQUIDs and other Multichannel Junctions}\label{SQUIDSec}

Various mechanisms can give rise to nonzero values of \(I_1\) and \(J_2\), or to other higher-order harmonics required for the JDE. These mechanisms are often nontrivial and form the focus of the remainder of this review. The simplest and historically first example of the JDE is provided by the asymmetric SQUID, as originally demonstrated in Refs.~\cite{DeWaele67, DeWaele69, FultonDynes72}.
The CPR of a SQUID composed of two JJs in parallel is given simply by 
\(J(\varphi_1,\varphi_2)=J_1(\varphi_1)+J_2(\varphi_2)\), where \(J_n\) and \(\varphi_n\) are the CPRs and phases of each JJ, respectively. More generally, one can consider a multichannel JJ with \(J(\varphi)=\sum_n J_n(\varphi_n)\), including multi-loop devices and wide JJs. In the case of a SQUID, the magnetic flux \(\Phi\) through the loop determines the phase difference \(\varphi_1-\varphi_2=2\pi\Phi/\Phi_0\) where \(\Phi_0=h/(2e)\) is the flux quantum. Assuming \(J_n(\varphi_n)=J_{cn}\sin\varphi_n\), the flux thus generates the \(I_1\) term in \(J(\varphi,\Phi)\), which is enough the generate AJE but not JDE.

While second and higher order harmonics may be generated in a single JJ due to MAR processes \cite{ZazunovEgger24}, the SQUID provides an alternative mechanism if the inductance \(L\) of the loop is accounted for, as done in \cite{DeWaele67, DeWaele69, FultonDynes72}. This gives rise to an additional magnetic field due to the current circulating in the loop, such that \(\Phi=\Phi_{ext}+L(J_1-J_2)\). Assuming \(L\) is small, one can then expand the CPR to linear order in \(L\):
\[J(\varphi,\phi)\approx J_{1c}\sin(\varphi+\phi)+ J_{2c}\sin\varphi-\frac{\pi L}{\Phi_0}J_{2c}\left(J_{1c}\sin\varphi_0(1+\cos2\varphi)+(J_{1c}\cos\varphi_0-J_{2c})\sin2\varphi\right)\]
where \(\varphi=\varphi_1-\phi\) and \(\phi=2\pi\Phi_{ext}/\Phi_0\). In particular, both the second harmonic and the constant (zeroth-order) term, \(I_0\)), can be generated by the circuit inductance, even when the individual JJs exhibit standard CPRs containing only the first harmonic. The constant term -- noted, for instance, in Ref.~\cite{CuozzoShabani24} -- must remain bounded so that 
\(J(\varphi,\phi)\) changes sign, thereby ensuring the stability of the SQUID. This condition, however, no longer holds once the system is driven out of equilibrium, as discussed in Sec.~\ref{NonEqSec}.

This inductance-driven mechanism for generating higher-order harmonics and realizing the JDE -- recently found to dominate over multiple Andreev reflection (MAR) processes in experiments~\cite{Kim25} -- has, in fact, been recognized for decades~\cite{DeWaele67, DeWaele69, FultonDynes72, Peterson79, Likharev22}. Following the discovery of the intrinsic SDE in 2020, several theoretical proposals for the JDE have invoked the asymmetric SQUID geometry, as well as related multiloop JJ arrays such as Andreev molecules, or configurations introducing additional asymmetry between the two arms~\cite{FominovMikhailov22, SeoaneSouto22, Haenel22, Pillet23, HodtLinder23, SongBezryadin23, BozkurtFatemi23, LeggLossKlinovaja23, SeleznevFominov24, DeSimoniGiazotto24, GrecoGiazotto24, Kotetes24, YerinGiazotto25, Zalom25}. A number of these designs have already been realized experimentally~\cite{MayerShabani20, PaolucciGiazotto23, ReinhardtManfra24, Wu25, ChieppaGiazotto25, CiacciaManfra23, GrecoGiazotto23, Leblanc24, Yu24, Kim25, BanszerusManfra25, Li25, ReinhardtGlazman25, SongBezryadin23, Kudriashov25, MatsuoManfra25}.

A broad class of systems known as vortex or fluxon ratchets -- also termed Josephson vortex ratchets, fluxon rectifiers, or fluxonic diodes -- likewise exploit the JDE in SQUIDs and other multiloop JJ arrays when driven by an AC signal~\cite{Zapata96, Falo99, Carapella01, Lee03, Marconi07, Spiechowicz14, Semenov15, TriasFalo00, Weiss00, CarapellaCostabile01, Sterck02, CarapellaFilatrella02, Majer03, Shalom05, BeckGoldobin05, Streck05, Streck09, KnufinkeGoldobin12, Chesca17}. We discuss these systems in more detail in Sec.~\ref{VDESec}.

The same inductance effect can occur not only in SQUIDs but also wide JJs, which was proposed by Goldman and Kreisman as an explanation their observation of JDE in a single junction in 1967 \cite{GoldmanKreisman67}. Such inductance, or screening self-field, effects are well-known to result in skewed Fraunhofer patterns with  \(J_{c\pm}(\mathbf{B})\neq J_{c\pm}(-\mathbf{B})\) that demonstrate JDE at nonzero magnetic fields \cite{VanHarlingen95, KrasnovPedersen97, Monaco13, VarvaStrunk13, GolodKrasnov22, ChirolliCuocoGiazotto24} (recall that global TRS implies that \(J_{c-}(\mathbf{B})=-J_{c+}(-\mathbf{B})\)); see Fig. \ref{NonEqfig} (d) below for an illustration. Such such effects have generally been considered undesirable, especially in the context of identifying unconventional SC states with TRSB/ISB in JJ experiments that from a modern perspective would be using intrinsic AJE/JDE \cite{GeshkenbeinLarkin86, VanHarlingen95, YangAgterberg00, Nelson04, KaurAgterbergSigrist05}.

A related inductive mechanism, in which current-dependent Zeeman spin splitting leads to nonreciprocal transport, has also been shown to produce JDE in JJs containing magnetic impurities or chiral quantum dots~\cite{ChengSun23, SunMaoSun23, SunMaoSun25, DebnathDutta24, DebnathDutta25}. An effective SQUID description has likewise been invoked to interpret recent JDE observations in two-dimensional topological insulators, where the two ``arms" of the SQUID correspond to proximitized superconducting edge states~\cite{Li24, NikodemFuAndo25} (see Sec.~\ref{TopoSec}). Earlier, in 2010, a network of effective SQUIDs comprising JJs between superconducting puddles was proposed as a model for the bulk superconducting diode effect observed in Sr\(_2\)RuO\(_4\) with Ru inclusions~\cite{KaneyasuSigrist10}.

In this way, the asymmetric SQUID mechanism provides a unified framework for understanding both the Josephson and superconducting diode effects, including intrinsic realizations. More broadly, the same conclusion applies across all diode phenomena in superconductors: while broken TRS and ISB are necessary ingredients, they are not sufficient on their own. The emergence of a diode response additionally requires nontrivial interference between at least two current-carrying channels or bands -- for example, the two helical bands in noncentrosymmetric superconductors -- analogous to the interference produced by the circulating loop current in an asymmetric SQUID.

\subsubsection{Multiterminal Junctions}\label{GLJDEmultiSec}

Another generalization of JDE has been considered in multiterminal devices, i.e. multiple JJs connected either via a common normal/insulating region, or to a common SC island. The former had been proposed theoretically in the form of a Josephson triode (with three SC terminals) by Likharev already in 1975 \cite{Likharev79}, and have been used to experimentally realize JDE both in the former \cite{PankratovaVavilovShabani20, GuptaPribiag23, Coraiola24, Behner25}, and latter \cite{GolodKrasnov22, MatsuoManfra23, ChilesFinkelstein23, ZhangKayyalha24, ArnaultFinkelstein25} cases.
Theoretically, JDE in multiterminal JJs had been noted already in 2001 in Ref. \cite{AminOmelyanchouk01}, as well as in more recent studies \cite{VirtanenHeikkila24, HuamaniCorrea24, ZalomNovotny25, Takeuchi25, SahooSoori25, SahooSoori25_2}.
The free energies for such devices can then be constructed, respectively, with a vector of order parameters \(\boldsymbol{\Psi}=(\Psi_0,\Psi_1,\dots,\Psi_{N-1})\):
\begin{align}\label{Fi}
    \mathcal{F}[\boldsymbol{\Psi}]&=\sum_n\alpha_n\Psi_n^*\Psi_{n+1}+\text{c.c.},\qquad    \mathcal{F}[\boldsymbol{\Psi}]=\sum_n\alpha_n\Psi_n^*\Psi_{0}+\text{c.c.}
\end{align}
with \(n=0,1,\dots,N-1\) and in the former case defined modulo \(N\), the number of SC terminals.
More complicated circuits and junction arrays are of course possible, with more complicated couplings \(\alpha_{nm}\Psi_n^*\Psi_{m}\) as well as higher-order nonlinear terms.
One can consider either phase- or current-biased setups. With \(\phi_j\) corresponding to the phase of each region, one can write the condensation energy \(F(\phi_0,\dots,\phi_{N-1})\) corresponding to \(\mathcal{F}\) optimized over the order parameter magnitudes \(|\Psi_n|\)
(see Ref. \cite{PankratovaVavilovShabani20}), and the currents can be defined as \(J_n=2e\partial_{\varphi_n}F\) (for \(n\neq 0\) in the case of an SC island) with \(\varphi_n=\phi_n-\phi_{n-1}\) or \(\varphi_n=\phi_n-\phi_{0}\) for the normal/SC island cases, respectively, with slightly different physical meanings. For the case of an SC island, \(J_n\) is the current flowing into the island; for a normal island, \(J_n\) is the current flowing from the \(n\) terminal to the \(n+1\) terminal (in this case, one can more generally define \(J_{nm}\) for currents running between \(n\) and \(m\) terminals, satisfying \(J_{nm}+J_{ml}=J_{nl}\)). By current conservation, one has \(\sum_n J_n=0\) in both cases.

The main advantage of multiterminal devices in the context JDE is that they can be operated as an open system with a single two-terminal JJ connected to additional sources. Interestingly, in these configurations there is no bound on the diode efficiency \(\eta\) even in equilibrium. This can be illustrated in the simplest example of a three-terminal device with a central SC island and ordinary CPRs \(J_n=J_{nc} \sin\varphi_n\) (i.e., without intrinsic TRSB or ISB), considered in Ref. \cite{ChilesFinkelstein23}. One of the currents, \(J_3\), was fixed in the experiment, while \(J_1\) was measured (we assume \(J_{2c}\geq J_{1c}\); the other case is similar). The critical values of the \(J_1\) current, \(J_{c\pm}\), depend on how \(|J_3|\) compares to \(|J_{2c}-J_{1c}|\): if \(|J_3|\leq |J_{2c}-J_{1c}|\), \(J_{c\pm}=\pm J_{1c}\), and there is no diode effect; if, on the other hand, \(J_{2c}-J_{1c}\leq|J_3|\leq J_{2c}+J_{1c}\), \(J_{c,\sgn[J_3]}=\pm J_{2c}-J_3\) and \(J_{c,-\sgn[J_3]}=\mp J_{1c}\). The corresponding JDE efficiency in the latter case is \(\eta=(\sgn[J_3](J_{2c}-J_{1c})-J_3)/(J_{2c}+J_{1c} - |J_3|)\), which note is unbounded and may even diverge (as \(|J_3|\rightarrow J_{1c}+J_{2c}\)).
The meaning of \(|\eta|>1\) is that not only does the JJ pass a supercurrent in only one direction, but it does so only for a sufficiently strong current above a threshold value.
In this situation, a more appropriate figure of merit may be \(\eta'=(|J_{c+}|-|J_{c-}|)(|J_{c+}|+|J_{c-}|)=\eta^{\sgn[1-|\eta|]}\), which goes to zero when \(\eta\rightarrow\infty\). In practice, depending on applications, both \(\eta\) and \(\eta'\) may be useful quantities to optimize, and devices with \(|\eta|>1\) may be useful for diode applications. Observe that \(J_3\) is sufficient for breaking both TRS and IS in the multi-terminal case, even if all the junctions are identical (in that case, \(|\eta|\leq 1\), however).

\subsection{GL Theory of Intrinsic Bulk SDE}\label{GLSDESec}

The intrinsic bulk SDE (here understood in a sense of intrinsic ISB) was anticipated on symmetry grounds by Levitov et al. (1985)~\cite{LevitovNazarovEliashberg85} and later addressed within a microscopically derived GL framework by Edelstein (1996)~\cite{Edelstein96}. The phenomenology of the intrinsic bulk SDE is, however, subtle: Edelstein's original calculation was later shown to be in-part erroneous, and subsequent corrections in the literature were required before a consistent microscopic description emerged. Tracing this sequence of results and misunderstandings is instructive: it clarifies the origin of the effect and highlights several parallels with the JDE phenomenology discussed above.

Any SC system can be described phenomenologically using the standard GL theory, which proceeds as an expansion of the free energy in powers of the SC order parameter and its gradient \cite{Tinkham04}:
\[F[\Psi(\mathbf{r})] =\sum_n\left(\Psi^*(\mathbf{r})\alpha_n\mathbf{D}^n\Psi(\mathbf{r})+(\Psi^2(\mathbf{r}))^*\beta_n\mathbf{D}^n\Psi^2(\mathbf{r})+\text{c.c.}\right)+\frac{B^2}{8\pi}\label{FGL}\]
where \(\mathbf{D}=\boldsymbol{\nabla}+2ie\mathbf{A}\) is the covariant derivative, \(\mathbf{A}\) is the vector potential of the magnetic field \(\mathbf{B}=\boldsymbol{\nabla}\times\mathbf{A}\). The GL coefficients \(\alpha_n\) and \(\beta_n\) should be understood as tensors with spatial direction indices \(i_1,\dots,i_n=x,y,z\), with the shorthand notation \(\alpha_n\mathbf{D}^n = \alpha_n^{(i_1,\dots,i_n)} D_{i_1} \dots D_{i_n}\), but below we will mostly restrict our attention to a single spatial direction for simplicity. Close to the SC transition temperature \(T_c\) we can take \(\alpha_0=\alpha'_0(T-T_{c0})\) and the rest of the coefficients can be taken to be constant in temperature (\(T_{c0}\) is the critical temperature in the absence of TRSB and ISB, see below). The GL equations are obtained by minimizing the free energy with respect to \(\Psi^*\) and \(\mathbf{A}\), and the supercurrent is obtained as \(\delta F/\delta\mathbf{A}=0=\mathbf{J}+\boldsymbol{\nabla}\times\mathbf{B}/(4\pi)\).

To simplify the analysis and following Refs. \cite{DaidoYanase22, DaidoYanase22_2}, we Fourier transform the order parameter, i.e., consider a plane wave decomposition and take \(\Psi(\mathbf{r})=\Psi(\mathbf{q})e^{i\mathbf{q}\cdot\mathbf{r}}\) where \(\mathbf{q}\) is the Cooper pair momentum. The free energy then becomes
\[F=\alpha(q)|\Psi(\mathbf{\mathbf{q}})|^2+\beta(\mathbf{q})|\Psi(\mathbf{q})|^4+\frac{B^2}{8\pi}\label{FGL}\]
with \(\alpha(\mathbf{q})=\sum_n\alpha_n\mathbf{q}^n\) and \(\beta(\mathbf{q})=\sum_n\beta_n\mathbf{q}^n\). Optimizing over the order parameter, one finds \(|\Psi(\mathbf{q})|^2=-\alpha(\mathbf{q})/(2\beta(\mathbf{q}))\), with corresponding condensation energy
\(f(q)=-\alpha^2(\mathbf{q})/4\beta(\mathbf{q})\equiv-\tilde{\alpha}^2(\mathbf{q})/4\,,\label{f}\)
where for later convenience we further define \(\tilde{\alpha}(\mathbf{q})=\sum_n\tilde{\alpha}_n \mathbf{q}^n\). Because of minimal coupling, the supercurrent is given simply by \(\mathbf{J}(\mathbf{q})=2\partial_\mathbf{q}f(\mathbf{q})\). Note that in equilibrium \(\mathbf{J}(\mathbf{q})=0\) by definition. There are some critical \(\mathbf{q}_{c\pm}\) above and below which no nontrivial solution for \(\Psi\) exist. Between those values, \(\mathbf{J}(\mathbf{q})\) reaches its maximum and minimum values \(J_{c\pm}\) at some \(\mathbf{q}_{\pm}\). The diode efficiency coefficient is then defined as \(\eta=(J_{c+}+J_{c-})/(J_{c+}-J_{c-})\). As in the case of JDE,  \(\eta\) is bounded between \(\pm1\) as long as \(f(q)\) is bounded from below.

\subsubsection{The Linear Lifshitz invariant and Nonuniform SC}

In the presence of either TRS or IS, the GL coefficients with odd \(n\) vanish. SDE thus arises due to these `anomalous' terms, as noted by Edelstein in Ref. \cite{Edelstein96}.
The first such term in the GL expansion linear in \(\mathbf{q}\), \(\alpha_1\), is known as the linear Lifshitz invariant, first introduced by E. M. Lifshitz \cite{Lifshitz41, Dzyaloshinskii64}.\footnote{Higher order anomalous \(\alpha_n\) are sometimes also referred to as higher order Lifshitz invariants; unless otherwise specified, we reserve the term Lifshitz invariant for the linear term.} In superconductors, the effect of the Lifshitz invariant was studied early on in the context of magnets with toroidal order by Gorbatsevich \cite{Gorbatsevich89} (see Sec. \ref{MagSec}), who already identified its main effect on the SC order parameter, which is to make it nonuniform or inhomogeneous, i.e. to induce finite momentum Cooper pairing with \(\mathbf{q}\neq0\). The Lifshitz invariant can also appear due to Josephson coupling between SCs of different symmetries \cite{KaurAgterberg03, ShafferChichinadzeLevchenko24}, and in NCSs with external magnetic fields, where the nonuniform SC phase is known as the helical SC state \cite{Mineev93, MineevSamokhin94, GorkovRashba01, BarzykinGorkov02, FrigeriAgterbergSigrist04, BauerSigrist12, Yip14, SmidmanAgterberg17, DimitrovaFeigelman03, AgterbergKaur07, DimitrovaFeigelman07}.

Indeed, keeping also \(\alpha_2\) for stability, one finds from \(\mathbf{J}(\mathbf{q})=0\) that in the ground state \(\mathbf{q}_0=-\alpha_2^{-1}\alpha_1/2\) (again using the shorthand tensor notation: \(\alpha_1\) and \(\alpha_2\) are a vector and a \(2\times2\) matrix, respectively); see Fig. \ref{SDEfig}(b).
However, no SDE occurs due to the Lifshitz invariant alone, as
pointed out by Agterberg in Ref. \cite{BauerSigrist12} (see also later Refs. \cite{KapustinRadzihovsky22, WangHao25}).
In particular, taking \(\delta\mathbf{q}=\mathbf{q}-\mathbf{q}_0\), one finds by completing the square that \(\alpha(q)=\alpha_0+\alpha_2q_0^2+\alpha_2\delta\mathbf{q}^2\), eliminating the anomalous term; this can also be achieved by a gauge transformation \(\Psi(\mathbf{r})\rightarrow \Psi(\mathbf{r})e^{i\mathbf{q}_0\cdot\mathbf{r}}\). As a result, the free energy is still symmetric under an inversion-like symmetry \(\mathcal{I}':\mathbf{q}-\mathbf{q}_0\rightarrow-\mathbf{q}-\mathbf{q}_0\), and no SDE occurs. This argument invalidates Edelstein's calculation in Ref. \cite{Edelstein96}!\footnote{Edelstein's mistake was a result of an incorrect definition of the critical current, which he identified with the value of the current at the critical Cooper pair momentum \(\mathbf{q}\) at which the GL equations cease to have a nontrivial solution. The current at that momentum, however is zero! Edelstein expresses his final result in Eq. (1.6) as \(J=J_0(1+\delta j)\), where \(\delta j\) is the nonreciprocal correction to the critical current \(J_0\) in the absence of ISB. But by Edelstein's definition, \(J_0=0\).}

The Lifshitz invariant is nevertheless sufficient to account for other signatures of the helical SC state, including spontaneous currents in a ring geometry \cite{Gorbatsevich89,  BauerSigrist12}, and AJE \cite{Gorbatsevich89,  BauerSigrist12} and JDE \cite{YangAgterberg00, KaurAgterberg05} in a JJ between a helical and a conventional SC.
It has also been pointed out that the Lifshitz invariant alone can give rise to SDE if boundary effects are included \cite{KochanZutic23}, but such SDE could not be purely intrinsic as it would be sensitive to extrinsic boundary conditions.

\subsubsection{SDE from Higher Order Gradient Terms}

\begin{figure}[t]
\includegraphics[width=0.99\textwidth]{SDEfig.pdf}
\caption{Condensation energies \(f(q)\) (top, blue) and supercurrents \(J(  q)\) (bottom, orange) as functions of the Cooper pair momentum \(q\): (a) without anomalous terms and no FFLO or SDE; (b) with the TRS and IS breaking (linear) Lifshitz invariant resulting in a shift of the condensation energy by \(q_0\) and a ground state with finite momentum pairing \(q_0\neq0\), but no SDE; and (c) with additional anomalous terms yielding the bulk SDE. Note the similarity to Fig. \ref{JDEfig}.}
\label{SDEfig}
\end{figure}

The absence of SDE with the Lifshitz invariant alone is analogous to the fact that no JDE occurs in JJs with CPRs containing only the first harmonics, even if the anomalous \(\cos\varphi\) term is present. Those are only sufficient to produce the AJE, but higher harmonics are needed to produce the necessary interference that gives rise to JDE. Similarly, after the intrinsic SDE was reported in 2020 in Ref. \cite{AndoYanase20}, it was quickly realized that higher order terms in the GL theory do in fact produce an SDE, as shown (independently from Edelstein's work) in Refs. \cite{DaidoYanase22, YuanFu22, HeNagaosa22, IlicBergeret22} (see Fig. \ref{SDEfig}(c)). However, there was no initial agreement on which higher order terms have to be kept: Ref. \cite{YuanFu22} kept \(\alpha_3\), Ref. \cite{DaidoYanase22} added \(\alpha_3\) and \(\beta_1\), Ref. \cite{HeNagaosa22} added \(\alpha_3, \alpha_4,\beta_1\) and \(\beta_2\), while Ref. \cite{IlicBergeret22} kept all terms up to fourth order in \(\mathbf{q}\). This raises the question of the minimal number of terms have to be kept to obtain the correct result in such models (within appropriate approximations). As we shall see in Sec. \ref{NCSSec}, keeping an incorrect combination of terms can yield qualitatively wrong results.

To answer this question, it is simplest to compute the critical currents. Here we follow Ref. \cite{HasanShafferKhodasLevchenko24} (see also Ref. \cite{DaidoYanase22_2}), and fix the direction of \(\mathbf{q}\) to avoid tensor quantities.
We work close to \(T_c\) where the GL theory is most valid and also assume, as is the case in most microscopic models, that the anomalous \(\alpha_n\) and \(\beta_n\) with odd \(n\) are of the same order in TRSB and ISB field, whatever its physical character. We then work to leading order in those terms.
It is natural to expand \(\tilde{\alpha}(\mathbf{q})\) around its minimum at \(\mathbf{q}_0\) (also computed to third order in \(\alpha_n\)):
\[\tilde{\alpha}(q)=a_0+a_2\delta q^2+a_3\delta q^3+\dots\]
Importantly, note that if we only keep terms up to quadratic order, \(f(q)\) (\(J(q)\)) is even (odd) under \(\mathcal{I}_{q_0}:\delta q\rightarrow -\delta q\), such that \(J_{c+}=-J_{c-}\) (and \(q_{c+}+q_0=-q_{c-}+q_0\)), and the SDE vanishes, consistent with our discussion above.
To capture the SDE, it is therefore necessary (and sufficient) to work up to the cubic order in \(\delta q\).

The critical currents are maxima and minima of \(J(q)\) and are therefore roots of \(\partial_q J(q)=0\). To third order in \(\delta q\), one finds
\[J_{c\pm}=\frac{4a_0^2a_3}{9a_2}\pm\frac{4(-a_0)^{3/2}\sqrt{a_2}}{3\sqrt{3}},\label{JcGL}\]
and the superconducting diode coefficient is given by \(\eta=\sqrt{-a_0/3}\, a_3/a_2^{3/2}\).
Note that these expressions are valid to any order in the anomalous terms. To linear order in the anomalous terms and the leading order in \(a_0=\alpha_0/\sqrt{\beta_0}\), 
\begin{equation}
\eta=\frac{2 \alpha_2 \alpha_3 \beta_0-4 \alpha_1 \alpha_4 \beta_0-\alpha_2^2 \beta_1+\alpha_1 \alpha_2 \beta_2}{2\sqrt{3} \alpha_2^{\frac{5}{2}}\beta_0}\sqrt{-\alpha_0}\label{etaGL}
\end{equation}
in terms of the original GL coefficients. To leading order, in other words, one must compute \(\alpha(q)\) to fourth order in \(q\) while it is sufficient to expand \(\beta(q)\) to second order.
Note in particular that the Lifshitz invariant alone does produce an SDE if either \(\alpha_4\) or \(\beta_2\) terms are nonzero, in which case \(\eta\propto\alpha_1(\alpha_2\beta_2-4\alpha_4\beta_0)\).
Because \(a_3\) contains \(\alpha_n\) with \(n=0\) through \(4\) and \(\beta_n\) with \(n=1\) through \(2\), all of those terms must be kept, as properly done for the first time in Ref. \cite{IlicBergeret22}
(Ref. \cite{HeNagaosa22} kept the correct terms, but appear to have an algebraic mistake in their analysis). In all cases, to leading order \(\eta\propto\sqrt{T-T_c}\), as noted 
in Refs. \cite{LevitovNazarovEliashberg85, DaidoYanase22, YuanFu22, HeNagaosa22, IlicBergeret22}.

Not only is the Lifshitz invariant (and finite momentum pairing) not sufficient to produce SDE, it is also not necessary in some rare cases: in particular, in  2D systems with three-fold rotation symmetry \(C_{3}\) (assuming it also not broken by the magnetic field), the linear in \(q\) terms are not allowed; see, e.g., Ref. \cite{Zhai22}. In that case, \(q_0=0\) by symmetry (assuming the ground state does not spontaneously break the \(C_3\) symmetry), but the SDE coefficient is nevertheless nonzero: \(\eta=\alpha_3\sqrt{-\alpha_0}/(\sqrt{3} \alpha_2^{3/2})\). In this case a cubic expansion of \(\alpha(q)\) in \(q\) is sufficient.

\subsubsection{FFLO with Spontaneous TRSB/ISB, Critical Points, and Perfect SDE}\label{GLFFLOSec}

\begin{figure}[t]
\includegraphics[width=0.99\textwidth]{FFLOfig.pdf}
\caption{Typical condensation energy (blue) and supercurrent (orange) as functions of \(q\) with ISB or TRSB: as \(\alpha_2\) decreases (from left to right) and becomes negative, the minimum at \(q=0\) splits into two minima at \(\pm q_0\). The resulting FFLO ground state spontaneously breaks TRS and/or IS.}
\label{FFLOfig}
\end{figure}

The GL theory above assumes that both TRSB and ISB occur explicitly. SDE, however, is also possible when TRSB and ISB occur spontaneously. The simplest example is the FFLO state driven by depairing due to a Zeeman splitting field \cite{FF, LO}, and more generally other PDW states without magnetic fields \cite{Agterberg20}; in principle, such states may preserve both TRS and IS in the LO-type (stripe) phases with \(\Psi(\mathbf{r})\sim\cos\mathbf{q\cdot r}\), but both are broken in FF-type phases with \(\Psi(\mathbf{r})\sim e^{\pm i\mathbf{q\cdot r}}\). The free energy \(F(\mathbf{q})\) and the condensation energy \(f(\mathbf{q})\) are in that case globally invariant under \(\mathbf{q}\rightarrow-\mathbf{q}\), but the ground state occurs at some finite momenta \(\pm\mathbf{q}_0\neq 0\). The simplest form of the free energy exhibiting such FFLO states is given by \(\alpha(q)=\alpha_0+\alpha_2q^2+\alpha_4q^4\) with \(\alpha_0,\alpha_2<0\) and \(\alpha_4>0\), and constant \(\beta(q)=\beta_0>0\), see Fig. \ref{FFLOfig} \cite{SamokhinTruong17}.

By symmetry, the ground state must be degenerate, with at least one other minimum at \(-\mathbf{q}_0\). This can lead to domain formation and more complicated dynamics that can affect the SDE: as the application of a current favors one minimum over the other, so increasing and decreasing the current can lead to hysteresis. Since the (at least) two ground states are separated by an energy barrier, assuming reasonable damping in the system implies that if one starts in one of the minima, however, the critical currents \(J_{c\pm}\) defined by the current at which dissipation starts to take place most likely correspond to the local minimum and maximum closest to the minimum of the condensation energy (see right panel in Fig. \ref{FFLOfig}): the system has to evolve across higher energy states in order to reach the true minimum of the free energy, so dissipation is necessary. If the free energy barrier becomes sufficiently small, however, the argument above will clearly fail due to thermal or quantum fluctuations. Identifying the energy scale that determines what sufficiently small means requires a more sophisticated microscopic theory accounting for nonequilibrium processes, or at least a phenomenological time-dependent GL (TDGL) analysis that has not yet been developed and is thus beyond the scope of this review.

SDE due to FF-type FFLO states was first proposed in Ref. \cite{YuanFu22}, who found, moreover, that the diode efficiency \(|\eta|\) reaches its maximum value of \(1\) at the tricritical point at which the second-order transition line of the uniform SC phase intersects the first-order transition line of the FFLO phase. Similar `perfect' SDE was found in Refs. \cite{ShafferChichinadzeLevchenko24, ChakrabortyBlackSchaffer25} and generally occurs at a second-order phase transition between a uniform and FFLO phases: as the global minimum at \(q=0\) splits into two at the phase transition, the two new critical points of \(J(q)\) that determine the \(J_{c\pm}\) of the FFLO states appear close to \(J=0\). Approaching the phase transition from the FFLO state thus implies that one of \(J_{c\pm}\) approaches zero, which yields \(|\eta|=1\). The caveats from the previous paragraph apply especially in this situation, since the free energy potential barrier also vanishes at the phase transition: one would therefore expect the system to evolve continuously at \(q=0\) at the transition and no diode effect to occur. On physical grounds, most likely \(|\eta|\) is maximized at the phase transition but not at its perfect value. In that respect, a first-order phase transition may be favorable for realizing larger \(|\eta|\), as in that case the free-energy barrier does not vanish. Additionally, Refs. \cite{YuanFu22,ChakrabortyBlackSchaffer25} noted a sign change in \(\eta\) at lower temperatures.

\subsection{Vortex Diode Effect}\label{VDESec}

Most proposals for SDE and JDE involve magnetic fields in order to break TRS. We have, with the notable exception of JDE in SQUIDs, largely neglected the resulting orbital effects (i.e., effects due to non-zero \(\mathbf{A}\) in the GL free energy). This can potentially be justified under some circumstances, for example for bulk SDE in type I SCs, or in truly 2D SCs. In most realistic situations, however, orbital effects cannot be neglected, which significantly complicates the identification of intrinsic SDE and JDE because orbital effects can themselves lead extrinsic SC diode effects, as emphasized in Refs. \cite{MollGeshkenbein23, HouMoodera23, GaggioliGeshkenbein24, GaggioliMoodera25}. Indeed, as mentioned above the original observations of SDE in Refs. \cite{EdwardsNewhouse62, SwartzHart67} likely originate from one such effect, the VDE.
With modern theoretical hindsight, VDE can be understood in relatively simple terms. In presence of a current, a single vortex carrying a magnetic flux quantum \(\Phi_0=hc/(2e)\) experiences a Lorentz force \(\mathbf{F}_L=\mathbf{J}\times\boldsymbol{\Phi}_0\). When a vortex moves in response to this force, since the magnetic flux moves with them they induce an electric field \(\mathbf{E}\sim\boldsymbol{\Phi}_0\times\mathbf{v}\), where \(\mathbf{v}\) is the velocity of the vortex. The electric field is felt by quasiparticles in the normal core of the vortex, resulting in a voltage and dissipation: assuming that vortices experience drag \(\mathbf{F}_D=\eta_D\mathbf{v}\), the resistivity is \(\rho\sim H\Phi_0/\eta_D\). In general, vortices also experience pinning forces \(\mathbf{F}_P\). If these are sufficiently strong, the Lorentz force must first exceed them before vortices start to move, which determines the critical current in this scenario: \(J_c\sim F_p/\Phi_0\).
The microscopic theories of vortex pinning, drag and flow in presence of disorder are quite complicated (see, e.g., Refs. \cite{BlatterLarkinRMP94, LarkinVarlamov05}), but in general it is natural for the pinning forces not to respect IS, and consequently the nonreciprocity of the critical currents can then be understood as due to the asymmetry of the pinning forces: \(J_{c\pm}\sim F_{p\pm}/\Phi_0\). 

Different observations of VDE in years prior to and since the discovery of the intrinsic SDE in 2020 in Refs. \cite{Broussard88, Roas90, Jiang94, PlourdeVodolazov01, Touitou04, MorelleMoshchalkov06, Papon08, Harrington09, Adami13, CerbuMoshchalkov13, Sivakov18, SuriStrunk22, GutfreundBuzdin23} have been attributed to such surface pinning effects.
At the turn of the 20\(^{\text{th}}\) century, VDE has mostly been studied in the context of the vortex ratchet effect, referring to a net DC vortex flow (and thus DC voltage) under an AC current drive.
This can occur also in JJs and were originally proposed \cite{Zapata96, Falo99, Carapella01, Lee03, Marconi07, Spiechowicz14, Semenov15, FominovMikhailov22, ReinhardtGlazman25} and realized \cite{TriasFalo00, Weiss00, CarapellaCostabile01, Sterck02, CarapellaFilatrella02, Majer03, Shalom05, BeckGoldobin05, Streck05, Streck09, KnufinkeGoldobin12, Chesca17, ReinhardtGlazman25} in SQUIDs and other multiloop JJ arrays based on the inductance mechanism discussed in Sec. \ref{SQUIDSec}, realizing a VDE-based JDE.
Vortex ratchets in wide JJs -- the continuum limit version of 1D JJ arrays -- have also been proposed \cite{KrasnovPedersen97, Goldobin01, Shaju03, ChirolliCuocoGiazotto24} and realized \cite{KrasnovPedersen97, WangGoldobin09, ChenYacoby24, Ghosh24}.\footnote{Ref. \cite{KrasnovPedersen97} was possibly the first to specifically refer to a superconducting diode application in 1997, while the first use of the term vortex diode is possibly Ref. \cite{OlsonReichhardt01} in 2001, but both are predated by the term `fluxonic diode' introduced by Kadin in 1990 in Ref. \cite{Kadin90} and  `Josephson fluxonic diode' in 1994 in Ref. \cite{RaissiNordman94}.}
Other implementations of JDE using fluxons or vortices include JJs with a control current \cite{GuarcelloFilatrella24} or a pinned vortex \cite{FukayaOrtixCuoco24} in one SC region have been proposed recently (note that these break both TRS and IS by themselves).
An early proposal for a fluxonic diode in 1990 relied on a junction between a region of vortices and antivortices (dual to a pn junction) \cite{Kadin90}, experimentally realized in 1994 \cite{RaissiNordman94}.
Finally, asymmetric 2D pinning potentials -- analogous to 2D JJ arrays -- have been proposed \cite{LeeBarabasi99, OlsonReichhardt01, Shklovskij14, Reichhardt15, He19, Bogush25, Koshelev25} and subsequently realized \cite{Villegas03, Wordenweber04, VanDeVondel05, PryadunMoshchalkov06, deSouzaSilvaMoshchalkov07, GillinjnsMoshchalkovReichhardt07, Lyu21, Dobrovolskiy22} the vortex ratchet in junction-free SCs equivalent to a VDE-based bulk SDE. Bulk VDE due to surface Bean-Livingston barriers have also been studied theoretically in Refs. \cite{VodolazovPeeters05, Kubo23, GaggioliGeshkenbein24}.

Given the ubiquity of vortex motion and asymmetric pinning barriers one may expect in realistic samples, it has been pointed out soon after the claim of the experimental observation of an intrinsic SDE in Ref. \cite{AndoYanase20} that it can be difficult to differentiate such an intrinsic SDE from the extrinsic VDE \cite{HouMoodera23, MollGeshkenbein23, GaggioliGeshkenbein24, GaggioliMoodera25}.
Since VDE is, by its nature, sensitive to various geometric effects like the presence of corners, wedges, curvature, adjacent currents, etc. (see Refs. \cite{WambaughReichhardt99, Shaju03, ClemPeeters12, Adami13, Jiang21, Cadorim24, Aguirre25}), one possible way to identify whether the SDE is intrinsic or extrinsic is to vary the sample geometry.
This may be challenging to do in novel materials if sample quality is highly variable, but using external `pinch' currents in the setup used in Ref. \cite{EdwardsNewhouse62} may be used to control the surface pinning barriers.
Identifying the critical current anisotropy for multiple directions and matching it with crystallographic axes is another possible strategy used to argue against intrinsic SDE for example in \cite{SundareshVayrynen23}, but it may not work in small samples whose geometry is also determined by the crystallographic axes (a false negative, on the other hand, may also occur due to self-field effects, etc.). More detailed GL and microscopic models are needed to identify incontrovertible signatures of VDE in the temperature and field dependence of the critical currents.
The issue is complicated by the fact that intrinsic SC diodes are also sources of VDE when vortices are present: an `intrinsic' VDE, for example, was suggested to take place in NCSs in Ref. \cite{HoshinoNagaosa18}. ISB in this case has also been shown to lead to the deformation of the shape of the vortices \cite{FuchsKochanManfra22, PutilovBuzdin25}, an effect that can be identified in vortex imaging. Note that vortex fluctuations are also generally present even in monolayer NCSs in in-plane fields, especially near the BKT transition, and their effect on SDE has been studied in Ref. \cite{NunchotYanase25}.

To complicate things even more, one may also expect more complicated vortex lattice effects to lead to ISB intrinsic to the vortex lattice, for example in meron vortex lattice proposed to possibly occur in CeRh\(_2\)As\(_2\) \cite{MinamideYanase25}. Such effects may generally be anticipated in coexistence regions of multiphase SCs that may host multiple vortex species, and can subsequently lead to another version of an intrinsic VDE \cite{Savelev02, ShafferChichinadzeLevchenko24, WangAgterbergBabaev24}.
In 3D, vortex structures can be quite complicated, for example twisted vortices, including twisted double core vortices, can form \cite{Dubrovskii89, Eltsov06, RantanenEltsov24}. 
Such twisted vortices, which can be pinned to columnar defects \cite{BlatterLarkinRMP94}, break IS and moreover have a chirality, which on symmetry grounds allows for VDE with \(\mathbf{J}\parallel\mathbf{H}\) (see end of Sec. \ref{GLSDESec}).
In the presence of twisted vortices, this effect can be realized if the vortex flow conductivity includes a nonzero Hall component corresponding to vortex motion parallel to the current direction \cite{BardeenStephen65, Kim65, Dorsey92, Tinkham04}. We thus note that VDE can also generically give rise to a transverse (or Hall) SC diode effect. 

\section{Intrinsic SDE and JDE in Noncentrosymmetric Materials}\label{NCSSec}

Having established the need for higher order harmonics in JJ CPRs and higher order anomalous GL terms in addition to TRSB and ISB in order to generate JDE and SDE respectively, we now review the microscopic mechanisms for generating them in this and following sections. As in the prototypical SQUID example in Sec. \ref{SQUIDSec}, some form of interference between two or more channels is generally required to produce the higher order terms. In this section we turn to the first, and most studied, intrinsic such mechanism: ISB due to SOC in NCSs. 
This was the mechanism considered in the initial theoretical predictions of both intrinsic SDE \cite{LevitovNazarovEliashberg85, Edelstein96} and JDE \cite{YangAgterberg00, KaurAgterbergSigrist05, ReynosoAvignon08}, as well as the one invoked to explain their first observations in Ref. \cite{AndoYanase20} and Refs. \cite{DartiailhZutic21, BaumgartnerManfra22}, respectively. 

In the presence of Zeeman splitting (neglecting orbital effects, which is particularly justified for 2D materials in in-plane magnetic fields), in the normal state each helical band is shifted in momentum space perpendicular to the applied field, see Fig. \ref{NCSfig} (a).
As we explain in more detail below, the main effect of this shift on the SC state is the realization of the so-called helical phases with finite-momentum pairing \cite{Mineev93, MineevSamokhin94, BarzykinGorkov02, DimitrovaFeigelman03, KaurAgterbergSigrist05, AgterbergKaur07, DimitrovaFeigelman07}.
Both intrinsic SDE and JDE were, in fact, initially predicted as experimental signatures of helical NCSs, which remain the most studied platforms for realizing intrinsic SDE. In addition, many proposed realizations of JDE with uniform SCs rely on non-centrosymmetric materials being used as tunneling barriers in the JJs. Given the importance of NCS and non-centrosymmetric materials for realizing both SDE and JDE, especially the Rashba-Zeeman model, we review them in some detail below. Despite its relative simplicity, the analytical computation of SDE in this model turns out to be surprisingly nontrivial, and therefore particularly instructive.

\subsection{Basic Theory of SC in Noncentrosymmetric Materials}

Noncentrosymmetric materials including NCSs have been extensively studied for a long time \cite{AndersonBlount65,Bulaevskii76, Edelstein95}, especially after the discovery of heavy fermion NCSs like CePt\(_3\)Si, as reviewed in Refs.  \cite{BauerSigrist12, Yip14, SmidmanAgterberg17}.
Dresselhaus or Rashba SOC \cite{Dresselhaus55, Rashba59, BychkovRashba84} that lifts the spin degeneracy resulting in two spin-split helical bands, leading to the magnetoelectric effect \cite{Dzyaloshinskii59, Edelstein90} (noncentrosymmetric materials can thus also be referred to as magnetoelectric materials). In the SC state, this results in mixing between singlet and triplet paired states \cite{ Edelstein89, GorkovRashba01, FrigeriAgterbergSigrist04}.
We first recall the normal state properties of systems with generic SOC and Zeeman terms. The corresponding single-body Hamiltonian is
\[H=\sum_{\mathbf{k}ss'}\mathcal{H}_{ss'}(\mathbf{k})c^\dagger_{\mathbf{k}s}c_{\mathbf{k}s'}=\sum_{\mathbf{k}ss'}\left[\varepsilon(\mathbf{k})\delta_{ss'}+\mathbf{g}(\mathbf{k})\cdot\boldsymbol{\sigma}_{ss'}\right]c^\dagger_{\mathbf{k}s}c_{\mathbf{k}s'}=\sum_{\mathbf{k}\lambda}\xi_\lambda(\mathbf{k})d^\dagger_{\mathbf{k}\lambda}d_{\mathbf{k}\lambda}\label{H0}\]
where \(s,s'=\uparrow,\downarrow\) (also \(\pm1\) respectively when convenient below) are spin indices, \(\varepsilon(\mathbf{k})\)
is the dispersion in the absence of spin-splitting fields and
\(\mathbf{g}(\mathbf{k})=\boldsymbol{\beta}(\mathbf{k})+\mathbf{h}\)
includes both the SOC \(\boldsymbol{\beta}(\mathbf{k})=-\boldsymbol{\beta}(-\mathbf{k})\) and the Zeeman term with magnetic field \(\mathbf{h}\) measured in units of the Bohr magneton. In the simplest case, we will take the dispersion to be that of free electrons: \(\varepsilon(\mathbf{k})=\frac{k^2}{2m}-\mu\) with electron mass \(m\) and chemical potential \(\mu\).
Rashba SOC corresponds to \(\boldsymbol{\beta}(\mathbf{k})=\alpha_R \mathbf{p}\times \hat{\mathbf{z}}\), but the same Hamiltonian can be used for various forms of SOC including Dresselhaus SOC such as Ising SOC \cite{Lu15,Xi16,HsuKim17,MockliKhodas19,ShafferBurnellFernandes20,WickramaratneAgterberg20,HamillShaffer21,IlicMeyerHouzet23,ShafferBurnellFernandes23,SieglUgeda25}; by symmetry considerations, however, Ising SOC by itself does not generate SDE in 2D systems, so we will mostly restrict our attention to the Rashba case.

The Hamiltonian is diagonalized by the operators (henceforth Einstein summation convention is assumed unless otherwise stated) \(d_{\mathbf{k}\lambda}=U^s_\lambda(\mathbf{k})c_{\mathbf{k}s}\) that form the so-called helical or band basis with helical/band indices \(\lambda=\pm1\) (see, e.g., Appendix E in Ref. \cite{HasanShafferKhodasLevchenko24} for explicit general expressions).
In general, the two helical bands have dispersion
\(\xi_\lambda(\mathbf{k})=\varepsilon(\mathbf{k})+\lambda g(\mathbf{k})\)
where \(g(\mathbf{k})=|\mathbf{g}(\mathbf{k})|\).
For Rashba SOC at zero magnetic field and free electron dispersion \(\varepsilon(\mathbf{k})=\frac{k^2}{2m}-\mu\) with Fermi momentum \(p_F=\sqrt{2m\mu}\) and Fermi velocity \(v_F=p_F/m\), one finds in particular that the helical bands from two circular Fermi surfaces with Fermi momenta \(p_{F\lambda}=\sqrt{p_F^2+\alpha_R^2m^2}-\lambda\alpha_Rm\) and dispersions \(\xi_\lambda(\mathbf{k})=v\delta k_\lambda+\delta k_\lambda^2/(2m)\) with \(\delta k_\lambda=|k-p_{F\lambda}|\) and \(v=\sqrt{v_F^2+\alpha_R^2}\).
In this case the Fermi velocities of the two helical bands are therefore equal, but the band-resolved DOSs \(\nu_\lambda=p_{F\lambda}/(2\pi v)\) are not.  Typically the Fermi energy \(\varepsilon_{F}=p_Fv_F\) is the largest energy scale, and to leading order in \(\alpha_Rp_F/\varepsilon_F=\alpha_R/v_F\), we have \(v=v_F\) and \(\nu_\lambda\approx\nu_0(1-\lambda \alpha_R/v_F)\), with \(\nu_0=m/(2\pi)\) being the DOS in the absence of SOC.
Note that when treating the magnetic field perturbatively, some authors refer to a helical basis that diagonalizes the SOC part of the Hamiltonian only and excluding the Zeeman term \cite{HouzetMeyer15, IlicBergeret22}; this is often reasonable but can lead to incorrect results when considering the form of the gap function, as we discuss below. To avoid confusion, we refer to the \(\lambda\)-bases as the band basis below.
In general, the main goal of diagonalizing the Hamiltonian is to work in a basis labeled by conserved quantities (operators commuting with the Hamiltonian): in the presence of SOC or magnetic fields the electron spin is no longer conserved and is not a good quantum number, whereas the band indices are good quantum numbers.

An important consequence of Rashba SOC and Zeeman fields being simultaneously present is that the Fermi surfaces formed by the two helical bands have shifted centers of mass \(\mathbf{Q}_\lambda\) and are deformed from perfect inversion symmetry: \(\xi_\lambda(\mathbf{k+q})\neq\xi_\lambda(-\mathbf{k+q})\) for any fixed \(\mathbf{q}\); see Fig. \ref{NCSfig} (a).  
For small magnetic fields \(h\ll\alpha_Rp_F\) (taking \(\mathbf{h}\) to be along the \(\hat{\mathbf{x}}\) direction), to leading order one finds \(\xi_\lambda(\mathbf{k})=v_F\delta k_\lambda+\lambda h\sin\theta_\mathbf{k}\approx v_F(|\mathbf{k-Q}_\lambda|-p_{F\lambda})\) with \(\mathbf{Q}_\lambda=\lambda \hat{\mathbf{z}}\times \mathbf{h}/v_F\) (working to linear order in \(h\)). To linear order in \(h\), therefore, the two Fermi surfaces remain circular but shift in opposite direction perpendicular to the in-plane magnetic field; this is the origin of finite-momentum pairing being favored over uniform SC. Second-order terms result in the deformation of the Fermi surfaces, breaking their symmetry under \(\mathbf{k-Q}_\lambda\rightarrow -\mathbf{k-Q}_\lambda\). As \(h\) increases further, the two Fermi surfaces move towards each other until they touch at a single point when \(h=\alpha_R p_F\) (due to the cancellation of the Zeeman and SOC fields resulting in spin degeneracy at that point), and then move apart in opposite directions (again becoming circular in the limit \(h\gg \alpha_R p_F\). At even higher magnetic fields (or small chemical potentials), topological SC has been predicted \cite{Kitaev01, FuKane08, Fujimoto08, ZhangDasSarma08, SauDasSarma10, OregRefaelvonOppen10, LutchynSauDasSarma10, CookFranz11,  WangZhang15}, but we will not consider that parameter regime in this section, which we reserve for Sec. \ref{TopoSec}.

\subsubsection{Bogolyubov-de Gennes Theory of NCSs}

Superconductivity in the presence of Rashba SOC and Zeeman splitting has been studied extensively \cite{MineevSamokhin94, BarzykinGorkov02, BauerSigrist12, Yip14,  SmidmanAgterberg17, DimitrovaFeigelman03, KaurAgterberg05, AgterbergKaur07, DimitrovaFeigelman07}, and can be described by the Hamiltonian \(H=\sum_{s_1s_2}\bar{\Psi}_{\mathbf{k}s_1}\mathcal{H}^{(BdG)}_{s_1s_2}(\mathbf{k;q})\Psi_{\mathbf{k}s_2}\)
where \(\Psi_{\mathbf{k}s}(\mathbf{q})=(c_{\mathbf{k+q}/2,s},\bar{c}_{-\mathbf{k+q}/2,s})^T\) are the Nambu spinors and the Bogolyubov-de Gennes (BdG) Hamiltonian
\[\mathcal{H}^{(BdG)}(\mathbf{k;q})=\left(\begin{array}{cc}
\mathcal{H}(\mathbf{k+q}/2) & \widehat{\Delta}(\mathbf{k;q}) \\
\widehat{\Delta}^\dagger(\mathbf{k;q}) & -\mathcal{H}^T(-\mathbf{k+q}/2)
\end{array}\right)\label{HBdG}\]
is a \(4\times4\) matrix, each block being a \(2\times2\) matrix with spin indices.
The gap functions thus correspond to pairing terms of the form \(\widehat{\Delta}_{ss'}(\mathbf{k;q})c^\dagger_{\mathbf{k+q}/2,s}c^\dagger_{-\mathbf{k+q}/2,s'}\) with total momentum \(\mathbf{q}\). We have assumed for simplicity that \(\widehat{\Delta}\) with only one \(\mathbf{q}\) is present at a time (i.e. no mixing between different momenta). Particle-holy symmetry due to fermionic anticommutation relations implies we can take \(\widehat{\Delta}(\mathbf{k;q})=-\widehat{\Delta}^T(-\mathbf{k;q})\).

The form of the gap function is determined by the pairing interactions from which the BdG Hamiltonian can formally be obtained via a Hubbard-Stratonovich transformation (see, e.g., Refs. \cite{ShafferBurnellFernandes23,HasanShafferKhodasLevchenko24}, including appendices, for details). In the spin basis, these can include singlet or triplet pairing channels, giving rise to singlet/triplet gap functions \(\widehat{\Delta}_{ss'}(\mathbf{k;q})=d_\mu(\mathbf{k;q})(\sigma^\mu i\sigma^y)_{ss'}\) with \(\mu=0\) for singlet and \(\mu=x,y,z\) for triplet terms. However, in the presence of SOC (and magnetic field) singlet and triplet channels do not, however, in general decouple in the gap equation and it is more appropriate to convert the gap function to the band basis , which is done via
\(\sum_{\mathbf{k}ss'}\widehat{\Delta}_{ss'}(\mathbf{k;q})c^\dagger_{\mathbf{k+q}/2,s} c^\dagger_{-\mathbf{k+q}/2,s'}=\sum_{\mathbf{k}\lambda\lambda'}\widehat{\Delta}_{\lambda\lambda'}(\mathbf{k;q})d^\dagger_{\mathbf{k+q}/2,\lambda} d^\dagger_{-\mathbf{k+q}/2,\lambda'}\).
Conventional \(s\)-wave spin-singlet pairing with \(\widehat{\Delta}=\Delta_0 i\sigma^y\) that is typically assumed
becomes, in the band basis (for in-plane magnetic field and Rashba SOC):
\[\widehat{\Delta}^{(s)}=\frac{\Delta_0}{2}\left[\tilde{\sigma}^z(e^{-i\varphi(\mathbf{k-q}/2)}-e^{-i\varphi(-\mathbf{k-q}/2)})+i\tilde{\sigma}^y(e^{-i\varphi(\mathbf{k-q}/2)}+e^{-i\varphi(-\mathbf{k-q}/2)})\right]\label{DeltaS}\] 
where
\(e^{i\varphi(\mathbf{k})}=(g_x(\mathbf{k})+ig_y(\mathbf{k}))/\sqrt{g_x^2(\mathbf{k})+g_y^2(\mathbf{k})}\).
In the absence of a magnetic field, this simplifies to \(\widehat{\Delta}^{(s)}=\Delta_0e^{-i\varphi(\mathbf{k})}\tilde{\sigma}^z\), which is a fully intraband band-triplet term diagonal in the band-basis,  significantly simplifying the gap equation.
Observe that even though the Cooper pairs in the presence of SOC are equal mixtures of spin-singlet and spin-triplet components due to opposite spin polarization on opposite sides of the Fermi surface of a given band, due to TRS the singlet channel nevertheless decouples from all spin-triplet channels. In the absence of SOC but with magnetic field,
the gap function also simplifies to \(\widehat{\Delta}^{(s)}=\Delta_0e^{-\vartheta}i\tilde{\sigma}^y\) (\(\vartheta\) is the angle specifying the direction of the magnetic field), but in this case has a fully interband band-singlet form. The absence of intraband terms in particular implies the absence of a weak coupling logarithmic instability, and leads to strong spin-limiting and the resulting Pauli limit. The spin-singlet channel still decouples from the spin-triplet channels due to the inversion symmetry.

Importantly, when both SOC and Zeeman terms are present, however, the spin-singlet gap function contains both interband (band-triplet) and intraband (band-singlet) terms, and is no longer a well-defined channel despite being a self-consistent solution. 
This reflects the fact that spin-singlet and spin-triplet terms are mixed by the simultaneous presence of SOC and Zeeman splitting and therefore the spin-singlet channel is no longer well-defined \cite{Edelstein89}.
As we shall see, the intraband terms play a crucial role in the bulk SDE.

\subsubsection{Derivation of the GL Theory and SC Phases of the Rashba-Zeeman Model}

\begin{figure}[t]
\includegraphics[width=0.99\textwidth]{NCSfig.pdf}
\caption{(a) Fermi surfaces in the normal state of the Rashba-Zeeman Hamiltonian formed by the two helical bands \(\lambda=\pm\) in the presence of an in-plane magnetic field (spin orientation shown by arrows). The centers of the two Fermi surfaces are shifted by \(\mathbf{Q}_\lambda\) in opposite directions perpendicular to \(\mathbf{h}\).
(b) Schematic SC phase diagram of the Rashba-Zeeman model.  Color scale shows the value of the Cooper pair momentum \(q\). Black solid line indicates the second order phase transition into the normal state (white); the dashed line indicates the Lifshitz transition (crossover) between the weak and strong helical phases (with \(q\sim\varsigma h\) and \(q\sim h\), respectively), marked by the formation of the BFS in the strong helical phase. For some parameter values the crossover becomes a first order phase transition (solid blue line) at a critical end point (blue dot) at low temperatures. The stripe phase with coexisting \(\mathbf{q}=2\mathbf{Q}_\lambda\) may occur at higher magnetic fields. (c) The BdG spectrum in the weak (bottom) and strong (top) helical phases; pink surface indicates zero energy (units are arbitrary). The spectrum is gapless in the strong helical phase as the BdG bands cross zero energy, forming the BFS (highlighted).
}
\label{NCSfig}
\end{figure}

The GL coefficients needed for computing the SDE can be obtained via the standard expansion of the SC free energy (in the Matsubara formalism), formally obtained by integrating out the fermions and expanding the resulting trace logarithm:
\[\mathcal{F}[\widehat{\Delta},\widehat{\Delta}^\dagger]=-T\sum_{n\mathbf{k} j} \frac{1}{2j}\text{Tr}\left[\left(\widehat{\Delta}^\dagger(\mathbf{k;q})G^{(0)}(i\omega_n,\mathbf{k+q}/2)\widehat{\Delta}(\mathbf{k;q})G^{(0,h)}(i\omega_n,\mathbf{k-q}/2)\right)^j\right]+H_{\Delta^2}\label{Fexpansion}\]
where \(j\) is summed over all positive integers, \(\omega_n=\pi T(2n+1)\) are the Matsubara frequencies, \(G^{(0)}(i\omega,\mathbf{k})=(i\omega-\mathcal{H}(\mathbf{k}))^{-1}\) is the bare Green's function of the single-body Hamiltonian Eq. \ref{H0}, and the hole normal state Green's function \(G^{(0,h)}(i\omega,\mathbf{k})=-G^T(-i\omega,-\mathbf{k})\) (odd \(j\) terms vanish by symmetry). \(H_{\Delta^2}\sim |\Delta|^2/V\) includes the pairing interactions.

The phase diagram, shown schematically in Fig. \ref{NCSfig} (b), can be constructed by minimizing the free energy with respect to both the order parameter 
\(\widehat{\Delta}^\dagger\) and the pair momentum \(\mathbf{q}\). Minimization with respect to \(\widehat{\Delta}^\dagger\) yields the self-consistency condition -- namely, the gap equation -- while variation with respect to \(\mathbf{q}\) determines the supercurrent \(\mathbf{J}(\mathbf{q})\). The equilibrium state corresponds to the value \(\mathbf{q}=\mathbf{q}_0\)	
  at which the current vanishes, \(J(\mathbf{q}_0)=0\). Alternatively, one may derive the GL functional directly by evaluating the momentum and frequency sums and truncating the harmonic expansion over 
\(j\), typically at  \(j=2\).

Assuming \(s\)-wave singlet interactions, the self-consistent phase diagrams of helical Rashba SCs have been studied in several works.
Though it was known based on phenomenological models that the helical phase should exist due to the presence of the Lifshtiz invariant \cite{MineevSamokhin94, Agterberg03}, in a microscopic interaction model it was first established in Ref. \cite{BarzykinGorkov02}, which was soon verified in Ref. \cite{DimitrovaFeigelman03} where a \(T-H\) phase diagram was constructed. Ref. \cite{DimitrovaFeigelman03}, however, neglected the difference in the DOSs of the two helical bands, and as a result found a uniform SC state at low fields, with a helical phase only emerging after a second order phase transition at higher fields, followed by another second order phase transition into the so-called stripe phase. In the stripe phase it becomes energetically preferable for electrons on band \(\lambda\) to pair with momentum \(2\mathbf{Q}_\lambda\), at the cost of interactions between Cooper pairs from different bands being forbidden. The stripe phase corresponds to the usual FFLO phase in the absence of SOC.
Rectifying the assumption of equal DOSs, Refs. \cite{DimitrovaFeigelman07, AgterbergKaur07} found that the helical phase emerges continuously at arbitrarily small fields with Cooper pair momentum \(\mathbf{q}\sim \delta\nu/\nu_0 \mathbf{Q}_-\) with \(\delta\nu=\nu_+-\nu_-=-2\nu_0(\alpha_R/v_F)\) (hence assuming \(\delta\nu=0\) resulted in \(\mathbf{q}=0\)). This is essentially a compromise between the shifted momenta of the two Fermi surfaces.
The phase transition between the uniform and helical phases at equal DOSs at \(\delta\nu=0\) becomes instead a crossover from the (weak) helical phase to the strong helical phase in which the \(\lambda=-1\) band with the higher DOS dominates and \(\mathbf{q}\sim  2\mathbf{Q}_-\). This crossover can become a first-order phase transition at low temperatures provided the SOC is not too strong. Both Refs. \cite{DimitrovaFeigelman07, AgterbergKaur07} also found that there is a Lifshitz transition into a gapless SC state determined by the field being equal to the direct gap between the BdG bands, \(\Delta=h\), resulting in the formation of Bogoliubov Fermi surfaces \cite{TimmAgterberg17, YuanFu18, LinkBoettcherHerbut20, ShafferBurnellFernandes20, AkbariThalmeier22, BanerjeeSchnyder22, BabkinSerbyn24, Mateos24, Sano25, Wei25} (see  Fig. \ref{NCSfig} (c)). The presence of BFSs, which form predominantly on the lower-DOS band, serves as a more robust marker of the strong helical phase. Finally, it was noted in Ref. \cite{AgterbergKaur07} that the region occupied by the stripe phase generally shrinks with increasing \(\delta\nu\).

\subsection{SDE in Helical NCSs}\label{NCSSDESec}

As noted in Sec.~\ref{GLSDESec}, it took a sequence of theoretical developments~\cite{LevitovNazarovEliashberg85, Edelstein96, BauerSigrist12, DevizoravaBuzdin21, DaidoYanase22, HeNagaosa22, YuanFu22, IlicBergeret22, HasanShafferKhodasLevchenko24}, along with renewed motivation from the 2020 experimental observation of the SDE in a NCS heterostructure~\cite{AndoYanase20}, to arrive at a consistent GL description of the effect.
Yet even after the correct GL formalism was established in Ref.~\cite{IlicBergeret22}, the story contained one final twist. The properly derived analytical expression for the SDE efficiency, $\eta$, in the Rashba–Zeeman model with $s$-wave pairing -- valid to linear order in the magnetic field and in the limit of strong spin–orbit coupling -- was again found to vanish, $\eta=0$. This unexpected result invalidated earlier findings reported in Refs.~\cite{HeNagaosa22, YuanFu22}. (As discussed below, the conclusions of Ref.~\cite{DaidoYanase22} remain qualitatively correct, although the corresponding analytical expressions are not quantitatively accurate for the microscopic model.) To clarify when and why the effect cancels, and, more importantly, under what conditions it survives, we now turn to a detailed examination of this computation.

\subsubsection{SDE in Rashba-Zeeman Model with \(s\)-wave Pairing}

To find the critical currents and the associated SDE, we now proceed to derive the GL theory from Eq. \ref{Fexpansion} and assuming a singlet \(s\)-wave gap function from Eq. \ref{DeltaS}. We can then use the formulas from Sec. \ref{GLSDESec}.
Without some further approximations, this can only be done numerically with some tight-binding regularization for the band dispersions, as done in Refs. \cite{DaidoYanase22, DaidoYanase22_2, IkedaDaidoYanase22, IlicBergeret22}. In most approaches, weak coupling is assumed, and, moreover, typically the Fermi energy \(\varepsilon_F=v_Fp_F\) is assumed to be the largest (effectively infinite) energy scale. In that case, one can make the quasiclassical approximation, e.g., in the Eilenberger approach as done in Refs. \cite{IlicBergeret22}, or used to evaluate the momentum integrals in a more direct approach \cite{Edelstein96, YuanFu22, HasanShafferKhodasLevchenko24, HasanShafferKhodasLevchenko25}.
However, it should be noted that the quasiclassical approximation is not valid if the Rashba SOC strength \(\alpha_Rp_F\) is comparable to the Fermi energy (as considered in Ref. \cite{HeNagaosa22}) if interband pairing is important, since the linearization of \(\xi_\lambda\) around \(p_{F\lambda}\) may not be valid at \(p_{F,-\lambda}\). For this reason, most calculations assume \(\alpha_Rp_F\ll \varepsilon_F\), or equivalently that the parameter \(\varsigma=\alpha_R/v_F\ll1\) is small and work to linear order in \(\varsigma\) in analytical calculations \cite{Edelstein96, IlicBergeret22, HasanShafferKhodasLevchenko24, HasanShafferKhodasLevchenko25}.

Additional energy scales of the problem are associated with the pairing momentum (\(v_Fq\)), magnetic field (\(h\)), the pairing strength (\(T_c\)), and the proximity to the phase transition line (\(\alpha_0=\nu_0t\) with \(t=(T-T_c)/T_c\) being the reduced temperature). A weak-coupling cutoff \(\Lambda\) is also formally needed to avoid the logarithmic divergence. For the applicability of the GL theory, \(t\ll1\) must be assumed, and in general all other energy scales must be assumed to be small. Analytical results can be found in the limit of small \(h\), in particular \(h\ll(\alpha_Rp_F,T_c)\) and working to linear order in \(h\).
Since \(v_Fq_0\) is at most of the order of \(h\) (\(\varsigma h\) in the weak-coupling theory) and close to \(T_c\) the range of \(q\) is constrained to be within \(v_Fq\sim T_c\), which is thus also assumed to be small and only leading terms in powers of \(q/p_F\) are kept.
Some works further assume in addition that Rashba SOC is strong compared to \(T_c\), \(\alpha_R p_F\gg T_c\). In that case interband pairing is strongly suppressed and can usually be neglected, as done explicitly or implicitly in Refs. \cite{YuanFu22, IlicBergeret22}. However, as found in Ref. \cite{IlicBergeret22}, in that limit no SDE occurs at all! As shown in Ref. \cite{HasanShafferKhodasLevchenko24}, it is therefore important to keep the interband terms.

It turns out that nearly analytic expressions can be obtained for arbitrary values of \(\kappa = \alpha_Rp_F/(\pi T_c)\) and \(q\), as long as \(h\ll\alpha_Rp_F\). In particular,
\begin{align}
\alpha(\mathbf{q})&=\alpha_\infty(\mathbf{q})+\nu_0\int\left[\digamma\left(\frac{v_F q \cos(\theta_\mathbf{k}-\theta_\mathbf{q})+2\alpha_Rp_F}{2\pi T}\right)-\digamma\left(\frac{v_F q \cos(\theta_\mathbf{k}-\theta_\mathbf{q})}{2\pi T}\right)\right]d\theta_\mathbf{k}\label{alpha}\\
\beta(\mathbf{q})&=\beta_\infty(\mathbf{q})+\delta\beta(\mathbf{q})=\int\sum_{\lambda}\frac{\nu_\lambda}{128\pi^3T^2}\psi^{(2)}\left(\frac{1}{2}-\frac{i\delta\xi_{\lambda}}{4\pi T}\right)d\theta_\mathbf{k}+\delta\beta(\mathbf{q})\label{beta}
\end{align} 
with
\[\alpha_\infty(\mathbf{q})=\left[\sum_{\lambda}\int \frac{\nu_{\lambda}}{4\pi}\left[-\ln\frac{1.13\Lambda}{T}+\digamma\left(\frac{\delta\xi_{\lambda}}{\pi T}\right)\right]d\theta_\mathbf{k}\right]^{T_c,\mathbf{q}=0}_T,\qquad\digamma\left(x\right)=\text{Re}\left[\psi\left(\frac{1+ix}{2}\right)-\psi\left(\frac{1}{2}\right)\right]\,,\]
where \(\psi(x)\) is the digamma function, \(\psi^{(n)}(x)=\partial_x^n\psi(x)\),
\(\delta\xi_{\lambda}(\mathbf{\theta_k;q})=2\lambda h \sin\theta_\mathbf{k}+v_F q \cos(\theta_\mathbf{k}-\theta_\mathbf{q})\), 
and \(\theta_\mathbf{q}\) is the angle of vector \(\mathbf{q}\) measured from the direction of \(\mathbf{h}\) (which recall we take to be the \(\hat{\mathbf{x}}\) direction) \cite{HasanShafferKhodasLevchenko24}. \(\alpha_\infty(\mathbf{q})\) and \(\beta_\infty(\mathbf{q})\) are the values of \(\alpha(\mathbf{q})\) and \(\beta(\mathbf{q})\) in the limit \(\kappa\rightarrow\infty\) (i.e., leading terms in the strong SOC limit). The correction \(\delta\beta(\mathbf{q})\) can in principle also be computed in terms of angular integrals of the digamma function, but the expression is complicated.

Analytic expressions for the GL coefficients \(\alpha_n\) and \(\beta_n\) are obtained by expanding the digamma functions in powers of \(q\) and evaluating the elementary angular integrals. The resulting coefficients can be found in Eqs. (D3-D4) in Ref. \cite{HasanShafferKhodasLevchenko24}. From Eq. \ref{etaGL}, the resulting SC diode efficiency coefficient reads:
\[\label{eq:eta}
\eta=\frac{\varsigma \sqrt{-t}}{7\sqrt{42\zeta^3(3)}} \frac{h\sin\theta_\mathbf{q}}{\pi T_c}\left(16\kappa^{-4}\digamma(\kappa)+8\kappa^{-3}\digamma^{(1)}(\kappa)+14\kappa^{-2}\zeta\left(3\right)-\kappa^{-2}\digamma^{(2)}(\kappa)\right)\]
and \(\zeta(n)\) is the Riemann zeta function and \(\digamma^{(n)}(x)=\partial_x^n\digamma(x)\).
This has simple expressions for strong and weak SOC limits: \(
\eta= 2
\varsigma \sqrt{-t}\,h\sin\theta_\mathbf{q}/(\pi\sqrt{21\zeta(3)} T_c\kappa^2)\label{etaLargeSOC}\)
and
\(\eta=635\zeta(7) \, \varsigma  \sqrt{-t}\,\kappa^2 
h\sin\theta_\mathbf{q}/(56\sqrt{42\zeta^3(3)}\,\pi T_c)\), respectively. In particular, \(\eta\) vanishes both in the absence of SOC (as expected by symmetry) and, more surprisingly, in the limit of strong SOC.
The assumption of pure interband pairing in Refs. \cite{YuanFu22, IlicBergeret22}, which is justified for strong SOC, amounts precisely to the limit \(\kappa\rightarrow\infty\), i.e., taking \(\alpha(\mathbf{q})=\alpha_\infty(\mathbf{q})\) and \(\beta(\mathbf{q})=\beta_\infty(\mathbf{q})\). As shown in Ref. \cite{IlicBergeret22} and can be seen from Eq. \ref{eq:eta}, this means that \(\eta=0\) to order \(\sqrt{-t}\). It was proven in Ref. \cite{HasanShafferKhodasLevchenko24} that if one ignores interband interactions, \(\eta=0\) to all orders in \(t\) and \(q\), to linear order in \(h\).

It is instructive to consider why SDE vanishes in this limit, which is again somewhat analogous to the vanishing of JDE in a SQUID in the absence of higher harmonics discussed in Sed. \ref{SQUIDSec}. In particular, the absence of interference in the \(s\)-wave Rashba-Zeeman model is in this case due to the shifted Fermi surfaces remaining circular, to linear order in \(h\).
As shown in Ref. \cite{HasanShafferKhodasLevchenko24}, in that case the condensation energy defined in Eq. \ref{f} satisfies an approximate (to linear order in \(h\)) inversion like symmetry \(\mathbf{q+q}_0\rightarrow-\mathbf{q+q}_0\), which implies no SDE (to that order in \(h\)).
Additional distortion of the Fermi surface is needed, which originates from cubic and higher order terms in \(h\) in the simplest model in the absence of interband pairing. Alternatively, interband pairing also provides the interference needed to produce the SDE in analogy to higher harmonics producing JDE in a SQUID.
In more realistic models, there are many other possible perturbations that break the approximate symmetry and thus also produce an SDE, for example noncircular Fermi surfaces, non-\(s\)-wave pairing, disorder \cite{HasanShafferKhodasLevchenko25}, etc., can all induce a nonzero \(\eta\). The Rashba-Zeeman SC model thus ironically turns out to be highly fine-tuned to minimize the SDE. On the other hand, this implies that a plethora of mechanisms for realizing and enhancing SDE in NCSs are possible by considering various ways of breaking the approximate low-field symmetry.

At higher fields and low temperatures the SDE has to be evaluated numerically, as done in Refs. \cite{DaidoYanase22, DaidoYanase22_2, IlicBergeret22, Aoyama24, IkedaDaidoYanase22}. Since lattice regularization is generally used, \(\eta\) in principal contains a small linear in \(h\) term, but it is generally found that it grows quickly with \(h\) within the weak helical phase. More interesting qualitative features are found near the crossover of phase transition between the weak and strong helical phases: it is found that \(\eta\) is strongly enhanced in this region of the phase diagram, with different numerical studies finding maximal \(\eta\) to be between 20\% and 50\% in this region. Moreover, \(\eta\) changes sign in the strong helical phase (sometimes multiple sign changes are seen). Qualitatively, one reason for the sign change is that the Fermi surface deformation at \(p_x=0\) (assuming the field is along the \(\hat{\mathbf{x}}\) direction, such that \(\mathbf{Q}_\lambda\) is along the \(\hat{\mathbf{y}}\) direction) has the opposite sense relative to \(\mathbf{Q}_\lambda\) itself. However, since interference between the two bands remains important and multiple sign changes are seen, the sign of \(\eta\) is in general not as simple to explain.
The enhancement of the diode efficiency can be understood as a consequence of the proximity to a critical end point \cite{ZhuangShafferHasanLevchenko25} (see  Fig. \ref{NCSfig} (b)). The first order phase transition implies an existence of multiple minima of \(f(q)\), whereas at a cross over the two minima merge. As discussed in Sec. \ref{GLFFLOSec}, \(\eta\) formally approaches its perfect value of \(\pm100\%\) at such points as a critical point of \(J(q)\) becomes an inflection point. Finally, the stripe phase with multiple \(q\) order parameters has also been studied numerically in Ref. \cite{Aoyama24}; though one may expect that SDE is reduced in a stripe phase as no SDE takes place in the LO-type/stripe phase in the absence of SOC, SDE is found to persist  due to the DOS and Fermi surface asymmetries due to ISB, and no sharp discontinuity was seen in \(\eta\) across the phase transition.  Moreover, \(\eta\) is found to be actually slightly enhanced in the stripe phase in comparison to the helical phase, possibly due to the increased competition between the two helical bands.

\subsubsection{Variations on the \(s\)-wave Rashba-Zeeman Model}

In the above discussion, several effects that can take place within the Rashba-Zeeman model have been neglected.
Rashba SOC specifically due to interfacial effects has been discussed in the context of SDE \cite{DevizoravaBuzdin21, PutilovBuzdin24, PlastovetsBuzdin24, MazanikBergeret25}, and the orbital Rashba effect (affecting orbital angular momentum instead of the spin) \cite{Park11} has been proposed for realizing SDE \cite{Saunderson25}.
In this section we also neglected other orbital effects discussed in Sec. \ref{VDESec}.
We had also assumed equal Fermi velocities of the two helical bands, but this need not be the case in general, for example in the presence of a confining potential in the 1D Rashba-Zeeman model, which can lead to SDE even if \(\delta\nu\) is assumed to vanish. \cite{dePicolliVayrynen23, MeyerHouzet24}.
Smaller chemical potentials have also been considered in 1D systems, and \(\eta\) appears to be maximized at near \(50\%\) close to \(\mu=0\)  \cite{BhowmikGhosh25}. A similar effect of low chemical potentials close to \(T_c\) had been noted in 2D in Ref. \cite{HeNagaosa22} (with much lower \(\eta\) due to the proximity to \(T_c\)), but a formula for \(\eta\) inconsistent with that in Ref. \cite{IlicBergeret22, HasanShafferKhodasLevchenko24} was used. Note that at low chemical potentials topological SC states may be realized with further implications for SDE considered in Sec. \ref{TopoSec}.

Another important consideration that can be important for SDE is disorder, which introduces another dimensionless parameter \(T_c\tau\) to the problem, with \(\tau\) being the electron scattering time on impurities. On symmetry grounds, since the disorder itself is typically inversion symmetric on average, it can be expected to be detrimental to intrinsic SDE. This has indeed been found to be the case in numerical studies in the strong disorder regime \(T_c\tau\ll1\) studied in Ref. \cite{KokkelerTokatlyBergeret24, IlicBergeret24, MazanikBergeret25} by solving the Usadel equation (the additional suppression of the critical current due to Coulomb interactions was also considered in Ref. \cite{NunchotYanase25_2}), as well as in weak-to-moderate disorder regime in the intraband limit studied using the Eilenberger approach close to \(T_c\) in Ref. \cite{IlicBergeret22}. In Refs. \cite{IlicBergeret22, IlicBergeret24, MazanikBergeret25}, the main detrimental effect was attributed to the suppression of the strong helical phase due to disorder-induced hybridization between the two helical bands, since \(\eta\) is maximized near the weak-to-strong helical phase crossover on which these works focused.
Interestingly, while the suppression of the strong helical phase was also found numerically in Ref. \cite{IkedaDaidoYanase22} using a self-consistent Born approximation on a square lattice, the maximum diode efficiency close to the weak-to-strong helical phase crossover was actually found to increase with disorder for weak to moderate disorder and at low temperature (again in the intraband limit): while both \(|J_{c\pm}|\) are found to decrease with disorder, the smaller of the two is found to decrease faster.
The counterintuitive increase of \(\eta\) with increasing disorder was explained in the low-field weak helical phase limit close to \(T_c\) using an analytical calculation employing a self-consistent Born approximation in Ref. \cite{HasanShafferKhodasLevchenko25}. In particular, it was found that disorder breaks the low-field approximate symmetry \(\mathcal{I}'\) that holds in the intraband limit considered in Refs. \cite{IlicBergeret22, IkedaDaidoYanase22}: as a result, \(\eta\) no longer vanishes to linear order in \(h\) and is therefore enhanced at lower fields and weak disorder before reaching a peak at moderate disorder and vanishing in the strong disorder limit. Additionally, in the case of weak SOC (far from the intraband limit), it was found that while even weak disorder generally suppresses SDE, \(\eta\) can change sign within the weak helical phase at moderate disorder and is generally a nonmonotonic function of \(T_c\tau\).

Beyond such variations within the \(s\)-wave Rashba-Zeeman model,
given the subtle vanishing of SDE in the intraband limit discussed above one can furthermore anticipate that small modifications to the model have large qualitative effect, as found in various studies.
This includes numerical results for lattice realizations of the Rashba-Zeeman model \cite{DaidoYanase22}, and  studies that found \(d\)-wave pairing to give a moderate enhancement of \(\eta\) over the \(s\)-wave case \cite{DaidoYanase22_2, OhNagaosa24}.
The Zeeman field can also be substituted by other TRSB fields such as altermagnetic orders as considered in Ref. \cite{BanerjeeScheurer24alt}, but we postpone the discussion of this possibility to Sec. \ref{MagSec}.
More generally, other NCSs that are not described by the Rashba-Zeeman model can be considered, including chiral and polar/ferroelectric systems.
These often have other forms of SOC, some of which have been considered in the literature, including radial Rashba SOC \cite{KangKochan24}, cubic Rashba SOC in 1D \cite{BhowmikGhosh25}, and Ising SOC \cite{BankierLevchenkoKhodas25}.
The latter was found to enhance the SDE: though Ising SOC by itself is not sufficient to realize SDE for in-plane magnetic fields, it results in a linear in \(\mathbf{h}\) term contributing to \(\eta\) even in the intraband limit. 
In particular, Ising SOC breaks the approximate symmetry \(\mathcal{I}'\) that otherwise suppresses the SDE in the low field limit.
The trigonal lattice has also been considered in the case of the 2D polar/ferroelectric material CuNb\(_2\)Se\(_4\): in this case no helical SC state is formed due to the \(C_3\) symmetry, but SDE is still possible \cite{Zhai22}.
More generally, point group symmetries also dictate the direction of the current along which the SDE takes place: in the Rashba-Zeeman model and polar point groups\footnote{I.e., leaving more than a single point invariant under the the symmetry action. These are \(C_s\), \(C_n\), \(C_{nv}\) and \(D_{nv}\). In particular, ferroelectric materials are polar.} SDE occurs for \(\mathbf{J,q}\perp\mathbf{h}\), but it can occur for \(\mathbf{J,q}\parallel\mathbf{h}\) in chiral\footnote{I.e., only containing pure rotation symmetries. These are \(C_n\), \(D_n\), \(O\) and \(T\), also called enantiomorphic.} systems including those with radial Rashba SOC.\footnote{Or some more general direction, including for noncentrosymmetric point groups that are neither polar nor chiral: \(D_{2d}\), \(C_{3h}\), \(D_{3h}\), \(S_4\), \(T_d\). All chiral and polar groups are non-centrosymmetric.}
The case \(\mathbf{J}\perp\mathbf{h}\) is more typical as it naturally arises from relativistic SOC, and the microscopic origins of \(\mathbf{J}\parallel\mathbf{H}\) are less studied, though known to occur in experiments \cite{KealhofferBalents23, HouMoodera23, GaggioliMoodera25}. In theory, SDE with \(\mathbf{J}\parallel\mathbf{h}\) has also been found to occur due to Dzyaloshinskii-Moriya interactions \cite{NunchotYanase24} and due to the Aharonov-Bohm phase in the normal state of chiral nanotubes \cite{HeNagaosa23, LiHe25}; in the latter, JDE with \(\mathbf{J}\parallel\mathbf{h}\) has also been found \cite{Cuozzo25}.

\subsection{JDE in Noncentrosymmetric Materials}

It is likely that the first instance of intrinsic JDE was also noted in the context of experimental signatures of helical NCSs, in particular skewed Fraunhofer patterns (equivalent to JDE, as discussed in Sec. \ref{SQUIDSec}) \cite{YangAgterberg00, KaurAgterbergSigrist05}. 
Helical NCSs have subsequently been predicted to realize AJE \cite{Buzdin08, DolciniMeyerHouzet15, HasanSongciLevchenko22} and there has in general been much interest in studying JJs with SOC \cite{AmundsenZutic24}, but surprisingly few works considered them in the context of JDE \cite{Roig24, MeyerHouzet24, OsinLevchenkoKhodas24, ZazunovEgger24, ZhuangShafferHasanLevchenko25}, despite the recognition that linear gradient terms and finite-momentum Cooper pairs generically lead to JDE \cite{DavydovaFu22, ZinklSigrist22, KochanZutic23, Zhang24, HuangVayrynen24, KarabassovBobkovaVasenko24}.
Most theoretical studies of JDE do, however, invoke the Rashba-Zeeman model to describe ISB and TRSB in the normal region of JJs \cite{ReynosoAvignon08, ZazunovEggerMartin09, ReynosoAvignon12, YokoyamaNazarov13, YokoyamaNazarov14, NesterovHouzetMeyer16, PekertenMatosAbiague22, WangWangWu22, ZhangJiang22, ChengSun23, CostaKochan23, CostaManfraKochanParadisoStrunk23, Huang23, LiuAndreevSpivak24,  DebnathDutta24, PekertenMatosAbiagueZutic24, MeyerHouzet24, WangChen24, Soori24, IlicBergeret24, RuizStrunk25}. 

\subsubsection{JDE with Rashba SOC and \(s\)-wave Pairing}\label{NCSJDESec}

Several methods exist for computing CPRs of JJs and have been used for computing JDE in noncentrosymmetric materials. The most direct approach is to find the Andreev bound states and their spectrum by obtaining the eigenvectors and eigenvalues of the BdG Hamiltonian in Eq. \ref{HBdG} in real space, allowing for spatial variation of the gap function (and other parameters of the Hamiltonian), fixing its phase difference across the junction as a boundary condition. Typically, tight-binding models are used for regularization in that case, as done numerically in Refs. \cite{ReynosoAvignon08, ZazunovEggerMartin09, ReynosoAvignon12, YokoyamaNazarov13, YokoyamaNazarov14, PekertenMatosAbiague22, WangWangWu22, ZhangJiang22, CostaKochan23, Roig24, ZazunovEgger24,  PekertenMatosAbiagueZutic24, WangChen24, Soori24}. The GL theory of a bulk SC can also be used to compute JDE, again allowing for the spatial variation of the order parameter \(\Psi\) in Eq. \ref{FGL}, fixing its phase difference across the junction. Refs. \cite{KochanZutic23, HasanSongciLevchenko22} used this approach for junctions with identical S regions, but it was first applied to  JJs between a uniform and nonuniform SCs (which includes helical SCs) in Ref. \cite{YangAgterberg00}.
Alternatively, the quasiclassical Eilenberger approach is well suited for computing JDE either in the ballistic \cite{ZhuangShafferHasanLevchenko25} or disordered \cite{LiuAndreevSpivak24, IlicBergeret24} cases (in the latter, the Usadel equation can be derived), and allows for a more analytical treatment that we outline here.

The length of the Josephson junction \(L\), to be compared to the coherence length \(\xi\sim v_F/T_c\), provides an additional scale in the problem, with the corresponding energy scale being the Thouless energy \(E_T=v_F/L\)  (for a ballistic junction without disorder).
Analytical results can be obtained in the limit of short and long junctions -- \(L\ll\xi\) or \(L\gg\xi\), respectively -- at zero temperature and small magnetic fields, and neglecting interband pairing. The two bands then contribute independently to the current, \(J(\varphi)=\sum_\lambda J_\lambda=\sum_\lambda (1-\lambda\varsigma) J_0(q-2Q_\lambda,\varphi-2Q_\lambda L)\), where \(J_0\) is the current evaluated in the absence of SOC and magnetic fields.
Here we further assumed a quasi-1D junction with the magnetic field perpendicular to the current direction.
Note the formal analogy to a multichannel JJ or a SQUID and the intrinsic JDE mechanism thus has the same phenomenology as discussed in Sec. \ref{SQUIDSec}.
In the short junction limit,
\[J_0(q,\varphi)=\Delta_q\sin\frac{\varphi}{2}\sgn\left[\cos\frac{\varphi}{2}-\frac{qv_F}{\Delta_q}\right]+\frac{2qv_F}{\pi}\label{eq:J0-JDE}\]
as first found using the scattering matrix approach in Ref. \cite{DavydovaFu22} for the case of a general nonuniform SC, and specifically for helical SCs in Refs. \cite{ZazunovEgger24, MeyerHouzet24} (with \(q=0\), this has the same form as the clean limit Kulik-Omelyanchuk CPR at zero temperature \cite{GolubovKupriyanov04}). The first term in Eq. \eqref{eq:J0-JDE} originates from the Andreev bound states, whereas the last term is a contribution from the continuum states.
In the weak helical phase, \(\Delta_q\) is independent of \(q\) to leading approximation (which justifies neglecting its self-consistency in the weak helical phase), while \(q v_F \approx\varsigma h\).
In the short junction, therefore, the JDE originates from the finite momentum of the Cooper pairs. Consequently, no JDE can be found in the short junction limit \(L=0\) without accounting for helical SC, as noted in Ref. \cite{MeyerHouzet24} and consistent with results found in numerical studies. It is also crucial that the band-resolved CPRs \(J_\lambda\) contain higher harmonics, as discussed in Sec. \ref{GLJDESec}.

In a long junction, on the other hand, one finds \(J_0(q,\varphi-2Q_\lambda L)\approx J_0(0,\varphi-2Q_\lambda L+q\xi)\), i.e., the shift of the Fermi surfaces amounts simply to an additional relative phase shift. In the limit \(L\gg\xi\), this shift can be neglected, so that the self-consistency condition for \(q\), which also accounts for current conservation at junction interfaces \cite{Krekels25}, is not as crucial (except in the strong helical phase in which the coherence length can be enhanced), and the phase shift between the two bands can be taken to be \(\delta\varphi\approx4Q_-L\approx 4h/E_T\), originating from interference due to momentum shifts of the two helical bands. For this reason a finite length of the JJ is necessary to observe JDE (and AJE) in the absence of finite-momentum pairing.
At zero temperature the CPR is a simple sawtooth function: \(J_0(0,\varphi)= \sigma\varphi\) for some constant \(\sigma\) and \(\varphi\in(-\pi,\pi)\), repeating modulo \(2\pi\).
The critical currents are then given by
\[J_{c\pm}=\pi\sigma\left\{\begin{array}{ll}
     \pm 2(1-x) + 2\varsigma x, & x <(1-\varsigma)/2\\
     \pm (1+\varsigma)+(1-\varsigma)(1-2x), & (1-\varsigma)/2 < x <(1+\varsigma)/2\\
     \pm 2x + 2\varsigma (1-x), & x >(1+\varsigma)/2
\end{array}\right.\]
The SC diode efficiency coefficient is thus
\[\eta(x,\varsigma)=\left\{\begin{array}{ll}
     \varsigma x/(1-x), & x <(1-\varsigma)/2\\
     (1-\varsigma)(1-2x)/(1+\varsigma), & (1-\varsigma)/2 < x <(1+\varsigma)/2\\
     \varsigma(1-x)/x, & x >(1+\varsigma)/2
\end{array}\right.\]
Note that because \(x\approx 4h/(\pi E_T) \mod 2\), in the long junction \(\eta\) is an oscillatory function of \(h\) and \(L\), so that the field can be used to change the sign of the effect.
The maximum efficiency is found to be \(\eta=3-2\sqrt2\approx17\%\). This is comparable to most reported maximal values in numerical studies of JDE in either short or long junctions, with values ranging up to approximately 40\%. Higher values have been found when additional sources are considered, such as quantum point contact potentials \cite{ReynosoAvignon08}.
Importantly, note that \(\eta=0\) when \(\varsigma=0\) or \(1\), i.e., if the two helical bands have equal DOSs or only one of them contributes to the supercurrent.
This again demonstrates the fact that TRSB and ISB are together not sufficient to produce JDE: an interference between at least two channels is required. This conclusion holds for a more general single-channel CPR \(J_0(0,\varphi)\) as long as it itself does not exhibit JDE, as noted in Ref. \cite{YokoyamaNazarov13,YokoyamaNazarov14}.

Other junction parameters can have significant effects on \(\eta\).
In some parameter ranges topological SC is famously realized, but we postpone the discussion of the effects of topology on JDE, studied in Refs. \cite{KopasovMelnikov21, LeggLossKlinovaja23,  CayaoNagaosaTanaka24, LiuWang24, MondalCayao25, GuanAn25}, to Sec. \ref{TopoSec}.
Effects of imperfect transmission in short nontransparent junctions have been studied in Ref. \cite{ZazunovEgger24}, with results consistent with those for JDE with generic nonuniform SC found in Ref. \cite{DavydovaFu22}, namely that reduced transparency is detrimental to the diode efficiency. Note, however, that those works treated \(q\) as a tunable parameter and did not solve for it self-consistently.
Junctions in the strongly disordered regime (\(T_c\tau\ll1\)) have been studied using the Usadel equation in Refs. \cite{LiuAndreevSpivak24, IlicBergeret24}, but no clear effects on JDE have been identified.
Temperature effects cannot be properly studied without a full self-consistent treatment of the gap function, which was done in Ref. \cite{ZhuangShafferHasanLevchenko25}; as one might expect, it was found that lower temperature are generally favorable for the diode effect.
In 1D junctions, the confining potential can play an important role and give rise to unequal Fermi velocities of the two helical bands, as considered in Ref. \cite{MeyerHouzet24}.
Counter-positively, the effect of the width in a 2D planar JJ has been considered in Ref. \cite{IlicBergeret24} for strongly disordered junctions of moderate length \(L/\xi\sim 2\), observing a sign change of \(\eta\) with \(W\) in certain cases. The sign change originates from the same oscillatory behavior of \(\eta\) noted above.

Finally, all the results discussed above pertain only to the weak helical phase. In the strong helical phase at higher magnetic fields, some qualitative differences occur due to the presence of the Bogolyubov Fermi Surface. In that case it becomes important to account for finite Cooper pair momentum even in long junctions, and additionally for the direction of \(\mathbf{Q}_\lambda\) relative to the current direction. Because the BFSs are anisotropic and centered along the direction of \(\mathbf{Q}_\lambda\), the spectral gap is effectively dependent on the current direction and is greatly suppressed when the current flows parallel to the BFSs. Moreover, the BFSs form first on the band with the lower DOS, such that only one band contributes to the total current for such current directions, and the JDE is also found to be strongly suppressed as a result \cite{ZhuangShafferHasanLevchenko25}. When the current is perpendicular to the BFS, on the other hand, the supercurrent is not suppressed, resulting in a strongly 2-fold anisotropic response (with respect to current or magnetic field direction). Such an anisotropy can, moreover, serve as a signature for BFSs and strong helical phases.

\subsubsection{Variations on the \(s\)-wave Rashba-Zeeman Model}\label{NCSJDEvarSec}

In close analogy with the situation for bulk SDE, JDE has been analyzed in a variety of noncentrosymmetric systems beyond the Rashba–Zeeman model with 
$s$-wave superconductivity. However, to date, the possible role of helical or finite-momentum superconducting order has not been systematically investigated to the best of our knowledge.

A natural extension of these ideas arises in superconductor–ferromagnet–superconductor (SFS) junctions~\cite{Buzdin05, GengBergeretHeikkila23, Mitrovic25}, where the ferromagnetic barrier enables a field-free realization of the Josephson diode effect (JDE), i.e., without the need for an external magnetic field~\cite{MinutilloCampagnano18, CostaKochan23, HessLeggLossKlinovaja23, Hikino25}. Consistent with results for short Rashba–Zeeman junctions, it has been found that in the short-junction limit (\(L/\xi\to0\)) multiple conducting channels~\cite{CostaKochan23} or multilayer heterostructures~\cite{MinutilloCampagnano18} are required to observe a finite JDE in systems with uniform $s$-wave superconductors. In contrast, long diffusive SFS junctions (i.e., those with strong disorder) exhibit only very small efficiencies, 
\(\eta\sim0.2\%\)~\cite{Hikino25}, although interestingly, short junctions with lower interface transparency can enhance the effect~\cite{CostaKochan23}.

JDE can also arise in the absence of Rashba spin–orbit coupling (SOC) if an inversion-symmetry-breaking magnetic texture -- such as a domain wall or skyrmion -- or a magnetization gradient is present. However, in such cases the efficiency remains small, 
\(\eta\sim0.25\%\)~\cite{HessLeggLossKlinovaja23, Roig24}. A significantly larger effect has been predicted when asymmetric transmission amplitudes are induced by interface magnetization, with efficiencies up to 
\(\eta\sim36\%\) inferred from measured critical-current asymmetries~\cite{GreinSchon09, MargarisFlytzanis10}. Finally, altermagnets -- which intrinsically break time-reversal symmetry -- can serve as viable alternatives to conventional ferromagnetic barriers in realizing JDEs~\cite{ChengMaoSun24, Jiang25} (see Sec.~\ref{MagSec}).

Just as the Zeeman term, the Rashba term may also be replaced by other forms of SOC to describe JDE in broader classes of noncentrosymmetric materials. For instance, Ref.~\cite{WangChen25} studied JDE in the polar compound \(T_d\)-MoTe\(_2\), reporting a large efficiency of \(\eta\sim67\%\) for certain parameter regimes. Ising SOC has likewise been explored in both SFS junctions~\cite{PatilBelzig24} and in junctions involving altermagnetic superconductors~\cite{Boruah25}. Because Ising SOC pins spins out of plane, the (alter)magnetic field must also possess an out-of-plane component for the JDE to occur.

Similar behavior has been found in systems with radial Rashba SOC, which polarizes spins perpendicular to the Fermi surface: JDE appears for currents aligned parallel to the magnetic field~\cite{CostaFabian24}. Moreover, JDE is implicitly present -- but was not emphasized -- in studies of the anomalous Josephson effect (AJE) with cubic Rashba SOC~\cite{AlidoustZutic21}.

Beyond conventional $s$-wave singlet superconductors, other pairing symmetries have been considered, including 
$d$-wave singlet~\cite{ZhangJiang22, VakiliKovalev24} and $p$-wave triplet states~\cite{Soori24, SharmaThakurathi25}. Notably, field-free JDEs can arise when the superconducting order parameter itself breaks time-reversal symmetry, as in 
\(d+id\) or \(d+is\) pairing~\cite{ZhangJiang22, VakiliKovalev24}, or through singlet–triplet mixing~\cite{Soori24, SharmaThakurathi25}. Additional extrinsic geometric mechanisms have also been proposed, for instance due to kinks or bends in the junction geometry~\cite{KopasovMelnikov21, MaiellaroCitro24}.

Across these diverse realizations, the resulting diode efficiencies do not differ dramatically from those predicted in the Rashba–Zeeman model, with typical maximum values in the range \(\eta\sim30-40\%\).

\section{Other Mechanisms}\label{OtherSec}

Since SDE and JDE can arise in any superconducting system where time-reversal symmetry (TRS) and inversion symmetry (IS) are simultaneously broken, there exist as many possible mechanisms for superconducting diodes as there are distinct ways to break these symmetries together. At the same time, the interplay of TRSB and ISB is central to many other areas of modern condensed matter physics beyond noncentrosymmetric superconductors. These include superconductors with finite-momentum pairing and intrinsic TRSB, such as FFLO and pair-density-wave (PDW) states (Sec. \ref{FFLOSec}); nonsuperconducting orders that break TRS and IS, including magnetic phases such as altermagnets (Sec. \ref{MagSec}) and valley-polarized states (Sec. \ref{FFLOSec}); and topological systems in which SOC and magnetic fields play a defining role (Sec. \ref{TopoSec}).

Nonequilibrium superconducting states also generally violate TRS and can additionally break IS, potentially modifying or even generating both SDE and JDE (Sec. \ref{NonEqSec}). Below, we review how these various symmetry-broken states of matter have been theoretically proposed to realize superconducting diode effects. Conversely, one may view these proposals from the opposite perspective -- as demonstrations that superconducting diodes can serve as sensitive probes of TRSB and ISB, thereby providing a diagnostic tool for identifying exotic symmetry-broken phases.

\subsection{FFLO, Pair Density Waves, Other Spontaneous Symmetry Breaking} \label{FFLOSec}

As discussed phenomenologically in Sec. \ref{GLFFLOSec} and briefly in Sec. \ref{NCSJDESec}, both bulk SDE \cite{YuanFu22} and JDE \cite{DavydovaFu22} can be realized from spontaneous TRSB and ISB in FFLO states. Besides the classical FFLO mechanism due to Zeeman splitting \cite{FF,LO, SamokhinTruong17} considered in Ref. \cite{YuanFu22}, other microscopic mechanisms that realize FFLO states have been considered in the context of SDE. Screening Meissner currents were considered in Refs. \cite{DavydovaFu22, LevichevVodolazov23}, for example, and several works consider FFLO states originating from altermagentism \cite{BanerjeeScheurer24alt, SimKnolle25, ChakrabortyBlackSchaffer25, MukasaMasaki25}, with the altermagnetic field resulting in spin-splitting similar to that due to Zeeman splitting mechanism of the classical FFLO mechanism (see Sec. \ref{MagSec} for more details on SDE in altermagnets). Below we review two other mechanisms for SDE via FFLO states that have been proposed in the literature: from FFLO due to multiphase/multicomponent SCs and due to coexisting non-SC orders with ISB such as valley polarization.

\subsubsection{Multiphase SDE and Orbital FFLO}

Nonreciprocal transport due to a subcategory of FFLO/PDW states has been considered in multiphase \cite{KaneyasuSigrist10, ChazonoYanase23, ShafferChichinadzeLevchenko24, MatsumotoYanase25} and multicomponent SCs \cite{NunchotYanase24}, including in bilayer SCs \cite{XieLaw23, NakamuraYanase24, MatsumotoYanase25}. 
The main modification to the GL theory in these cases is an additional degree of freedom \(j=1,2\):
\[F=\sum_j\left(\alpha_j(q)|\Psi_j(\mathbf{\mathbf{q}})|^2+\beta_j(\mathbf{q})|\Psi_j(\mathbf{q})|^4\right)+\left[\alpha_{12}(\mathbf{q})\Psi_1^*(\mathbf{\mathbf{q}})\Psi_2(\mathbf{\mathbf{q}})+c.c.\right]+\dots\]
(additional mixed \(\beta\)-terms are also possible).\footnote{More generally, each order parameter \(\Psi_j\) may have its own momentum \(\mathbf{\mathbf{q}}_j\); the difference  \(\mathbf{\mathbf{q}}_1-\mathbf{\mathbf{q}}_2\), however, can be optimized and absorbed into the phase difference between the two, while the current couples only to the sum, \(\mathcal{F}_J=\mathbf{J}\cdot(\mathbf{q}_1+\mathbf{q}_2)\); see, e.g. \cite{DaidoYanaseLaw25}.} 

Note the similarity of this free energy to that of a multichannel JJ or a SQUID. The essential idea is that while the individual components  \(\Psi_1\) and \(\Psi_2\) do not produce a SDE on their own, their combination can do so through interference mediated by the mixing term \(\alpha_{12}\), which plays the role of an effective Josephson coupling. Free energies of this type have long been considered in theoretical studies of JJs involving unconventional superconductors with mixed symmetries \cite{VanHarlingen95, AgterbergSigrist98, YangAgterberg00, KaurAgterberg03, KaurAgterberg05, KaurAgterbergSigrist05, LeridonVarma07}.\footnote{Interestingly, while current nonreciprocity was not explicitly noted in these earlier works, the presence of skewed Fraunhofer patterns -- equivalently implying a Josephson diode effect -- was; see also Sec.~\ref{TopoSec}.} More recently, JDE has been studied explicitly in multicomponent JJs and SQUIDs \cite{YerinGiazotto24, YerinGiazotto25}.

The relevant degrees of freedom can correspond to pairing channels of opposite parity in so-called anapole superconductors \cite{ChazonoYanase23}, suggested to be relevant to UTe$_2$ \cite{KanasugiYanase22}; to singlet–triplet mixing \cite{NunchotYanase24, AzambujMockli25}; to the components of a two-dimensional irreducible representation (irrep) \cite{KaneyasuSigrist10}; or to a layer index in multilayer systems \cite{XieLaw23, NakamuraYanase24, Yuan25}. More generally, as pointed out in Ref.~\cite{ShafferChichinadzeLevchenko24}, \(\Psi_1\) and \(\Psi_2\) can represent any two order parameters belonging to distinct irreps -- not necessarily of opposite parity. In such cases, symmetry requires  \(\alpha_{12}(\mathbf{q})\) to vanish at $q=0$, implying that a finite-momentum state \(\mathbf{q}\neq0\) may be energetically favored in the mixed phase even if it is not in either pure state.

In the bilayer case, the so-called orbital FFLO states have also been considered more recently
\cite{Liu17, MockliYanaseSigrsit18, XieLaw23orb, Wan23, Zhao23, NakamuraYanase24, Clepkens24, NagJain24, Cao24, Yan24, MatsumotoYanase25, Zhao25, ZhuDasSarma25, ChazonoYanase25}: for an in-plane magnetic field with \(\mathbf{A}=\mathbf{B}\times\mathbf{z}\), by minimal coupling the vector potential can be absorbed into relative shift of momenta in each layer by \(\lambda\mathbf{q}_0\) with layer index \(\lambda=\pm\). The microscopic problem is then completely analogous to the case of helical NCSs discussed in Sec. \ref{NCSSDESec}: if the interlayer pairing is predominant, FF-like pairing with momentum \(2\mathbf{q}_0\) is favored. On the other hand, if intraband pairing is dominant, a state analogous to the LO-like stripe phase in helical SCs is realized with \(\Psi_\lambda\) carrying opposite momenta \(\lambda\mathbf{q}_0\) (see previous footnote). This corresponds to a Josephson vortex state, which is favored at higher fields as found in Ref. \cite{NakamuraYanase24}; the transition to the Josephson vortex state was also found to be accompanied by a sign change of \(\eta\). Note that the orbital effect is not sufficient to produce the SDE if each layer is inversion symmetric, as the normal state is then symmetric under a layer-dependent inversion-like symmetry and additional terms need to be introduced to break it in order to realize a diode effect. Similar finite-momentum states that can be expected to give rise to SDE have also been considered in bilayers with SOC \cite{YoshidaSigristYanase12, YoshidaSigristYanase15, NakamuraYanase17, LiuDevereaux24}.

\subsubsection{SDE and JDE with Coexisting Orders: Valley Polarization}

An alternative possibility for realizing SDE is that the spontaneous TRSB and/or ISB takes place already in the normal state due to a non-SC order that then coexists with the SC order. For example, TRSB can be spontaneously broken by magnetic order, which leads to SDE with explicit ISB; we discuss this case in the next Sec. \ref{MagSec}. ISB, in turn, can be spontaneously broken by ferroelectric order, a situation considered in Ref. \cite{Zhai22}; in that case an external magnetic field is required to realize the SDE. In both cases, the SDE can be `trained' by an external field that couples to the non-SC order, typically magnetic or electric fields, resulting in hysteretic behavior similar to that for SDE from FF-type pairing (for which the training field is the applied current itself).

A final possibility is a non-SC order parameter that breaks both TRS and IS. An example of recent interest is valley polarization (VP), observed in several moir\'{e} \cite{BurgMacDonald19, ZondinerVonOppen20, Zhang20, LiFuMak21, SaitoYoung21, YuFeldman22, XieMakLaw22} 
and non-moir\'{e} \cite{ZhouYoung21, delaBarrera22, Seiler22, PantaleonGuinea23, LiLi24, HolleisYoung25, Kumawat25} 
multilayer structures (including spin-polarized quarter metal states).
Many theories have been developed to study VP order \cite{PoSenthil18, LiuDai19, BurgMacDonald19, ChichinadzeClassenChubukov20, XuXu20, HsuDasSarma20, LianBernevig21, ChichinadzeClassenChubukov22, ChristosSachdev22, MandalFernandes23, DongLevitovChubukov23, WangZaletel24, LeeChichinadzeChubukov24, RainesGlazmanChubukov24, Friedlan25, MayrhoferChubukov25}, including VP-induced PDWs and finite-momentum SC that may lead to SDE \cite{HanKivelson22, WuWuWu23, CastroShafferWuSantos23, GilBerg25, ParraMartinezGuinea25}. 
Since SDE has in particular been observed in twisted trilayer graphene \cite{LinScheurerLi22}, there have been several theoretical studies of both bulk SDE \cite{ScammellScheurer22, BanerjeeScheurer24, BanerjeeScheurer25, Zhuang25, DaidoYanaseLaw25, ChenSchrade25, ChenScheurerSchrade25, Yoon25} (additional symmetries may need to be broken, with many possible corresponding orders analyzed in detail in Ref. \cite{ScammellScheurer22}); JDE has also been considered with valley polarization in the normal barrier \cite{WeiLiu22, XieLaw23, HuLaw23}. The (bulk) free energy in this case has the form \(\mathcal{F}[\Psi,\Phi]=\mathcal{F}_{SC}[\Psi]+\mathcal{F}_{VP}[\Phi]+\mathcal{F}_{I}[\Psi,\Phi]\) where \(\Phi\) is the VP order parameter, \(\mathcal{F}_{SC}\) is as in Eq. \ref{FGL}, \(\mathcal{F}_{VP}[\Phi]=a\Phi^2+b\Phi^4\), and the two field interact via \(\mathcal{F}_{I}[\Psi,\Phi]=v(q)\Phi|\Psi|^2+\dots\). By symmetry, \(v(q)\) has to be odd under \(q\) in this case, which induces an effective Lifshitz invariant to linear order in \(q\), as well as the higher order Lifshitz invariants that lead to SDE. Moreover, VP order is known to couple directly to the current with terms of the form \(cJ\Phi\) allowed in \(\mathcal{F}_{VP}\), where \(J\) can be either a normal or a supercurrent, leading to interesting backaction \cite{BanerjeeScheurer24} and nonequilibrium effects \cite{BanerjeeScheurer25, Zhuang25, DaidoYanaseLaw25} that can greatly enhance the SDE efficiency \(\eta\), up to near perfect values. A particularly dramatic example is the possibility of a supra-perfect SDE observed in experiment \cite{LinScheurerLi22} and studied theoretically in Ref. \cite{DaidoYanaseLaw25}, where it was attributed to coexistence of multiple domains of three degenerate FF-like states (as discussed in Sec. \ref{GLJDESec}, the supra-perfect effect can only take place in open systems/out of equilibrium). 

\subsection{Altermagnetism}\label{MagSec}

Time-reversal symmetry breaking, required for both SDE and JDE, is most commonly realized via magnetic fields. These fields can be externally applied or intrinsic to the system, as in materials that exhibit spontaneous magnetization. Conventional ferromagnets (FMs) are most natural for this purpose (e.g., realizing JDE in SFS junction) see Sec. \ref{NCSJDEvarSec}), but any other magnetic order such as unconventional ferromagnetism (e.g., \(p\)-wave) partially compensated ferrimagnetism, or antiferromagnetism (AFMs), can also be considered. This also includes the newly classified altermagnets (AMs) that, like AFMs, are characterized by TRSB and zero net magnetization, but, unlike AFMs, preserve both inversion and translation symmetries (under this classification, AFMs are thus restricted to systems that break translation symmetry). The net zero magnetization in altermagnets results instead from a combination of TRS with a point group symmetry, such as rotation or mirror symmetries, a possibility that has been considered in the classification of magnetic symmetry groups in the context of piezomagnetism and anitferromagnetic magnetoelectric materials \cite{TavgerZaitsev56, Dzyaloshinskii58, Dzyaloshinskii60, Eerenstein06, Mineev25}. 
AMs as translation symmetry-preserving phases of matter distinct from AFMs, however, have been recognized only recently \cite{SmejkalJungwirth22, Mazin24, RoigAgterberg24, Jiang24, Mineev24, MaedaCayao25, ParshukovSchnyder25, JungwirthFernandes25, FukayaCayao25}, and as a result have received particular attention as a source of TRSB in the context of both SDE \cite{BanerjeeScheurer24alt, SimKnolle25, ChakrabortyBlackSchaffer25, MukasaMasaki25, SamantaGhosh25, Froldi25} and JDE \cite{ChengMaoSun24,SharmaThakurathi25, Boruah25, Jiang25}, as well as AJE \cite{ChengSun24, WeiWang24} and SC phenomena more broadly \cite{KokkelerTokatlyBergeret25, FukayaCayao25}. Relatively high diode efficiencies \(\eta\sim40-50\%\) have been found for both SDE and JDE.
Rashba SOC has typically been considered as the additional source of ISB needed for realizing the bulk SDE \cite{BanerjeeScheurer24alt, MukasaMasaki25, Froldi25} and JDE \cite{Jiang25,SharmaThakurathi25}, as has Ising SOC for JDE \cite{Boruah25}, in which case the AM magnetization must have components parallel to the SOC field to produce the diode effect (as for SDE driven by an external magnetic field). Junction asymmetry can also act as a source of ISB \cite{ChengMaoSun24}. 
Alternatively, AMs have been theorized to lead to SC with finite-momentum pairing and spontaneous ISB \cite{SimKnolle25, ChakrabortyBlackSchaffer25, Froldi25}, analogous to FFLO states driven by Zeeman splitting \cite{FF,LO}, in which case SDE is realized without SOC. In the latter case, since translation symmetry is also broken by the finite-momentum paired state, it is interesting to consider whether the resulting state should be classified as an AM FFLO or an AFM FFLO state.
At the time of publication (and to the best of our knowledge), JDE with altermagnetic SCs has not been considered. AFM have also not been considered as sources of SDE or JDE, though they are known to induce AJE in presence of Rashba SOC \cite{RabinovichBobkovs19}.

Of more direct relevance to SC diode effects are a subset of noncentrosymmetric (or magnetoelectric) magnets in which the inversion symmetry is broken by the magnetic order itself. These include multiferroic materials (that are both ferromagnetic and ferroelectric) \cite{Eerenstein06}; noncollinear FMs, AFMs, and AMs like magnetic helices and chiral magnets that additionally break mirror symmetries \cite{Coey87, CheongXu22, CheongHuang24}, along with some frustrated magnets \cite{Starykh15}; as well as toroidal AFMs with a magnetic toroidal (also called anapole \cite{Zeldovich58}) moment and that preserve the product \(\mathcal{TI}\) of TRS and IS that enforces the vanishing of the net magnetization \cite{Gorbatsevich89, Dubovik90, GorbatsevichKopaev94, Spaldin08, Kopaev09, Hayami14, Pourovskii25}. 
Though systems with toroidal and other multipole magnetic orders have been theoretically found to host nonuniform SC \cite{Gorbatsevich89, SumitaYanase16, SumitaYanase17, SumitaYanase20, ChazonoYanase23},
there have been relatively few works that explored such intrinsic magnetic orders as possible mechanisms for realizing SDE and JDE: only conical FMs \cite{KamraFu24, NikolicBuzdin25} and frustrated AFMs \cite{FrazierLi25} have been proposed for realizing JDE, while anapole/toroidal AFMs had been suggested \cite{WuAgterberg24} and investigated \cite{ChazonoYanase23} as possible sources of SDE. 
Note in the latter case that the \(\mathcal{TI}\) symmetry does not reverse the current and so does not need to be broken for SDE to be realized (it also need not be preserved in the SC state).

Notably, SC in toroidal AFMs had already been studied as early as 1989 in Ref. \cite{Gorbatsevich89}, where the Lifshitz invariant and many of its consequences -- including the non-uniform SC and characteristics of AJE such as phase battery operation and spontaneous current generation in a ring geometry -- had been identified likely for the first time; no SDE had been identified, however, since higher order gradient terms were not considered.
Refs. \cite{Gorbatsevich89, GorbatsevichKopaev94} in particular studied orbital AFMs with magnetization due to orbital currents, which more recently have been investigated in the context of loop current states (sometimes also called imaginary charge density waves) \cite{Varma14, Bourges21, Fernandes25, SchultzFernandesSchmalian25}. Since such orbital magnetic states also break TRS (and in principle can also break inversion), they provide another potential mechanism for realizing SDE and JDE, but have so far received little attention in the literature  \cite{Varma25, ShenZhang25}. 

An extrinsic magnetic field can also achieve ISB and thus lead to SDE and JDE. Such SDE has been computed in a 1D helical Shiba chain (with an additional AM) \cite{BhowmikSaha25} 
and 2D conical Shiba lattice \cite{BhowmikGhosh25shiba2D} formed due to magnetic impurities. JDE, in turn, has been found in SFS junctions with domain walls or skyrmion textures in the FM regions \cite{HessLeggLossKlinovaja23,SinnerChotorlishvili24} and due to magnetization gradients \cite{Silaev14, Roig24}.
Ref. \cite{Roig24} noted in particular that since SOC acts as a momentum-dependent magnetic field, a spatially varying magnetic field can act as a synthetic SOC.
Larger efficiencies have been found in S/F/F/F/S junctions with non-collinear magnetization in the FM layers, with a maximum \(\eta\sim36\%\) \cite{GreinSchon09, MargarisFlytzanis10, PalBenjamin19, Halterman22, Schulz25I, Schulz25II} (a similar non-reciprocity was noted but not computed in an S/F/AFM/F/S junction \cite{Malshukov24}).

\subsection{Topology and Quantum Geometry}\label{TopoSec}

Given that SDE and JDE require TRSB and ISB, and that both TRSB and ISB play an important role in the classification of topological phases of matter, there are unsurprisingly many intersections between the two subjects. Presence or absence of TRS, along with particle hole symmetry (PHS) and their product, underlies the 10-fold classification of both gapped and gapless symmetry protected phases of matter \cite{ChiuRyu16, SatoAndo17, Bernevig22}. For example, TRSB is required to realize Chern insulators \cite{Haldane88} and chiral topological SCs (TSCs) \cite{ReadGreen00}, while either TRSB or ISB is necessary to realize Weyl semimetals (and both must be broken to realize Weyl SCs).
Moreover, ISB plays a crucial indirect role in many experimental proposals and realizations of topological phases, as those often rely on SOC to lift the spin degeneracy, including both 2D \cite{KaneMele05} and 3D \cite{HasanMoore11} TRS-preserving topological insulators (TIs).
SOC is also instrumental in proposed realizations of 1D TSCs with Majorana zero modes (MZMs) \cite{Kitaev01, OregRefaelvonOppen10, LutchynSauDasSarma10, CookFranz11} and chiral 2D TSCs \cite{FuKane08, Fujimoto08, ZhangDasSarma08, SauDasSarma10, WangZhang15} using heterostructures with proximitized conventional \(s\)-wave SCs, which remain a promising platform for realizing topological quantum computers \cite{NayakSimonDasSarma08, Alicea12, Beenakker13, CaiZutic23, SchielaShabani24}. 

The relation between topology and SC diode effects, however, is not straightforward since neither SDE nor JDE require topology. Topology has, nevertheless, been found to affect both in nontrivial ways.
The most direct connection has been considered for topological phases that also break TRS or IS and thus can serve as a source of the diode effects: this includes  ISB due to SOC in the helical surface states on 3D TIs, again proposed for both SDE \cite{KarabassovBobkovaVasenko22, Kotetes23, KarabassovBobkovaVasenko23} and JDE \cite{TanakaLuNagaosa22, LuTanakaNagaosa23, LiuAndreevSpivak24, KarabassovBobkovaVasenko24}; and
TRSB in chiral SCs proposed for realization of both SDE \cite{ZinklSigrist22, Varma25} and JDE \cite{ZinklSigrist22, LiuWang24, VolkovFranz24, YerinGiazotto24, VakiliKovalev24} and in a Chern insulator considered in a JDE study \cite{ShenZhang25} (in this case ISB has been assumed from extrinsic sources).
The helical states of 3D TIs can be described by the same Hamiltonian with SOC as the Rashba-Zeeman model Eq. \ref{H0}, but with \(\varepsilon(\mathbf{k})=-\mu\), with only a single helical band being occupied at a time.
Since both SDE and JDE are enhanced by the interference between different helical bands, helical surface states are thus relatively disadvantageous compared to NCSs for realizing diode effects, and consequently relatively small efficiencies \(\eta\) have been found for SDE, while additional ISB terms such as misaligned \(d\)-wave pairing \cite{TanakaLuNagaosa22} or magnetic fields \cite{KarabassovBobkovaVasenko24} have been considered for JDE. Large JDE efficiencies, \(\eta\sim40\%\), has, however, been reported in Ref. \cite{LuTanakaNagaosa23} with no additional ISB even in a short junction limit and with uniform SC (although JDE is known to vanish in this limit in the Rashba-Zeeman model, as discussed in Sec. \ref{NCSJDESec}). 
On the other hand, topology has been found to be beneficial to JDE in TSCs in Refs. \cite{LiuWang24, VakiliKovalev24}. Ref. \cite{LiuWang24} in particular studies chiral SC realized in a 3D TI with a ferromagnet proximitized to an \(s\)-wave SC on one of its surfaces, the model considered in \cite{WangZhang15} (the normal state in this case is strictly speaking a higher-order axion TI with \emph{surface} quantum anomalous Hall effect, sometimes referred to as a (3D) magnetic TI \cite{Bernevig22}). 
Ref. \cite{LiuWang24} found that \(\eta\) was nearly vanishing in the topologically trivial state despite both TRS and IS being broken, but was significant in states with Chern numbers \(C=1\) and \(C=2\). Ref. \cite{VakiliKovalev24} similarly considered JDE in a chiral \(d+id\) SC (with \(C=2\)), with a relative rotation of the order parameter between the two leads breaking IS, and studied the contribution of the chiral edge modes to JDE. Interestingly, it was found that while edge modes contribute most of the Josephson current for junctions between SCs with equal chirality, the JDE appears to originate mostly from the bulk states; on the other hand, the edge states appeared to contribute most of the JDE in JJs with SCs of opposite chirality, but the diode effect itself was much weaker than in the case of equal chirality (\(\eta\sim4\%\) vs \(\eta\sim30\%\), respectively).
We also note that many early studies of unconventional SCs with TRSB and ISB in JJs proposed skewed Fraunhofer patters as their signature in JJs including SQUIDs \cite{VanHarlingen95, AgterbergSigrist98, YangAgterberg00, KaurAgterberg03, KaurAgterberg05, KaurAgterbergSigrist05, LeridonVarma07}, but had not noted the JDE that it implies.

Relatively greater attention has been devoted to the influence of edge states on the JDE in two-dimensional topological insulators (2D TIs), namely quantum spin Hall insulators \cite{DolciniMeyerHouzet15, ChenLaw18, WeiWangWang23, WangLiu24, FracassiSassetti24, HuangVayrynen24, DingWang24, ScharfKochanMatosAbiague24, Guo25}. Similar to 3D TIs, they do not break TRS but the pairs of helical edge states on each side of the 2D TI (say top and bottom) do typically break IS locally (this is the case for the Kane-Mele model \cite{KaneMele05}, but in general inversion-symmetric TIs are also possible \cite{FuKane07, HughesProdanBernevig11}). Consequently, JDE is realized in a single pair of helical edge states when a Zeeman field is applied, as already noted in Ref. \cite{DolciniMeyerHouzet15}. However, the effect is usually opposite on opposite sides such that \(J_t(\varphi)=-J_b(-\varphi)\) (assuming equal magnetic fields on both sides), and if both contribute equally to the total current, then \(J(\varphi)=J_t(\varphi)+J_b(\varphi)=-J(-\varphi)\) and no JDE occurs. As in the case of 1D NCSs or SQUIDs, some additional interference due to an asymmetry between the top and bottom edge states -- e.g. due to different top and bottom Fermi velocities \cite{ChenLaw18, WeiWangWang23, DingWang24} or magnetic fields \cite{WangLiu24, FracassiSassetti24, HuangVayrynen24} -- is then needed to realize the JDE. Alternatively, coupling the SCs in the JJ to a single side of the 2D TI can also give rise to JDE \cite{ScharfKochanMatosAbiague24}, but it has been noted that the interference between the two sides can in fact enhance the diode efficiency \cite{WangLiu24} (from \(\eta\sim30\%\) to \(\sim70\%\)).

SDE and JDE in 1D TSCs with MZMs have also received a lot of attention, in particular realized in Rashba nanowires \cite{ KopasovMelnikov21, PekertenMatosAbiague22, LeggLossKlinovaja22, LeggLossKlinovaja23, PekertenMatosAbiagueZutic24, MonroeZutic24,  MaiellaroCitro24,  CayaoNagaosaTanaka24, CuozzoShabani24, LiuWang24, MondalCayao25, GuanAn25, Soori23II, WangLiu25} and 2D TI nanotubes \cite{LeggLossKlinovaja22}, following the proposals in Refs. \cite{OregRefaelvonOppen10, LutchynSauDasSarma10, CookFranz11} that rely on TRSB due to magnetic fields in addition to SOC that breaks IS. SDE in a 1D TSC with finite momentum pairing in a helical Shiba chain has also been considered \cite{BhowmikSaha25}. From symmetry considerations, 1D TSCs are thus natural platforms for realizing SDE, as noted in Ref. \cite{LeggLossKlinovaja22}, especially since they can be effectively described by the Rashba-Zeeman Hamiltonian in Eq. \ref{H0} in the limit \(\Delta\gg\mu\), though relatively small SDE efficiencies (up to \(\eta\sim7\%\)) have been found. Since the bulk SDE is not sensitive to edge effects, most of the focus has been on JDE and the role of MZMs in its realization \cite{KopasovMelnikov21, PekertenMatosAbiague22, LeggLossKlinovaja23, PekertenMatosAbiagueZutic24, MonroeZutic24,  MaiellaroCitro24,  CayaoNagaosaTanaka24, LiuWang24, MondalCayao25, GuanAn25, Soori23II} (Refs. \cite{KopasovMelnikov21, MaiellaroCitro24} additionally consider kinks in the JJs, which affects the JDE in a nontrivial fashion). Generally, it has been found that the efficiency \(\eta\) is peaked in the vicinity of the topological phase transition \cite{KopasovMelnikov21, CayaoNagaosaTanaka24, LiuWang24, MondalCayao25, GuanAn25}, and several works claim that JDE is maximized in the topological phase \cite{LeggLossKlinovaja23, CayaoNagaosaTanaka24, GuanAn25}, especially in the case of low transparency JJs \cite{MondalCayao25}. Ref. \cite{LiuWang24}, on the other hand, found that while JDE is sometimes enhanced in the topological phase, it is not the case for all parameter regimes. One possible source of the enhancement of JDE is the fractional (so-called \(4\pi\)-periodic) Josephson effect, one of the predicted signatures of MZMs in 1D TSCs \cite{Kitaev01, Alicea12}. This effect relies on fermionic parity conservation, in which case the CPR is doubly valued at certain \(\varphi\) and switches between the two branches as the phase winds by \(2\pi\) (hence resulting in the \(4\pi\) periodicity). The \(4\pi\)-periodic CPR has been found to enhance JDE, especially in SQUIDs, in Refs. \cite{LeggLossKlinovaja23,CuozzoShabani24,Kotetes24} (also in 2D TIs in Ref. \cite{ScharfKochanMatosAbiague24}).

SDE \cite{ChenHosur24} and JDE \cite{ChenLaw18, LiuSun25} in Weyl and Dirac semimetals have also been considered, as well as in nodal \(d\)-wave SCs \cite{DaidoYanase22_2, Aoyama24, TanakaLuNagaosa22, VakiliKovalev24}. Tilted Weyl semimetals in particular can break both TRS and IS, but otherwise some extrinsic symmetry breaking is required in this case.
Finally, beyond topological phases characterized by the Berry curvature, which is the imaginary part of the quantum geometric tensor (given by \(Q^{ij}=\langle\partial_{k_i}u|[1-|u\rangle\langle u|]|\partial_{k_j}u\rangle\) for a Bloch state \(u(\mathbf{k})\)), the importance of the quantum (Fubini-Study) metric given by the real part of the quantum geometric tensor has been recognized more recently \cite{YuBernevigTorma25}. The quantum metric is especially important in SCs in flat band systems, in which it contributes to the superfluid stiffness (corresponding to \(\alpha_2\) in Eq. \ref{FGL}) that would otherwise vanish and that is necessary to stabilize the SC state \cite{PeottaTorma15}. Consequently, the quantum metric also contributes to SDE both via various GL coefficients, including \(\alpha_2\) and \(\alpha_3\) (through the quantum metric dipole) \cite{HuLaw24}. The quantum metric was also found to contribute to the Lifshitz invariant \(\alpha_1\) \cite{DunbrackHeikkila25}. There are also likely contributions to \(\alpha_4\) and \(\beta_n\) that have not yet been reported in the literature. 

To summarize, the relationship between topology and diode effects is not clear cut: while topological phase, and in particular topological edge states, have been found to give rise to SDE and JDE due to TRSB and ISB, the diode efficiencies may or may not be very sensitive to the topology.
Nevertheless, SDE and JDE may be useful as sensors for identifying topological phases: if SC can be restricted to surfaces of topological materials, the diodicity can be used to confirm the presence of edge states via TRSB or ISB, as proposed for example for detecting chiral SC phases \cite{VanHarlingen95, AgterbergSigrist98, YangAgterberg00, KaurAgterberg03, KaurAgterberg05, KaurAgterbergSigrist05, LeridonVarma07, ZinklSigrist22} and 2D \cite{DolciniMeyerHouzet15, ChenLaw18, WeiWangWang23, WangLiu24, FracassiSassetti24, HuangVayrynen24, DingWang24, ScharfKochanMatosAbiague24, Guo25} and 3D TIs \cite{KarabassovBobkovaVasenko22, Kotetes23, KarabassovBobkovaVasenko23, TanakaLuNagaosa22, LuTanakaNagaosa23, LiuAndreevSpivak24, KarabassovBobkovaVasenko24}.
In general, other TRS and IS breaking chiral and helical edge modes that have not yet been considered, for example in fractional quantum Hall states, can also be expected to give rise to SDE and JDE, which can in turn be used to detect the edge modes.

\subsection{Nonequilibrium effects}\label{NonEqSec}

\begin{figure}[t]
\includegraphics[width=0.99\textwidth]{NonEqfig.pdf}
\caption{Nonequilibrium effects in JJs. (a) Hysteresis and non-reciprocal retrapping currents in an IVC illustrating pseudo-JDE; adopted from Ref. \cite{SteinerVonOppen23}. (b) Shapiro JDE in a JJ driven by AC current with amplitude \(J_{AC}\), resulting in non-reciprocal Shapiro steps in DC voltage \(\bar{V}\); adopted from Ref. \cite{FominovMikhailov22}. (c) Multiterminal Andreev interferometer proposed for realizing proper JDE from non-equilibrium Andreev scattering processes (indicated in purple); adopted from Ref. \cite{ShafferLiLevchenko25}. (d) Skewed Fraunhofer pattern measured in a multiterminal JJ with three SC terminals and a common normal region, adopted from Ref. \cite{GuptaPribiag23}.}
\label{NonEqfig}
\end{figure}

While the superconducting diode coefficient \(|\eta|\) is bounded in a closed system to be at most \(100\%\), this is not the case in open systems, as discussed already in Sec. \ref{GLJDEmultiSec}. As a consequence, non-equilibrium effects have been found to be generally beneficial for increasing \(|\eta|\), even above its nominally perfect value, though generally at the cost of dissipation. For example, perfect bulk SDE has been theoretically found using time-dependent GL (TDGL) models for systems with valley polarization \cite{BanerjeeScheurer25} and bilayers in perpendicular electric fields or current \cite{DaidoYanase25}. Several works considered chiral light as a source of TRSB to achieve SDE \cite{Matsyshyn25, Arora25},
but with the exception of bulk VDE and the vortex ratchet effect discussed in Sec. \ref{VDESec}, there have been few other works studying SDE out of equilibrium.

Non-equilibrium effects have been considered more widely in JJs with JDE that can be studied using RCSJ models. Such models assume that the total external current is assumed to be a sum of three terms:
\(J_{ext}=J(\varphi)+J_d(V)+J_C(\dot{V})\),
where \(J_d(V)=V/R\) and \(J_C(\dot{V})=C\dot{V}\) are the dissipative and capacitive current respectively with the dot denoting the time derivative (a fourth noise current can also be added). The current equation is complemented by Josephson's second relation, \(\dot{\varphi}=2eV\). This amounts to considering a particle moving in an effective `washboard' potential \(U(\varphi)=F(\varphi)-J_{ext}\varphi/2e\), and consequently one can consider the effective CPR to be \(J(\varphi)-J_{ext}\).
In the DC driven case one finds that in underdamped JJs (with large \(RC\))
exhibit hysteretic behavior: in addition to the critical currents \(J_{c\pm}\) (usually identified by sweeping the current up/down from zero), there is a retrapping current \(J_{r\pm}\) with \(|J_{r\pm}|<|J_{c\pm}|\) at which the JJ switched back to the SC state from the resistive state; see Fig. \ref{NonEqfig}(a). Because the retrapping currents are found in dissipative nonequilibrium states that by themselves break TRS, they may be nonreciprocal even if the ground state of the JJ does not break TRS (see Ref. \cite{ZhangJiang22}). ISB is thus enough to observe nonreciprocity of \(J_{r\pm}\), just as it is for sufficient for realizing non-SC diodes.
Indeed, the non-reciprocity of \(J_{r\pm}\) can be realized if the dissipative part of the current is itself non-reciprocal, i.e. \(J_d(-V)\neq -J_d(V)\) (which generally requires nonlinear terms, as is the case for diodes), as shown in Ref. \cite{SteinerVonOppen23}.
Such a retrapping diode effect is thus of the same physical origin as the regular diode effect, and is in that sense not a true SC diode effect sometimes referred to as a pseudo-JDE \cite{WangWangWu25}.
In earlier literature, however, the pseudo-JDE is often simply referred to as an SC diode effect \cite{HuWuDai07, MisakiNagaosa21}.

Another JDE-like effect occurs if the JJ is driven by a combination of DC and AC currents, \(J_{ext}=J_0+J_1\cos\omega t\). In that case Shapiro steps appear in IVCs, with the first Shapiro step at which the DC component of the voltage jumps from zero to a non-zero value generally occurring when the DC component of the current reaches \(|J_0|=|J_{c\pm}^{(Sh)}|<|J_{c\pm}|\). In the absence of TRS and IS, the Shapiro steps are generally nonreciprocal, and thus also exhibit a diode-like effect \(J_{c+}^{(Sh)}\neq-J_{c-}^{(Sh)}\) (see Fig. \ref{NonEqfig}(b)). Such a AC-driven Shapiro diode effect has been observed in several experiments in Refs. \cite{MendittoGoldobin16, SchmidGoldobin24, Su24, MatsuoManfra25, BorgonginoGiazotto25,Wang25} and studied theoretically in Refs. \cite{SeoaneSoutoDanon24, ScheerDanon25, BorgonginoGiazotto25, MonroeZutic24, Soori23I, Soori23II, OrtegaTaberner23, LiuAndreevSpivak24, ZazunovEgger24, TsarevFominov25, QiXie25}.
Multiterminal devices such as SQUIDs have also been studied under similar AC-driving in Refs. \cite{Zapata96, Goldobin01, Carapella01, Lee03, GolodKrasnov22, SeoaneSouto22, SeoaneSoutoDanon24, GuarcelloFilatrella24, ValentiniDanon24, CuozzoShabani24}, and a related effect in a bulk SC with SDE exposed to microwave radiation resulting in a DC current has also been considered \cite{MironovBuzdin24}.
It is generally found that the diode efficiency \(\eta_{Sh}=(J_{c+}^{(Sh)}+J_{c-}^{(Sh)})/(J_{c+}^{(Sh)}-J_{c-}^{(Sh)})\) is not bounded to \(|\eta_{Sh}|\leq 1\), since \(J_{c+}^{(Sh)}\) and \(J_{c-}^{(Sh)}\) need not have opposite signs.
Cases with \(|\eta_{Sh}| = 1\) as well as \(|\eta_{Sh}| > 1\) have both been predicted theoretically in Refs. \cite{Zapata96, OrtegaTaberner23, SeoaneSoutoDanon24, BorgonginoGiazotto25} and more recently observed in experiments in Refs. \cite{ValentiniDanon24, Su24, BorgonginoGiazotto25, Wang25}. However, it is important to emphasize that
the Shapiro diode effect only takes place in the presence of AC voltages and dissipation, and it is thus not an entirely bona fide SC diode effect.

A true JDE due to nonequilibrium effects is nevertheless possible, as shown in Ref. \cite{ShafferLiLevchenko25}, for example, in multiterminal Andreev interferometer devices shown in Fig. \ref{NonEqfig}(c). With two terminals being superconducting and the rest being in the normal state such a device can be considered as an SNS junction, as studied, for example, in Refs. \cite{Volkov95, GolubovWilhelmZaikin97, WilhelmZaikin98, VirtanenHeikkila04, Titov08, DolgirevKalenkovZaikin18, KalenkovZaikin21, Baselmans99, ShaikhaidarovVolkov00, Bezuglyi03, SunLinder24}.
Phenomenologically, the observation is based on the fact that in such a device the total current through the superconductors is given by
\(J(\varphi)=J_0(\varphi)+cJ_{ext}\), where \(J_0\) is the usual equilibrium Josephson current, \(J_{ext}\) is an external dissipative current running between the normal terminals, and \(c\) is some constant.
If \(J_{0}\) is itself reciprocal with \(J_{c0+}=-J_{c0-}=J_{c0}\), the diode efficiency is simply given by \(\eta = c J_{ext}/J_{c0}\). There is no a priori reason why \(c J_{ext}\) has to be less than \(J_{c0}\), and indeed it was found in Ref. \cite{ShafferLiLevchenko25}, based on earlier work Ref. \cite{Titov08}, that this can indeed be the case. The crucial and simple observation is that the external dissipative current \(J_{ext}\) is itself a TRSB and ISB quantity, and can thus lead to a true JDE. For the same reason, a dissipative can lead to AJE, as noted in Refs. \cite{DolgirevKalenkovZaikin18, HijanoIlicBergeret21, MarginedaCheckley23}. Another simple example of this is an SDE consisting of a regular SC in parallel with a current source \cite{TianZhang25}.

\section{Outlook}

Much current interest in superconducting diodes is driven by their potential applications in superconducting electronics, including many realizations of quantum computers \cite{Clarke08, Devoret13, Huang20, Abughanem25}. This technology is still in its infancy in terms of practical realization, despite many efforts \cite{LikharevSemenov91, Likharev12, Rasmussen21, KimNguyen25} to build superconducting computers that go back to at least the invention of the cryotron by Buck in 1956 \cite{Buck56}. One of the stumbling blocks in this development had been the absence of fast and efficient switches and memory elements, which SC diodes promise to overcome. As reviewed here, both SDE and JDE also have a long history, but the early realizations had not resulted in useful devices, likely at least in part due to their low diode efficiencies, requirements for large magnetic fields, and other limitations of extrinsic SDE and JDE mechanisms. The discovery of intrinsic SDE in novel materials may, therefore, finally lead to more practical devices. Further progress in the development of quantum materials and heterostructures is likely to result in SC diodes with even higher efficiencies. More theoretical efforts identifying novel SDE and JDE mechanisms based on new ways of breaking time-reversal and inversion symmetries beyond those reviewed here are thus also highly desirable.

There are other related effects we did not have space to review here, including: the magneto-chiral anisotropy and the accompanying second harmonic generation that presaged the discovery of intrinsic SDE \cite{Rikken01, WakatsukiNagaosa17, WakatsukiNagaosa18, HoshinoNagaosa18, DaidoYanase24, AttiasKhodas24}; Josephson LEDs \cite{Suemune06, RecherNazarovKouwenhoven10}; spin JDE \cite{Zhang23, WeiWangWang23, MaoSun24, SunLinder25, Schulz25I, Schulz25II}; thermal JDE \cite{MartinezPerezGiazotto13, TrupianoGiazotto25, DebnathSaha25}; transverse JDE and rectification effects \cite{Zeng25, Fu25} as well as the ratchet Hall effect in SDE \cite{Parafilo25}; and nonreciprocal tunneling effects in SN junctions \cite{DavydovaFu24}.
Finally, we note that there have been recent efforts to design and build circuits that incorporate SC diodes \cite{Hosur24,UpadhyayGolubev24, InglaAynesMoodera25, Castellani25}. These are only the first steps necessary to design more complicated SC circuits with SC diodes capable of (classical or quantum) computation, especially competitive with existing technologies. Much work remains to be done in that direction.

\section*{DISCLOSURE STATEMENT}
The authors are not aware of any affiliations, memberships, funding, or financial holdings that
might be perceived as affecting the objectivity of this review. 

\section*{ACKNOWLEDGMENTS}

We thank Lotan Attias, Itai Bankier, Dmitry Chichinadze, Jaglul Hasan, Manuel Houzet, Maxim Khodas, Songci Li, Tony Liu, Julia Meyer, Konstantin Nesterov, Alexander Osin, Mikhail Titov, and Zekun Zhuang for prior collaboration on the topics of anomalous Josephson effect, the superconducting diode effect, and the Josephson diode effect, which have greatly shaped our understanding of the essential physics presented in this review. We are grateful to Daniel Agterberg, Anton Andreev, Alexander Eaton, Dmitri Efetov, Gleb Finkelstein, Uri Vool, and Matthew Yankowitz for enlightening discussions. 
This work was financially supported by the National Science Foundation (NSF), Quantum Leap Challenge Institute for Hybrid Quantum Architectures and Networks Grant No. OMA-2016136 (D. S.); NSF Grant No. DMR-2452658 (A. L.) and H. I. Romnes Faculty Fellowship provided by the University of Wisconsin-Madison Office of the Vice Chancellor for Research and Graduate Education with funding from the Wisconsin Alumni Research Foundation. This review was written in part 
at Aspen Center for Physics, which is supported by the National Science Foundation grant PHY-2210452. 
%


\bibliographystyle{apsrev}
\renewcommand{\bibname}{}
\bibliography{bibliography_url}

\begin{thebibliography}{564}
\expandafter\ifx\csname natexlab\endcsname\relax\def\natexlab#1{#1}\fi
\expandafter\ifx\csname bibnamefont\endcsname\relax
  \def\bibnamefont#1{#1}\fi
\expandafter\ifx\csname bibfnamefont\endcsname\relax
  \def\bibfnamefont#1{#1}\fi
\expandafter\ifx\csname citenamefont\endcsname\relax
  \def\citenamefont#1{#1}\fi
\expandafter\ifx\csname url\endcsname\relax
  \def\url#1{\texttt{#1}}\fi
\expandafter\ifx\csname urlprefix\endcsname\relax\def\urlprefix{URL }\fi
\providecommand{\bibinfo}[2]{#2}
\providecommand{\eprint}[2][]{\url{#2}}

\bibitem[{\citenamefont{Nadeem et~al.}(2023)\citenamefont{Nadeem, Fuhrer, and
  Wang}}]{Nadeem23}
\bibinfo{author}{\bibfnamefont{M.}~\bibnamefont{Nadeem}},
  \bibinfo{author}{\bibfnamefont{M.~S.} \bibnamefont{Fuhrer}},
  \bibnamefont{and} \bibinfo{author}{\bibfnamefont{X.}~\bibnamefont{Wang}},
  \bibinfo{journal}{Nature Reviews Physics} \textbf{\bibinfo{volume}{5}},
  \bibinfo{pages}{558} (\bibinfo{year}{2023}),
  \urlprefix\url{https://www.nature.com/articles/s42254-023-00632-w}.

\bibitem[{\citenamefont{Moll and Geshkenbein}(2023)}]{MollGeshkenbein23}
\bibinfo{author}{\bibfnamefont{P.~J.~W.} \bibnamefont{Moll}} \bibnamefont{and}
  \bibinfo{author}{\bibfnamefont{V.~B.} \bibnamefont{Geshkenbein}},
  \bibinfo{journal}{Nature Physics} pp. \bibinfo{pages}{1--2}
  (\bibinfo{year}{2023}), ISSN \bibinfo{issn}{1745-2481},
  \bibinfo{note}{publisher: Nature Publishing Group},
  \urlprefix\url{https://www.nature.com/articles/s41567-023-02229-7}.

\bibitem[{\citenamefont{Nagaosa and Yanase}(2024)}]{NagaosaYanase24}
\bibinfo{author}{\bibfnamefont{N.}~\bibnamefont{Nagaosa}} \bibnamefont{and}
  \bibinfo{author}{\bibfnamefont{Y.}~\bibnamefont{Yanase}},
  \bibinfo{journal}{Annual Review of Condensed Matter Physics}
  \textbf{\bibinfo{volume}{15}}, \bibinfo{pages}{63} (\bibinfo{year}{2024}),
  ISSN \bibinfo{issn}{1947-5454, 1947-5462}, \bibinfo{note}{publisher: Annual
  Reviews},
  \urlprefix\url{https://www.annualreviews.org/content/journals/10.1146/annurev-conmatphys-032822-033734}.

\bibitem[{\citenamefont{Ma et~al.}(2025)\citenamefont{Ma, Zhan, and
  Lin}}]{Ma25}
\bibinfo{author}{\bibfnamefont{J.}~\bibnamefont{Ma}},
  \bibinfo{author}{\bibfnamefont{R.}~\bibnamefont{Zhan}}, \bibnamefont{and}
  \bibinfo{author}{\bibfnamefont{X.}~\bibnamefont{Lin}},
  \bibinfo{journal}{Advanced Physics Research} \textbf{\bibinfo{volume}{4}},
  \bibinfo{pages}{2400180} (\bibinfo{year}{2025}),
  \eprint{https://advanced.onlinelibrary.wiley.com/doi/pdf/10.1002/apxr.202400180},
  \urlprefix\url{https://advanced.onlinelibrary.wiley.com/doi/abs/10.1002/apxr.202400180}.

\bibitem[{\citenamefont{Daido et~al.}(2022)\citenamefont{Daido, Ikeda, and
  Yanase}}]{DaidoYanase22}
\bibinfo{author}{\bibfnamefont{A.}~\bibnamefont{Daido}},
  \bibinfo{author}{\bibfnamefont{Y.}~\bibnamefont{Ikeda}}, \bibnamefont{and}
  \bibinfo{author}{\bibfnamefont{Y.}~\bibnamefont{Yanase}},
  \bibinfo{journal}{Physical Review Letters} \textbf{\bibinfo{volume}{128}},
  \bibinfo{pages}{037001} (\bibinfo{year}{2022}), \bibinfo{note}{publisher:
  American Physical Society},
  \urlprefix\url{https://link.aps.org/doi/10.1103/PhysRevLett.128.037001}.

\bibitem[{\citenamefont{Levitov et~al.}(1985)\citenamefont{Levitov, Nazarov,
  and Eliashberg}}]{LevitovNazarovEliashberg85}
\bibinfo{author}{\bibfnamefont{L.~S.} \bibnamefont{Levitov}},
  \bibinfo{author}{\bibfnamefont{Y.~V.} \bibnamefont{Nazarov}},
  \bibnamefont{and} \bibinfo{author}{\bibfnamefont{G.~M.}
  \bibnamefont{Eliashberg}}, \bibinfo{journal}{JETP Lett. (Engl. Transl.);
  (United States)} \textbf{\bibinfo{volume}{41:9}} (\bibinfo{year}{1985}),
  \bibinfo{note}{institution: L. D. Landau Institute of Theoretical Physics,
  Academy of Sciences of the USSR},
  \urlprefix\url{https://www.osti.gov/biblio/5259947}.

\bibitem[{\citenamefont{Ando et~al.}(2020)\citenamefont{Ando, Miyasaka, Li,
  Ishizuka, Arakawa, Shiota, Moriyama, Yanase, and Ono}}]{AndoYanase20}
\bibinfo{author}{\bibfnamefont{F.}~\bibnamefont{Ando}},
  \bibinfo{author}{\bibfnamefont{Y.}~\bibnamefont{Miyasaka}},
  \bibinfo{author}{\bibfnamefont{T.}~\bibnamefont{Li}},
  \bibinfo{author}{\bibfnamefont{J.}~\bibnamefont{Ishizuka}},
  \bibinfo{author}{\bibfnamefont{T.}~\bibnamefont{Arakawa}},
  \bibinfo{author}{\bibfnamefont{Y.}~\bibnamefont{Shiota}},
  \bibinfo{author}{\bibfnamefont{T.}~\bibnamefont{Moriyama}},
  \bibinfo{author}{\bibfnamefont{Y.}~\bibnamefont{Yanase}}, \bibnamefont{and}
  \bibinfo{author}{\bibfnamefont{T.}~\bibnamefont{Ono}},
  \bibinfo{journal}{Nature} \textbf{\bibinfo{volume}{584}},
  \bibinfo{pages}{373} (\bibinfo{year}{2020}), ISSN \bibinfo{issn}{1476-4687},
  \bibinfo{note}{number: 7821 Publisher: Nature Publishing Group},
  \urlprefix\url{https://www.nature.com/articles/s41586-020-2590-4}.

\bibitem[{\citenamefont{Scammell et~al.}(2022)\citenamefont{Scammell, Li, and
  Scheurer}}]{ScammellScheurer22}
\bibinfo{author}{\bibfnamefont{H.~D.} \bibnamefont{Scammell}},
  \bibinfo{author}{\bibfnamefont{J.~I.~A.} \bibnamefont{Li}}, \bibnamefont{and}
  \bibinfo{author}{\bibfnamefont{M.~S.} \bibnamefont{Scheurer}},
  \bibinfo{journal}{2D Materials} \textbf{\bibinfo{volume}{9}},
  \bibinfo{pages}{025027} (\bibinfo{year}{2022}),
  \urlprefix\url{https://dx.doi.org/10.1088/2053-1583/ac5b16}.

\bibitem[{\citenamefont{Chazono et~al.}(2023)\citenamefont{Chazono, Kanasugi,
  Kitamura, and Yanase}}]{ChazonoYanase23}
\bibinfo{author}{\bibfnamefont{M.}~\bibnamefont{Chazono}},
  \bibinfo{author}{\bibfnamefont{S.}~\bibnamefont{Kanasugi}},
  \bibinfo{author}{\bibfnamefont{T.}~\bibnamefont{Kitamura}}, \bibnamefont{and}
  \bibinfo{author}{\bibfnamefont{Y.}~\bibnamefont{Yanase}},
  \bibinfo{journal}{Physical Review B} \textbf{\bibinfo{volume}{107}},
  \bibinfo{pages}{214512} (\bibinfo{year}{2023}), \bibinfo{note}{publisher:
  American Physical Society},
  \urlprefix\url{https://link.aps.org/doi/10.1103/PhysRevB.107.214512}.

\bibitem[{\citenamefont{Banerjee and
  Scheurer}(2024{\natexlab{a}})}]{BanerjeeScheurer24}
\bibinfo{author}{\bibfnamefont{S.}~\bibnamefont{Banerjee}} \bibnamefont{and}
  \bibinfo{author}{\bibfnamefont{M.~S.} \bibnamefont{Scheurer}},
  \bibinfo{journal}{Physical Review Letters} \textbf{\bibinfo{volume}{132}},
  \bibinfo{pages}{046003} (\bibinfo{year}{2024}{\natexlab{a}}),
  \bibinfo{note}{publisher: American Physical Society},
  \urlprefix\url{https://link.aps.org/doi/10.1103/PhysRevLett.132.046003}.

\bibitem[{\citenamefont{Banerjee and Scheurer}(2025)}]{BanerjeeScheurer25}
\bibinfo{author}{\bibfnamefont{S.}~\bibnamefont{Banerjee}} \bibnamefont{and}
  \bibinfo{author}{\bibfnamefont{M.~S.} \bibnamefont{Scheurer}},
  \bibinfo{journal}{Phys. Rev. Appl.} \textbf{\bibinfo{volume}{24}},
  \bibinfo{pages}{L031004} (\bibinfo{year}{2025}),
  \urlprefix\url{https://link.aps.org/doi/10.1103/n681-k9dl}.

\bibitem[{\citenamefont{Chen et~al.}(2025{\natexlab{a}})\citenamefont{Chen, Xu,
  Zhang, and Schrade}}]{ChenSchrade25}
\bibinfo{author}{\bibfnamefont{Y.}~\bibnamefont{Chen}},
  \bibinfo{author}{\bibfnamefont{C.}~\bibnamefont{Xu}},
  \bibinfo{author}{\bibfnamefont{Y.}~\bibnamefont{Zhang}}, \bibnamefont{and}
  \bibinfo{author}{\bibfnamefont{C.}~\bibnamefont{Schrade}},
  \emph{\bibinfo{title}{Finite-momentum superconductivity from chiral bands in
  twisted mote$_2$}} (\bibinfo{year}{2025}{\natexlab{a}}), \eprint{2506.18886},
  \urlprefix\url{https://arxiv.org/abs/2506.18886}.

\bibitem[{\citenamefont{Yoon et~al.}(2025)\citenamefont{Yoon, Xu, Barlas, and
  Zhang}}]{Yoon25}
\bibinfo{author}{\bibfnamefont{C.}~\bibnamefont{Yoon}},
  \bibinfo{author}{\bibfnamefont{T.}~\bibnamefont{Xu}},
  \bibinfo{author}{\bibfnamefont{Y.}~\bibnamefont{Barlas}}, \bibnamefont{and}
  \bibinfo{author}{\bibfnamefont{F.}~\bibnamefont{Zhang}}
  (\bibinfo{year}{2025}), \eprint{2502.17555},
  \urlprefix\url{https://arxiv.org/abs/2502.17555}.

\bibitem[{\citenamefont{Chen et~al.}(2025{\natexlab{b}})\citenamefont{Chen,
  Scheurer, and Schrade}}]{ChenScheurerSchrade25}
\bibinfo{author}{\bibfnamefont{Y.}~\bibnamefont{Chen}},
  \bibinfo{author}{\bibfnamefont{M.~S.} \bibnamefont{Scheurer}},
  \bibnamefont{and} \bibinfo{author}{\bibfnamefont{C.}~\bibnamefont{Schrade}},
  \bibinfo{journal}{Phys. Rev. B} \textbf{\bibinfo{volume}{112}},
  \bibinfo{pages}{L060505} (\bibinfo{year}{2025}{\natexlab{b}}),
  \urlprefix\url{https://link.aps.org/doi/10.1103/zgnk-rw1p}.

\bibitem[{\citenamefont{Daido et~al.}(2025)\citenamefont{Daido, Yanase, and
  Law}}]{DaidoYanaseLaw25}
\bibinfo{author}{\bibfnamefont{A.}~\bibnamefont{Daido}},
  \bibinfo{author}{\bibfnamefont{Y.}~\bibnamefont{Yanase}}, \bibnamefont{and}
  \bibinfo{author}{\bibfnamefont{K.~T.} \bibnamefont{Law}}
  (\bibinfo{year}{2025}), \eprint{2503.16923},
  \urlprefix\url{https://arxiv.org/abs/2503.16923}.

\bibitem[{\citenamefont{Samanta and Ghosh}(2025)}]{SamantaGhosh25}
\bibinfo{author}{\bibfnamefont{D.}~\bibnamefont{Samanta}} \bibnamefont{and}
  \bibinfo{author}{\bibfnamefont{S.~K.} \bibnamefont{Ghosh}}
  (\bibinfo{year}{2025}), \eprint{2507.21446},
  \urlprefix\url{https://arxiv.org/abs/2507.21446}.

\bibitem[{\citenamefont{Bhowmik
  et~al.}(2025{\natexlab{a}})\citenamefont{Bhowmik, Samanta, Nandy, Saha, and
  Ghosh}}]{BhowmikGhosh25}
\bibinfo{author}{\bibfnamefont{S.}~\bibnamefont{Bhowmik}},
  \bibinfo{author}{\bibfnamefont{D.}~\bibnamefont{Samanta}},
  \bibinfo{author}{\bibfnamefont{A.~K.} \bibnamefont{Nandy}},
  \bibinfo{author}{\bibfnamefont{A.}~\bibnamefont{Saha}}, \bibnamefont{and}
  \bibinfo{author}{\bibfnamefont{S.~K.} \bibnamefont{Ghosh}},
  \bibinfo{journal}{Communications Physics} \textbf{\bibinfo{volume}{8}},
  \bibinfo{pages}{260} (\bibinfo{year}{2025}{\natexlab{a}}),
  \urlprefix\url{https://doi.org/10.1038/s42005-025-02044-x}.

\bibitem[{\citenamefont{Lin et~al.}(2022)\citenamefont{Lin, Siriviboon,
  Scammell, Liu, Rhodes, Watanabe, Taniguchi, Hone, Scheurer, and
  Li}}]{LinScheurerLi22}
\bibinfo{author}{\bibfnamefont{J.-X.} \bibnamefont{Lin}},
  \bibinfo{author}{\bibfnamefont{P.}~\bibnamefont{Siriviboon}},
  \bibinfo{author}{\bibfnamefont{H.~D.} \bibnamefont{Scammell}},
  \bibinfo{author}{\bibfnamefont{S.}~\bibnamefont{Liu}},
  \bibinfo{author}{\bibfnamefont{D.}~\bibnamefont{Rhodes}},
  \bibinfo{author}{\bibfnamefont{K.}~\bibnamefont{Watanabe}},
  \bibinfo{author}{\bibfnamefont{T.}~\bibnamefont{Taniguchi}},
  \bibinfo{author}{\bibfnamefont{J.}~\bibnamefont{Hone}},
  \bibinfo{author}{\bibfnamefont{M.~S.} \bibnamefont{Scheurer}},
  \bibnamefont{and} \bibinfo{author}{\bibfnamefont{J.~I.~A.} \bibnamefont{Li}},
  \bibinfo{journal}{Nature Physics} \textbf{\bibinfo{volume}{18}},
  \bibinfo{pages}{1221} (\bibinfo{year}{2022}),
  \urlprefix\url{https://doi.org/10.1038/s41567-022-01700-1}.

\bibitem[{\citenamefont{Narita et~al.}(2022)\citenamefont{Narita, Ishizuka,
  Kawarazaki, Kan, Shiota, Moriyama, Shimakawa, Ognev, Samardak, Yanase
  et~al.}}]{NaritaYanase22}
\bibinfo{author}{\bibfnamefont{H.}~\bibnamefont{Narita}},
  \bibinfo{author}{\bibfnamefont{J.}~\bibnamefont{Ishizuka}},
  \bibinfo{author}{\bibfnamefont{R.}~\bibnamefont{Kawarazaki}},
  \bibinfo{author}{\bibfnamefont{D.}~\bibnamefont{Kan}},
  \bibinfo{author}{\bibfnamefont{Y.}~\bibnamefont{Shiota}},
  \bibinfo{author}{\bibfnamefont{T.}~\bibnamefont{Moriyama}},
  \bibinfo{author}{\bibfnamefont{Y.}~\bibnamefont{Shimakawa}},
  \bibinfo{author}{\bibfnamefont{A.~V.} \bibnamefont{Ognev}},
  \bibinfo{author}{\bibfnamefont{A.~S.} \bibnamefont{Samardak}},
  \bibinfo{author}{\bibfnamefont{Y.}~\bibnamefont{Yanase}},
  \bibnamefont{et~al.}, \bibinfo{journal}{Nature Nanotechnology}
  \textbf{\bibinfo{volume}{17}}, \bibinfo{pages}{823} (\bibinfo{year}{2022}),
  \urlprefix\url{https://doi.org/10.1038/s41565-022-01159-4}.

\bibitem[{\citenamefont{Qi et~al.}(2025{\natexlab{a}})\citenamefont{Qi, Ge, Ji,
  Ai, Ma, Wang, Cui, Liu, Wang, and Wang}}]{Qi25}
\bibinfo{author}{\bibfnamefont{S.}~\bibnamefont{Qi}},
  \bibinfo{author}{\bibfnamefont{J.}~\bibnamefont{Ge}},
  \bibinfo{author}{\bibfnamefont{C.}~\bibnamefont{Ji}},
  \bibinfo{author}{\bibfnamefont{Y.}~\bibnamefont{Ai}},
  \bibinfo{author}{\bibfnamefont{G.}~\bibnamefont{Ma}},
  \bibinfo{author}{\bibfnamefont{Z.}~\bibnamefont{Wang}},
  \bibinfo{author}{\bibfnamefont{Z.}~\bibnamefont{Cui}},
  \bibinfo{author}{\bibfnamefont{Y.}~\bibnamefont{Liu}},
  \bibinfo{author}{\bibfnamefont{Z.}~\bibnamefont{Wang}}, \bibnamefont{and}
  \bibinfo{author}{\bibfnamefont{J.}~\bibnamefont{Wang}},
  \bibinfo{journal}{Nature Communications} \textbf{\bibinfo{volume}{16}},
  \bibinfo{pages}{531} (\bibinfo{year}{2025}{\natexlab{a}}), ISSN
  \bibinfo{issn}{2041-1723}, \bibinfo{note}{publisher: Nature Publishing
  Group}, \urlprefix\url{https://www.nature.com/articles/s41467-025-55880-4}.

\bibitem[{\citenamefont{Nagata et~al.}(2025)\citenamefont{Nagata, Aoki, Daido,
  Kasahara, Kasahara, Ohshima, Ando, Yanase, Matsuda, and
  Shiraishi}}]{NagataYanase25}
\bibinfo{author}{\bibfnamefont{U.}~\bibnamefont{Nagata}},
  \bibinfo{author}{\bibfnamefont{M.}~\bibnamefont{Aoki}},
  \bibinfo{author}{\bibfnamefont{A.}~\bibnamefont{Daido}},
  \bibinfo{author}{\bibfnamefont{S.}~\bibnamefont{Kasahara}},
  \bibinfo{author}{\bibfnamefont{Y.}~\bibnamefont{Kasahara}},
  \bibinfo{author}{\bibfnamefont{R.}~\bibnamefont{Ohshima}},
  \bibinfo{author}{\bibfnamefont{Y.}~\bibnamefont{Ando}},
  \bibinfo{author}{\bibfnamefont{Y.}~\bibnamefont{Yanase}},
  \bibinfo{author}{\bibfnamefont{Y.}~\bibnamefont{Matsuda}}, \bibnamefont{and}
  \bibinfo{author}{\bibfnamefont{M.}~\bibnamefont{Shiraishi}},
  \bibinfo{journal}{Phys. Rev. Lett.} \textbf{\bibinfo{volume}{134}},
  \bibinfo{pages}{236703} (\bibinfo{year}{2025}),
  \urlprefix\url{https://link.aps.org/doi/10.1103/PhysRevLett.134.236703}.

\bibitem[{\citenamefont{Zinkl et~al.}(2022)\citenamefont{Zinkl, Hamamoto, and
  Sigrist}}]{ZinklSigrist22}
\bibinfo{author}{\bibfnamefont{B.}~\bibnamefont{Zinkl}},
  \bibinfo{author}{\bibfnamefont{K.}~\bibnamefont{Hamamoto}}, \bibnamefont{and}
  \bibinfo{author}{\bibfnamefont{M.}~\bibnamefont{Sigrist}},
  \bibinfo{journal}{Physical Review Research} \textbf{\bibinfo{volume}{4}},
  \bibinfo{pages}{033167} (\bibinfo{year}{2022}), \bibinfo{note}{publisher:
  American Physical Society},
  \urlprefix\url{https://link.aps.org/doi/10.1103/PhysRevResearch.4.033167}.

\bibitem[{\citenamefont{Edwards and Newhouse}(1962)}]{EdwardsNewhouse62}
\bibinfo{author}{\bibfnamefont{H.~H.} \bibnamefont{Edwards}} \bibnamefont{and}
  \bibinfo{author}{\bibfnamefont{V.~L.} \bibnamefont{Newhouse}},
  \bibinfo{journal}{Journal of Applied Physics} \textbf{\bibinfo{volume}{33}},
  \bibinfo{pages}{868} (\bibinfo{year}{1962}), ISSN \bibinfo{issn}{0021-8979},
  \urlprefix\url{https://doi.org/10.1063/1.1777183}.

\bibitem[{\citenamefont{Swartz and Hart}(1967{\natexlab{a}})}]{SwartzHart65}
\bibinfo{author}{\bibfnamefont{P.~S.} \bibnamefont{Swartz}} \bibnamefont{and}
  \bibinfo{author}{\bibfnamefont{H.~R.} \bibnamefont{Hart}},
  \bibinfo{journal}{Phys. Rev.} \textbf{\bibinfo{volume}{156}},
  \bibinfo{pages}{412} (\bibinfo{year}{1967}{\natexlab{a}}),
  \urlprefix\url{https://link.aps.org/doi/10.1103/PhysRev.156.412}.

\bibitem[{\citenamefont{De~Waele et~al.}(1967)\citenamefont{De~Waele, Kraan,
  De~Bruyn~Ouboter, and Taconis}}]{DeWaele67}
\bibinfo{author}{\bibfnamefont{A.~T. A.~M.} \bibnamefont{De~Waele}},
  \bibinfo{author}{\bibfnamefont{W.~H.} \bibnamefont{Kraan}},
  \bibinfo{author}{\bibfnamefont{R.}~\bibnamefont{De~Bruyn~Ouboter}},
  \bibnamefont{and} \bibinfo{author}{\bibfnamefont{K.~W.}
  \bibnamefont{Taconis}}, \bibinfo{journal}{Physica}
  \textbf{\bibinfo{volume}{37}}, \bibinfo{pages}{114} (\bibinfo{year}{1967}),
  ISSN \bibinfo{issn}{0031-8914},
  \urlprefix\url{https://www.sciencedirect.com/science/article/pii/0031891467901103}.

\bibitem[{\citenamefont{Goldman and Kreisman}(1967{\natexlab{a}})}]{Goldman67}
\bibinfo{author}{\bibfnamefont{A.~M.} \bibnamefont{Goldman}} \bibnamefont{and}
  \bibinfo{author}{\bibfnamefont{P.~J.} \bibnamefont{Kreisman}},
  \bibinfo{journal}{Physical Review} \textbf{\bibinfo{volume}{164}},
  \bibinfo{pages}{544} (\bibinfo{year}{1967}{\natexlab{a}}),
  \bibinfo{note}{publisher: American Physical Society},
  \urlprefix\url{https://link.aps.org/doi/10.1103/PhysRev.164.544}.

\bibitem[{\citenamefont{Wang et~al.}(2025{\natexlab{a}})\citenamefont{Wang,
  Wang, and Wu}}]{WangWangWu25}
\bibinfo{author}{\bibfnamefont{D.}~\bibnamefont{Wang}},
  \bibinfo{author}{\bibfnamefont{Q.-H.} \bibnamefont{Wang}}, \bibnamefont{and}
  \bibinfo{author}{\bibfnamefont{C.}~\bibnamefont{Wu}},
  \emph{\bibinfo{title}{Josephson diode effect: a phenomenological
  perspective}} (\bibinfo{year}{2025}{\natexlab{a}}), \eprint{2506.23200},
  \urlprefix\url{https://arxiv.org/abs/2506.23200}.

\bibitem[{\citenamefont{Fulde and Ferrell}(1964)}]{FF}
\bibinfo{author}{\bibfnamefont{P.}~\bibnamefont{Fulde}} \bibnamefont{and}
  \bibinfo{author}{\bibfnamefont{R.~A.} \bibnamefont{Ferrell}},
  \bibinfo{journal}{Phys. Rev.} \textbf{\bibinfo{volume}{135}},
  \bibinfo{pages}{A550} (\bibinfo{year}{1964}),
  \urlprefix\url{https://link.aps.org/doi/10.1103/PhysRev.135.A550}.

\bibitem[{\citenamefont{Larkin}(1965)}]{LO}
\bibinfo{author}{\bibfnamefont{A.~I.} \bibnamefont{Larkin}},
  \bibinfo{journal}{Sov. Phys. JETP} \textbf{\bibinfo{volume}{20}},
  \bibinfo{pages}{762} (\bibinfo{year}{1965}).

\bibitem[{\citenamefont{Agterberg et~al.}(2020)\citenamefont{Agterberg, Davis,
  Edkins, Fradkin, Van~Harlingen, Kivelson, Lee, Radzihovsky, Tranquada, and
  Wang}}]{Agterberg20}
\bibinfo{author}{\bibfnamefont{D.~F.} \bibnamefont{Agterberg}},
  \bibinfo{author}{\bibfnamefont{J.~S.} \bibnamefont{Davis}},
  \bibinfo{author}{\bibfnamefont{S.~D.} \bibnamefont{Edkins}},
  \bibinfo{author}{\bibfnamefont{E.}~\bibnamefont{Fradkin}},
  \bibinfo{author}{\bibfnamefont{D.~J.} \bibnamefont{Van~Harlingen}},
  \bibinfo{author}{\bibfnamefont{S.~A.} \bibnamefont{Kivelson}},
  \bibinfo{author}{\bibfnamefont{P.~A.} \bibnamefont{Lee}},
  \bibinfo{author}{\bibfnamefont{L.}~\bibnamefont{Radzihovsky}},
  \bibinfo{author}{\bibfnamefont{J.~M.} \bibnamefont{Tranquada}},
  \bibnamefont{and} \bibinfo{author}{\bibfnamefont{Y.}~\bibnamefont{Wang}},
  \bibinfo{journal}{Annual Review of Condensed Matter Physics}
  \textbf{\bibinfo{volume}{11}}, \bibinfo{pages}{231} (\bibinfo{year}{2020}),
  ISSN \bibinfo{issn}{1947-5462},
  \urlprefix\url{https://www.annualreviews.org/content/journals/10.1146/annurev-conmatphys-031119-050711}.

\bibitem[{\citenamefont{Geshkenbein and Larkin}(1986)}]{GeshkenbeinLarkin86}
\bibinfo{author}{\bibfnamefont{V.}~\bibnamefont{Geshkenbein}} \bibnamefont{and}
  \bibinfo{author}{\bibfnamefont{A.}~\bibnamefont{Larkin}},
  \bibinfo{journal}{JETP Lett.} \textbf{\bibinfo{volume}{43}}
  (\bibinfo{year}{1986}).

\bibitem[{\citenamefont{Sickinger et~al.}(2012)\citenamefont{Sickinger, Lipman,
  Weides, Mints, Kohlstedt, Koelle, Kleiner, and
  Goldobin}}]{SickingerGoldobin12}
\bibinfo{author}{\bibfnamefont{H.}~\bibnamefont{Sickinger}},
  \bibinfo{author}{\bibfnamefont{A.}~\bibnamefont{Lipman}},
  \bibinfo{author}{\bibfnamefont{M.}~\bibnamefont{Weides}},
  \bibinfo{author}{\bibfnamefont{R.~G.} \bibnamefont{Mints}},
  \bibinfo{author}{\bibfnamefont{H.}~\bibnamefont{Kohlstedt}},
  \bibinfo{author}{\bibfnamefont{D.}~\bibnamefont{Koelle}},
  \bibinfo{author}{\bibfnamefont{R.}~\bibnamefont{Kleiner}}, \bibnamefont{and}
  \bibinfo{author}{\bibfnamefont{E.}~\bibnamefont{Goldobin}},
  \bibinfo{journal}{Phys. Rev. Lett.} \textbf{\bibinfo{volume}{109}},
  \bibinfo{pages}{107002} (\bibinfo{year}{2012}),
  \urlprefix\url{https://link.aps.org/doi/10.1103/PhysRevLett.109.107002}.

\bibitem[{\citenamefont{Szombati et~al.}(2016)\citenamefont{Szombati,
  Nadj-Perge, Car, Plissard, Bakkers, and Kouwenhoven}}]{SzombatiKouwenhoven16}
\bibinfo{author}{\bibfnamefont{D.~B.} \bibnamefont{Szombati}},
  \bibinfo{author}{\bibfnamefont{S.}~\bibnamefont{Nadj-Perge}},
  \bibinfo{author}{\bibfnamefont{D.}~\bibnamefont{Car}},
  \bibinfo{author}{\bibfnamefont{S.~R.} \bibnamefont{Plissard}},
  \bibinfo{author}{\bibfnamefont{E.~P. a.~M.} \bibnamefont{Bakkers}},
  \bibnamefont{and} \bibinfo{author}{\bibfnamefont{L.~P.}
  \bibnamefont{Kouwenhoven}}, \bibinfo{journal}{Nature Physics}
  \textbf{\bibinfo{volume}{12}}, \bibinfo{pages}{568} (\bibinfo{year}{2016}),
  ISSN \bibinfo{issn}{1745-2481}, \bibinfo{note}{number: 6 Publisher: Nature
  Publishing Group}, \urlprefix\url{https://www.nature.com/articles/nphys3742}.

\bibitem[{\citenamefont{Buzdin}(2005)}]{Buzdin05}
\bibinfo{author}{\bibfnamefont{A.~I.} \bibnamefont{Buzdin}},
  \bibinfo{journal}{Rev. Mod. Phys.} \textbf{\bibinfo{volume}{77}},
  \bibinfo{pages}{935} (\bibinfo{year}{2005}),
  \urlprefix\url{https://link.aps.org/doi/10.1103/RevModPhys.77.935}.

\bibitem[{\citenamefont{Shukrinov}(2022)}]{Shukrinov22}
\bibinfo{author}{\bibfnamefont{Y.~M.} \bibnamefont{Shukrinov}},
  \bibinfo{journal}{Physics-Uspekhi} \textbf{\bibinfo{volume}{65}},
  \bibinfo{pages}{317} (\bibinfo{year}{2022}),
  \urlprefix\url{https://doi.org/10.3367/UFNe.2020.11.038894}.

\bibitem[{\citenamefont{Edelstein}(1996)}]{Edelstein96}
\bibinfo{author}{\bibfnamefont{V.~M.} \bibnamefont{Edelstein}},
  \bibinfo{journal}{Journal of Physics: Condensed Matter}
  \textbf{\bibinfo{volume}{8}}, \bibinfo{pages}{339} (\bibinfo{year}{1996}),
  \urlprefix\url{https://dx.doi.org/10.1088/0953-8984/8/3/012}.

\bibitem[{\citenamefont{Swartz and Hart}(1967{\natexlab{b}})}]{SwartzHart67}
\bibinfo{author}{\bibfnamefont{P.~S.} \bibnamefont{Swartz}} \bibnamefont{and}
  \bibinfo{author}{\bibfnamefont{H.~R.} \bibnamefont{Hart}},
  \bibinfo{journal}{Phys. Rev.} \textbf{\bibinfo{volume}{156}},
  \bibinfo{pages}{412} (\bibinfo{year}{1967}{\natexlab{b}}),
  \urlprefix\url{https://link.aps.org/doi/10.1103/PhysRev.156.412}.

\bibitem[{\citenamefont{Saint-James and Gennes}(1963)}]{SaintJamesDeGennes63}
\bibinfo{author}{\bibfnamefont{D.}~\bibnamefont{Saint-James}} \bibnamefont{and}
  \bibinfo{author}{\bibfnamefont{P.}~\bibnamefont{Gennes}},
  \bibinfo{journal}{Physics Letters} \textbf{\bibinfo{volume}{7}},
  \bibinfo{pages}{306} (\bibinfo{year}{1963}), ISSN \bibinfo{issn}{0031-9163},
  \urlprefix\url{https://www.sciencedirect.com/science/article/pii/0031916363900477}.

\bibitem[{\citenamefont{Tinkham}(2004)}]{Tinkham04}
\bibinfo{author}{\bibfnamefont{M.}~\bibnamefont{Tinkham}},
  \emph{\bibinfo{title}{Introduction to superconductivity}}
  (\bibinfo{publisher}{Courier Corporation}, \bibinfo{year}{2004}).

\bibitem[{\citenamefont{Josephson}(1962)}]{josephson_possible_1962}
\bibinfo{author}{\bibfnamefont{B.~D.} \bibnamefont{Josephson}},
  \bibinfo{journal}{Physics Letters} \textbf{\bibinfo{volume}{1}},
  \bibinfo{pages}{251} (\bibinfo{year}{1962}), ISSN \bibinfo{issn}{0031-9163},
  \urlprefix\url{https://www.sciencedirect.com/science/article/pii/0031916362913690}.

\bibitem[{\citenamefont{Little and Parks}(1962)}]{LittleParks62}
\bibinfo{author}{\bibfnamefont{W.~A.} \bibnamefont{Little}} \bibnamefont{and}
  \bibinfo{author}{\bibfnamefont{R.~D.} \bibnamefont{Parks}},
  \bibinfo{journal}{Phys. Rev. Lett.} \textbf{\bibinfo{volume}{9}},
  \bibinfo{pages}{9} (\bibinfo{year}{1962}),
  \urlprefix\url{https://link.aps.org/doi/10.1103/PhysRevLett.9.9}.

\bibitem[{\citenamefont{Vodolazov and Peeters}(2005)}]{VodolazovPeeters05}
\bibinfo{author}{\bibfnamefont{D.~Y.} \bibnamefont{Vodolazov}}
  \bibnamefont{and} \bibinfo{author}{\bibfnamefont{F.~M.}
  \bibnamefont{Peeters}}, \bibinfo{journal}{Phys. Rev. B}
  \textbf{\bibinfo{volume}{72}}, \bibinfo{pages}{172508}
  (\bibinfo{year}{2005}),
  \urlprefix\url{https://link.aps.org/doi/10.1103/PhysRevB.72.172508}.

\bibitem[{\citenamefont{Bean and Livingston}(1964)}]{BeanLivingston64}
\bibinfo{author}{\bibfnamefont{C.~P.} \bibnamefont{Bean}} \bibnamefont{and}
  \bibinfo{author}{\bibfnamefont{J.~D.} \bibnamefont{Livingston}},
  \bibinfo{journal}{Phys. Rev. Lett.} \textbf{\bibinfo{volume}{12}},
  \bibinfo{pages}{14} (\bibinfo{year}{1964}),
  \urlprefix\url{https://link.aps.org/doi/10.1103/PhysRevLett.12.14}.

\bibitem[{\citenamefont{Bardeen and Stephen}(1965)}]{BardeenStephen65}
\bibinfo{author}{\bibfnamefont{J.}~\bibnamefont{Bardeen}} \bibnamefont{and}
  \bibinfo{author}{\bibfnamefont{M.~J.} \bibnamefont{Stephen}},
  \bibinfo{journal}{Phys. Rev.} \textbf{\bibinfo{volume}{140}},
  \bibinfo{pages}{A1197} (\bibinfo{year}{1965}),
  \urlprefix\url{https://link.aps.org/doi/10.1103/PhysRev.140.A1197}.

\bibitem[{\citenamefont{Kim et~al.}(1965)\citenamefont{Kim, Hempstead, and
  Strnad}}]{Kim65}
\bibinfo{author}{\bibfnamefont{Y.~B.} \bibnamefont{Kim}},
  \bibinfo{author}{\bibfnamefont{C.~F.} \bibnamefont{Hempstead}},
  \bibnamefont{and} \bibinfo{author}{\bibfnamefont{A.~R.}
  \bibnamefont{Strnad}}, \bibinfo{journal}{Phys. Rev.}
  \textbf{\bibinfo{volume}{139}}, \bibinfo{pages}{A1163}
  (\bibinfo{year}{1965}),
  \urlprefix\url{https://link.aps.org/doi/10.1103/PhysRev.139.A1163}.

\bibitem[{\citenamefont{Cribier et~al.}(1967)\citenamefont{Cribier, Jacrot,
  Rao, and Farnoux}}]{Cribier67}
\bibinfo{author}{\bibfnamefont{D.}~\bibnamefont{Cribier}},
  \bibinfo{author}{\bibfnamefont{B.}~\bibnamefont{Jacrot}},
  \bibinfo{author}{\bibfnamefont{L.~M.} \bibnamefont{Rao}}, \bibnamefont{and}
  \bibinfo{author}{\bibfnamefont{B.}~\bibnamefont{Farnoux}}
  (\bibinfo{publisher}{Elsevier}, \bibinfo{year}{1967}),
  vol.~\bibinfo{volume}{5} of \emph{\bibinfo{series}{Progress in Low
  Temperature Physics}}, pp. \bibinfo{pages}{161--180},
  \urlprefix\url{https://www.sciencedirect.com/science/article/pii/S0079641708601225}.

\bibitem[{\citenamefont{Essmann and Träuble}(1967)}]{Essmann67}
\bibinfo{author}{\bibfnamefont{U.}~\bibnamefont{Essmann}} \bibnamefont{and}
  \bibinfo{author}{\bibfnamefont{H.}~\bibnamefont{Träuble}},
  \bibinfo{journal}{Physics Letters A} \textbf{\bibinfo{volume}{24}},
  \bibinfo{pages}{526} (\bibinfo{year}{1967}), ISSN \bibinfo{issn}{0375-9601},
  \urlprefix\url{https://www.sciencedirect.com/science/article/pii/0375960167908195}.

\bibitem[{\citenamefont{Abrikosov}(2004)}]{Abrikosov04}
\bibinfo{author}{\bibfnamefont{A.~A.} \bibnamefont{Abrikosov}},
  \bibinfo{journal}{Rev. Mod. Phys.} \textbf{\bibinfo{volume}{76}},
  \bibinfo{pages}{975} (\bibinfo{year}{2004}),
  \urlprefix\url{https://link.aps.org/doi/10.1103/RevModPhys.76.975}.

\bibitem[{\citenamefont{Miller and Cody}(1968)}]{Miller68}
\bibinfo{author}{\bibfnamefont{R.~E.} \bibnamefont{Miller}} \bibnamefont{and}
  \bibinfo{author}{\bibfnamefont{G.~D.} \bibnamefont{Cody}},
  \bibinfo{journal}{Phys. Rev.} \textbf{\bibinfo{volume}{173}},
  \bibinfo{pages}{494} (\bibinfo{year}{1968}),
  \urlprefix\url{https://link.aps.org/doi/10.1103/PhysRev.173.494}.

\bibitem[{\citenamefont{De~Waele and de~Bruyn~Ouboter}(1969)}]{DeWaele69}
\bibinfo{author}{\bibfnamefont{A.~T.~A.} \bibnamefont{De~Waele}}
  \bibnamefont{and}
  \bibinfo{author}{\bibfnamefont{R.}~\bibnamefont{de~Bruyn~Ouboter}},
  \bibinfo{journal}{Physica} \textbf{\bibinfo{volume}{41}},
  \bibinfo{pages}{225} (\bibinfo{year}{1969}).

\bibitem[{\citenamefont{Goldman and
  Kreisman}(1967{\natexlab{b}})}]{GoldmanKreisman67}
\bibinfo{author}{\bibfnamefont{A.~M.} \bibnamefont{Goldman}} \bibnamefont{and}
  \bibinfo{author}{\bibfnamefont{P.~J.} \bibnamefont{Kreisman}},
  \bibinfo{journal}{Physical Review} \textbf{\bibinfo{volume}{164}},
  \bibinfo{pages}{544} (\bibinfo{year}{1967}{\natexlab{b}}),
  \bibinfo{note}{publisher: American Physical Society},
  \urlprefix\url{https://link.aps.org/doi/10.1103/PhysRev.164.544}.

\bibitem[{\citenamefont{Golubov et~al.}(2004)\citenamefont{Golubov, Kupriyanov,
  and Il'ichev}}]{GolubovKupriyanov04}
\bibinfo{author}{\bibfnamefont{A.~A.} \bibnamefont{Golubov}},
  \bibinfo{author}{\bibfnamefont{M.~Y.} \bibnamefont{Kupriyanov}},
  \bibnamefont{and} \bibinfo{author}{\bibfnamefont{E.}~\bibnamefont{Il'ichev}},
  \bibinfo{journal}{Rev. Mod. Phys.} \textbf{\bibinfo{volume}{76}},
  \bibinfo{pages}{411} (\bibinfo{year}{2004}),
  \urlprefix\url{https://link.aps.org/doi/10.1103/RevModPhys.76.411}.

\bibitem[{\citenamefont{Zazunov et~al.}(2024)\citenamefont{Zazunov, Rech,
  Jonckheere, Gr\'emaud, Martin, and Egger}}]{ZazunovEgger24}
\bibinfo{author}{\bibfnamefont{A.}~\bibnamefont{Zazunov}},
  \bibinfo{author}{\bibfnamefont{J.}~\bibnamefont{Rech}},
  \bibinfo{author}{\bibfnamefont{T.}~\bibnamefont{Jonckheere}},
  \bibinfo{author}{\bibfnamefont{B.}~\bibnamefont{Gr\'emaud}},
  \bibinfo{author}{\bibfnamefont{T.}~\bibnamefont{Martin}}, \bibnamefont{and}
  \bibinfo{author}{\bibfnamefont{R.}~\bibnamefont{Egger}},
  \bibinfo{journal}{Phys. Rev. B} \textbf{\bibinfo{volume}{109}},
  \bibinfo{pages}{024504} (\bibinfo{year}{2024}),
  \urlprefix\url{https://link.aps.org/doi/10.1103/PhysRevB.109.024504}.

\bibitem[{\citenamefont{Fominov and Mikhailov}(2022)}]{FominovMikhailov22}
\bibinfo{author}{\bibfnamefont{Y.~V.} \bibnamefont{Fominov}} \bibnamefont{and}
  \bibinfo{author}{\bibfnamefont{D.~S.} \bibnamefont{Mikhailov}},
  \bibinfo{journal}{Phys. Rev. B} \textbf{\bibinfo{volume}{106}},
  \bibinfo{pages}{134514} (\bibinfo{year}{2022}),
  \urlprefix\url{https://link.aps.org/doi/10.1103/PhysRevB.106.134514}.

\bibitem[{\citenamefont{Jeon et~al.}(2022)\citenamefont{Jeon, Kim, Yoon, Jeon,
  Han, Cottet, Kontos, and Parkin}}]{JeonParkin22}
\bibinfo{author}{\bibfnamefont{K.-R.} \bibnamefont{Jeon}},
  \bibinfo{author}{\bibfnamefont{J.-K.} \bibnamefont{Kim}},
  \bibinfo{author}{\bibfnamefont{J.}~\bibnamefont{Yoon}},
  \bibinfo{author}{\bibfnamefont{J.-C.} \bibnamefont{Jeon}},
  \bibinfo{author}{\bibfnamefont{H.}~\bibnamefont{Han}},
  \bibinfo{author}{\bibfnamefont{A.}~\bibnamefont{Cottet}},
  \bibinfo{author}{\bibfnamefont{T.}~\bibnamefont{Kontos}}, \bibnamefont{and}
  \bibinfo{author}{\bibfnamefont{S.~S.} \bibnamefont{Parkin}},
  \bibinfo{journal}{Nature Materials} \textbf{\bibinfo{volume}{21}},
  \bibinfo{pages}{1008} (\bibinfo{year}{2022}).

\bibitem[{\citenamefont{Zhao et~al.}(2023{\natexlab{a}})\citenamefont{Zhao,
  Cui, Volkov, Yoo, Lee, Gardener, Akey, Engelke, Ronen, Zhong
  et~al.}}]{ZhaoKim23}
\bibinfo{author}{\bibfnamefont{S.~F.} \bibnamefont{Zhao}},
  \bibinfo{author}{\bibfnamefont{X.}~\bibnamefont{Cui}},
  \bibinfo{author}{\bibfnamefont{P.~A.} \bibnamefont{Volkov}},
  \bibinfo{author}{\bibfnamefont{H.}~\bibnamefont{Yoo}},
  \bibinfo{author}{\bibfnamefont{S.}~\bibnamefont{Lee}},
  \bibinfo{author}{\bibfnamefont{J.~A.} \bibnamefont{Gardener}},
  \bibinfo{author}{\bibfnamefont{A.~J.} \bibnamefont{Akey}},
  \bibinfo{author}{\bibfnamefont{R.}~\bibnamefont{Engelke}},
  \bibinfo{author}{\bibfnamefont{Y.}~\bibnamefont{Ronen}},
  \bibinfo{author}{\bibfnamefont{R.}~\bibnamefont{Zhong}},
  \bibnamefont{et~al.}, \bibinfo{journal}{Science}
  \textbf{\bibinfo{volume}{382}}, \bibinfo{pages}{1422}
  (\bibinfo{year}{2023}{\natexlab{a}}).

\bibitem[{\citenamefont{Zhu et~al.}(2023)\citenamefont{Zhu, Wang, Wang, Hu, Gu,
  Zhu, Zhang, and Xue}}]{Zhu23}
\bibinfo{author}{\bibfnamefont{Y.}~\bibnamefont{Zhu}},
  \bibinfo{author}{\bibfnamefont{H.}~\bibnamefont{Wang}},
  \bibinfo{author}{\bibfnamefont{Z.}~\bibnamefont{Wang}},
  \bibinfo{author}{\bibfnamefont{S.}~\bibnamefont{Hu}},
  \bibinfo{author}{\bibfnamefont{G.}~\bibnamefont{Gu}},
  \bibinfo{author}{\bibfnamefont{J.}~\bibnamefont{Zhu}},
  \bibinfo{author}{\bibfnamefont{D.}~\bibnamefont{Zhang}}, \bibnamefont{and}
  \bibinfo{author}{\bibfnamefont{Q.-K.} \bibnamefont{Xue}},
  \bibinfo{journal}{Phys. Rev. B} \textbf{\bibinfo{volume}{108}},
  \bibinfo{pages}{174508} (\bibinfo{year}{2023}),
  \urlprefix\url{https://link.aps.org/doi/10.1103/PhysRevB.108.174508}.

\bibitem[{\citenamefont{Ghosh et~al.}(2024)\citenamefont{Ghosh, Patil, Basu,
  Kuldeep, Dutta, Jangade, Kulkarni, Thamizhavel, Steiner, von Oppen
  et~al.}}]{Ghosh24}
\bibinfo{author}{\bibfnamefont{S.}~\bibnamefont{Ghosh}},
  \bibinfo{author}{\bibfnamefont{V.}~\bibnamefont{Patil}},
  \bibinfo{author}{\bibfnamefont{A.}~\bibnamefont{Basu}},
  \bibinfo{author}{\bibnamefont{Kuldeep}},
  \bibinfo{author}{\bibfnamefont{A.}~\bibnamefont{Dutta}},
  \bibinfo{author}{\bibfnamefont{D.~A.} \bibnamefont{Jangade}},
  \bibinfo{author}{\bibfnamefont{R.}~\bibnamefont{Kulkarni}},
  \bibinfo{author}{\bibfnamefont{A.}~\bibnamefont{Thamizhavel}},
  \bibinfo{author}{\bibfnamefont{J.~F.} \bibnamefont{Steiner}},
  \bibinfo{author}{\bibfnamefont{F.}~\bibnamefont{von Oppen}},
  \bibnamefont{et~al.}, \bibinfo{journal}{Nature Materials} pp.
  \bibinfo{pages}{1--7} (\bibinfo{year}{2024}), ISSN \bibinfo{issn}{1476-4660},
  \bibinfo{note}{publisher: Nature Publishing Group},
  \urlprefix\url{https://www.nature.com/articles/s41563-024-01804-4}.

\bibitem[{\citenamefont{Volkov et~al.}(2024)\citenamefont{Volkov,
  Lantagne-Hurtubise, Tummuru, Plugge, Pixley, and Franz}}]{VolkovFranz24}
\bibinfo{author}{\bibfnamefont{P.~A.} \bibnamefont{Volkov}},
  \bibinfo{author}{\bibfnamefont{E.}~\bibnamefont{Lantagne-Hurtubise}},
  \bibinfo{author}{\bibfnamefont{T.}~\bibnamefont{Tummuru}},
  \bibinfo{author}{\bibfnamefont{S.}~\bibnamefont{Plugge}},
  \bibinfo{author}{\bibfnamefont{J.~H.} \bibnamefont{Pixley}},
  \bibnamefont{and} \bibinfo{author}{\bibfnamefont{M.}~\bibnamefont{Franz}},
  \bibinfo{journal}{Phys. Rev. B} \textbf{\bibinfo{volume}{109}},
  \bibinfo{pages}{094518} (\bibinfo{year}{2024}),
  \urlprefix\url{https://link.aps.org/doi/10.1103/PhysRevB.109.094518}.

\bibitem[{\citenamefont{Goldobin et~al.}(2011)\citenamefont{Goldobin, Koelle,
  Kleiner, and Mints}}]{Goldobin11}
\bibinfo{author}{\bibfnamefont{E.}~\bibnamefont{Goldobin}},
  \bibinfo{author}{\bibfnamefont{D.}~\bibnamefont{Koelle}},
  \bibinfo{author}{\bibfnamefont{R.}~\bibnamefont{Kleiner}}, \bibnamefont{and}
  \bibinfo{author}{\bibfnamefont{R.~G.} \bibnamefont{Mints}},
  \bibinfo{journal}{Phys. Rev. Lett.} \textbf{\bibinfo{volume}{107}},
  \bibinfo{pages}{227001} (\bibinfo{year}{2011}),
  \urlprefix\url{https://link.aps.org/doi/10.1103/PhysRevLett.107.227001}.

\bibitem[{\citenamefont{Goldobin et~al.}(2007)\citenamefont{Goldobin, Koelle,
  Kleiner, and Buzdin}}]{GoldobinBuzdin07}
\bibinfo{author}{\bibfnamefont{E.}~\bibnamefont{Goldobin}},
  \bibinfo{author}{\bibfnamefont{D.}~\bibnamefont{Koelle}},
  \bibinfo{author}{\bibfnamefont{R.}~\bibnamefont{Kleiner}}, \bibnamefont{and}
  \bibinfo{author}{\bibfnamefont{A.}~\bibnamefont{Buzdin}},
  \bibinfo{journal}{Physical Review B} \textbf{\bibinfo{volume}{76}},
  \bibinfo{pages}{224523} (\bibinfo{year}{2007}), \bibinfo{note}{publisher:
  American Physical Society},
  \urlprefix\url{https://link.aps.org/doi/10.1103/PhysRevB.76.224523}.

\bibitem[{\citenamefont{Fulton et~al.}(1972)\citenamefont{Fulton, Dunkleberger,
  and Dynes}}]{FultonDynes72}
\bibinfo{author}{\bibfnamefont{T.~A.} \bibnamefont{Fulton}},
  \bibinfo{author}{\bibfnamefont{L.~N.} \bibnamefont{Dunkleberger}},
  \bibnamefont{and} \bibinfo{author}{\bibfnamefont{R.~C.} \bibnamefont{Dynes}},
  \bibinfo{journal}{Physical Review B} \textbf{\bibinfo{volume}{6}},
  \bibinfo{pages}{855} (\bibinfo{year}{1972}), \bibinfo{note}{publisher:
  American Physical Society},
  \urlprefix\url{https://link.aps.org/doi/10.1103/PhysRevB.6.855}.

\bibitem[{\citenamefont{Cuozzo et~al.}(2024)\citenamefont{Cuozzo, Pan, Shabani,
  and Rossi}}]{CuozzoShabani24}
\bibinfo{author}{\bibfnamefont{J.~J.} \bibnamefont{Cuozzo}},
  \bibinfo{author}{\bibfnamefont{W.}~\bibnamefont{Pan}},
  \bibinfo{author}{\bibfnamefont{J.}~\bibnamefont{Shabani}}, \bibnamefont{and}
  \bibinfo{author}{\bibfnamefont{E.}~\bibnamefont{Rossi}},
  \bibinfo{journal}{Phys. Rev. Res.} \textbf{\bibinfo{volume}{6}},
  \bibinfo{pages}{023011} (\bibinfo{year}{2024}),
  \urlprefix\url{https://link.aps.org/doi/10.1103/PhysRevResearch.6.023011}.

\bibitem[{\citenamefont{Kim et~al.}(2025)\citenamefont{Kim, Hays, Rosen, An,
  Zhang, Goswami, Azar, Gertler, Niedzielski, Schwartz et~al.}}]{Kim25}
\bibinfo{author}{\bibfnamefont{J.}~\bibnamefont{Kim}},
  \bibinfo{author}{\bibfnamefont{M.}~\bibnamefont{Hays}},
  \bibinfo{author}{\bibfnamefont{I.~T.} \bibnamefont{Rosen}},
  \bibinfo{author}{\bibfnamefont{J.}~\bibnamefont{An}},
  \bibinfo{author}{\bibfnamefont{H.}~\bibnamefont{Zhang}},
  \bibinfo{author}{\bibfnamefont{A.}~\bibnamefont{Goswami}},
  \bibinfo{author}{\bibfnamefont{K.}~\bibnamefont{Azar}},
  \bibinfo{author}{\bibfnamefont{J.~M.} \bibnamefont{Gertler}},
  \bibinfo{author}{\bibfnamefont{B.~M.} \bibnamefont{Niedzielski}},
  \bibinfo{author}{\bibfnamefont{M.~E.} \bibnamefont{Schwartz}},
  \bibnamefont{et~al.}, \emph{\bibinfo{title}{Emergent harmonics in josephson
  tunnel junctions due to series inductance}} (\bibinfo{year}{2025}),
  \eprint{2507.08171}, \urlprefix\url{https://arxiv.org/abs/2507.08171}.

\bibitem[{\citenamefont{Peterson and Hamilton}(1979)}]{Peterson79}
\bibinfo{author}{\bibfnamefont{R.~L.} \bibnamefont{Peterson}} \bibnamefont{and}
  \bibinfo{author}{\bibfnamefont{C.~A.} \bibnamefont{Hamilton}},
  \bibinfo{journal}{Journal of Applied Physics} \textbf{\bibinfo{volume}{50}},
  \bibinfo{pages}{8135} (\bibinfo{year}{1979}), ISSN \bibinfo{issn}{0021-8979},
  \eprint{https://pubs.aip.org/aip/jap/article-pdf/50/12/8135/18385428/8135_1_online.pdf},
  \urlprefix\url{https://doi.org/10.1063/1.325954}.

\bibitem[{\citenamefont{Likharev}(2022)}]{Likharev22}
\bibinfo{author}{\bibfnamefont{K.~K.} \bibnamefont{Likharev}},
  \emph{\bibinfo{title}{Dynamics of Josephson junctions and circuits}}
  (\bibinfo{publisher}{Routledge}, \bibinfo{year}{2022}).

\bibitem[{\citenamefont{Souto et~al.}(2022)\citenamefont{Souto, Leijnse, and
  Schrade}}]{SeoaneSouto22}
\bibinfo{author}{\bibfnamefont{R.~S.} \bibnamefont{Souto}},
  \bibinfo{author}{\bibfnamefont{M.}~\bibnamefont{Leijnse}}, \bibnamefont{and}
  \bibinfo{author}{\bibfnamefont{C.}~\bibnamefont{Schrade}},
  \bibinfo{journal}{Phys. Rev. Lett.} \textbf{\bibinfo{volume}{129}},
  \bibinfo{pages}{267702} (\bibinfo{year}{2022}),
  \urlprefix\url{https://link.aps.org/doi/10.1103/PhysRevLett.129.267702}.

\bibitem[{\citenamefont{Haenel and Can}(2022)}]{Haenel22}
\bibinfo{author}{\bibfnamefont{R.}~\bibnamefont{Haenel}} \bibnamefont{and}
  \bibinfo{author}{\bibfnamefont{O.}~\bibnamefont{Can}},
  \emph{\bibinfo{title}{Superconducting diode from flux biased josephson
  junction arrays}} (\bibinfo{year}{2022}), \eprint{2212.02657},
  \urlprefix\url{https://arxiv.org/abs/2212.02657}.

\bibitem[{\citenamefont{Pillet et~al.}(2023)\citenamefont{Pillet, Annabi,
  Peugeot, Riechert, Arrighi, Griesmar, and Bretheau}}]{Pillet23}
\bibinfo{author}{\bibfnamefont{J.-D.} \bibnamefont{Pillet}},
  \bibinfo{author}{\bibfnamefont{S.}~\bibnamefont{Annabi}},
  \bibinfo{author}{\bibfnamefont{A.}~\bibnamefont{Peugeot}},
  \bibinfo{author}{\bibfnamefont{H.}~\bibnamefont{Riechert}},
  \bibinfo{author}{\bibfnamefont{E.}~\bibnamefont{Arrighi}},
  \bibinfo{author}{\bibfnamefont{J.}~\bibnamefont{Griesmar}}, \bibnamefont{and}
  \bibinfo{author}{\bibfnamefont{L.}~\bibnamefont{Bretheau}},
  \bibinfo{journal}{Phys. Rev. Res.} \textbf{\bibinfo{volume}{5}},
  \bibinfo{pages}{033199} (\bibinfo{year}{2023}),
  \urlprefix\url{https://link.aps.org/doi/10.1103/PhysRevResearch.5.033199}.

\bibitem[{\citenamefont{Hodt and Linder}(2023)}]{HodtLinder23}
\bibinfo{author}{\bibfnamefont{E.~W.} \bibnamefont{Hodt}} \bibnamefont{and}
  \bibinfo{author}{\bibfnamefont{J.}~\bibnamefont{Linder}},
  \bibinfo{journal}{Phys. Rev. B} \textbf{\bibinfo{volume}{108}},
  \bibinfo{pages}{174502} (\bibinfo{year}{2023}),
  \urlprefix\url{https://link.aps.org/doi/10.1103/PhysRevB.108.174502}.

\bibitem[{\citenamefont{Song et~al.}(2023)\citenamefont{Song, Suresh~Babu, Bai,
  Golubev, Burkova, Romanov, Ilin, Eckstein, and Bezryadin}}]{SongBezryadin23}
\bibinfo{author}{\bibfnamefont{X.}~\bibnamefont{Song}},
  \bibinfo{author}{\bibfnamefont{S.}~\bibnamefont{Suresh~Babu}},
  \bibinfo{author}{\bibfnamefont{Y.}~\bibnamefont{Bai}},
  \bibinfo{author}{\bibfnamefont{D.~S.} \bibnamefont{Golubev}},
  \bibinfo{author}{\bibfnamefont{I.}~\bibnamefont{Burkova}},
  \bibinfo{author}{\bibfnamefont{A.}~\bibnamefont{Romanov}},
  \bibinfo{author}{\bibfnamefont{E.}~\bibnamefont{Ilin}},
  \bibinfo{author}{\bibfnamefont{J.~N.} \bibnamefont{Eckstein}},
  \bibnamefont{and}
  \bibinfo{author}{\bibfnamefont{A.}~\bibnamefont{Bezryadin}},
  \bibinfo{journal}{Communications Physics} \textbf{\bibinfo{volume}{6}},
  \bibinfo{pages}{177} (\bibinfo{year}{2023}).

\bibitem[{\citenamefont{Bozkurt et~al.}(2023)\citenamefont{Bozkurt, Brookman,
  Fatemi, and Akhmerov}}]{BozkurtFatemi23}
\bibinfo{author}{\bibfnamefont{A.~M.} \bibnamefont{Bozkurt}},
  \bibinfo{author}{\bibfnamefont{J.}~\bibnamefont{Brookman}},
  \bibinfo{author}{\bibfnamefont{V.}~\bibnamefont{Fatemi}}, \bibnamefont{and}
  \bibinfo{author}{\bibfnamefont{A.~R.} \bibnamefont{Akhmerov}},
  \bibinfo{journal}{SciPost Physics} \textbf{\bibinfo{volume}{15}},
  \bibinfo{pages}{204} (\bibinfo{year}{2023}).

\bibitem[{\citenamefont{Legg et~al.}(2023)\citenamefont{Legg, Laubscher, Loss,
  and Klinovaja}}]{LeggLossKlinovaja23}
\bibinfo{author}{\bibfnamefont{H.~F.} \bibnamefont{Legg}},
  \bibinfo{author}{\bibfnamefont{K.}~\bibnamefont{Laubscher}},
  \bibinfo{author}{\bibfnamefont{D.}~\bibnamefont{Loss}}, \bibnamefont{and}
  \bibinfo{author}{\bibfnamefont{J.}~\bibnamefont{Klinovaja}},
  \bibinfo{journal}{Physical Review B} \textbf{\bibinfo{volume}{108}},
  \bibinfo{pages}{214520} (\bibinfo{year}{2023}), \bibinfo{note}{publisher:
  American Physical Society},
  \urlprefix\url{https://link.aps.org/doi/10.1103/PhysRevB.108.214520}.

\bibitem[{\citenamefont{Seleznev and Fominov}(2024)}]{SeleznevFominov24}
\bibinfo{author}{\bibfnamefont{G.~S.} \bibnamefont{Seleznev}} \bibnamefont{and}
  \bibinfo{author}{\bibfnamefont{Y.~V.} \bibnamefont{Fominov}},
  \bibinfo{journal}{Phys. Rev. B} \textbf{\bibinfo{volume}{110}},
  \bibinfo{pages}{104508} (\bibinfo{year}{2024}),
  \urlprefix\url{https://link.aps.org/doi/10.1103/PhysRevB.110.104508}.

\bibitem[{\citenamefont{De~Simoni and Giazotto}(2024)}]{DeSimoniGiazotto24}
\bibinfo{author}{\bibfnamefont{G.}~\bibnamefont{De~Simoni}} \bibnamefont{and}
  \bibinfo{author}{\bibfnamefont{F.}~\bibnamefont{Giazotto}},
  \bibinfo{journal}{Phys. Rev. Appl.} \textbf{\bibinfo{volume}{21}},
  \bibinfo{pages}{064058} (\bibinfo{year}{2024}),
  \urlprefix\url{https://link.aps.org/doi/10.1103/PhysRevApplied.21.064058}.

\bibitem[{\citenamefont{Greco et~al.}(2024)\citenamefont{Greco, Pichard,
  Strambini, and Giazotto}}]{GrecoGiazotto24}
\bibinfo{author}{\bibfnamefont{A.}~\bibnamefont{Greco}},
  \bibinfo{author}{\bibfnamefont{Q.}~\bibnamefont{Pichard}},
  \bibinfo{author}{\bibfnamefont{E.}~\bibnamefont{Strambini}},
  \bibnamefont{and} \bibinfo{author}{\bibfnamefont{F.}~\bibnamefont{Giazotto}},
  \bibinfo{journal}{Applied Physics Letters} \textbf{\bibinfo{volume}{125}},
  \bibinfo{pages}{072601} (\bibinfo{year}{2024}), ISSN
  \bibinfo{issn}{0003-6951},
  \eprint{https://pubs.aip.org/aip/apl/article-pdf/doi/10.1063/5.0211021/20303950/072601_1_5.0211021.pdf},
  \urlprefix\url{https://doi.org/10.1063/5.0211021}.

\bibitem[{\citenamefont{Kotetes et~al.}(2024)\citenamefont{Kotetes, Roig, and
  Andersen}}]{Kotetes24}
\bibinfo{author}{\bibfnamefont{P.}~\bibnamefont{Kotetes}},
  \bibinfo{author}{\bibfnamefont{M.}~\bibnamefont{Roig}}, \bibnamefont{and}
  \bibinfo{author}{\bibfnamefont{B.~M.} \bibnamefont{Andersen}},
  \emph{\bibinfo{title}{Nonreciprocal equilibrium 4$\pi$-periodic josephson
  effect from poor man's majorana zero modes}} (\bibinfo{year}{2024}),
  \eprint{2409.13027}, \urlprefix\url{https://arxiv.org/abs/2409.13027}.

\bibitem[{\citenamefont{Yerin et~al.}(2025)\citenamefont{Yerin, Drechsler,
  Varlamov, Giazotto, and Cuoco}}]{YerinGiazotto25}
\bibinfo{author}{\bibfnamefont{Y.}~\bibnamefont{Yerin}},
  \bibinfo{author}{\bibfnamefont{S.-L.} \bibnamefont{Drechsler}},
  \bibinfo{author}{\bibfnamefont{A.~A.} \bibnamefont{Varlamov}},
  \bibinfo{author}{\bibfnamefont{F.}~\bibnamefont{Giazotto}}, \bibnamefont{and}
  \bibinfo{author}{\bibfnamefont{M.}~\bibnamefont{Cuoco}},
  \emph{\bibinfo{title}{Supercurrent diode effect in josephson interferometers
  with multiband superconductors}} (\bibinfo{year}{2025}),
  \urlprefix\url{https://doi.org/10.1038/s42005-025-02253-4}.

\bibitem[{\citenamefont{Zalom et~al.}(2025)\citenamefont{Zalom, Rolih, and
  Žitko}}]{Zalom25}
\bibinfo{author}{\bibfnamefont{P.}~\bibnamefont{Zalom}},
  \bibinfo{author}{\bibfnamefont{D.}~\bibnamefont{Rolih}}, \bibnamefont{and}
  \bibinfo{author}{\bibfnamefont{R.}~\bibnamefont{Žitko}},
  \emph{\bibinfo{title}{Andreev bound state spectroscopy of a quantum-dot-based
  aharonov-bohm interferometer with superconducting terminals}}
  (\bibinfo{year}{2025}), \eprint{2507.03614},
  \urlprefix\url{https://arxiv.org/abs/2507.03614}.

\bibitem[{\citenamefont{Mayer et~al.}(2020)\citenamefont{Mayer, Dartiailh,
  Yuan, Wickramasinghe, Rossi, and Shabani}}]{MayerShabani20}
\bibinfo{author}{\bibfnamefont{W.}~\bibnamefont{Mayer}},
  \bibinfo{author}{\bibfnamefont{M.~C.} \bibnamefont{Dartiailh}},
  \bibinfo{author}{\bibfnamefont{J.}~\bibnamefont{Yuan}},
  \bibinfo{author}{\bibfnamefont{K.~S.} \bibnamefont{Wickramasinghe}},
  \bibinfo{author}{\bibfnamefont{E.}~\bibnamefont{Rossi}}, \bibnamefont{and}
  \bibinfo{author}{\bibfnamefont{J.}~\bibnamefont{Shabani}},
  \bibinfo{journal}{Nature Communications} \textbf{\bibinfo{volume}{11}},
  \bibinfo{pages}{212} (\bibinfo{year}{2020}),
  \urlprefix\url{https://doi.org/10.1038/s41467-019-14094-1}.

\bibitem[{\citenamefont{Paolucci et~al.}(2023)\citenamefont{Paolucci,
  De~Simoni, and Giazotto}}]{PaolucciGiazotto23}
\bibinfo{author}{\bibfnamefont{F.}~\bibnamefont{Paolucci}},
  \bibinfo{author}{\bibfnamefont{G.}~\bibnamefont{De~Simoni}},
  \bibnamefont{and} \bibinfo{author}{\bibfnamefont{F.}~\bibnamefont{Giazotto}},
  \bibinfo{journal}{Applied Physics Letters} \textbf{\bibinfo{volume}{122}},
  \bibinfo{pages}{042601} (\bibinfo{year}{2023}), ISSN
  \bibinfo{issn}{0003-6951},
  \eprint{https://pubs.aip.org/aip/apl/article-pdf/doi/10.1063/5.0136709/20272933/042601_1_5.0136709.pdf},
  \urlprefix\url{https://doi.org/10.1063/5.0136709}.

\bibitem[{\citenamefont{Reinhardt et~al.}(2024)\citenamefont{Reinhardt,
  Ascherl, Costa, Berger, Gronin, Gardner, Lindemann, Manfra, Fabian, Kochan
  et~al.}}]{ReinhardtManfra24}
\bibinfo{author}{\bibfnamefont{S.}~\bibnamefont{Reinhardt}},
  \bibinfo{author}{\bibfnamefont{T.}~\bibnamefont{Ascherl}},
  \bibinfo{author}{\bibfnamefont{A.}~\bibnamefont{Costa}},
  \bibinfo{author}{\bibfnamefont{J.}~\bibnamefont{Berger}},
  \bibinfo{author}{\bibfnamefont{S.}~\bibnamefont{Gronin}},
  \bibinfo{author}{\bibfnamefont{G.~C.} \bibnamefont{Gardner}},
  \bibinfo{author}{\bibfnamefont{T.}~\bibnamefont{Lindemann}},
  \bibinfo{author}{\bibfnamefont{M.~J.} \bibnamefont{Manfra}},
  \bibinfo{author}{\bibfnamefont{J.}~\bibnamefont{Fabian}},
  \bibinfo{author}{\bibfnamefont{D.}~\bibnamefont{Kochan}},
  \bibnamefont{et~al.}, \bibinfo{journal}{Nature Communications}
  \textbf{\bibinfo{volume}{15}}, \bibinfo{pages}{4413} (\bibinfo{year}{2024}),
  \urlprefix\url{https://doi.org/10.1038/s41467-024-48741-z}.

\bibitem[{\citenamefont{Wu et~al.}(2025)\citenamefont{Wu, Wang, Su, Yan, Pan,
  Zhao, Zhang, and Xu}}]{Wu25}
\bibinfo{author}{\bibfnamefont{X.}~\bibnamefont{Wu}},
  \bibinfo{author}{\bibfnamefont{J.-Y.} \bibnamefont{Wang}},
  \bibinfo{author}{\bibfnamefont{H.}~\bibnamefont{Su}},
  \bibinfo{author}{\bibfnamefont{S.}~\bibnamefont{Yan}},
  \bibinfo{author}{\bibfnamefont{D.}~\bibnamefont{Pan}},
  \bibinfo{author}{\bibfnamefont{J.}~\bibnamefont{Zhao}},
  \bibinfo{author}{\bibfnamefont{P.}~\bibnamefont{Zhang}}, \bibnamefont{and}
  \bibinfo{author}{\bibfnamefont{H.}~\bibnamefont{Xu}}, \bibinfo{journal}{New
  Journal of Physics} \textbf{\bibinfo{volume}{27}}, \bibinfo{pages}{023031}
  (\bibinfo{year}{2025}).

\bibitem[{\citenamefont{Chieppa et~al.}(2025)\citenamefont{Chieppa, Shukla,
  Traverso, Bucci, Zannier, Fracassi, Ziani, Sassetti, Carrega, Beltram
  et~al.}}]{ChieppaGiazotto25}
\bibinfo{author}{\bibfnamefont{A.}~\bibnamefont{Chieppa}},
  \bibinfo{author}{\bibfnamefont{G.}~\bibnamefont{Shukla}},
  \bibinfo{author}{\bibfnamefont{S.}~\bibnamefont{Traverso}},
  \bibinfo{author}{\bibfnamefont{G.}~\bibnamefont{Bucci}},
  \bibinfo{author}{\bibfnamefont{V.}~\bibnamefont{Zannier}},
  \bibinfo{author}{\bibfnamefont{S.}~\bibnamefont{Fracassi}},
  \bibinfo{author}{\bibfnamefont{N.~T.} \bibnamefont{Ziani}},
  \bibinfo{author}{\bibfnamefont{M.}~\bibnamefont{Sassetti}},
  \bibinfo{author}{\bibfnamefont{M.}~\bibnamefont{Carrega}},
  \bibinfo{author}{\bibfnamefont{F.}~\bibnamefont{Beltram}},
  \bibnamefont{et~al.}, \emph{\bibinfo{title}{Superconducting quantum
  interference devices based on insb nanoflag josephson junctions}}
  (\bibinfo{year}{2025}), \eprint{2504.18965},
  \urlprefix\url{https://arxiv.org/abs/2504.18965}.

\bibitem[{\citenamefont{Ciaccia et~al.}(2023)\citenamefont{Ciaccia, Haller,
  Drachmann, Lindemann, Manfra, Schrade, and
  Sch\"onenberger}}]{CiacciaManfra23}
\bibinfo{author}{\bibfnamefont{C.}~\bibnamefont{Ciaccia}},
  \bibinfo{author}{\bibfnamefont{R.}~\bibnamefont{Haller}},
  \bibinfo{author}{\bibfnamefont{A.~C.~C.} \bibnamefont{Drachmann}},
  \bibinfo{author}{\bibfnamefont{T.}~\bibnamefont{Lindemann}},
  \bibinfo{author}{\bibfnamefont{M.~J.} \bibnamefont{Manfra}},
  \bibinfo{author}{\bibfnamefont{C.}~\bibnamefont{Schrade}}, \bibnamefont{and}
  \bibinfo{author}{\bibfnamefont{C.}~\bibnamefont{Sch\"onenberger}},
  \bibinfo{journal}{Phys. Rev. Res.} \textbf{\bibinfo{volume}{5}},
  \bibinfo{pages}{033131} (\bibinfo{year}{2023}),
  \urlprefix\url{https://link.aps.org/doi/10.1103/PhysRevResearch.5.033131}.

\bibitem[{\citenamefont{Greco et~al.}(2023)\citenamefont{Greco, Pichard, and
  Giazotto}}]{GrecoGiazotto23}
\bibinfo{author}{\bibfnamefont{A.}~\bibnamefont{Greco}},
  \bibinfo{author}{\bibfnamefont{Q.}~\bibnamefont{Pichard}}, \bibnamefont{and}
  \bibinfo{author}{\bibfnamefont{F.}~\bibnamefont{Giazotto}},
  \bibinfo{journal}{Applied Physics Letters} \textbf{\bibinfo{volume}{123}}
  (\bibinfo{year}{2023}).

\bibitem[{\citenamefont{Leblanc et~al.}(2024)\citenamefont{Leblanc,
  Tangchingchai, Momtaz, Kiyooka, Hartmann, Fernandez-Bada, Scher\"ubl, Brun,
  Schmitt, Zihlmann et~al.}}]{Leblanc24}
\bibinfo{author}{\bibfnamefont{A.}~\bibnamefont{Leblanc}},
  \bibinfo{author}{\bibfnamefont{C.}~\bibnamefont{Tangchingchai}},
  \bibinfo{author}{\bibfnamefont{Z.~S.} \bibnamefont{Momtaz}},
  \bibinfo{author}{\bibfnamefont{E.}~\bibnamefont{Kiyooka}},
  \bibinfo{author}{\bibfnamefont{J.-M.} \bibnamefont{Hartmann}},
  \bibinfo{author}{\bibfnamefont{G.~T.} \bibnamefont{Fernandez-Bada}},
  \bibinfo{author}{\bibfnamefont{Z.}~\bibnamefont{Scher\"ubl}},
  \bibinfo{author}{\bibfnamefont{B.}~\bibnamefont{Brun}},
  \bibinfo{author}{\bibfnamefont{V.}~\bibnamefont{Schmitt}},
  \bibinfo{author}{\bibfnamefont{S.}~\bibnamefont{Zihlmann}},
  \bibnamefont{et~al.}, \bibinfo{journal}{Phys. Rev. Res.}
  \textbf{\bibinfo{volume}{6}}, \bibinfo{pages}{033281} (\bibinfo{year}{2024}),
  \urlprefix\url{https://link.aps.org/doi/10.1103/PhysRevResearch.6.033281}.

\bibitem[{\citenamefont{Yu et~al.}(2024)\citenamefont{Yu, Cuozzo, Sapkota,
  Rossi, Rademacher, Nenoff, and Pan}}]{Yu24}
\bibinfo{author}{\bibfnamefont{W.}~\bibnamefont{Yu}},
  \bibinfo{author}{\bibfnamefont{J.~J.} \bibnamefont{Cuozzo}},
  \bibinfo{author}{\bibfnamefont{K.}~\bibnamefont{Sapkota}},
  \bibinfo{author}{\bibfnamefont{E.}~\bibnamefont{Rossi}},
  \bibinfo{author}{\bibfnamefont{D.~X.} \bibnamefont{Rademacher}},
  \bibinfo{author}{\bibfnamefont{T.~M.} \bibnamefont{Nenoff}},
  \bibnamefont{and} \bibinfo{author}{\bibfnamefont{W.}~\bibnamefont{Pan}},
  \bibinfo{journal}{Phys. Rev. B} \textbf{\bibinfo{volume}{110}},
  \bibinfo{pages}{104510} (\bibinfo{year}{2024}),
  \urlprefix\url{https://link.aps.org/doi/10.1103/PhysRevB.110.104510}.

\bibitem[{\citenamefont{Banszerus et~al.}(2025)\citenamefont{Banszerus,
  Andersson, Marshall, Lindemann, Manfra, Marcus, and
  Vaitiek\ifmmode~\dot{e}\else \.{e}\fi{}nas}}]{BanszerusManfra25}
\bibinfo{author}{\bibfnamefont{L.}~\bibnamefont{Banszerus}},
  \bibinfo{author}{\bibfnamefont{C.~W.} \bibnamefont{Andersson}},
  \bibinfo{author}{\bibfnamefont{W.}~\bibnamefont{Marshall}},
  \bibinfo{author}{\bibfnamefont{T.}~\bibnamefont{Lindemann}},
  \bibinfo{author}{\bibfnamefont{M.~J.} \bibnamefont{Manfra}},
  \bibinfo{author}{\bibfnamefont{C.~M.} \bibnamefont{Marcus}},
  \bibnamefont{and}
  \bibinfo{author}{\bibfnamefont{S.}~\bibnamefont{Vaitiek\ifmmode~\dot{e}\else
  \.{e}\fi{}nas}}, \bibinfo{journal}{Phys. Rev. X}
  \textbf{\bibinfo{volume}{15}}, \bibinfo{pages}{011021}
  (\bibinfo{year}{2025}),
  \urlprefix\url{https://link.aps.org/doi/10.1103/PhysRevX.15.011021}.

\bibitem[{\citenamefont{Li et~al.}(2025)\citenamefont{Li, Sato, Tanaka, and
  Fujimaki}}]{Li25}
\bibinfo{author}{\bibfnamefont{F.}~\bibnamefont{Li}},
  \bibinfo{author}{\bibfnamefont{T.}~\bibnamefont{Sato}},
  \bibinfo{author}{\bibfnamefont{M.}~\bibnamefont{Tanaka}}, \bibnamefont{and}
  \bibinfo{author}{\bibfnamefont{A.}~\bibnamefont{Fujimaki}},
  \bibinfo{journal}{Applied Physics Letters} \textbf{\bibinfo{volume}{127}}
  (\bibinfo{year}{2025}).

\bibitem[{\citenamefont{Reinhardt et~al.}(2025)\citenamefont{Reinhardt, Penner,
  Berger, Baumgartner, Gronin, Gardner, Lindemann, Manfra, Glazman, von Oppen
  et~al.}}]{ReinhardtGlazman25}
\bibinfo{author}{\bibfnamefont{S.}~\bibnamefont{Reinhardt}},
  \bibinfo{author}{\bibfnamefont{A.-G.} \bibnamefont{Penner}},
  \bibinfo{author}{\bibfnamefont{J.}~\bibnamefont{Berger}},
  \bibinfo{author}{\bibfnamefont{C.}~\bibnamefont{Baumgartner}},
  \bibinfo{author}{\bibfnamefont{S.}~\bibnamefont{Gronin}},
  \bibinfo{author}{\bibfnamefont{G.~C.} \bibnamefont{Gardner}},
  \bibinfo{author}{\bibfnamefont{T.}~\bibnamefont{Lindemann}},
  \bibinfo{author}{\bibfnamefont{M.~J.} \bibnamefont{Manfra}},
  \bibinfo{author}{\bibfnamefont{L.~I.} \bibnamefont{Glazman}},
  \bibinfo{author}{\bibfnamefont{F.}~\bibnamefont{von Oppen}},
  \bibnamefont{et~al.} (\bibinfo{year}{2025}), \eprint{2406.13819},
  \urlprefix\url{https://arxiv.org/abs/2406.13819}.

\bibitem[{\citenamefont{Kudriashov et~al.}(2025)\citenamefont{Kudriashov, Zhou,
  Hovhannisyan, Frolov, Elesin, Wang, Zharkova, Taniguchi, Watanabe, Yashina
  et~al.}}]{Kudriashov25}
\bibinfo{author}{\bibfnamefont{A.}~\bibnamefont{Kudriashov}},
  \bibinfo{author}{\bibfnamefont{X.}~\bibnamefont{Zhou}},
  \bibinfo{author}{\bibfnamefont{R.~A.} \bibnamefont{Hovhannisyan}},
  \bibinfo{author}{\bibfnamefont{A.}~\bibnamefont{Frolov}},
  \bibinfo{author}{\bibfnamefont{L.}~\bibnamefont{Elesin}},
  \bibinfo{author}{\bibfnamefont{Y.}~\bibnamefont{Wang}},
  \bibinfo{author}{\bibfnamefont{E.~V.} \bibnamefont{Zharkova}},
  \bibinfo{author}{\bibfnamefont{T.}~\bibnamefont{Taniguchi}},
  \bibinfo{author}{\bibfnamefont{K.}~\bibnamefont{Watanabe}},
  \bibinfo{author}{\bibfnamefont{L.~A.} \bibnamefont{Yashina}},
  \bibnamefont{et~al.}, \emph{\bibinfo{title}{Non-reciprocal current-phase
  relation and superconducting diode effect in topological-insulator-based
  josephson junctions}} (\bibinfo{year}{2025}), \eprint{2502.08527},
  \urlprefix\url{https://arxiv.org/abs/2502.08527}.

\bibitem[{\citenamefont{Matsuo et~al.}(2025)\citenamefont{Matsuo, Deacon,
  Kobayashi, Sato, Yokoyama, Lindemann, Gronin, Gardner, Ishibashi, Manfra
  et~al.}}]{MatsuoManfra25}
\bibinfo{author}{\bibfnamefont{S.}~\bibnamefont{Matsuo}},
  \bibinfo{author}{\bibfnamefont{R.~S.} \bibnamefont{Deacon}},
  \bibinfo{author}{\bibfnamefont{S.}~\bibnamefont{Kobayashi}},
  \bibinfo{author}{\bibfnamefont{Y.}~\bibnamefont{Sato}},
  \bibinfo{author}{\bibfnamefont{T.}~\bibnamefont{Yokoyama}},
  \bibinfo{author}{\bibfnamefont{T.}~\bibnamefont{Lindemann}},
  \bibinfo{author}{\bibfnamefont{S.}~\bibnamefont{Gronin}},
  \bibinfo{author}{\bibfnamefont{G.~C.} \bibnamefont{Gardner}},
  \bibinfo{author}{\bibfnamefont{K.}~\bibnamefont{Ishibashi}},
  \bibinfo{author}{\bibfnamefont{M.~J.} \bibnamefont{Manfra}},
  \bibnamefont{et~al.}, \bibinfo{journal}{Physical Review B}
  \textbf{\bibinfo{volume}{111}}, \bibinfo{pages}{094512}
  (\bibinfo{year}{2025}), \bibinfo{note}{publisher: American Physical Society},
  \urlprefix\url{https://link.aps.org/doi/10.1103/PhysRevB.111.094512}.

\bibitem[{\citenamefont{Zapata et~al.}(1996)\citenamefont{Zapata, Bartussek,
  Sols, and H\"anggi}}]{Zapata96}
\bibinfo{author}{\bibfnamefont{I.}~\bibnamefont{Zapata}},
  \bibinfo{author}{\bibfnamefont{R.}~\bibnamefont{Bartussek}},
  \bibinfo{author}{\bibfnamefont{F.}~\bibnamefont{Sols}}, \bibnamefont{and}
  \bibinfo{author}{\bibfnamefont{P.}~\bibnamefont{H\"anggi}},
  \bibinfo{journal}{Phys. Rev. Lett.} \textbf{\bibinfo{volume}{77}},
  \bibinfo{pages}{2292} (\bibinfo{year}{1996}),
  \urlprefix\url{https://link.aps.org/doi/10.1103/PhysRevLett.77.2292}.

\bibitem[{\citenamefont{Falo et~al.}(1999)\citenamefont{Falo, Martinez, Mazo,
  and Cilla}}]{Falo99}
\bibinfo{author}{\bibfnamefont{F.}~\bibnamefont{Falo}},
  \bibinfo{author}{\bibfnamefont{P.~J.} \bibnamefont{Martinez}},
  \bibinfo{author}{\bibfnamefont{J.~J.} \bibnamefont{Mazo}}, \bibnamefont{and}
  \bibinfo{author}{\bibfnamefont{S.}~\bibnamefont{Cilla}},
  \bibinfo{journal}{Europhysics Letters} \textbf{\bibinfo{volume}{45}},
  \bibinfo{pages}{700} (\bibinfo{year}{1999}).

\bibitem[{\citenamefont{Carapella}(2001)}]{Carapella01}
\bibinfo{author}{\bibfnamefont{G.}~\bibnamefont{Carapella}},
  \bibinfo{journal}{Phys. Rev. B} \textbf{\bibinfo{volume}{63}},
  \bibinfo{pages}{054515} (\bibinfo{year}{2001}),
  \urlprefix\url{https://link.aps.org/doi/10.1103/PhysRevB.63.054515}.

\bibitem[{\citenamefont{Lee}(2003)}]{Lee03}
\bibinfo{author}{\bibfnamefont{K.~H.} \bibnamefont{Lee}},
  \bibinfo{journal}{Applied Physics Letters} \textbf{\bibinfo{volume}{83}},
  \bibinfo{pages}{117} (\bibinfo{year}{2003}), ISSN \bibinfo{issn}{0003-6951},
  \urlprefix\url{https://doi.org/10.1063/1.1591244}.

\bibitem[{\citenamefont{Marconi}(2007)}]{Marconi07}
\bibinfo{author}{\bibfnamefont{V.~I.} \bibnamefont{Marconi}},
  \bibinfo{journal}{Phys. Rev. Lett.} \textbf{\bibinfo{volume}{98}},
  \bibinfo{pages}{047006} (\bibinfo{year}{2007}),
  \urlprefix\url{https://link.aps.org/doi/10.1103/PhysRevLett.98.047006}.

\bibitem[{\citenamefont{Spiechowicz et~al.}(2014)\citenamefont{Spiechowicz,
  H\"anggi, and \L{}uczka}}]{Spiechowicz14}
\bibinfo{author}{\bibfnamefont{J.}~\bibnamefont{Spiechowicz}},
  \bibinfo{author}{\bibfnamefont{P.}~\bibnamefont{H\"anggi}}, \bibnamefont{and}
  \bibinfo{author}{\bibfnamefont{J.}~\bibnamefont{\L{}uczka}},
  \bibinfo{journal}{Phys. Rev. B} \textbf{\bibinfo{volume}{90}},
  \bibinfo{pages}{054520} (\bibinfo{year}{2014}),
  \urlprefix\url{https://link.aps.org/doi/10.1103/PhysRevB.90.054520}.

\bibitem[{\citenamefont{Semenov et~al.}(2015)\citenamefont{Semenov, Polyakov,
  and Tolpygo}}]{Semenov15}
\bibinfo{author}{\bibfnamefont{V.~K.} \bibnamefont{Semenov}},
  \bibinfo{author}{\bibfnamefont{Y.~A.} \bibnamefont{Polyakov}},
  \bibnamefont{and} \bibinfo{author}{\bibfnamefont{S.~K.}
  \bibnamefont{Tolpygo}}, \bibinfo{journal}{IEEE Transactions on Applied
  Superconductivity} \textbf{\bibinfo{volume}{25}}, \bibinfo{pages}{1}
  (\bibinfo{year}{2015}).

\bibitem[{\citenamefont{Tr\'{\i}as et~al.}(2000)\citenamefont{Tr\'{\i}as, Mazo,
  Falo, and Orlando}}]{TriasFalo00}
\bibinfo{author}{\bibfnamefont{E.}~\bibnamefont{Tr\'{\i}as}},
  \bibinfo{author}{\bibfnamefont{J.~J.} \bibnamefont{Mazo}},
  \bibinfo{author}{\bibfnamefont{F.}~\bibnamefont{Falo}}, \bibnamefont{and}
  \bibinfo{author}{\bibfnamefont{T.~P.} \bibnamefont{Orlando}},
  \bibinfo{journal}{Phys. Rev. E} \textbf{\bibinfo{volume}{61}},
  \bibinfo{pages}{2257} (\bibinfo{year}{2000}),
  \urlprefix\url{https://link.aps.org/doi/10.1103/PhysRevE.61.2257}.

\bibitem[{\citenamefont{Weiss et~al.}(2000)\citenamefont{Weiss, Koelle,
  Müller, Gross, and Barthel}}]{Weiss00}
\bibinfo{author}{\bibfnamefont{S.}~\bibnamefont{Weiss}},
  \bibinfo{author}{\bibfnamefont{D.}~\bibnamefont{Koelle}},
  \bibinfo{author}{\bibfnamefont{J.}~\bibnamefont{Müller}},
  \bibinfo{author}{\bibfnamefont{R.}~\bibnamefont{Gross}}, \bibnamefont{and}
  \bibinfo{author}{\bibfnamefont{K.}~\bibnamefont{Barthel}},
  \bibinfo{journal}{Europhysics Letters} \textbf{\bibinfo{volume}{51}},
  \bibinfo{pages}{499} (\bibinfo{year}{2000}),
  \urlprefix\url{https://doi.org/10.1209/epl/i2000-00365-x}.

\bibitem[{\citenamefont{Carapella and Costabile}(2001)}]{CarapellaCostabile01}
\bibinfo{author}{\bibfnamefont{G.}~\bibnamefont{Carapella}} \bibnamefont{and}
  \bibinfo{author}{\bibfnamefont{G.}~\bibnamefont{Costabile}},
  \bibinfo{journal}{Phys. Rev. Lett.} \textbf{\bibinfo{volume}{87}},
  \bibinfo{pages}{077002} (\bibinfo{year}{2001}),
  \urlprefix\url{https://link.aps.org/doi/10.1103/PhysRevLett.87.077002}.

\bibitem[{\citenamefont{Sterck et~al.}(2002)\citenamefont{Sterck, Weiss, and
  Koelle}}]{Sterck02}
\bibinfo{author}{\bibfnamefont{A.}~\bibnamefont{Sterck}},
  \bibinfo{author}{\bibfnamefont{S.}~\bibnamefont{Weiss}}, \bibnamefont{and}
  \bibinfo{author}{\bibfnamefont{D.}~\bibnamefont{Koelle}},
  \bibinfo{journal}{Applied Physics A} \textbf{\bibinfo{volume}{75}},
  \bibinfo{pages}{253} (\bibinfo{year}{2002}), ISSN \bibinfo{issn}{1432-0630},
  \bibinfo{note}{company: Springer Distributor: Springer Institution: Springer
  Label: Springer Number: 2 Publisher: Springer-Verlag},
  \urlprefix\url{https://link-springer-com.ezproxy.library.wisc.edu/article/10.1007/s003390201326}.

\bibitem[{\citenamefont{Carapella et~al.}(2002)\citenamefont{Carapella,
  Costabile, Martucciello, Cirillo, Latempa, Polcari, and
  Filatrella}}]{CarapellaFilatrella02}
\bibinfo{author}{\bibfnamefont{G.}~\bibnamefont{Carapella}},
  \bibinfo{author}{\bibfnamefont{G.}~\bibnamefont{Costabile}},
  \bibinfo{author}{\bibfnamefont{N.}~\bibnamefont{Martucciello}},
  \bibinfo{author}{\bibfnamefont{M.}~\bibnamefont{Cirillo}},
  \bibinfo{author}{\bibfnamefont{R.}~\bibnamefont{Latempa}},
  \bibinfo{author}{\bibfnamefont{A.}~\bibnamefont{Polcari}}, \bibnamefont{and}
  \bibinfo{author}{\bibfnamefont{G.}~\bibnamefont{Filatrella}},
  \bibinfo{journal}{Physica C: Superconductivity}
  \textbf{\bibinfo{volume}{382}}, \bibinfo{pages}{337} (\bibinfo{year}{2002}),
  ISSN \bibinfo{issn}{0921-4534},
  \urlprefix\url{https://www.sciencedirect.com/science/article/pii/S0921453402012327}.

\bibitem[{\citenamefont{Majer et~al.}(2003)\citenamefont{Majer, Peguiron,
  Grifoni, Tusveld, and Mooij}}]{Majer03}
\bibinfo{author}{\bibfnamefont{J.~B.} \bibnamefont{Majer}},
  \bibinfo{author}{\bibfnamefont{J.}~\bibnamefont{Peguiron}},
  \bibinfo{author}{\bibfnamefont{M.}~\bibnamefont{Grifoni}},
  \bibinfo{author}{\bibfnamefont{M.}~\bibnamefont{Tusveld}}, \bibnamefont{and}
  \bibinfo{author}{\bibfnamefont{J.~E.} \bibnamefont{Mooij}},
  \bibinfo{journal}{Phys. Rev. Lett.} \textbf{\bibinfo{volume}{90}},
  \bibinfo{pages}{056802} (\bibinfo{year}{2003}),
  \urlprefix\url{https://link.aps.org/doi/10.1103/PhysRevLett.90.056802}.

\bibitem[{\citenamefont{Shal\'om and Pastoriza}(2005)}]{Shalom05}
\bibinfo{author}{\bibfnamefont{D.~E.} \bibnamefont{Shal\'om}} \bibnamefont{and}
  \bibinfo{author}{\bibfnamefont{H.}~\bibnamefont{Pastoriza}},
  \bibinfo{journal}{Phys. Rev. Lett.} \textbf{\bibinfo{volume}{94}},
  \bibinfo{pages}{177001} (\bibinfo{year}{2005}),
  \urlprefix\url{https://link.aps.org/doi/10.1103/PhysRevLett.94.177001}.

\bibitem[{\citenamefont{Beck et~al.}(2005)\citenamefont{Beck, Goldobin,
  Neuhaus, Siegel, Kleiner, and Koelle}}]{BeckGoldobin05}
\bibinfo{author}{\bibfnamefont{M.}~\bibnamefont{Beck}},
  \bibinfo{author}{\bibfnamefont{E.}~\bibnamefont{Goldobin}},
  \bibinfo{author}{\bibfnamefont{M.}~\bibnamefont{Neuhaus}},
  \bibinfo{author}{\bibfnamefont{M.}~\bibnamefont{Siegel}},
  \bibinfo{author}{\bibfnamefont{R.}~\bibnamefont{Kleiner}}, \bibnamefont{and}
  \bibinfo{author}{\bibfnamefont{D.}~\bibnamefont{Koelle}},
  \bibinfo{journal}{Phys. Rev. Lett.} \textbf{\bibinfo{volume}{95}},
  \bibinfo{pages}{090603} (\bibinfo{year}{2005}),
  \urlprefix\url{https://link.aps.org/doi/10.1103/PhysRevLett.95.090603}.

\bibitem[{\citenamefont{Sterck et~al.}(2005)\citenamefont{Sterck, Kleiner, and
  Koelle}}]{Streck05}
\bibinfo{author}{\bibfnamefont{A.}~\bibnamefont{Sterck}},
  \bibinfo{author}{\bibfnamefont{R.}~\bibnamefont{Kleiner}}, \bibnamefont{and}
  \bibinfo{author}{\bibfnamefont{D.}~\bibnamefont{Koelle}},
  \bibinfo{journal}{Phys. Rev. Lett.} \textbf{\bibinfo{volume}{95}},
  \bibinfo{pages}{177006} (\bibinfo{year}{2005}),
  \urlprefix\url{https://link.aps.org/doi/10.1103/PhysRevLett.95.177006}.

\bibitem[{\citenamefont{Sterck et~al.}(2009)\citenamefont{Sterck, Koelle, and
  Kleiner}}]{Streck09}
\bibinfo{author}{\bibfnamefont{A.}~\bibnamefont{Sterck}},
  \bibinfo{author}{\bibfnamefont{D.}~\bibnamefont{Koelle}}, \bibnamefont{and}
  \bibinfo{author}{\bibfnamefont{R.}~\bibnamefont{Kleiner}},
  \bibinfo{journal}{Phys. Rev. Lett.} \textbf{\bibinfo{volume}{103}},
  \bibinfo{pages}{047001} (\bibinfo{year}{2009}),
  \urlprefix\url{https://link.aps.org/doi/10.1103/PhysRevLett.103.047001}.

\bibitem[{\citenamefont{Knufinke et~al.}(2012)\citenamefont{Knufinke, Ilin,
  Siegel, Koelle, Kleiner, and Goldobin}}]{KnufinkeGoldobin12}
\bibinfo{author}{\bibfnamefont{M.}~\bibnamefont{Knufinke}},
  \bibinfo{author}{\bibfnamefont{K.}~\bibnamefont{Ilin}},
  \bibinfo{author}{\bibfnamefont{M.}~\bibnamefont{Siegel}},
  \bibinfo{author}{\bibfnamefont{D.}~\bibnamefont{Koelle}},
  \bibinfo{author}{\bibfnamefont{R.}~\bibnamefont{Kleiner}}, \bibnamefont{and}
  \bibinfo{author}{\bibfnamefont{E.}~\bibnamefont{Goldobin}},
  \bibinfo{journal}{Phys. Rev. E} \textbf{\bibinfo{volume}{85}},
  \bibinfo{pages}{011122} (\bibinfo{year}{2012}),
  \urlprefix\url{https://link.aps.org/doi/10.1103/PhysRevE.85.011122}.

\bibitem[{\citenamefont{Chesca et~al.}(2017)\citenamefont{Chesca, John,
  Pollett, Gaifullin, Cox, Mellor, and Savel'ev}}]{Chesca17}
\bibinfo{author}{\bibfnamefont{B.}~\bibnamefont{Chesca}},
  \bibinfo{author}{\bibfnamefont{D.}~\bibnamefont{John}},
  \bibinfo{author}{\bibfnamefont{R.}~\bibnamefont{Pollett}},
  \bibinfo{author}{\bibfnamefont{M.}~\bibnamefont{Gaifullin}},
  \bibinfo{author}{\bibfnamefont{J.}~\bibnamefont{Cox}},
  \bibinfo{author}{\bibfnamefont{C.~J.} \bibnamefont{Mellor}},
  \bibnamefont{and} \bibinfo{author}{\bibfnamefont{S.}~\bibnamefont{Savel'ev}},
  \bibinfo{journal}{Applied Physics Letters} \textbf{\bibinfo{volume}{111}}
  (\bibinfo{year}{2017}).

\bibitem[{\citenamefont{Van~Harlingen}(1995)}]{VanHarlingen95}
\bibinfo{author}{\bibfnamefont{D.~J.} \bibnamefont{Van~Harlingen}},
  \bibinfo{journal}{Rev. Mod. Phys.} \textbf{\bibinfo{volume}{67}},
  \bibinfo{pages}{515} (\bibinfo{year}{1995}),
  \urlprefix\url{https://link.aps.org/doi/10.1103/RevModPhys.67.515}.

\bibitem[{\citenamefont{Krasnov et~al.}(1997)\citenamefont{Krasnov, Oboznov,
  and Pedersen}}]{KrasnovPedersen97}
\bibinfo{author}{\bibfnamefont{V.~M.} \bibnamefont{Krasnov}},
  \bibinfo{author}{\bibfnamefont{V.~A.} \bibnamefont{Oboznov}},
  \bibnamefont{and} \bibinfo{author}{\bibfnamefont{N.~F.}
  \bibnamefont{Pedersen}}, \bibinfo{journal}{Physical Review B}
  \textbf{\bibinfo{volume}{55}}, \bibinfo{pages}{14486} (\bibinfo{year}{1997}),
  \bibinfo{note}{publisher: American Physical Society},
  \urlprefix\url{https://link.aps.org/doi/10.1103/PhysRevB.55.14486}.

\bibitem[{\citenamefont{Monaco et~al.}(2013)\citenamefont{Monaco, Koshelets,
  Mukhortova, and Mygind}}]{Monaco13}
\bibinfo{author}{\bibfnamefont{R.}~\bibnamefont{Monaco}},
  \bibinfo{author}{\bibfnamefont{V.~P.} \bibnamefont{Koshelets}},
  \bibinfo{author}{\bibfnamefont{A.}~\bibnamefont{Mukhortova}},
  \bibnamefont{and} \bibinfo{author}{\bibfnamefont{J.}~\bibnamefont{Mygind}},
  \bibinfo{journal}{Superconductor Science and Technology}
  \textbf{\bibinfo{volume}{26}}, \bibinfo{pages}{055021}
  (\bibinfo{year}{2013}),
  \urlprefix\url{https://dx.doi.org/10.1088/0953-2048/26/5/055021}.

\bibitem[{\citenamefont{V\'{a}vra et~al.}(2013)\citenamefont{V\'{a}vra, Pfaff,
  Monaco, Aprili, and Strunk}}]{VarvaStrunk13}
\bibinfo{author}{\bibfnamefont{O.}~\bibnamefont{V\'{a}vra}},
  \bibinfo{author}{\bibfnamefont{W.}~\bibnamefont{Pfaff}},
  \bibinfo{author}{\bibfnamefont{R.}~\bibnamefont{Monaco}},
  \bibinfo{author}{\bibfnamefont{M.}~\bibnamefont{Aprili}}, \bibnamefont{and}
  \bibinfo{author}{\bibfnamefont{C.}~\bibnamefont{Strunk}},
  \bibinfo{journal}{Applied Physics Letters} \textbf{\bibinfo{volume}{102}},
  \bibinfo{pages}{072602} (\bibinfo{year}{2013}), ISSN
  \bibinfo{issn}{0003-6951},
  \eprint{https://pubs.aip.org/aip/apl/article-pdf/doi/10.1063/1.4792213/14281559/072602\_1\_online.pdf},
  \urlprefix\url{https://doi.org/10.1063/1.4792213}.

\bibitem[{\citenamefont{Golod and Krasnov}(2022)}]{GolodKrasnov22}
\bibinfo{author}{\bibfnamefont{T.}~\bibnamefont{Golod}} \bibnamefont{and}
  \bibinfo{author}{\bibfnamefont{V.~M.} \bibnamefont{Krasnov}},
  \bibinfo{journal}{Nature Communications} \textbf{\bibinfo{volume}{13}},
  \bibinfo{pages}{3658} (\bibinfo{year}{2022}), ISSN \bibinfo{issn}{2041-1723},
  \bibinfo{note}{number: 1 Publisher: Nature Publishing Group},
  \urlprefix\url{https://www.nature.com/articles/s41467-022-31256-w}.

\bibitem[{\citenamefont{Chirolli et~al.}(2024)\citenamefont{Chirolli, Greco,
  Crippa, Strambini, Cuoco, Amico, and Giazotto}}]{ChirolliCuocoGiazotto24}
\bibinfo{author}{\bibfnamefont{L.}~\bibnamefont{Chirolli}},
  \bibinfo{author}{\bibfnamefont{A.}~\bibnamefont{Greco}},
  \bibinfo{author}{\bibfnamefont{A.}~\bibnamefont{Crippa}},
  \bibinfo{author}{\bibfnamefont{E.}~\bibnamefont{Strambini}},
  \bibinfo{author}{\bibfnamefont{M.}~\bibnamefont{Cuoco}},
  \bibinfo{author}{\bibfnamefont{L.}~\bibnamefont{Amico}}, \bibnamefont{and}
  \bibinfo{author}{\bibfnamefont{F.}~\bibnamefont{Giazotto}}
  (\bibinfo{year}{2024}), \bibinfo{note}{arXiv:2411.19338 [cond-mat,
  physics:quant-ph]}, \urlprefix\url{http://arxiv.org/abs/2411.19338}.

\bibitem[{\citenamefont{Yang and Agterberg}(2000)}]{YangAgterberg00}
\bibinfo{author}{\bibfnamefont{K.}~\bibnamefont{Yang}} \bibnamefont{and}
  \bibinfo{author}{\bibfnamefont{D.~F.} \bibnamefont{Agterberg}},
  \bibinfo{journal}{Physical Review Letters} \textbf{\bibinfo{volume}{84}},
  \bibinfo{pages}{4970} (\bibinfo{year}{2000}), \bibinfo{note}{publisher:
  American Physical Society},
  \urlprefix\url{https://link.aps.org/doi/10.1103/PhysRevLett.84.4970}.

\bibitem[{\citenamefont{Nelson et~al.}(2004)\citenamefont{Nelson, Mao, Maeno,
  and Liu}}]{Nelson04}
\bibinfo{author}{\bibfnamefont{K.~D.} \bibnamefont{Nelson}},
  \bibinfo{author}{\bibfnamefont{Z.~Q.} \bibnamefont{Mao}},
  \bibinfo{author}{\bibfnamefont{Y.}~\bibnamefont{Maeno}}, \bibnamefont{and}
  \bibinfo{author}{\bibfnamefont{Y.}~\bibnamefont{Liu}},
  \bibinfo{journal}{Science} \textbf{\bibinfo{volume}{306}},
  \bibinfo{pages}{1151} (\bibinfo{year}{2004}),
  \eprint{https://www.science.org/doi/pdf/10.1126/science.1103881},
  \urlprefix\url{https://www.science.org/doi/abs/10.1126/science.1103881}.

\bibitem[{\citenamefont{Kaur et~al.}(2005{\natexlab{a}})\citenamefont{Kaur,
  Agterberg, and Sigrist}}]{KaurAgterbergSigrist05}
\bibinfo{author}{\bibfnamefont{R.~P.} \bibnamefont{Kaur}},
  \bibinfo{author}{\bibfnamefont{D.~F.} \bibnamefont{Agterberg}},
  \bibnamefont{and} \bibinfo{author}{\bibfnamefont{M.}~\bibnamefont{Sigrist}},
  \bibinfo{journal}{Physical Review Letters} \textbf{\bibinfo{volume}{94}},
  \bibinfo{pages}{137002} (\bibinfo{year}{2005}{\natexlab{a}}),
  \bibinfo{note}{publisher: American Physical Society},
  \urlprefix\url{https://link.aps.org/doi/10.1103/PhysRevLett.94.137002}.

\bibitem[{\citenamefont{Cheng and Sun}(2023)}]{ChengSun23}
\bibinfo{author}{\bibfnamefont{Q.}~\bibnamefont{Cheng}} \bibnamefont{and}
  \bibinfo{author}{\bibfnamefont{Q.-F.} \bibnamefont{Sun}},
  \bibinfo{journal}{Physical Review B} \textbf{\bibinfo{volume}{107}},
  \bibinfo{pages}{184511} (\bibinfo{year}{2023}), \bibinfo{note}{publisher:
  American Physical Society},
  \urlprefix\url{https://link.aps.org/doi/10.1103/PhysRevB.107.184511}.

\bibitem[{\citenamefont{Sun et~al.}(2023)\citenamefont{Sun, Mao, and
  Sun}}]{SunMaoSun23}
\bibinfo{author}{\bibfnamefont{Y.-F.} \bibnamefont{Sun}},
  \bibinfo{author}{\bibfnamefont{Y.}~\bibnamefont{Mao}}, \bibnamefont{and}
  \bibinfo{author}{\bibfnamefont{Q.-F.} \bibnamefont{Sun}},
  \bibinfo{journal}{Physical Review B} \textbf{\bibinfo{volume}{108}},
  \bibinfo{pages}{214519} (\bibinfo{year}{2023}), \bibinfo{note}{publisher:
  American Physical Society},
  \urlprefix\url{https://link.aps.org/doi/10.1103/PhysRevB.108.214519}.

\bibitem[{\citenamefont{Sun et~al.}(2025{\natexlab{a}})\citenamefont{Sun, Mao,
  and Sun}}]{SunMaoSun25}
\bibinfo{author}{\bibfnamefont{Y.-F.} \bibnamefont{Sun}},
  \bibinfo{author}{\bibfnamefont{Y.}~\bibnamefont{Mao}}, \bibnamefont{and}
  \bibinfo{author}{\bibfnamefont{Q.-F.} \bibnamefont{Sun}},
  \bibinfo{journal}{Phys. Rev. B} \textbf{\bibinfo{volume}{111}},
  \bibinfo{pages}{054515} (\bibinfo{year}{2025}{\natexlab{a}}),
  \urlprefix\url{https://link.aps.org/doi/10.1103/PhysRevB.111.054515}.

\bibitem[{\citenamefont{Debnath and Dutta}(2024)}]{DebnathDutta24}
\bibinfo{author}{\bibfnamefont{D.}~\bibnamefont{Debnath}} \bibnamefont{and}
  \bibinfo{author}{\bibfnamefont{P.}~\bibnamefont{Dutta}}
  (\bibinfo{year}{2024}), \bibinfo{note}{arXiv:2402.00817 [cond-mat]},
  \urlprefix\url{http://arxiv.org/abs/2402.00817}.

\bibitem[{\citenamefont{Debnath and Dutta}(2025)}]{DebnathDutta25}
\bibinfo{author}{\bibfnamefont{D.}~\bibnamefont{Debnath}} \bibnamefont{and}
  \bibinfo{author}{\bibfnamefont{P.}~\bibnamefont{Dutta}},
  \bibinfo{journal}{Journal of Physics: Condensed Matter}
  \textbf{\bibinfo{volume}{37}}, \bibinfo{pages}{175301}
  (\bibinfo{year}{2025}).

\bibitem[{\citenamefont{Li et~al.}(2024{\natexlab{a}})\citenamefont{Li, Yan,
  Hong, Sheng, Wang, Dou, Guo, Shi, Su, Lyu et~al.}}]{Li24}
\bibinfo{author}{\bibfnamefont{Y.}~\bibnamefont{Li}},
  \bibinfo{author}{\bibfnamefont{D.}~\bibnamefont{Yan}},
  \bibinfo{author}{\bibfnamefont{Y.}~\bibnamefont{Hong}},
  \bibinfo{author}{\bibfnamefont{H.}~\bibnamefont{Sheng}},
  \bibinfo{author}{\bibfnamefont{A.}~\bibnamefont{Wang}},
  \bibinfo{author}{\bibfnamefont{Z.}~\bibnamefont{Dou}},
  \bibinfo{author}{\bibfnamefont{X.}~\bibnamefont{Guo}},
  \bibinfo{author}{\bibfnamefont{X.}~\bibnamefont{Shi}},
  \bibinfo{author}{\bibfnamefont{Z.}~\bibnamefont{Su}},
  \bibinfo{author}{\bibfnamefont{Z.}~\bibnamefont{Lyu}}, \bibnamefont{et~al.},
  \bibinfo{journal}{Nature Communications} \textbf{\bibinfo{volume}{15}},
  \bibinfo{pages}{9031} (\bibinfo{year}{2024}{\natexlab{a}}),
  \urlprefix\url{https://doi.org/10.1038/s41467-024-53383-2}.

\bibitem[{\citenamefont{Nikodem et~al.}(2025)\citenamefont{Nikodem, Schluck,
  Geier, Papaj, Legg, Feng, Bagchi, Fu, and Ando}}]{NikodemFuAndo25}
\bibinfo{author}{\bibfnamefont{E.}~\bibnamefont{Nikodem}},
  \bibinfo{author}{\bibfnamefont{J.}~\bibnamefont{Schluck}},
  \bibinfo{author}{\bibfnamefont{M.}~\bibnamefont{Geier}},
  \bibinfo{author}{\bibfnamefont{M.}~\bibnamefont{Papaj}},
  \bibinfo{author}{\bibfnamefont{H.~F.} \bibnamefont{Legg}},
  \bibinfo{author}{\bibfnamefont{J.}~\bibnamefont{Feng}},
  \bibinfo{author}{\bibfnamefont{M.}~\bibnamefont{Bagchi}},
  \bibinfo{author}{\bibfnamefont{L.}~\bibnamefont{Fu}}, \bibnamefont{and}
  \bibinfo{author}{\bibfnamefont{Y.}~\bibnamefont{Ando}},
  \bibinfo{journal}{Science Advances} \textbf{\bibinfo{volume}{11}},
  \bibinfo{pages}{eadw4898} (\bibinfo{year}{2025}),
  \eprint{https://www.science.org/doi/pdf/10.1126/sciadv.adw4898},
  \urlprefix\url{https://www.science.org/doi/abs/10.1126/sciadv.adw4898}.

\bibitem[{\citenamefont{Kaneyasu et~al.}(2010)\citenamefont{Kaneyasu, Hayashi,
  Gut, Makoshi, and Sigrist}}]{KaneyasuSigrist10}
\bibinfo{author}{\bibfnamefont{H.}~\bibnamefont{Kaneyasu}},
  \bibinfo{author}{\bibfnamefont{N.}~\bibnamefont{Hayashi}},
  \bibinfo{author}{\bibfnamefont{B.}~\bibnamefont{Gut}},
  \bibinfo{author}{\bibfnamefont{K.}~\bibnamefont{Makoshi}}, \bibnamefont{and}
  \bibinfo{author}{\bibfnamefont{M.}~\bibnamefont{Sigrist}},
  \bibinfo{journal}{Journal of the Physical Society of Japan}
  \textbf{\bibinfo{volume}{79}}, \bibinfo{pages}{104705}
  (\bibinfo{year}{2010}), \eprint{https://doi.org/10.1143/JPSJ.79.104705},
  \urlprefix\url{https://doi.org/10.1143/JPSJ.79.104705}.

\bibitem[{\citenamefont{Likharev}(1979)}]{Likharev79}
\bibinfo{author}{\bibfnamefont{K.~K.} \bibnamefont{Likharev}},
  \bibinfo{journal}{Rev. Mod. Phys.} \textbf{\bibinfo{volume}{51}},
  \bibinfo{pages}{101} (\bibinfo{year}{1979}),
  \urlprefix\url{https://link.aps.org/doi/10.1103/RevModPhys.51.101}.

\bibitem[{\citenamefont{Pankratova et~al.}(2020)\citenamefont{Pankratova, Lee,
  Kuzmin, Wickramasinghe, Mayer, Yuan, Vavilov, Shabani, and
  Manucharyan}}]{PankratovaVavilovShabani20}
\bibinfo{author}{\bibfnamefont{N.}~\bibnamefont{Pankratova}},
  \bibinfo{author}{\bibfnamefont{H.}~\bibnamefont{Lee}},
  \bibinfo{author}{\bibfnamefont{R.}~\bibnamefont{Kuzmin}},
  \bibinfo{author}{\bibfnamefont{K.}~\bibnamefont{Wickramasinghe}},
  \bibinfo{author}{\bibfnamefont{W.}~\bibnamefont{Mayer}},
  \bibinfo{author}{\bibfnamefont{J.}~\bibnamefont{Yuan}},
  \bibinfo{author}{\bibfnamefont{M.~G.} \bibnamefont{Vavilov}},
  \bibinfo{author}{\bibfnamefont{J.}~\bibnamefont{Shabani}}, \bibnamefont{and}
  \bibinfo{author}{\bibfnamefont{V.~E.} \bibnamefont{Manucharyan}},
  \bibinfo{journal}{Phys. Rev. X} \textbf{\bibinfo{volume}{10}},
  \bibinfo{pages}{031051} (\bibinfo{year}{2020}),
  \urlprefix\url{https://link.aps.org/doi/10.1103/PhysRevX.10.031051}.

\bibitem[{\citenamefont{Gupta et~al.}(2023)\citenamefont{Gupta, Graziano,
  Pendharkar, Dong, Dempsey, Palmstr{\o}m, and Pribiag}}]{GuptaPribiag23}
\bibinfo{author}{\bibfnamefont{M.}~\bibnamefont{Gupta}},
  \bibinfo{author}{\bibfnamefont{G.~V.} \bibnamefont{Graziano}},
  \bibinfo{author}{\bibfnamefont{M.}~\bibnamefont{Pendharkar}},
  \bibinfo{author}{\bibfnamefont{J.~T.} \bibnamefont{Dong}},
  \bibinfo{author}{\bibfnamefont{C.~P.} \bibnamefont{Dempsey}},
  \bibinfo{author}{\bibfnamefont{C.}~\bibnamefont{Palmstr{\o}m}},
  \bibnamefont{and} \bibinfo{author}{\bibfnamefont{V.~S.}
  \bibnamefont{Pribiag}}, \bibinfo{journal}{Nature Communications}
  \textbf{\bibinfo{volume}{14}}, \bibinfo{pages}{3078} (\bibinfo{year}{2023}),
  \urlprefix\url{https://doi.org/10.1038/s41467-023-38856-0}.

\bibitem[{\citenamefont{Coraiola et~al.}(2024)\citenamefont{Coraiola,
  Svetogorov, Haxell, Sabonis, Hinderling, Ten~Kate, Cheah, Krizek, Schott,
  Wegscheider et~al.}}]{Coraiola24}
\bibinfo{author}{\bibfnamefont{M.}~\bibnamefont{Coraiola}},
  \bibinfo{author}{\bibfnamefont{A.~E.} \bibnamefont{Svetogorov}},
  \bibinfo{author}{\bibfnamefont{D.~Z.} \bibnamefont{Haxell}},
  \bibinfo{author}{\bibfnamefont{D.}~\bibnamefont{Sabonis}},
  \bibinfo{author}{\bibfnamefont{M.}~\bibnamefont{Hinderling}},
  \bibinfo{author}{\bibfnamefont{S.~C.} \bibnamefont{Ten~Kate}},
  \bibinfo{author}{\bibfnamefont{E.}~\bibnamefont{Cheah}},
  \bibinfo{author}{\bibfnamefont{F.}~\bibnamefont{Krizek}},
  \bibinfo{author}{\bibfnamefont{R.}~\bibnamefont{Schott}},
  \bibinfo{author}{\bibfnamefont{W.}~\bibnamefont{Wegscheider}},
  \bibnamefont{et~al.}, \bibinfo{journal}{ACS nano}
  \textbf{\bibinfo{volume}{18}}, \bibinfo{pages}{9221} (\bibinfo{year}{2024}).

\bibitem[{\citenamefont{Behner et~al.}(2025)\citenamefont{Behner, Jalil, Rupp,
  Lüth, Grützmacher, and Schäpers}}]{Behner25}
\bibinfo{author}{\bibfnamefont{G.}~\bibnamefont{Behner}},
  \bibinfo{author}{\bibfnamefont{A.~R.} \bibnamefont{Jalil}},
  \bibinfo{author}{\bibfnamefont{A.}~\bibnamefont{Rupp}},
  \bibinfo{author}{\bibfnamefont{H.}~\bibnamefont{Lüth}},
  \bibinfo{author}{\bibfnamefont{D.}~\bibnamefont{Grützmacher}},
  \bibnamefont{and}
  \bibinfo{author}{\bibfnamefont{T.}~\bibnamefont{Schäpers}},
  \bibinfo{journal}{ACS nano} \textbf{\bibinfo{volume}{19}},
  \bibinfo{pages}{3878} (\bibinfo{year}{2025}).

\bibitem[{\citenamefont{Matsuo et~al.}(2023)\citenamefont{Matsuo, Imoto,
  Yokoyama, Sato, Lindemann, Gronin, Gardner, Manfra, and
  Tarucha}}]{MatsuoManfra23}
\bibinfo{author}{\bibfnamefont{S.}~\bibnamefont{Matsuo}},
  \bibinfo{author}{\bibfnamefont{T.}~\bibnamefont{Imoto}},
  \bibinfo{author}{\bibfnamefont{T.}~\bibnamefont{Yokoyama}},
  \bibinfo{author}{\bibfnamefont{Y.}~\bibnamefont{Sato}},
  \bibinfo{author}{\bibfnamefont{T.}~\bibnamefont{Lindemann}},
  \bibinfo{author}{\bibfnamefont{S.}~\bibnamefont{Gronin}},
  \bibinfo{author}{\bibfnamefont{G.~C.} \bibnamefont{Gardner}},
  \bibinfo{author}{\bibfnamefont{M.~J.} \bibnamefont{Manfra}},
  \bibnamefont{and} \bibinfo{author}{\bibfnamefont{S.}~\bibnamefont{Tarucha}},
  \bibinfo{journal}{Nature Physics} \textbf{\bibinfo{volume}{19}},
  \bibinfo{pages}{1636} (\bibinfo{year}{2023}).

\bibitem[{\citenamefont{Chiles et~al.}(2023)\citenamefont{Chiles, Arnault,
  Chen, Larson, Zhao, Watanabe, Taniguchi, Amet, and
  Finkelstein}}]{ChilesFinkelstein23}
\bibinfo{author}{\bibfnamefont{J.}~\bibnamefont{Chiles}},
  \bibinfo{author}{\bibfnamefont{E.~G.} \bibnamefont{Arnault}},
  \bibinfo{author}{\bibfnamefont{C.-C.} \bibnamefont{Chen}},
  \bibinfo{author}{\bibfnamefont{T.~F.~Q.} \bibnamefont{Larson}},
  \bibinfo{author}{\bibfnamefont{L.}~\bibnamefont{Zhao}},
  \bibinfo{author}{\bibfnamefont{K.}~\bibnamefont{Watanabe}},
  \bibinfo{author}{\bibfnamefont{T.}~\bibnamefont{Taniguchi}},
  \bibinfo{author}{\bibfnamefont{F.}~\bibnamefont{Amet}}, \bibnamefont{and}
  \bibinfo{author}{\bibfnamefont{G.}~\bibnamefont{Finkelstein}},
  \bibinfo{journal}{Nano Letters} \textbf{\bibinfo{volume}{23}},
  \bibinfo{pages}{5257} (\bibinfo{year}{2023}), ISSN \bibinfo{issn}{1530-6984},
  \bibinfo{note}{publisher: American Chemical Society},
  \urlprefix\url{https://doi.org/10.1021/acs.nanolett.3c01276}.

\bibitem[{\citenamefont{Zhang et~al.}(2024)\citenamefont{Zhang, Rashid,
  Tanhayi~Ahari, de~Coster, Taniguchi, Watanabe, Gilbert, Samarth, and
  Kayyalha}}]{ZhangKayyalha24}
\bibinfo{author}{\bibfnamefont{F.}~\bibnamefont{Zhang}},
  \bibinfo{author}{\bibfnamefont{A.~S.} \bibnamefont{Rashid}},
  \bibinfo{author}{\bibfnamefont{M.}~\bibnamefont{Tanhayi~Ahari}},
  \bibinfo{author}{\bibfnamefont{G.~J.} \bibnamefont{de~Coster}},
  \bibinfo{author}{\bibfnamefont{T.}~\bibnamefont{Taniguchi}},
  \bibinfo{author}{\bibfnamefont{K.}~\bibnamefont{Watanabe}},
  \bibinfo{author}{\bibfnamefont{M.~J.} \bibnamefont{Gilbert}},
  \bibinfo{author}{\bibfnamefont{N.}~\bibnamefont{Samarth}}, \bibnamefont{and}
  \bibinfo{author}{\bibfnamefont{M.}~\bibnamefont{Kayyalha}},
  \bibinfo{journal}{Phys. Rev. Appl.} \textbf{\bibinfo{volume}{21}},
  \bibinfo{pages}{034011} (\bibinfo{year}{2024}),
  \urlprefix\url{https://link.aps.org/doi/10.1103/PhysRevApplied.21.034011}.

\bibitem[{\citenamefont{Arnault et~al.}(2025)\citenamefont{Arnault, Chiles,
  Larson, Chen, Zhao, Watanabe, Taniguchi, Amet, and
  Finkelstein}}]{ArnaultFinkelstein25}
\bibinfo{author}{\bibfnamefont{E.~G.} \bibnamefont{Arnault}},
  \bibinfo{author}{\bibfnamefont{J.}~\bibnamefont{Chiles}},
  \bibinfo{author}{\bibfnamefont{T.~F.~Q.} \bibnamefont{Larson}},
  \bibinfo{author}{\bibfnamefont{C.-C.} \bibnamefont{Chen}},
  \bibinfo{author}{\bibfnamefont{L.}~\bibnamefont{Zhao}},
  \bibinfo{author}{\bibfnamefont{K.}~\bibnamefont{Watanabe}},
  \bibinfo{author}{\bibfnamefont{T.}~\bibnamefont{Taniguchi}},
  \bibinfo{author}{\bibfnamefont{F.~m.~c.} \bibnamefont{Amet}},
  \bibnamefont{and}
  \bibinfo{author}{\bibfnamefont{G.}~\bibnamefont{Finkelstein}},
  \bibinfo{journal}{Phys. Rev. Lett.} \textbf{\bibinfo{volume}{134}},
  \bibinfo{pages}{067001} (\bibinfo{year}{2025}),
  \urlprefix\url{https://link.aps.org/doi/10.1103/PhysRevLett.134.067001}.

\bibitem[{\citenamefont{Amin et~al.}(2001)\citenamefont{Amin, Omelyanchouk, and
  Zagoskin}}]{AminOmelyanchouk01}
\bibinfo{author}{\bibfnamefont{M.}~\bibnamefont{Amin}},
  \bibinfo{author}{\bibfnamefont{A.}~\bibnamefont{Omelyanchouk}},
  \bibnamefont{and} \bibinfo{author}{\bibfnamefont{A.}~\bibnamefont{Zagoskin}},
  \bibinfo{journal}{Low Temperature Physics} \textbf{\bibinfo{volume}{27}},
  \bibinfo{pages}{616} (\bibinfo{year}{2001}).

\bibitem[{\citenamefont{Virtanen and Heikkil\"a}(2024)}]{VirtanenHeikkila24}
\bibinfo{author}{\bibfnamefont{P.}~\bibnamefont{Virtanen}} \bibnamefont{and}
  \bibinfo{author}{\bibfnamefont{T.~T.} \bibnamefont{Heikkil\"a}},
  \bibinfo{journal}{Phys. Rev. Lett.} \textbf{\bibinfo{volume}{132}},
  \bibinfo{pages}{046002} (\bibinfo{year}{2024}),
  \urlprefix\url{https://link.aps.org/doi/10.1103/PhysRevLett.132.046002}.

\bibitem[{\citenamefont{Huamani~Correa and Nowak}(2024)}]{HuamaniCorrea24}
\bibinfo{author}{\bibfnamefont{J.~L.} \bibnamefont{Huamani~Correa}}
  \bibnamefont{and} \bibinfo{author}{\bibfnamefont{M.~P.} \bibnamefont{Nowak}},
  \bibinfo{journal}{SciPost Physics} \textbf{\bibinfo{volume}{17}},
  \bibinfo{pages}{037} (\bibinfo{year}{2024}).

\bibitem[{\citenamefont{Zalom et~al.}(2024)\citenamefont{Zalom,
  \ifmmode~\check{Z}\else \v{Z}\fi{}onda, and Novotn\'y}}]{ZalomNovotny25}
\bibinfo{author}{\bibfnamefont{P.}~\bibnamefont{Zalom}},
  \bibinfo{author}{\bibfnamefont{M.}~\bibnamefont{\ifmmode~\check{Z}\else
  \v{Z}\fi{}onda}}, \bibnamefont{and}
  \bibinfo{author}{\bibfnamefont{T.}~\bibnamefont{Novotn\'y}},
  \bibinfo{journal}{Phys. Rev. Lett.} \textbf{\bibinfo{volume}{132}},
  \bibinfo{pages}{126505} (\bibinfo{year}{2024}),
  \urlprefix\url{https://link.aps.org/doi/10.1103/PhysRevLett.132.126505}.

\bibitem[{\citenamefont{Takeuchi and Eto}(2025)}]{Takeuchi25}
\bibinfo{author}{\bibfnamefont{G.}~\bibnamefont{Takeuchi}} \bibnamefont{and}
  \bibinfo{author}{\bibfnamefont{M.}~\bibnamefont{Eto}},
  \bibinfo{journal}{Journal of the Physical Society of Japan}
  \textbf{\bibinfo{volume}{94}}, \bibinfo{pages}{054701}
  (\bibinfo{year}{2025}).

\bibitem[{\citenamefont{Sahoo and Soori}(2025{\natexlab{a}})}]{SahooSoori25}
\bibinfo{author}{\bibfnamefont{B.~K.} \bibnamefont{Sahoo}} \bibnamefont{and}
  \bibinfo{author}{\bibfnamefont{A.}~\bibnamefont{Soori}},
  \bibinfo{journal}{Journal of Physics: Condensed Matter}
  \textbf{\bibinfo{volume}{37}}, \bibinfo{pages}{305302}
  (\bibinfo{year}{2025}{\natexlab{a}}),
  \urlprefix\url{https://doi.org/10.1088/1361-648X/adf1d0}.

\bibitem[{\citenamefont{Sahoo and Soori}(2025{\natexlab{b}})}]{SahooSoori25_2}
\bibinfo{author}{\bibfnamefont{B.~K.} \bibnamefont{Sahoo}} \bibnamefont{and}
  \bibinfo{author}{\bibfnamefont{A.}~\bibnamefont{Soori}}
  (\bibinfo{year}{2025}{\natexlab{b}}), \eprint{2509.14109},
  \urlprefix\url{https://arxiv.org/abs/2509.14109}.

\bibitem[{\citenamefont{Daido and Yanase}(2022)}]{DaidoYanase22_2}
\bibinfo{author}{\bibfnamefont{A.}~\bibnamefont{Daido}} \bibnamefont{and}
  \bibinfo{author}{\bibfnamefont{Y.}~\bibnamefont{Yanase}},
  \bibinfo{journal}{Phys. Rev. B} \textbf{\bibinfo{volume}{106}},
  \bibinfo{pages}{205206} (\bibinfo{year}{2022}),
  \urlprefix\url{https://link.aps.org/doi/10.1103/PhysRevB.106.205206}.

\bibitem[{\citenamefont{Lifshitz}(1941)}]{Lifshitz41}
\bibinfo{author}{\bibfnamefont{E.~M.} \bibnamefont{Lifshitz}},
  \bibinfo{journal}{Zh. Eksp. Teor. Fiz} \textbf{\bibinfo{volume}{11}},
  \bibinfo{pages}{269} (\bibinfo{year}{1941}).

\bibitem[{\citenamefont{Dzyaloshinskii}(1964)}]{Dzyaloshinskii64}
\bibinfo{author}{\bibfnamefont{I.~E.} \bibnamefont{Dzyaloshinskii}},
  \bibinfo{journal}{Sov. Phys. JETP} \textbf{\bibinfo{volume}{19}},
  \bibinfo{pages}{960} (\bibinfo{year}{1964}).

\bibitem[{\citenamefont{Gorbatsevich}(1989)}]{Gorbatsevich89}
\bibinfo{author}{\bibfnamefont{A.}~\bibnamefont{Gorbatsevich}},
  \bibinfo{journal}{Sov Phys JETP} \textbf{\bibinfo{volume}{68}},
  \bibinfo{pages}{847} (\bibinfo{year}{1989}).

\bibitem[{\citenamefont{Kaur and Agterberg}(2003)}]{KaurAgterberg03}
\bibinfo{author}{\bibfnamefont{R.~P.} \bibnamefont{Kaur}} \bibnamefont{and}
  \bibinfo{author}{\bibfnamefont{D.~F.} \bibnamefont{Agterberg}},
  \bibinfo{journal}{Phys. Rev. B} \textbf{\bibinfo{volume}{68}},
  \bibinfo{pages}{100506} (\bibinfo{year}{2003}),
  \urlprefix\url{https://link.aps.org/doi/10.1103/PhysRevB.68.100506}.

\bibitem[{\citenamefont{Shaffer et~al.}(2024)\citenamefont{Shaffer,
  Chichinadze, and Levchenko}}]{ShafferChichinadzeLevchenko24}
\bibinfo{author}{\bibfnamefont{D.}~\bibnamefont{Shaffer}},
  \bibinfo{author}{\bibfnamefont{D.~V.} \bibnamefont{Chichinadze}},
  \bibnamefont{and}
  \bibinfo{author}{\bibfnamefont{A.}~\bibnamefont{Levchenko}},
  \bibinfo{journal}{Physical Review B} \textbf{\bibinfo{volume}{110}},
  \bibinfo{pages}{184509} (\bibinfo{year}{2024}), \bibinfo{note}{publisher:
  American Physical Society},
  \urlprefix\url{https://link.aps.org/doi/10.1103/PhysRevB.110.184509}.

\bibitem[{\citenamefont{Mineev}(1993)}]{Mineev93}
\bibinfo{author}{\bibfnamefont{V.~P.} \bibnamefont{Mineev}},
  \bibinfo{journal}{JETP LETTERS C/C OF PIS'MA V ZHURNAL EKSPERIMENTAL'NOI
  TEORETICHESKOI FIZIKI} \textbf{\bibinfo{volume}{57}}, \bibinfo{pages}{680}
  (\bibinfo{year}{1993}).

\bibitem[{\citenamefont{Mineev and Samokhin}(1994)}]{MineevSamokhin94}
\bibinfo{author}{\bibfnamefont{V.}~\bibnamefont{Mineev}} \bibnamefont{and}
  \bibinfo{author}{\bibfnamefont{K.}~\bibnamefont{Samokhin}},
  \bibinfo{journal}{Journal of Experimental and Theoretical Physics}
  \textbf{\bibinfo{volume}{78}}, \bibinfo{pages}{401} (\bibinfo{year}{1994}),
  ISSN \bibinfo{issn}{1063-7761}, \bibinfo{note}{place: United States INIS
  Reference Number: 26032477}.

\bibitem[{\citenamefont{Gor'kov and Rashba}(2001)}]{GorkovRashba01}
\bibinfo{author}{\bibfnamefont{L.~P.} \bibnamefont{Gor'kov}} \bibnamefont{and}
  \bibinfo{author}{\bibfnamefont{E.~I.} \bibnamefont{Rashba}},
  \bibinfo{journal}{Phys. Rev. Lett.} \textbf{\bibinfo{volume}{87}},
  \bibinfo{pages}{037004} (\bibinfo{year}{2001}),
  \urlprefix\url{https://link.aps.org/doi/10.1103/PhysRevLett.87.037004}.

\bibitem[{\citenamefont{Barzykin and Gor'kov}(2002)}]{BarzykinGorkov02}
\bibinfo{author}{\bibfnamefont{V.}~\bibnamefont{Barzykin}} \bibnamefont{and}
  \bibinfo{author}{\bibfnamefont{L.~P.} \bibnamefont{Gor'kov}},
  \bibinfo{journal}{Phys. Rev. Lett.} \textbf{\bibinfo{volume}{89}},
  \bibinfo{pages}{227002} (\bibinfo{year}{2002}),
  \urlprefix\url{https://link.aps.org/doi/10.1103/PhysRevLett.89.227002}.

\bibitem[{\citenamefont{Frigeri et~al.}(2004)\citenamefont{Frigeri, Agterberg,
  Koga, and Sigrist}}]{FrigeriAgterbergSigrist04}
\bibinfo{author}{\bibfnamefont{P.~A.} \bibnamefont{Frigeri}},
  \bibinfo{author}{\bibfnamefont{D.~F.} \bibnamefont{Agterberg}},
  \bibinfo{author}{\bibfnamefont{A.}~\bibnamefont{Koga}}, \bibnamefont{and}
  \bibinfo{author}{\bibfnamefont{M.}~\bibnamefont{Sigrist}},
  \bibinfo{journal}{Phys. Rev. Lett.} \textbf{\bibinfo{volume}{92}},
  \bibinfo{pages}{097001} (\bibinfo{year}{2004}),
  \urlprefix\url{https://link.aps.org/doi/10.1103/PhysRevLett.92.097001}.

\bibitem[{\citenamefont{Bauer and Sigrist}(2012)}]{BauerSigrist12}
\bibinfo{author}{\bibfnamefont{E.}~\bibnamefont{Bauer}} \bibnamefont{and}
  \bibinfo{author}{\bibfnamefont{M.}~\bibnamefont{Sigrist}},
  \emph{\bibinfo{title}{Non-centrosymmetric superconductors: introduction and
  overview}}, vol. \bibinfo{volume}{847} (\bibinfo{publisher}{Springer Science
  \& Business Media}, \bibinfo{year}{2012}).

\bibitem[{\citenamefont{Yip}(2014)}]{Yip14}
\bibinfo{author}{\bibfnamefont{S.}~\bibnamefont{Yip}}, \bibinfo{journal}{Annual
  Review of Condensed Matter Physics} \textbf{\bibinfo{volume}{5}},
  \bibinfo{pages}{15} (\bibinfo{year}{2014}), ISSN \bibinfo{issn}{1947-5462},
  \urlprefix\url{https://www.annualreviews.org/content/journals/10.1146/annurev-conmatphys-031113-133912}.

\bibitem[{\citenamefont{Smidman et~al.}(2017)\citenamefont{Smidman, Salamon,
  Yuan, and Agterberg}}]{SmidmanAgterberg17}
\bibinfo{author}{\bibfnamefont{M.}~\bibnamefont{Smidman}},
  \bibinfo{author}{\bibfnamefont{M.~B.} \bibnamefont{Salamon}},
  \bibinfo{author}{\bibfnamefont{H.~Q.} \bibnamefont{Yuan}}, \bibnamefont{and}
  \bibinfo{author}{\bibfnamefont{D.~F.} \bibnamefont{Agterberg}},
  \bibinfo{journal}{Reports on Progress in Physics}
  \textbf{\bibinfo{volume}{80}}, \bibinfo{pages}{036501}
  (\bibinfo{year}{2017}), ISSN \bibinfo{issn}{0034-4885},
  \bibinfo{note}{publisher: IOP Publishing},
  \urlprefix\url{https://doi.org/10.1088/1361-6633/80/3/036501}.

\bibitem[{\citenamefont{Dimitrova and Feigel'man}(2003)}]{DimitrovaFeigelman03}
\bibinfo{author}{\bibfnamefont{O.}~\bibnamefont{Dimitrova}} \bibnamefont{and}
  \bibinfo{author}{\bibfnamefont{M.~V.} \bibnamefont{Feigel'man}},
  \bibinfo{journal}{Journal of Experimental and Theoretical Physics Letters}
  \textbf{\bibinfo{volume}{78}}, \bibinfo{pages}{637} (\bibinfo{year}{2003}),
  ISSN \bibinfo{issn}{1090-6487},
  \urlprefix\url{https://doi.org/10.1134/1.1644308}.

\bibitem[{\citenamefont{Agterberg and Kaur}(2007)}]{AgterbergKaur07}
\bibinfo{author}{\bibfnamefont{D.~F.} \bibnamefont{Agterberg}}
  \bibnamefont{and} \bibinfo{author}{\bibfnamefont{R.~P.} \bibnamefont{Kaur}},
  \bibinfo{journal}{Physical Review B} \textbf{\bibinfo{volume}{75}},
  \bibinfo{pages}{064511} (\bibinfo{year}{2007}), \bibinfo{note}{publisher:
  American Physical Society},
  \urlprefix\url{https://link.aps.org/doi/10.1103/PhysRevB.75.064511}.

\bibitem[{\citenamefont{Dimitrova and Feigel'man}(2007)}]{DimitrovaFeigelman07}
\bibinfo{author}{\bibfnamefont{O.}~\bibnamefont{Dimitrova}} \bibnamefont{and}
  \bibinfo{author}{\bibfnamefont{M.~V.} \bibnamefont{Feigel'man}},
  \bibinfo{journal}{Phys. Rev. B} \textbf{\bibinfo{volume}{76}},
  \bibinfo{pages}{014522} (\bibinfo{year}{2007}),
  \urlprefix\url{https://link.aps.org/doi/10.1103/PhysRevB.76.014522}.

\bibitem[{\citenamefont{Kapustin and
  Radzihovsky}(2022)}]{KapustinRadzihovsky22}
\bibinfo{author}{\bibfnamefont{A.}~\bibnamefont{Kapustin}} \bibnamefont{and}
  \bibinfo{author}{\bibfnamefont{L.}~\bibnamefont{Radzihovsky}},
  \bibinfo{journal}{Phys. Rev. B} \textbf{\bibinfo{volume}{105}},
  \bibinfo{pages}{134514} (\bibinfo{year}{2022}),
  \urlprefix\url{https://link.aps.org/doi/10.1103/PhysRevB.105.134514}.

\bibitem[{\citenamefont{Wang and Hao}(2025)}]{WangHao25}
\bibinfo{author}{\bibfnamefont{R.}~\bibnamefont{Wang}} \bibnamefont{and}
  \bibinfo{author}{\bibfnamefont{N.}~\bibnamefont{Hao}},
  \emph{\bibinfo{title}{Universal diagnostic criterion for intrinsic
  superconducting diode effect}} (\bibinfo{year}{2025}), \eprint{2507.04876},
  \urlprefix\url{https://arxiv.org/abs/2507.04876}.

\bibitem[{\citenamefont{Kaur et~al.}(2005{\natexlab{b}})\citenamefont{Kaur,
  Agterberg, and Sigrist}}]{KaurAgterberg05}
\bibinfo{author}{\bibfnamefont{R.~P.} \bibnamefont{Kaur}},
  \bibinfo{author}{\bibfnamefont{D.~F.} \bibnamefont{Agterberg}},
  \bibnamefont{and} \bibinfo{author}{\bibfnamefont{M.}~\bibnamefont{Sigrist}},
  \bibinfo{journal}{Physical Review Letters} \textbf{\bibinfo{volume}{94}},
  \bibinfo{pages}{137002} (\bibinfo{year}{2005}{\natexlab{b}}),
  \bibinfo{note}{publisher: American Physical Society},
  \urlprefix\url{https://link.aps.org/doi/10.1103/PhysRevLett.94.137002}.

\bibitem[{\citenamefont{Kochan et~al.}(2023)\citenamefont{Kochan, Costa,
  Zhumagulov, and ?uti?}}]{KochanZutic23}
\bibinfo{author}{\bibfnamefont{D.}~\bibnamefont{Kochan}},
  \bibinfo{author}{\bibfnamefont{A.}~\bibnamefont{Costa}},
  \bibinfo{author}{\bibfnamefont{I.}~\bibnamefont{Zhumagulov}},
  \bibnamefont{and} \bibinfo{author}{\bibfnamefont{I.}~\bibnamefont{?uti?}}
  (\bibinfo{year}{2023}), \eprint{2303.11975},
  \urlprefix\url{https://arxiv.org/abs/2303.11975}.

\bibitem[{\citenamefont{Yuan and Fu}(2022)}]{YuanFu22}
\bibinfo{author}{\bibfnamefont{N.~F.~Q.} \bibnamefont{Yuan}} \bibnamefont{and}
  \bibinfo{author}{\bibfnamefont{L.}~\bibnamefont{Fu}},
  \bibinfo{journal}{Proceedings of the National Academy of Sciences}
  \textbf{\bibinfo{volume}{119}}, \bibinfo{pages}{e2119548119}
  (\bibinfo{year}{2022}), \bibinfo{note}{publisher: Proceedings of the National
  Academy of Sciences},
  \urlprefix\url{https://www.pnas.org/doi/10.1073/pnas.2119548119}.

\bibitem[{\citenamefont{He et~al.}(2022)\citenamefont{He, Tanaka, and
  Nagaosa}}]{HeNagaosa22}
\bibinfo{author}{\bibfnamefont{J.~J.} \bibnamefont{He}},
  \bibinfo{author}{\bibfnamefont{Y.}~\bibnamefont{Tanaka}}, \bibnamefont{and}
  \bibinfo{author}{\bibfnamefont{N.}~\bibnamefont{Nagaosa}},
  \bibinfo{journal}{New Journal of Physics} \textbf{\bibinfo{volume}{24}},
  \bibinfo{pages}{053014} (\bibinfo{year}{2022}),
  \urlprefix\url{https://dx.doi.org/10.1088/1367-2630/ac6766}.

\bibitem[{\citenamefont{Ili\ifmmode~\acute{c}\else \'{c}\fi{} and
  Bergeret}(2022)}]{IlicBergeret22}
\bibinfo{author}{\bibfnamefont{S.}~\bibnamefont{Ili\ifmmode~\acute{c}\else
  \'{c}\fi{}}} \bibnamefont{and} \bibinfo{author}{\bibfnamefont{F.~S.}
  \bibnamefont{Bergeret}}, \bibinfo{journal}{Phys. Rev. Lett.}
  \textbf{\bibinfo{volume}{128}}, \bibinfo{pages}{177001}
  (\bibinfo{year}{2022}),
  \urlprefix\url{https://link.aps.org/doi/10.1103/PhysRevLett.128.177001}.

\bibitem[{\citenamefont{Hasan et~al.}(2024)\citenamefont{Hasan, Shaffer,
  Khodas, and Levchenko}}]{HasanShafferKhodasLevchenko24}
\bibinfo{author}{\bibfnamefont{J.}~\bibnamefont{Hasan}},
  \bibinfo{author}{\bibfnamefont{D.}~\bibnamefont{Shaffer}},
  \bibinfo{author}{\bibfnamefont{M.}~\bibnamefont{Khodas}}, \bibnamefont{and}
  \bibinfo{author}{\bibfnamefont{A.}~\bibnamefont{Levchenko}},
  \bibinfo{journal}{Physical Review B} \textbf{\bibinfo{volume}{110}},
  \bibinfo{pages}{024508} (\bibinfo{year}{2024}), \bibinfo{note}{publisher:
  American Physical Society},
  \urlprefix\url{https://link.aps.org/doi/10.1103/PhysRevB.110.024508}.

\bibitem[{\citenamefont{Zhai et~al.}(2022)\citenamefont{Zhai, Li, Wen, Wu, and
  He}}]{Zhai22}
\bibinfo{author}{\bibfnamefont{B.}~\bibnamefont{Zhai}},
  \bibinfo{author}{\bibfnamefont{B.}~\bibnamefont{Li}},
  \bibinfo{author}{\bibfnamefont{Y.}~\bibnamefont{Wen}},
  \bibinfo{author}{\bibfnamefont{F.}~\bibnamefont{Wu}}, \bibnamefont{and}
  \bibinfo{author}{\bibfnamefont{J.}~\bibnamefont{He}}, \bibinfo{journal}{Phys.
  Rev. B} \textbf{\bibinfo{volume}{106}}, \bibinfo{pages}{L140505}
  (\bibinfo{year}{2022}),
  \urlprefix\url{https://link.aps.org/doi/10.1103/PhysRevB.106.L140505}.

\bibitem[{\citenamefont{Samokhin and Truong}(2017)}]{SamokhinTruong17}
\bibinfo{author}{\bibfnamefont{K.~V.} \bibnamefont{Samokhin}} \bibnamefont{and}
  \bibinfo{author}{\bibfnamefont{B.~P.} \bibnamefont{Truong}},
  \bibinfo{journal}{Phys. Rev. B} \textbf{\bibinfo{volume}{96}},
  \bibinfo{pages}{214501} (\bibinfo{year}{2017}),
  \urlprefix\url{https://link.aps.org/doi/10.1103/PhysRevB.96.214501}.

\bibitem[{\citenamefont{Chakraborty and
  Black-Schaffer}(2025)}]{ChakrabortyBlackSchaffer25}
\bibinfo{author}{\bibfnamefont{D.}~\bibnamefont{Chakraborty}} \bibnamefont{and}
  \bibinfo{author}{\bibfnamefont{A.~M.} \bibnamefont{Black-Schaffer}},
  \bibinfo{journal}{Phys. Rev. Lett.} \textbf{\bibinfo{volume}{135}},
  \bibinfo{pages}{026001} (\bibinfo{year}{2025}),
  \urlprefix\url{https://link.aps.org/doi/10.1103/cv8s-tk4c}.

\bibitem[{\citenamefont{Hou et~al.}(2023)\citenamefont{Hou, Nichele, Chi,
  Lodesani, Wu, Ritter, Haxell, Davydova, Ili\ifmmode~\acute{c}\else
  \'{c}\fi{}, Glezakou-Elbert et~al.}}]{HouMoodera23}
\bibinfo{author}{\bibfnamefont{Y.}~\bibnamefont{Hou}},
  \bibinfo{author}{\bibfnamefont{F.}~\bibnamefont{Nichele}},
  \bibinfo{author}{\bibfnamefont{H.}~\bibnamefont{Chi}},
  \bibinfo{author}{\bibfnamefont{A.}~\bibnamefont{Lodesani}},
  \bibinfo{author}{\bibfnamefont{Y.}~\bibnamefont{Wu}},
  \bibinfo{author}{\bibfnamefont{M.~F.} \bibnamefont{Ritter}},
  \bibinfo{author}{\bibfnamefont{D.~Z.} \bibnamefont{Haxell}},
  \bibinfo{author}{\bibfnamefont{M.}~\bibnamefont{Davydova}},
  \bibinfo{author}{\bibfnamefont{S.}~\bibnamefont{Ili\ifmmode~\acute{c}\else
  \'{c}\fi{}}},
  \bibinfo{author}{\bibfnamefont{O.}~\bibnamefont{Glezakou-Elbert}},
  \bibnamefont{et~al.}, \bibinfo{journal}{Phys. Rev. Lett.}
  \textbf{\bibinfo{volume}{131}}, \bibinfo{pages}{027001}
  (\bibinfo{year}{2023}),
  \urlprefix\url{https://link.aps.org/doi/10.1103/PhysRevLett.131.027001}.

\bibitem[{\citenamefont{Gaggioli et~al.}(2024)\citenamefont{Gaggioli, Blatter,
  Novoselov, and Geshkenbein}}]{GaggioliGeshkenbein24}
\bibinfo{author}{\bibfnamefont{F.}~\bibnamefont{Gaggioli}},
  \bibinfo{author}{\bibfnamefont{G.}~\bibnamefont{Blatter}},
  \bibinfo{author}{\bibfnamefont{K.~S.} \bibnamefont{Novoselov}},
  \bibnamefont{and} \bibinfo{author}{\bibfnamefont{V.~B.}
  \bibnamefont{Geshkenbein}}, \bibinfo{journal}{Phys. Rev. Res.}
  \textbf{\bibinfo{volume}{6}}, \bibinfo{pages}{023190} (\bibinfo{year}{2024}),
  \urlprefix\url{https://link.aps.org/doi/10.1103/PhysRevResearch.6.023190}.

\bibitem[{\citenamefont{Gaggioli et~al.}(2025)\citenamefont{Gaggioli, Hou,
  Moodera, and Kamra}}]{GaggioliMoodera25}
\bibinfo{author}{\bibfnamefont{F.}~\bibnamefont{Gaggioli}},
  \bibinfo{author}{\bibfnamefont{Y.}~\bibnamefont{Hou}},
  \bibinfo{author}{\bibfnamefont{J.~S.} \bibnamefont{Moodera}},
  \bibnamefont{and} \bibinfo{author}{\bibfnamefont{A.}~\bibnamefont{Kamra}}
  (\bibinfo{year}{2025}), \eprint{2405.05306},
  \urlprefix\url{https://arxiv.org/abs/2405.05306}.

\bibitem[{\citenamefont{Blatter et~al.}(1994)\citenamefont{Blatter, Feigel'man,
  Geshkenbein, Larkin, and Vinokur}}]{BlatterLarkinRMP94}
\bibinfo{author}{\bibfnamefont{G.}~\bibnamefont{Blatter}},
  \bibinfo{author}{\bibfnamefont{M.~V.} \bibnamefont{Feigel'man}},
  \bibinfo{author}{\bibfnamefont{V.~B.} \bibnamefont{Geshkenbein}},
  \bibinfo{author}{\bibfnamefont{A.~I.} \bibnamefont{Larkin}},
  \bibnamefont{and} \bibinfo{author}{\bibfnamefont{V.~M.}
  \bibnamefont{Vinokur}}, \bibinfo{journal}{Rev. Mod. Phys.}
  \textbf{\bibinfo{volume}{66}}, \bibinfo{pages}{1125} (\bibinfo{year}{1994}),
  \urlprefix\url{https://link.aps.org/doi/10.1103/RevModPhys.66.1125}.

\bibitem[{\citenamefont{Larkin and Varlamov}(2005)}]{LarkinVarlamov05}
\bibinfo{author}{\bibfnamefont{A.}~\bibnamefont{Larkin}} \bibnamefont{and}
  \bibinfo{author}{\bibfnamefont{A.}~\bibnamefont{Varlamov}},
  \emph{\bibinfo{title}{Theory of fluctuations in superconductors}}, vol.
  \bibinfo{volume}{127} (\bibinfo{publisher}{OUP Oxford},
  \bibinfo{year}{2005}).

\bibitem[{\citenamefont{Broussard and Geballe}(1988)}]{Broussard88}
\bibinfo{author}{\bibfnamefont{P.~R.} \bibnamefont{Broussard}}
  \bibnamefont{and} \bibinfo{author}{\bibfnamefont{T.~H.}
  \bibnamefont{Geballe}}, \bibinfo{journal}{Phys. Rev. B}
  \textbf{\bibinfo{volume}{37}}, \bibinfo{pages}{68} (\bibinfo{year}{1988}),
  \urlprefix\url{https://link.aps.org/doi/10.1103/PhysRevB.37.68}.

\bibitem[{\citenamefont{Roas et~al.}(1990)\citenamefont{Roas, Schultz, and
  Saemann-Ischenko}}]{Roas90}
\bibinfo{author}{\bibfnamefont{B.}~\bibnamefont{Roas}},
  \bibinfo{author}{\bibfnamefont{L.}~\bibnamefont{Schultz}}, \bibnamefont{and}
  \bibinfo{author}{\bibfnamefont{G.}~\bibnamefont{Saemann-Ischenko}},
  \bibinfo{journal}{Phys. Rev. Lett.} \textbf{\bibinfo{volume}{64}},
  \bibinfo{pages}{479} (\bibinfo{year}{1990}),
  \urlprefix\url{https://link.aps.org/doi/10.1103/PhysRevLett.64.479}.

\bibitem[{\citenamefont{Jiang et~al.}(1994)\citenamefont{Jiang, Connolly,
  Hagen, and Lobb}}]{Jiang94}
\bibinfo{author}{\bibfnamefont{X.}~\bibnamefont{Jiang}},
  \bibinfo{author}{\bibfnamefont{P.~J.} \bibnamefont{Connolly}},
  \bibinfo{author}{\bibfnamefont{S.~J.} \bibnamefont{Hagen}}, \bibnamefont{and}
  \bibinfo{author}{\bibfnamefont{C.~J.} \bibnamefont{Lobb}},
  \bibinfo{journal}{Phys. Rev. B} \textbf{\bibinfo{volume}{49}},
  \bibinfo{pages}{9244} (\bibinfo{year}{1994}),
  \urlprefix\url{https://link.aps.org/doi/10.1103/PhysRevB.49.9244}.

\bibitem[{\citenamefont{Plourde et~al.}(2001)\citenamefont{Plourde,
  Van~Harlingen, Vodolazov, Besseling, Hesselberth, and
  Kes}}]{PlourdeVodolazov01}
\bibinfo{author}{\bibfnamefont{B.~L.~T.} \bibnamefont{Plourde}},
  \bibinfo{author}{\bibfnamefont{D.~J.} \bibnamefont{Van~Harlingen}},
  \bibinfo{author}{\bibfnamefont{D.~Y.} \bibnamefont{Vodolazov}},
  \bibinfo{author}{\bibfnamefont{R.}~\bibnamefont{Besseling}},
  \bibinfo{author}{\bibfnamefont{M.~B.~S.} \bibnamefont{Hesselberth}},
  \bibnamefont{and} \bibinfo{author}{\bibfnamefont{P.~H.} \bibnamefont{Kes}},
  \bibinfo{journal}{Phys. Rev. B} \textbf{\bibinfo{volume}{64}},
  \bibinfo{pages}{014503} (\bibinfo{year}{2001}),
  \urlprefix\url{https://link.aps.org/doi/10.1103/PhysRevB.64.014503}.

\bibitem[{\citenamefont{Touitou et~al.}(2004)\citenamefont{Touitou, Bernstein,
  Hamet, Simon, M{\'e}chin, Contour, and Jacquet}}]{Touitou04}
\bibinfo{author}{\bibfnamefont{N.}~\bibnamefont{Touitou}},
  \bibinfo{author}{\bibfnamefont{P.}~\bibnamefont{Bernstein}},
  \bibinfo{author}{\bibfnamefont{J.}~\bibnamefont{Hamet}},
  \bibinfo{author}{\bibfnamefont{C.}~\bibnamefont{Simon}},
  \bibinfo{author}{\bibfnamefont{L.}~\bibnamefont{M{\'e}chin}},
  \bibinfo{author}{\bibfnamefont{J.}~\bibnamefont{Contour}}, \bibnamefont{and}
  \bibinfo{author}{\bibfnamefont{E.}~\bibnamefont{Jacquet}},
  \bibinfo{journal}{Applied physics letters} \textbf{\bibinfo{volume}{85}},
  \bibinfo{pages}{1742} (\bibinfo{year}{2004}).

\bibitem[{\citenamefont{Morelle and Moshchalkov}(2006)}]{MorelleMoshchalkov06}
\bibinfo{author}{\bibfnamefont{M.}~\bibnamefont{Morelle}} \bibnamefont{and}
  \bibinfo{author}{\bibfnamefont{V.~V.} \bibnamefont{Moshchalkov}},
  \bibinfo{journal}{Applied Physics Letters} \textbf{\bibinfo{volume}{88}},
  \bibinfo{pages}{172507} (\bibinfo{year}{2006}), ISSN
  \bibinfo{issn}{0003-6951},
  \eprint{https://pubs.aip.org/aip/apl/article-pdf/doi/10.1063/1.2199468/14656756/172507\_1\_online.pdf},
  \urlprefix\url{https://doi.org/10.1063/1.2199468}.

\bibitem[{\citenamefont{Papon et~al.}(2008)\citenamefont{Papon, Senapati, and
  Barber}}]{Papon08}
\bibinfo{author}{\bibfnamefont{A.}~\bibnamefont{Papon}},
  \bibinfo{author}{\bibfnamefont{K.}~\bibnamefont{Senapati}}, \bibnamefont{and}
  \bibinfo{author}{\bibfnamefont{Z.}~\bibnamefont{Barber}},
  \bibinfo{journal}{Applied Physics Letters} \textbf{\bibinfo{volume}{93}}
  (\bibinfo{year}{2008}).

\bibitem[{\citenamefont{Harrington et~al.}(2009)\citenamefont{Harrington,
  MacManus-Driscoll, and Durrell}}]{Harrington09}
\bibinfo{author}{\bibfnamefont{S.~A.} \bibnamefont{Harrington}},
  \bibinfo{author}{\bibfnamefont{J.~L.} \bibnamefont{MacManus-Driscoll}},
  \bibnamefont{and} \bibinfo{author}{\bibfnamefont{J.~H.}
  \bibnamefont{Durrell}}, \bibinfo{journal}{Applied Physics Letters}
  \textbf{\bibinfo{volume}{95}}, \bibinfo{pages}{022518}
  (\bibinfo{year}{2009}), ISSN \bibinfo{issn}{0003-6951},
  \eprint{https://pubs.aip.org/aip/apl/article-pdf/doi/10.1063/1.3182735/13668412/022518\_1\_online.pdf},
  \urlprefix\url{https://doi.org/10.1063/1.3182735}.

\bibitem[{\citenamefont{Adami et~al.}(2013)\citenamefont{Adami, Cerbu,
  Cabosart, Motta, Cuppens, Ortiz, Moshchalkov, Hackens, Delamare, Van~de
  Vondel et~al.}}]{Adami13}
\bibinfo{author}{\bibfnamefont{O.-A.} \bibnamefont{Adami}},
  \bibinfo{author}{\bibfnamefont{D.}~\bibnamefont{Cerbu}},
  \bibinfo{author}{\bibfnamefont{D.}~\bibnamefont{Cabosart}},
  \bibinfo{author}{\bibfnamefont{M.}~\bibnamefont{Motta}},
  \bibinfo{author}{\bibfnamefont{J.}~\bibnamefont{Cuppens}},
  \bibinfo{author}{\bibfnamefont{W.~A.} \bibnamefont{Ortiz}},
  \bibinfo{author}{\bibfnamefont{V.}~\bibnamefont{Moshchalkov}},
  \bibinfo{author}{\bibfnamefont{B.}~\bibnamefont{Hackens}},
  \bibinfo{author}{\bibfnamefont{R.}~\bibnamefont{Delamare}},
  \bibinfo{author}{\bibfnamefont{J.}~\bibnamefont{Van~de Vondel}},
  \bibnamefont{et~al.}, \bibinfo{journal}{Applied Physics Letters}
  \textbf{\bibinfo{volume}{102}} (\bibinfo{year}{2013}).

\bibitem[{\citenamefont{Cerbu et~al.}(2013)\citenamefont{Cerbu, Gladilin,
  Cuppens, Fritzsche, Tempere, Devreese, Moshchalkov, Silhanek, and Van~de
  Vondel}}]{CerbuMoshchalkov13}
\bibinfo{author}{\bibfnamefont{D.}~\bibnamefont{Cerbu}},
  \bibinfo{author}{\bibfnamefont{V.}~\bibnamefont{Gladilin}},
  \bibinfo{author}{\bibfnamefont{J.}~\bibnamefont{Cuppens}},
  \bibinfo{author}{\bibfnamefont{J.}~\bibnamefont{Fritzsche}},
  \bibinfo{author}{\bibfnamefont{J.}~\bibnamefont{Tempere}},
  \bibinfo{author}{\bibfnamefont{J.}~\bibnamefont{Devreese}},
  \bibinfo{author}{\bibfnamefont{V.}~\bibnamefont{Moshchalkov}},
  \bibinfo{author}{\bibfnamefont{A.}~\bibnamefont{Silhanek}}, \bibnamefont{and}
  \bibinfo{author}{\bibfnamefont{J.}~\bibnamefont{Van~de Vondel}},
  \bibinfo{journal}{New Journal of Physics} \textbf{\bibinfo{volume}{15}},
  \bibinfo{pages}{063022} (\bibinfo{year}{2013}).

\bibitem[{\citenamefont{Sivakov et~al.}(2018)\citenamefont{Sivakov, Turutanov,
  Kolinko, and Pokhila}}]{Sivakov18}
\bibinfo{author}{\bibfnamefont{A.~G.} \bibnamefont{Sivakov}},
  \bibinfo{author}{\bibfnamefont{O.~G.} \bibnamefont{Turutanov}},
  \bibinfo{author}{\bibfnamefont{A.~E.} \bibnamefont{Kolinko}},
  \bibnamefont{and} \bibinfo{author}{\bibfnamefont{A.~S.}
  \bibnamefont{Pokhila}}, \bibinfo{journal}{Low Temperature Physics}
  \textbf{\bibinfo{volume}{44}}, \bibinfo{pages}{226} (\bibinfo{year}{2018}),
  ISSN \bibinfo{issn}{1063-777X},
  \eprint{https://pubs.aip.org/aip/ltp/article-pdf/44/3/226/15667842/226\_1\_online.pdf},
  \urlprefix\url{https://doi.org/10.1063/1.5024540}.

\bibitem[{\citenamefont{Suri et~al.}(2022)\citenamefont{Suri, Kamra, Meier,
  Kronseder, Belzig, Back, and Strunk}}]{SuriStrunk22}
\bibinfo{author}{\bibfnamefont{D.}~\bibnamefont{Suri}},
  \bibinfo{author}{\bibfnamefont{A.}~\bibnamefont{Kamra}},
  \bibinfo{author}{\bibfnamefont{T.~N.} \bibnamefont{Meier}},
  \bibinfo{author}{\bibfnamefont{M.}~\bibnamefont{Kronseder}},
  \bibinfo{author}{\bibfnamefont{W.}~\bibnamefont{Belzig}},
  \bibinfo{author}{\bibfnamefont{C.~H.} \bibnamefont{Back}}, \bibnamefont{and}
  \bibinfo{author}{\bibfnamefont{C.}~\bibnamefont{Strunk}},
  \bibinfo{journal}{Applied Physics Letters} \textbf{\bibinfo{volume}{121}}
  (\bibinfo{year}{2022}).

\bibitem[{\citenamefont{Gutfreund et~al.}(2023)\citenamefont{Gutfreund,
  Matsuki, Plastovets, Noah, Gorzawski, Fridman, Yang, Buzdin, Millo, Robinson
  et~al.}}]{GutfreundBuzdin23}
\bibinfo{author}{\bibfnamefont{A.}~\bibnamefont{Gutfreund}},
  \bibinfo{author}{\bibfnamefont{H.}~\bibnamefont{Matsuki}},
  \bibinfo{author}{\bibfnamefont{V.}~\bibnamefont{Plastovets}},
  \bibinfo{author}{\bibfnamefont{A.}~\bibnamefont{Noah}},
  \bibinfo{author}{\bibfnamefont{L.}~\bibnamefont{Gorzawski}},
  \bibinfo{author}{\bibfnamefont{N.}~\bibnamefont{Fridman}},
  \bibinfo{author}{\bibfnamefont{G.}~\bibnamefont{Yang}},
  \bibinfo{author}{\bibfnamefont{A.}~\bibnamefont{Buzdin}},
  \bibinfo{author}{\bibfnamefont{O.}~\bibnamefont{Millo}},
  \bibinfo{author}{\bibfnamefont{J.~W.~A.} \bibnamefont{Robinson}},
  \bibnamefont{et~al.}, \bibinfo{journal}{Nature Communications}
  \textbf{\bibinfo{volume}{14}}, \bibinfo{pages}{1630} (\bibinfo{year}{2023}),
  ISSN \bibinfo{issn}{2041-1723}, \bibinfo{note}{number: 1 Publisher: Nature
  Publishing Group},
  \urlprefix\url{https://www.nature.com/articles/s41467-023-37294-2}.

\bibitem[{\citenamefont{Goldobin et~al.}(2001)\citenamefont{Goldobin, Sterck,
  and Koelle}}]{Goldobin01}
\bibinfo{author}{\bibfnamefont{E.}~\bibnamefont{Goldobin}},
  \bibinfo{author}{\bibfnamefont{A.}~\bibnamefont{Sterck}}, \bibnamefont{and}
  \bibinfo{author}{\bibfnamefont{D.}~\bibnamefont{Koelle}},
  \bibinfo{journal}{Physical Review E} \textbf{\bibinfo{volume}{63}},
  \bibinfo{pages}{031111} (\bibinfo{year}{2001}), \bibinfo{note}{publisher:
  American Physical Society},
  \urlprefix\url{https://link.aps.org/doi/10.1103/PhysRevE.63.031111}.

\bibitem[{\citenamefont{Shaju and Kuriakose}(2003)}]{Shaju03}
\bibinfo{author}{\bibfnamefont{P.~D.} \bibnamefont{Shaju}} \bibnamefont{and}
  \bibinfo{author}{\bibfnamefont{V.~C.} \bibnamefont{Kuriakose}},
  \bibinfo{journal}{Superconductor Science and Technology}
  \textbf{\bibinfo{volume}{16}}, \bibinfo{pages}{L25} (\bibinfo{year}{2003}),
  \urlprefix\url{https://dx.doi.org/10.1088/0953-2048/16/5/103}.

\bibitem[{\citenamefont{Wang et~al.}(2009)\citenamefont{Wang, Zhu, G\"urlich,
  Ruoff, Kim, Hatano, Zhao, Zhao, Goldobin, Koelle et~al.}}]{WangGoldobin09}
\bibinfo{author}{\bibfnamefont{H.~B.} \bibnamefont{Wang}},
  \bibinfo{author}{\bibfnamefont{B.~Y.} \bibnamefont{Zhu}},
  \bibinfo{author}{\bibfnamefont{C.}~\bibnamefont{G\"urlich}},
  \bibinfo{author}{\bibfnamefont{M.}~\bibnamefont{Ruoff}},
  \bibinfo{author}{\bibfnamefont{S.}~\bibnamefont{Kim}},
  \bibinfo{author}{\bibfnamefont{T.}~\bibnamefont{Hatano}},
  \bibinfo{author}{\bibfnamefont{B.~R.} \bibnamefont{Zhao}},
  \bibinfo{author}{\bibfnamefont{Z.~X.} \bibnamefont{Zhao}},
  \bibinfo{author}{\bibfnamefont{E.}~\bibnamefont{Goldobin}},
  \bibinfo{author}{\bibfnamefont{D.}~\bibnamefont{Koelle}},
  \bibnamefont{et~al.}, \bibinfo{journal}{Phys. Rev. B}
  \textbf{\bibinfo{volume}{80}}, \bibinfo{pages}{224507}
  (\bibinfo{year}{2009}),
  \urlprefix\url{https://link.aps.org/doi/10.1103/PhysRevB.80.224507}.

\bibitem[{\citenamefont{Chen et~al.}(2024{\natexlab{a}})\citenamefont{Chen,
  Park, Vool, Maksimovic, Broadway, Flaks, Zhou, Maletinsky, Stern, Halperin
  et~al.}}]{ChenYacoby24}
\bibinfo{author}{\bibfnamefont{S.}~\bibnamefont{Chen}},
  \bibinfo{author}{\bibfnamefont{S.}~\bibnamefont{Park}},
  \bibinfo{author}{\bibfnamefont{U.}~\bibnamefont{Vool}},
  \bibinfo{author}{\bibfnamefont{N.}~\bibnamefont{Maksimovic}},
  \bibinfo{author}{\bibfnamefont{D.~A.} \bibnamefont{Broadway}},
  \bibinfo{author}{\bibfnamefont{M.}~\bibnamefont{Flaks}},
  \bibinfo{author}{\bibfnamefont{T.~X.} \bibnamefont{Zhou}},
  \bibinfo{author}{\bibfnamefont{P.}~\bibnamefont{Maletinsky}},
  \bibinfo{author}{\bibfnamefont{A.}~\bibnamefont{Stern}},
  \bibinfo{author}{\bibfnamefont{B.~I.} \bibnamefont{Halperin}},
  \bibnamefont{et~al.}, \bibinfo{journal}{Nature Communications}
  \textbf{\bibinfo{volume}{15}}, \bibinfo{pages}{8059}
  (\bibinfo{year}{2024}{\natexlab{a}}).

\bibitem[{\citenamefont{Olson et~al.}(2001)\citenamefont{Olson, Reichhardt,
  Jank\'o, and Nori}}]{OlsonReichhardt01}
\bibinfo{author}{\bibfnamefont{C.~J.} \bibnamefont{Olson}},
  \bibinfo{author}{\bibfnamefont{C.}~\bibnamefont{Reichhardt}},
  \bibinfo{author}{\bibfnamefont{B.}~\bibnamefont{Jank\'o}}, \bibnamefont{and}
  \bibinfo{author}{\bibfnamefont{F.}~\bibnamefont{Nori}},
  \bibinfo{journal}{Phys. Rev. Lett.} \textbf{\bibinfo{volume}{87}},
  \bibinfo{pages}{177002} (\bibinfo{year}{2001}),
  \urlprefix\url{https://link.aps.org/doi/10.1103/PhysRevLett.87.177002}.

\bibitem[{\citenamefont{Kadin}(1990)}]{Kadin90}
\bibinfo{author}{\bibfnamefont{A.~M.} \bibnamefont{Kadin}},
  \bibinfo{journal}{Journal of Applied Physics} \textbf{\bibinfo{volume}{68}},
  \bibinfo{pages}{5741} (\bibinfo{year}{1990}), ISSN \bibinfo{issn}{0021-8979},
  \eprint{https://pubs.aip.org/aip/jap/article-pdf/68/11/5741/18638064/5741\_1\_online.pdf},
  \urlprefix\url{https://doi.org/10.1063/1.346969}.

\bibitem[{\citenamefont{Raissi and Nordman}(1994)}]{RaissiNordman94}
\bibinfo{author}{\bibfnamefont{F.}~\bibnamefont{Raissi}} \bibnamefont{and}
  \bibinfo{author}{\bibfnamefont{J.~E.} \bibnamefont{Nordman}},
  \bibinfo{journal}{Applied Physics Letters} \textbf{\bibinfo{volume}{65}},
  \bibinfo{pages}{1838} (\bibinfo{year}{1994}), ISSN \bibinfo{issn}{0003-6951},
  \urlprefix\url{https://doi.org/10.1063/1.112859}.

\bibitem[{\citenamefont{Guarcello et~al.}(2024)\citenamefont{Guarcello, Pagano,
  and Filatrella}}]{GuarcelloFilatrella24}
\bibinfo{author}{\bibfnamefont{C.}~\bibnamefont{Guarcello}},
  \bibinfo{author}{\bibfnamefont{S.}~\bibnamefont{Pagano}}, \bibnamefont{and}
  \bibinfo{author}{\bibfnamefont{G.}~\bibnamefont{Filatrella}},
  \bibinfo{journal}{Applied Physics Letters} \textbf{\bibinfo{volume}{124}},
  \bibinfo{pages}{162601} (\bibinfo{year}{2024}), ISSN
  \bibinfo{issn}{0003-6951}, \urlprefix\url{https://doi.org/10.1063/5.0211230}.

\bibitem[{\citenamefont{Fukaya et~al.}(2024)\citenamefont{Fukaya, Mercaldo,
  Margineda, Crippa, Strambini, Giazotto, Ortix, and
  Cuoco}}]{FukayaOrtixCuoco24}
\bibinfo{author}{\bibfnamefont{Y.}~\bibnamefont{Fukaya}},
  \bibinfo{author}{\bibfnamefont{M.~T.} \bibnamefont{Mercaldo}},
  \bibinfo{author}{\bibfnamefont{D.}~\bibnamefont{Margineda}},
  \bibinfo{author}{\bibfnamefont{A.}~\bibnamefont{Crippa}},
  \bibinfo{author}{\bibfnamefont{E.}~\bibnamefont{Strambini}},
  \bibinfo{author}{\bibfnamefont{F.}~\bibnamefont{Giazotto}},
  \bibinfo{author}{\bibfnamefont{C.}~\bibnamefont{Ortix}}, \bibnamefont{and}
  \bibinfo{author}{\bibfnamefont{M.}~\bibnamefont{Cuoco}}
  (\bibinfo{year}{2024}), \bibinfo{note}{arXiv:2403.04421 [cond-mat]},
  \urlprefix\url{http://arxiv.org/abs/2403.04421}.

\bibitem[{\citenamefont{Lee et~al.}(1999)\citenamefont{Lee, Jank{\'o},
  Der{\'e}nyi, and Barab{\'a}si}}]{LeeBarabasi99}
\bibinfo{author}{\bibfnamefont{C.~S.} \bibnamefont{Lee}},
  \bibinfo{author}{\bibfnamefont{B.}~\bibnamefont{Jank{\'o}}},
  \bibinfo{author}{\bibfnamefont{I.}~\bibnamefont{Der{\'e}nyi}},
  \bibnamefont{and} \bibinfo{author}{\bibfnamefont{A.~L.}
  \bibnamefont{Barab{\'a}si}}, \bibinfo{journal}{Nature}
  \textbf{\bibinfo{volume}{400}}, \bibinfo{pages}{337} (\bibinfo{year}{1999}),
  \urlprefix\url{https://doi.org/10.1038/22485}.

\bibitem[{\citenamefont{Shklovskij et~al.}(2013)\citenamefont{Shklovskij,
  Sosedkin, and Dobrovolskiy}}]{Shklovskij14}
\bibinfo{author}{\bibfnamefont{V.~A.} \bibnamefont{Shklovskij}},
  \bibinfo{author}{\bibfnamefont{V.~V.} \bibnamefont{Sosedkin}},
  \bibnamefont{and} \bibinfo{author}{\bibfnamefont{O.~V.}
  \bibnamefont{Dobrovolskiy}}, \bibinfo{journal}{Journal of Physics: Condensed
  Matter} \textbf{\bibinfo{volume}{26}}, \bibinfo{pages}{025703}
  (\bibinfo{year}{2013}),
  \urlprefix\url{https://dx.doi.org/10.1088/0953-8984/26/2/025703}.

\bibitem[{\citenamefont{Reichhardt et~al.}(2015)\citenamefont{Reichhardt, Ray,
  and Reichhardt}}]{Reichhardt15}
\bibinfo{author}{\bibfnamefont{C.}~\bibnamefont{Reichhardt}},
  \bibinfo{author}{\bibfnamefont{D.}~\bibnamefont{Ray}}, \bibnamefont{and}
  \bibinfo{author}{\bibfnamefont{C.~J.~O.} \bibnamefont{Reichhardt}},
  \bibinfo{journal}{Phys. Rev. B} \textbf{\bibinfo{volume}{91}},
  \bibinfo{pages}{184502} (\bibinfo{year}{2015}),
  \urlprefix\url{https://link.aps.org/doi/10.1103/PhysRevB.91.184502}.

\bibitem[{\citenamefont{He et~al.}(2019)\citenamefont{He, Xue, and
  Zhou}}]{He19}
\bibinfo{author}{\bibfnamefont{A.}~\bibnamefont{He}},
  \bibinfo{author}{\bibfnamefont{C.}~\bibnamefont{Xue}}, \bibnamefont{and}
  \bibinfo{author}{\bibfnamefont{Y.-H.} \bibnamefont{Zhou}},
  \bibinfo{journal}{Applied Physics Letters} \textbf{\bibinfo{volume}{115}}
  (\bibinfo{year}{2019}).

\bibitem[{\citenamefont{Bogush et~al.}(2025)\citenamefont{Bogush, de~Bragança,
  Fomin, and Dobrovolskiy}}]{Bogush25}
\bibinfo{author}{\bibfnamefont{I.}~\bibnamefont{Bogush}},
  \bibinfo{author}{\bibfnamefont{R.~H.} \bibnamefont{de~Bragança}},
  \bibinfo{author}{\bibfnamefont{V.~M.} \bibnamefont{Fomin}}, \bibnamefont{and}
  \bibinfo{author}{\bibfnamefont{O.~V.} \bibnamefont{Dobrovolskiy}}
  (\bibinfo{year}{2025}), \eprint{2407.20780},
  \urlprefix\url{https://arxiv.org/abs/2407.20780}.

\bibitem[{\citenamefont{Koshelev}(2025)}]{Koshelev25}
\bibinfo{author}{\bibfnamefont{A.~E.} \bibnamefont{Koshelev}}
  (\bibinfo{year}{2025}), \eprint{2504.09321},
  \urlprefix\url{https://arxiv.org/abs/2504.09321}.

\bibitem[{\citenamefont{Villegas et~al.}(2003)\citenamefont{Villegas, Savel'ev,
  Nori, Gonzalez, Anguita, Garcia, and Vicent}}]{Villegas03}
\bibinfo{author}{\bibfnamefont{J.}~\bibnamefont{Villegas}},
  \bibinfo{author}{\bibfnamefont{S.}~\bibnamefont{Savel'ev}},
  \bibinfo{author}{\bibfnamefont{F.}~\bibnamefont{Nori}},
  \bibinfo{author}{\bibfnamefont{E.}~\bibnamefont{Gonzalez}},
  \bibinfo{author}{\bibfnamefont{J.}~\bibnamefont{Anguita}},
  \bibinfo{author}{\bibfnamefont{R.}~\bibnamefont{Garcia}}, \bibnamefont{and}
  \bibinfo{author}{\bibfnamefont{J.}~\bibnamefont{Vicent}},
  \bibinfo{journal}{science} \textbf{\bibinfo{volume}{302}},
  \bibinfo{pages}{1188} (\bibinfo{year}{2003}).

\bibitem[{\citenamefont{W\"ordenweber et~al.}(2004)\citenamefont{W\"ordenweber,
  Dymashevski, and Misko}}]{Wordenweber04}
\bibinfo{author}{\bibfnamefont{R.}~\bibnamefont{W\"ordenweber}},
  \bibinfo{author}{\bibfnamefont{P.}~\bibnamefont{Dymashevski}},
  \bibnamefont{and} \bibinfo{author}{\bibfnamefont{V.~R.} \bibnamefont{Misko}},
  \bibinfo{journal}{Phys. Rev. B} \textbf{\bibinfo{volume}{69}},
  \bibinfo{pages}{184504} (\bibinfo{year}{2004}),
  \urlprefix\url{https://link.aps.org/doi/10.1103/PhysRevB.69.184504}.

\bibitem[{\citenamefont{Van~de Vondel et~al.}(2005)\citenamefont{Van~de Vondel,
  de~Souza~Silva, Zhu, Morelle, and Moshchalkov}}]{VanDeVondel05}
\bibinfo{author}{\bibfnamefont{J.}~\bibnamefont{Van~de Vondel}},
  \bibinfo{author}{\bibfnamefont{C.~C.} \bibnamefont{de~Souza~Silva}},
  \bibinfo{author}{\bibfnamefont{B.~Y.} \bibnamefont{Zhu}},
  \bibinfo{author}{\bibfnamefont{M.}~\bibnamefont{Morelle}}, \bibnamefont{and}
  \bibinfo{author}{\bibfnamefont{V.~V.} \bibnamefont{Moshchalkov}},
  \bibinfo{journal}{Phys. Rev. Lett.} \textbf{\bibinfo{volume}{94}},
  \bibinfo{pages}{057003} (\bibinfo{year}{2005}),
  \urlprefix\url{https://link.aps.org/doi/10.1103/PhysRevLett.94.057003}.

\bibitem[{\citenamefont{Pryadun et~al.}(2006)\citenamefont{Pryadun, Sierra,
  Aliev, Golubovic, and Moshchalkov}}]{PryadunMoshchalkov06}
\bibinfo{author}{\bibfnamefont{V.~V.} \bibnamefont{Pryadun}},
  \bibinfo{author}{\bibfnamefont{J.}~\bibnamefont{Sierra}},
  \bibinfo{author}{\bibfnamefont{F.~G.} \bibnamefont{Aliev}},
  \bibinfo{author}{\bibfnamefont{D.~S.} \bibnamefont{Golubovic}},
  \bibnamefont{and} \bibinfo{author}{\bibfnamefont{V.~V.}
  \bibnamefont{Moshchalkov}}, \bibinfo{journal}{Applied Physics Letters}
  \textbf{\bibinfo{volume}{88}}, \bibinfo{pages}{062517}
  (\bibinfo{year}{2006}), ISSN \bibinfo{issn}{0003-6951},
  \eprint{https://pubs.aip.org/aip/apl/article-pdf/doi/10.1063/1.2171788/14652461/062517\_1\_online.pdf},
  \urlprefix\url{https://doi.org/10.1063/1.2171788}.

\bibitem[{\citenamefont{de~Souza~Silva
  et~al.}(2007)\citenamefont{de~Souza~Silva, Silhanek, Van~de Vondel, Gillijns,
  Metlushko, Ilic, and Moshchalkov}}]{deSouzaSilvaMoshchalkov07}
\bibinfo{author}{\bibfnamefont{C.~C.} \bibnamefont{de~Souza~Silva}},
  \bibinfo{author}{\bibfnamefont{A.~V.} \bibnamefont{Silhanek}},
  \bibinfo{author}{\bibfnamefont{J.}~\bibnamefont{Van~de Vondel}},
  \bibinfo{author}{\bibfnamefont{W.}~\bibnamefont{Gillijns}},
  \bibinfo{author}{\bibfnamefont{V.}~\bibnamefont{Metlushko}},
  \bibinfo{author}{\bibfnamefont{B.}~\bibnamefont{Ilic}}, \bibnamefont{and}
  \bibinfo{author}{\bibfnamefont{V.~V.} \bibnamefont{Moshchalkov}},
  \bibinfo{journal}{Phys. Rev. Lett.} \textbf{\bibinfo{volume}{98}},
  \bibinfo{pages}{117005} (\bibinfo{year}{2007}),
  \urlprefix\url{https://link.aps.org/doi/10.1103/PhysRevLett.98.117005}.

\bibitem[{\citenamefont{Gillijns et~al.}(2007)\citenamefont{Gillijns, Silhanek,
  Moshchalkov, Reichhardt, and Reichhardt}}]{GillinjnsMoshchalkovReichhardt07}
\bibinfo{author}{\bibfnamefont{W.}~\bibnamefont{Gillijns}},
  \bibinfo{author}{\bibfnamefont{A.~V.} \bibnamefont{Silhanek}},
  \bibinfo{author}{\bibfnamefont{V.~V.} \bibnamefont{Moshchalkov}},
  \bibinfo{author}{\bibfnamefont{C.~J.~O.} \bibnamefont{Reichhardt}},
  \bibnamefont{and}
  \bibinfo{author}{\bibfnamefont{C.}~\bibnamefont{Reichhardt}},
  \bibinfo{journal}{Phys. Rev. Lett.} \textbf{\bibinfo{volume}{99}},
  \bibinfo{pages}{247002} (\bibinfo{year}{2007}),
  \urlprefix\url{https://link.aps.org/doi/10.1103/PhysRevLett.99.247002}.

\bibitem[{\citenamefont{Lyu et~al.}(2021)\citenamefont{Lyu, Jiang, Wang, Xiao,
  Dong, Chen, Milo{\v{s}}evi{\'c}, Wang, Divan, Pearson et~al.}}]{Lyu21}
\bibinfo{author}{\bibfnamefont{Y.-Y.} \bibnamefont{Lyu}},
  \bibinfo{author}{\bibfnamefont{J.}~\bibnamefont{Jiang}},
  \bibinfo{author}{\bibfnamefont{Y.-L.} \bibnamefont{Wang}},
  \bibinfo{author}{\bibfnamefont{Z.-L.} \bibnamefont{Xiao}},
  \bibinfo{author}{\bibfnamefont{S.}~\bibnamefont{Dong}},
  \bibinfo{author}{\bibfnamefont{Q.-H.} \bibnamefont{Chen}},
  \bibinfo{author}{\bibfnamefont{M.~V.} \bibnamefont{Milo{\v{s}}evi{\'c}}},
  \bibinfo{author}{\bibfnamefont{H.}~\bibnamefont{Wang}},
  \bibinfo{author}{\bibfnamefont{R.}~\bibnamefont{Divan}},
  \bibinfo{author}{\bibfnamefont{J.~E.} \bibnamefont{Pearson}},
  \bibnamefont{et~al.}, \bibinfo{journal}{Nature communications}
  \textbf{\bibinfo{volume}{12}}, \bibinfo{pages}{2703} (\bibinfo{year}{2021}).

\bibitem[{\citenamefont{Dobrovolskiy and Chumak}(2022)}]{Dobrovolskiy22}
\bibinfo{author}{\bibfnamefont{O.~V.} \bibnamefont{Dobrovolskiy}}
  \bibnamefont{and} \bibinfo{author}{\bibfnamefont{A.~V.}
  \bibnamefont{Chumak}}, \bibinfo{journal}{Journal of Magnetism and Magnetic
  Materials} \textbf{\bibinfo{volume}{543}}, \bibinfo{pages}{168633}
  (\bibinfo{year}{2022}), ISSN \bibinfo{issn}{0304-8853},
  \urlprefix\url{https://www.sciencedirect.com/science/article/pii/S0304885321008726}.

\bibitem[{\citenamefont{Kubo}(2023)}]{Kubo23}
\bibinfo{author}{\bibfnamefont{T.}~\bibnamefont{Kubo}}, \bibinfo{journal}{Phys.
  Rev. Appl.} \textbf{\bibinfo{volume}{20}}, \bibinfo{pages}{034033}
  (\bibinfo{year}{2023}),
  \urlprefix\url{https://link.aps.org/doi/10.1103/PhysRevApplied.20.034033}.

\bibitem[{\citenamefont{Wambaugh et~al.}(1999)\citenamefont{Wambaugh,
  Reichhardt, Olson, Marchesoni, and Nori}}]{WambaughReichhardt99}
\bibinfo{author}{\bibfnamefont{J.~F.} \bibnamefont{Wambaugh}},
  \bibinfo{author}{\bibfnamefont{C.}~\bibnamefont{Reichhardt}},
  \bibinfo{author}{\bibfnamefont{C.~J.} \bibnamefont{Olson}},
  \bibinfo{author}{\bibfnamefont{F.}~\bibnamefont{Marchesoni}},
  \bibnamefont{and} \bibinfo{author}{\bibfnamefont{F.}~\bibnamefont{Nori}},
  \bibinfo{journal}{Phys. Rev. Lett.} \textbf{\bibinfo{volume}{83}},
  \bibinfo{pages}{5106} (\bibinfo{year}{1999}),
  \urlprefix\url{https://link.aps.org/doi/10.1103/PhysRevLett.83.5106}.

\bibitem[{\citenamefont{Clem et~al.}(2012)\citenamefont{Clem, Mawatari,
  Berdiyorov, and Peeters}}]{ClemPeeters12}
\bibinfo{author}{\bibfnamefont{J.~R.} \bibnamefont{Clem}},
  \bibinfo{author}{\bibfnamefont{Y.}~\bibnamefont{Mawatari}},
  \bibinfo{author}{\bibfnamefont{G.~R.} \bibnamefont{Berdiyorov}},
  \bibnamefont{and} \bibinfo{author}{\bibfnamefont{F.~M.}
  \bibnamefont{Peeters}}, \bibinfo{journal}{Phys. Rev. B}
  \textbf{\bibinfo{volume}{85}}, \bibinfo{pages}{144511}
  (\bibinfo{year}{2012}),
  \urlprefix\url{https://link.aps.org/doi/10.1103/PhysRevB.85.144511}.

\bibitem[{\citenamefont{Jiang et~al.}(2021)\citenamefont{Jiang, Wang,
  Milo\ifmmode \check{s}\else \v{s}\fi{}evi\ifmmode~\acute{c}\else \'{c}\fi{},
  Xiao, Peeters, and Chen}}]{Jiang21}
\bibinfo{author}{\bibfnamefont{J.}~\bibnamefont{Jiang}},
  \bibinfo{author}{\bibfnamefont{Y.-L.} \bibnamefont{Wang}},
  \bibinfo{author}{\bibfnamefont{M.~V.} \bibnamefont{Milo\ifmmode
  \check{s}\else \v{s}\fi{}evi\ifmmode~\acute{c}\else \'{c}\fi{}}},
  \bibinfo{author}{\bibfnamefont{Z.-L.} \bibnamefont{Xiao}},
  \bibinfo{author}{\bibfnamefont{F.~M.} \bibnamefont{Peeters}},
  \bibnamefont{and} \bibinfo{author}{\bibfnamefont{Q.-H.} \bibnamefont{Chen}},
  \bibinfo{journal}{Phys. Rev. B} \textbf{\bibinfo{volume}{103}},
  \bibinfo{pages}{014502} (\bibinfo{year}{2021}),
  \urlprefix\url{https://link.aps.org/doi/10.1103/PhysRevB.103.014502}.

\bibitem[{\citenamefont{Cadorim et~al.}(2024)\citenamefont{Cadorim, Sardella,
  and Silva}}]{Cadorim24}
\bibinfo{author}{\bibfnamefont{L.~R.} \bibnamefont{Cadorim}},
  \bibinfo{author}{\bibfnamefont{E.}~\bibnamefont{Sardella}}, \bibnamefont{and}
  \bibinfo{author}{\bibfnamefont{C.~C. d.~S.} \bibnamefont{Silva}},
  \bibinfo{journal}{Phys. Rev. Appl.} \textbf{\bibinfo{volume}{21}},
  \bibinfo{pages}{054040} (\bibinfo{year}{2024}),
  \urlprefix\url{https://link.aps.org/doi/10.1103/PhysRevApplied.21.054040}.

\bibitem[{\citenamefont{Aguirre et~al.}(2025)\citenamefont{Aguirre,
  Barba-Ortega, de~Arruda, and Faundez}}]{Aguirre25}
\bibinfo{author}{\bibfnamefont{C.~A.} \bibnamefont{Aguirre}},
  \bibinfo{author}{\bibfnamefont{J.}~\bibnamefont{Barba-Ortega}},
  \bibinfo{author}{\bibfnamefont{A.~S.} \bibnamefont{de~Arruda}},
  \bibnamefont{and} \bibinfo{author}{\bibfnamefont{J.}~\bibnamefont{Faundez}}
  (\bibinfo{year}{2025}), \eprint{2506.01797},
  \urlprefix\url{https://arxiv.org/abs/2506.01797}.

\bibitem[{\citenamefont{Sundaresh et~al.}(2023)\citenamefont{Sundaresh,
  V{\"a}yrynen, Lyanda-Geller, and Rokhinson}}]{SundareshVayrynen23}
\bibinfo{author}{\bibfnamefont{A.}~\bibnamefont{Sundaresh}},
  \bibinfo{author}{\bibfnamefont{J.~I.} \bibnamefont{V{\"a}yrynen}},
  \bibinfo{author}{\bibfnamefont{Y.}~\bibnamefont{Lyanda-Geller}},
  \bibnamefont{and} \bibinfo{author}{\bibfnamefont{L.~P.}
  \bibnamefont{Rokhinson}}, \bibinfo{journal}{Nature Communications}
  \textbf{\bibinfo{volume}{14}}, \bibinfo{pages}{1628} (\bibinfo{year}{2023}).

\bibitem[{\citenamefont{Hoshino et~al.}(2018)\citenamefont{Hoshino, Wakatsuki,
  Hamamoto, and Nagaosa}}]{HoshinoNagaosa18}
\bibinfo{author}{\bibfnamefont{S.}~\bibnamefont{Hoshino}},
  \bibinfo{author}{\bibfnamefont{R.}~\bibnamefont{Wakatsuki}},
  \bibinfo{author}{\bibfnamefont{K.}~\bibnamefont{Hamamoto}}, \bibnamefont{and}
  \bibinfo{author}{\bibfnamefont{N.}~\bibnamefont{Nagaosa}},
  \bibinfo{journal}{Phys. Rev. B} \textbf{\bibinfo{volume}{98}},
  \bibinfo{pages}{054510} (\bibinfo{year}{2018}),
  \urlprefix\url{https://link.aps.org/doi/10.1103/PhysRevB.98.054510}.

\bibitem[{\citenamefont{Fuchs et~al.}(2022)\citenamefont{Fuchs, Kochan,
  Schmidt, H\"uttner, Baumgartner, Reinhardt, Gronin, Gardner, Lindemann,
  Manfra et~al.}}]{FuchsKochanManfra22}
\bibinfo{author}{\bibfnamefont{L.}~\bibnamefont{Fuchs}},
  \bibinfo{author}{\bibfnamefont{D.}~\bibnamefont{Kochan}},
  \bibinfo{author}{\bibfnamefont{J.}~\bibnamefont{Schmidt}},
  \bibinfo{author}{\bibfnamefont{N.}~\bibnamefont{H\"uttner}},
  \bibinfo{author}{\bibfnamefont{C.}~\bibnamefont{Baumgartner}},
  \bibinfo{author}{\bibfnamefont{S.}~\bibnamefont{Reinhardt}},
  \bibinfo{author}{\bibfnamefont{S.}~\bibnamefont{Gronin}},
  \bibinfo{author}{\bibfnamefont{G.~C.} \bibnamefont{Gardner}},
  \bibinfo{author}{\bibfnamefont{T.}~\bibnamefont{Lindemann}},
  \bibinfo{author}{\bibfnamefont{M.~J.} \bibnamefont{Manfra}},
  \bibnamefont{et~al.}, \bibinfo{journal}{Phys. Rev. X}
  \textbf{\bibinfo{volume}{12}}, \bibinfo{pages}{041020}
  (\bibinfo{year}{2022}),
  \urlprefix\url{https://link.aps.org/doi/10.1103/PhysRevX.12.041020}.

\bibitem[{\citenamefont{Putilov et~al.}(2025)\citenamefont{Putilov, Zakharov,
  Kudlis, Mel'nikov, and Buzdin}}]{PutilovBuzdin25}
\bibinfo{author}{\bibfnamefont{A.~V.} \bibnamefont{Putilov}},
  \bibinfo{author}{\bibfnamefont{D.~V.} \bibnamefont{Zakharov}},
  \bibinfo{author}{\bibfnamefont{A.}~\bibnamefont{Kudlis}},
  \bibinfo{author}{\bibfnamefont{A.~S.} \bibnamefont{Mel'nikov}},
  \bibnamefont{and} \bibinfo{author}{\bibfnamefont{A.~I.}
  \bibnamefont{Buzdin}}, \bibinfo{journal}{Phys. Rev. B}
  \textbf{\bibinfo{volume}{112}}, \bibinfo{pages}{134507}
  (\bibinfo{year}{2025}),
  \urlprefix\url{https://link.aps.org/doi/10.1103/3nnt-b8xv}.

\bibitem[{\citenamefont{Nunchot and
  Yanase}(2025{\natexlab{a}})}]{NunchotYanase25}
\bibinfo{author}{\bibfnamefont{N.}~\bibnamefont{Nunchot}} \bibnamefont{and}
  \bibinfo{author}{\bibfnamefont{Y.}~\bibnamefont{Yanase}},
  \bibinfo{journal}{Phys. Rev. B} \textbf{\bibinfo{volume}{111}},
  \bibinfo{pages}{094515} (\bibinfo{year}{2025}{\natexlab{a}}),
  \urlprefix\url{https://link.aps.org/doi/10.1103/PhysRevB.111.094515}.

\bibitem[{\citenamefont{Minamide and Yanase}(2025)}]{MinamideYanase25}
\bibinfo{author}{\bibfnamefont{A.}~\bibnamefont{Minamide}} \bibnamefont{and}
  \bibinfo{author}{\bibfnamefont{Y.}~\bibnamefont{Yanase}},
  \bibinfo{journal}{Phys. Rev. Lett.} \textbf{\bibinfo{volume}{134}},
  \bibinfo{pages}{026002} (\bibinfo{year}{2025}),
  \urlprefix\url{https://link.aps.org/doi/10.1103/PhysRevLett.134.026002}.

\bibitem[{\citenamefont{Savel'Ev and Nori}(2002)}]{Savelev02}
\bibinfo{author}{\bibfnamefont{S.}~\bibnamefont{Savel'Ev}} \bibnamefont{and}
  \bibinfo{author}{\bibfnamefont{F.}~\bibnamefont{Nori}},
  \bibinfo{journal}{Nature Materials} \textbf{\bibinfo{volume}{1}},
  \bibinfo{pages}{179} (\bibinfo{year}{2002}).

\bibitem[{\citenamefont{Wang et~al.}(2024{\natexlab{a}})\citenamefont{Wang,
  Yao, Winyard, Broyles, Gould, He, Liang, Li, Zhang, Chen
  et~al.}}]{WangAgterbergBabaev24}
\bibinfo{author}{\bibfnamefont{Y.}~\bibnamefont{Wang}},
  \bibinfo{author}{\bibfnamefont{H.}~\bibnamefont{Yao}},
  \bibinfo{author}{\bibfnamefont{T.}~\bibnamefont{Winyard}},
  \bibinfo{author}{\bibfnamefont{C.}~\bibnamefont{Broyles}},
  \bibinfo{author}{\bibfnamefont{S.}~\bibnamefont{Gould}},
  \bibinfo{author}{\bibfnamefont{Q.}~\bibnamefont{He}},
  \bibinfo{author}{\bibfnamefont{K.}~\bibnamefont{Liang}},
  \bibinfo{author}{\bibfnamefont{Z.}~\bibnamefont{Li}},
  \bibinfo{author}{\bibfnamefont{P.}~\bibnamefont{Zhang}},
  \bibinfo{author}{\bibfnamefont{B.}~\bibnamefont{Chen}}, \bibnamefont{et~al.},
  \bibinfo{journal}{Nano Letters}  (\bibinfo{year}{2024}{\natexlab{a}}).

\bibitem[{\citenamefont{Dubrovskii}(1989)}]{Dubrovskii89}
\bibinfo{author}{\bibfnamefont{I.}~\bibnamefont{Dubrovskii}},
  \bibinfo{journal}{Soviet Journal of Low Temperature Physics}
  \textbf{\bibinfo{volume}{15}}, \bibinfo{pages}{197} (\bibinfo{year}{1989}).

\bibitem[{\citenamefont{Eltsov et~al.}(2006)\citenamefont{Eltsov, Finne,
  H\"anninen, Kopu, Krusius, Tsubota, and Thuneberg}}]{Eltsov06}
\bibinfo{author}{\bibfnamefont{V.~B.} \bibnamefont{Eltsov}},
  \bibinfo{author}{\bibfnamefont{A.~P.} \bibnamefont{Finne}},
  \bibinfo{author}{\bibfnamefont{R.}~\bibnamefont{H\"anninen}},
  \bibinfo{author}{\bibfnamefont{J.}~\bibnamefont{Kopu}},
  \bibinfo{author}{\bibfnamefont{M.}~\bibnamefont{Krusius}},
  \bibinfo{author}{\bibfnamefont{M.}~\bibnamefont{Tsubota}}, \bibnamefont{and}
  \bibinfo{author}{\bibfnamefont{E.~V.} \bibnamefont{Thuneberg}},
  \bibinfo{journal}{Phys. Rev. Lett.} \textbf{\bibinfo{volume}{96}},
  \bibinfo{pages}{215302} (\bibinfo{year}{2006}),
  \urlprefix\url{https://link.aps.org/doi/10.1103/PhysRevLett.96.215302}.

\bibitem[{\citenamefont{Rantanen and Eltsov}(2024)}]{RantanenEltsov24}
\bibinfo{author}{\bibfnamefont{R.}~\bibnamefont{Rantanen}} \bibnamefont{and}
  \bibinfo{author}{\bibfnamefont{V.}~\bibnamefont{Eltsov}},
  \bibinfo{journal}{Phys. Rev. Res.} \textbf{\bibinfo{volume}{6}},
  \bibinfo{pages}{043112} (\bibinfo{year}{2024}),
  \urlprefix\url{https://link.aps.org/doi/10.1103/PhysRevResearch.6.043112}.

\bibitem[{\citenamefont{Dorsey}(1992)}]{Dorsey92}
\bibinfo{author}{\bibfnamefont{A.~T.} \bibnamefont{Dorsey}},
  \bibinfo{journal}{Phys. Rev. B} \textbf{\bibinfo{volume}{46}},
  \bibinfo{pages}{8376} (\bibinfo{year}{1992}),
  \urlprefix\url{https://link.aps.org/doi/10.1103/PhysRevB.46.8376}.

\bibitem[{\citenamefont{Reynoso et~al.}(2008)\citenamefont{Reynoso, Usaj,
  Balseiro, Feinberg, and Avignon}}]{ReynosoAvignon08}
\bibinfo{author}{\bibfnamefont{A.~A.} \bibnamefont{Reynoso}},
  \bibinfo{author}{\bibfnamefont{G.}~\bibnamefont{Usaj}},
  \bibinfo{author}{\bibfnamefont{C.~A.} \bibnamefont{Balseiro}},
  \bibinfo{author}{\bibfnamefont{D.}~\bibnamefont{Feinberg}}, \bibnamefont{and}
  \bibinfo{author}{\bibfnamefont{M.}~\bibnamefont{Avignon}},
  \bibinfo{journal}{Physical Review Letters} \textbf{\bibinfo{volume}{101}},
  \bibinfo{pages}{107001} (\bibinfo{year}{2008}), \bibinfo{note}{publisher:
  American Physical Society},
  \urlprefix\url{https://link.aps.org/doi/10.1103/PhysRevLett.101.107001}.

\bibitem[{\citenamefont{Dartiailh et~al.}(2021)\citenamefont{Dartiailh, Mayer,
  Yuan, Wickramasinghe, Matos-Abiague, \ifmmode \check{Z}\else
  \v{Z}\fi{}uti\ifmmode~\acute{c}\else \'{c}\fi{}, and
  Shabani}}]{DartiailhZutic21}
\bibinfo{author}{\bibfnamefont{M.~C.} \bibnamefont{Dartiailh}},
  \bibinfo{author}{\bibfnamefont{W.}~\bibnamefont{Mayer}},
  \bibinfo{author}{\bibfnamefont{J.}~\bibnamefont{Yuan}},
  \bibinfo{author}{\bibfnamefont{K.~S.} \bibnamefont{Wickramasinghe}},
  \bibinfo{author}{\bibfnamefont{A.}~\bibnamefont{Matos-Abiague}},
  \bibinfo{author}{\bibfnamefont{I.}~\bibnamefont{\ifmmode \check{Z}\else
  \v{Z}\fi{}uti\ifmmode~\acute{c}\else \'{c}\fi{}}}, \bibnamefont{and}
  \bibinfo{author}{\bibfnamefont{J.}~\bibnamefont{Shabani}},
  \bibinfo{journal}{Phys. Rev. Lett.} \textbf{\bibinfo{volume}{126}},
  \bibinfo{pages}{036802} (\bibinfo{year}{2021}),
  \urlprefix\url{https://link.aps.org/doi/10.1103/PhysRevLett.126.036802}.

\bibitem[{\citenamefont{Baumgartner et~al.}(2022)\citenamefont{Baumgartner,
  Fuchs, Costa, Reinhardt, Gronin, Gardner, Lindemann, Manfra, Faria~Junior,
  Kochan et~al.}}]{BaumgartnerManfra22}
\bibinfo{author}{\bibfnamefont{C.}~\bibnamefont{Baumgartner}},
  \bibinfo{author}{\bibfnamefont{L.}~\bibnamefont{Fuchs}},
  \bibinfo{author}{\bibfnamefont{A.}~\bibnamefont{Costa}},
  \bibinfo{author}{\bibfnamefont{S.}~\bibnamefont{Reinhardt}},
  \bibinfo{author}{\bibfnamefont{S.}~\bibnamefont{Gronin}},
  \bibinfo{author}{\bibfnamefont{G.~C.} \bibnamefont{Gardner}},
  \bibinfo{author}{\bibfnamefont{T.}~\bibnamefont{Lindemann}},
  \bibinfo{author}{\bibfnamefont{M.~J.} \bibnamefont{Manfra}},
  \bibinfo{author}{\bibfnamefont{P.~E.} \bibnamefont{Faria~Junior}},
  \bibinfo{author}{\bibfnamefont{D.}~\bibnamefont{Kochan}},
  \bibnamefont{et~al.}, \bibinfo{journal}{Nature Nanotechnology}
  \textbf{\bibinfo{volume}{17}}, \bibinfo{pages}{39} (\bibinfo{year}{2022}),
  ISSN \bibinfo{issn}{1748-3395}, \bibinfo{note}{number: 1 Publisher: Nature
  Publishing Group},
  \urlprefix\url{https://www.nature.com/articles/s41565-021-01009-9}.

\bibitem[{\citenamefont{Anderson and Blount}(1965)}]{AndersonBlount65}
\bibinfo{author}{\bibfnamefont{P.~W.} \bibnamefont{Anderson}} \bibnamefont{and}
  \bibinfo{author}{\bibfnamefont{E.~I.} \bibnamefont{Blount}},
  \bibinfo{journal}{Phys. Rev. Lett.} \textbf{\bibinfo{volume}{14}},
  \bibinfo{pages}{217} (\bibinfo{year}{1965}),
  \urlprefix\url{https://link.aps.org/doi/10.1103/PhysRevLett.14.217}.

\bibitem[{\citenamefont{Bulaevskii et~al.}(1976)\citenamefont{Bulaevskii,
  Guseinov, and Rusinov}}]{Bulaevskii76}
\bibinfo{author}{\bibfnamefont{L.}~\bibnamefont{Bulaevskii}},
  \bibinfo{author}{\bibfnamefont{A.}~\bibnamefont{Guseinov}}, \bibnamefont{and}
  \bibinfo{author}{\bibfnamefont{A.}~\bibnamefont{Rusinov}},
  \bibinfo{journal}{Zh. Eksp. Teor. Fiz} \textbf{\bibinfo{volume}{71}},
  \bibinfo{pages}{2356} (\bibinfo{year}{1976}).

\bibitem[{\citenamefont{Edelstein}(1995)}]{Edelstein95}
\bibinfo{author}{\bibfnamefont{V.~M.} \bibnamefont{Edelstein}},
  \bibinfo{journal}{Physical Review Letters} \textbf{\bibinfo{volume}{75}},
  \bibinfo{pages}{2004} (\bibinfo{year}{1995}), \bibinfo{note}{publisher:
  American Physical Society},
  \urlprefix\url{https://link.aps.org/doi/10.1103/PhysRevLett.75.2004}.

\bibitem[{\citenamefont{Dresselhaus}(1955)}]{Dresselhaus55}
\bibinfo{author}{\bibfnamefont{G.}~\bibnamefont{Dresselhaus}},
  \bibinfo{journal}{Phys. Rev.} \textbf{\bibinfo{volume}{100}},
  \bibinfo{pages}{580} (\bibinfo{year}{1955}),
  \urlprefix\url{https://link.aps.org/doi/10.1103/PhysRev.100.580}.

\bibitem[{\citenamefont{Rashba and Sheka}(1959)}]{Rashba59}
\bibinfo{author}{\bibfnamefont{E.}~\bibnamefont{Rashba}} \bibnamefont{and}
  \bibinfo{author}{\bibfnamefont{V.}~\bibnamefont{Sheka}},
  \bibinfo{journal}{Fizika tverd. tela} \textbf{\bibinfo{volume}{1}},
  \bibinfo{pages}{162} (\bibinfo{year}{1959}).

\bibitem[{\citenamefont{Bychkov and Rashba}(1984)}]{BychkovRashba84}
\bibinfo{author}{\bibfnamefont{Y.~A.} \bibnamefont{Bychkov}} \bibnamefont{and}
  \bibinfo{author}{\bibfnamefont{{\'E}.~I.} \bibnamefont{Rashba}},
  \bibinfo{journal}{JETP lett} \textbf{\bibinfo{volume}{39}},
  \bibinfo{pages}{78} (\bibinfo{year}{1984}).

\bibitem[{\citenamefont{Dzyaloshinskii}(1959)}]{Dzyaloshinskii59}
\bibinfo{author}{\bibfnamefont{I.}~\bibnamefont{Dzyaloshinskii}},
  \bibinfo{journal}{J. Exp. Theor. Phys} \textbf{\bibinfo{volume}{37}},
  \bibinfo{pages}{881} (\bibinfo{year}{1959}).

\bibitem[{\citenamefont{Edelstein}(1990)}]{Edelstein90}
\bibinfo{author}{\bibfnamefont{V.~M.} \bibnamefont{Edelstein}},
  \bibinfo{journal}{Solid State Communications} \textbf{\bibinfo{volume}{73}},
  \bibinfo{pages}{233} (\bibinfo{year}{1990}).

\bibitem[{\citenamefont{Edel'shtein}(1989)}]{Edelstein89}
\bibinfo{author}{\bibfnamefont{V.}~\bibnamefont{Edel'shtein}},
  \bibinfo{journal}{Soviet Journal of Experimental and Theoretical Physics}
  \textbf{\bibinfo{volume}{68}}, \bibinfo{pages}{1244} (\bibinfo{year}{1989}).

\bibitem[{\citenamefont{Lu et~al.}(2015)\citenamefont{Lu, Zheliuk, Leermakers,
  Yuan, Zeitler, Law, and Ye}}]{Lu15}
\bibinfo{author}{\bibfnamefont{J.}~\bibnamefont{Lu}},
  \bibinfo{author}{\bibfnamefont{O.}~\bibnamefont{Zheliuk}},
  \bibinfo{author}{\bibfnamefont{I.}~\bibnamefont{Leermakers}},
  \bibinfo{author}{\bibfnamefont{N.~F.} \bibnamefont{Yuan}},
  \bibinfo{author}{\bibfnamefont{U.}~\bibnamefont{Zeitler}},
  \bibinfo{author}{\bibfnamefont{K.~T.} \bibnamefont{Law}}, \bibnamefont{and}
  \bibinfo{author}{\bibfnamefont{J.}~\bibnamefont{Ye}},
  \bibinfo{journal}{Science} \textbf{\bibinfo{volume}{350}},
  \bibinfo{pages}{1353} (\bibinfo{year}{2015}).

\bibitem[{\citenamefont{Xi et~al.}(2016)\citenamefont{Xi, Wang, Zhao, Park,
  Law, Berger, Forr{\'o}, Shan, and Mak}}]{Xi16}
\bibinfo{author}{\bibfnamefont{X.}~\bibnamefont{Xi}},
  \bibinfo{author}{\bibfnamefont{Z.}~\bibnamefont{Wang}},
  \bibinfo{author}{\bibfnamefont{W.}~\bibnamefont{Zhao}},
  \bibinfo{author}{\bibfnamefont{J.-H.} \bibnamefont{Park}},
  \bibinfo{author}{\bibfnamefont{K.~T.} \bibnamefont{Law}},
  \bibinfo{author}{\bibfnamefont{H.}~\bibnamefont{Berger}},
  \bibinfo{author}{\bibfnamefont{L.}~\bibnamefont{Forr{\'o}}},
  \bibinfo{author}{\bibfnamefont{J.}~\bibnamefont{Shan}}, \bibnamefont{and}
  \bibinfo{author}{\bibfnamefont{K.~F.} \bibnamefont{Mak}},
  \bibinfo{journal}{Nature Physics} \textbf{\bibinfo{volume}{12}},
  \bibinfo{pages}{139} (\bibinfo{year}{2016}).

\bibitem[{\citenamefont{Hsu et~al.}(2017)\citenamefont{Hsu, Vaezi, Fischer, and
  Kim}}]{HsuKim17}
\bibinfo{author}{\bibfnamefont{Y.-T.} \bibnamefont{Hsu}},
  \bibinfo{author}{\bibfnamefont{A.}~\bibnamefont{Vaezi}},
  \bibinfo{author}{\bibfnamefont{M.~H.} \bibnamefont{Fischer}},
  \bibnamefont{and} \bibinfo{author}{\bibfnamefont{E.-A.} \bibnamefont{Kim}},
  \bibinfo{journal}{Nature communications} \textbf{\bibinfo{volume}{8}},
  \bibinfo{pages}{14985} (\bibinfo{year}{2017}).

\bibitem[{\citenamefont{M\"ockli and Khodas}(2019)}]{MockliKhodas19}
\bibinfo{author}{\bibfnamefont{D.}~\bibnamefont{M\"ockli}} \bibnamefont{and}
  \bibinfo{author}{\bibfnamefont{M.}~\bibnamefont{Khodas}},
  \bibinfo{journal}{Phys. Rev. B} \textbf{\bibinfo{volume}{99}},
  \bibinfo{pages}{180505} (\bibinfo{year}{2019}),
  \urlprefix\url{https://link.aps.org/doi/10.1103/PhysRevB.99.180505}.

\bibitem[{\citenamefont{Shaffer et~al.}(2020)\citenamefont{Shaffer, Kang,
  Burnell, and Fernandes}}]{ShafferBurnellFernandes20}
\bibinfo{author}{\bibfnamefont{D.}~\bibnamefont{Shaffer}},
  \bibinfo{author}{\bibfnamefont{J.}~\bibnamefont{Kang}},
  \bibinfo{author}{\bibfnamefont{F.~J.} \bibnamefont{Burnell}},
  \bibnamefont{and} \bibinfo{author}{\bibfnamefont{R.~M.}
  \bibnamefont{Fernandes}}, \bibinfo{journal}{Physical Review B}
  \textbf{\bibinfo{volume}{101}}, \bibinfo{pages}{224503}
  (\bibinfo{year}{2020}), \bibinfo{note}{publisher: American Physical Society},
  \urlprefix\url{https://link.aps.org/doi/10.1103/PhysRevB.101.224503}.

\bibitem[{\citenamefont{Wickramaratne et~al.}(2020)\citenamefont{Wickramaratne,
  Khmelevskyi, Agterberg, and Mazin}}]{WickramaratneAgterberg20}
\bibinfo{author}{\bibfnamefont{D.}~\bibnamefont{Wickramaratne}},
  \bibinfo{author}{\bibfnamefont{S.}~\bibnamefont{Khmelevskyi}},
  \bibinfo{author}{\bibfnamefont{D.~F.} \bibnamefont{Agterberg}},
  \bibnamefont{and} \bibinfo{author}{\bibfnamefont{I.~I.} \bibnamefont{Mazin}},
  \bibinfo{journal}{Phys. Rev. X} \textbf{\bibinfo{volume}{10}},
  \bibinfo{pages}{041003} (\bibinfo{year}{2020}),
  \urlprefix\url{https://link.aps.org/doi/10.1103/PhysRevX.10.041003}.

\bibitem[{\citenamefont{Hamill et~al.}(2021)\citenamefont{Hamill, Heischmidt,
  Sohn, Shaffer, Tsai, Zhang, Xi, Suslov, Berger, Forr{\'o}
  et~al.}}]{HamillShaffer21}
\bibinfo{author}{\bibfnamefont{A.}~\bibnamefont{Hamill}},
  \bibinfo{author}{\bibfnamefont{B.}~\bibnamefont{Heischmidt}},
  \bibinfo{author}{\bibfnamefont{E.}~\bibnamefont{Sohn}},
  \bibinfo{author}{\bibfnamefont{D.}~\bibnamefont{Shaffer}},
  \bibinfo{author}{\bibfnamefont{K.-T.} \bibnamefont{Tsai}},
  \bibinfo{author}{\bibfnamefont{X.}~\bibnamefont{Zhang}},
  \bibinfo{author}{\bibfnamefont{X.}~\bibnamefont{Xi}},
  \bibinfo{author}{\bibfnamefont{A.}~\bibnamefont{Suslov}},
  \bibinfo{author}{\bibfnamefont{H.}~\bibnamefont{Berger}},
  \bibinfo{author}{\bibfnamefont{L.}~\bibnamefont{Forr{\'o}}},
  \bibnamefont{et~al.}, \bibinfo{journal}{Nature physics}
  \textbf{\bibinfo{volume}{17}}, \bibinfo{pages}{949} (\bibinfo{year}{2021}).

\bibitem[{\citenamefont{Ili\ifmmode~\acute{c}\else \'{c}\fi{}
  et~al.}(2023)\citenamefont{Ili\ifmmode~\acute{c}\else \'{c}\fi{}, Meyer, and
  Houzet}}]{IlicMeyerHouzet23}
\bibinfo{author}{\bibfnamefont{S.}~\bibnamefont{Ili\ifmmode~\acute{c}\else
  \'{c}\fi{}}}, \bibinfo{author}{\bibfnamefont{J.~S.} \bibnamefont{Meyer}},
  \bibnamefont{and} \bibinfo{author}{\bibfnamefont{M.}~\bibnamefont{Houzet}},
  \bibinfo{journal}{Phys. Rev. B} \textbf{\bibinfo{volume}{108}},
  \bibinfo{pages}{214510} (\bibinfo{year}{2023}),
  \urlprefix\url{https://link.aps.org/doi/10.1103/PhysRevB.108.214510}.

\bibitem[{\citenamefont{Shaffer et~al.}(2023)\citenamefont{Shaffer, Burnell,
  and Fernandes}}]{ShafferBurnellFernandes23}
\bibinfo{author}{\bibfnamefont{D.}~\bibnamefont{Shaffer}},
  \bibinfo{author}{\bibfnamefont{F.~J.} \bibnamefont{Burnell}},
  \bibnamefont{and} \bibinfo{author}{\bibfnamefont{R.~M.}
  \bibnamefont{Fernandes}}, \bibinfo{journal}{Phys. Rev. B}
  \textbf{\bibinfo{volume}{107}}, \bibinfo{pages}{224516}
  (\bibinfo{year}{2023}),
  \urlprefix\url{https://link.aps.org/doi/10.1103/PhysRevB.107.224516}.

\bibitem[{\citenamefont{Siegl et~al.}(2025)\citenamefont{Siegl, Bleibaum, Wan,
  Kurpas, Schliemann, Ugeda, Marganska, and Grifoni}}]{SieglUgeda25}
\bibinfo{author}{\bibfnamefont{J.}~\bibnamefont{Siegl}},
  \bibinfo{author}{\bibfnamefont{A.}~\bibnamefont{Bleibaum}},
  \bibinfo{author}{\bibfnamefont{W.}~\bibnamefont{Wan}},
  \bibinfo{author}{\bibfnamefont{M.}~\bibnamefont{Kurpas}},
  \bibinfo{author}{\bibfnamefont{J.}~\bibnamefont{Schliemann}},
  \bibinfo{author}{\bibfnamefont{M.~M.} \bibnamefont{Ugeda}},
  \bibinfo{author}{\bibfnamefont{M.}~\bibnamefont{Marganska}},
  \bibnamefont{and} \bibinfo{author}{\bibfnamefont{M.}~\bibnamefont{Grifoni}},
  \bibinfo{journal}{Nature Communications} \textbf{\bibinfo{volume}{16}},
  \bibinfo{pages}{8228} (\bibinfo{year}{2025}).

\bibitem[{\citenamefont{Houzet and Meyer}(2015)}]{HouzetMeyer15}
\bibinfo{author}{\bibfnamefont{M.}~\bibnamefont{Houzet}} \bibnamefont{and}
  \bibinfo{author}{\bibfnamefont{J.~S.} \bibnamefont{Meyer}},
  \bibinfo{journal}{Phys. Rev. B} \textbf{\bibinfo{volume}{92}},
  \bibinfo{pages}{014509} (\bibinfo{year}{2015}),
  \urlprefix\url{https://link.aps.org/doi/10.1103/PhysRevB.92.014509}.

\bibitem[{\citenamefont{Kitaev}(2001)}]{Kitaev01}
\bibinfo{author}{\bibfnamefont{A.~Y.} \bibnamefont{Kitaev}},
  \bibinfo{journal}{Physics-Uspekhi} \textbf{\bibinfo{volume}{44}},
  \bibinfo{pages}{131} (\bibinfo{year}{2001}),
  \urlprefix\url{https://dx.doi.org/10.1070/1063-7869/44/10S/S29}.

\bibitem[{\citenamefont{Fu and Kane}(2008)}]{FuKane08}
\bibinfo{author}{\bibfnamefont{L.}~\bibnamefont{Fu}} \bibnamefont{and}
  \bibinfo{author}{\bibfnamefont{C.~L.} \bibnamefont{Kane}},
  \bibinfo{journal}{Phys. Rev. Lett.} \textbf{\bibinfo{volume}{100}},
  \bibinfo{pages}{096407} (\bibinfo{year}{2008}),
  \urlprefix\url{https://link.aps.org/doi/10.1103/PhysRevLett.100.096407}.

\bibitem[{\citenamefont{Fujimoto}(2008)}]{Fujimoto08}
\bibinfo{author}{\bibfnamefont{S.}~\bibnamefont{Fujimoto}},
  \bibinfo{journal}{Phys. Rev. B} \textbf{\bibinfo{volume}{77}},
  \bibinfo{pages}{220501} (\bibinfo{year}{2008}),
  \urlprefix\url{https://link.aps.org/doi/10.1103/PhysRevB.77.220501}.

\bibitem[{\citenamefont{Zhang et~al.}(2008)\citenamefont{Zhang, Tewari,
  Lutchyn, and Das~Sarma}}]{ZhangDasSarma08}
\bibinfo{author}{\bibfnamefont{C.}~\bibnamefont{Zhang}},
  \bibinfo{author}{\bibfnamefont{S.}~\bibnamefont{Tewari}},
  \bibinfo{author}{\bibfnamefont{R.~M.} \bibnamefont{Lutchyn}},
  \bibnamefont{and}
  \bibinfo{author}{\bibfnamefont{S.}~\bibnamefont{Das~Sarma}},
  \bibinfo{journal}{Phys. Rev. Lett.} \textbf{\bibinfo{volume}{101}},
  \bibinfo{pages}{160401} (\bibinfo{year}{2008}),
  \urlprefix\url{https://link.aps.org/doi/10.1103/PhysRevLett.101.160401}.

\bibitem[{\citenamefont{Sau et~al.}(2010)\citenamefont{Sau, Lutchyn, Tewari,
  and Das~Sarma}}]{SauDasSarma10}
\bibinfo{author}{\bibfnamefont{J.~D.} \bibnamefont{Sau}},
  \bibinfo{author}{\bibfnamefont{R.~M.} \bibnamefont{Lutchyn}},
  \bibinfo{author}{\bibfnamefont{S.}~\bibnamefont{Tewari}}, \bibnamefont{and}
  \bibinfo{author}{\bibfnamefont{S.}~\bibnamefont{Das~Sarma}},
  \bibinfo{journal}{Phys. Rev. Lett.} \textbf{\bibinfo{volume}{104}},
  \bibinfo{pages}{040502} (\bibinfo{year}{2010}),
  \urlprefix\url{https://link.aps.org/doi/10.1103/PhysRevLett.104.040502}.

\bibitem[{\citenamefont{Oreg et~al.}(2010)\citenamefont{Oreg, Refael, and von
  Oppen}}]{OregRefaelvonOppen10}
\bibinfo{author}{\bibfnamefont{Y.}~\bibnamefont{Oreg}},
  \bibinfo{author}{\bibfnamefont{G.}~\bibnamefont{Refael}}, \bibnamefont{and}
  \bibinfo{author}{\bibfnamefont{F.}~\bibnamefont{von Oppen}},
  \bibinfo{journal}{Phys. Rev. Lett.} \textbf{\bibinfo{volume}{105}},
  \bibinfo{pages}{177002} (\bibinfo{year}{2010}),
  \urlprefix\url{https://link.aps.org/doi/10.1103/PhysRevLett.105.177002}.

\bibitem[{\citenamefont{Lutchyn et~al.}(2010)\citenamefont{Lutchyn, Sau, and
  Das~Sarma}}]{LutchynSauDasSarma10}
\bibinfo{author}{\bibfnamefont{R.~M.} \bibnamefont{Lutchyn}},
  \bibinfo{author}{\bibfnamefont{J.~D.} \bibnamefont{Sau}}, \bibnamefont{and}
  \bibinfo{author}{\bibfnamefont{S.}~\bibnamefont{Das~Sarma}},
  \bibinfo{journal}{Phys. Rev. Lett.} \textbf{\bibinfo{volume}{105}},
  \bibinfo{pages}{077001} (\bibinfo{year}{2010}),
  \urlprefix\url{https://link.aps.org/doi/10.1103/PhysRevLett.105.077001}.

\bibitem[{\citenamefont{Cook and Franz}(2011)}]{CookFranz11}
\bibinfo{author}{\bibfnamefont{A.}~\bibnamefont{Cook}} \bibnamefont{and}
  \bibinfo{author}{\bibfnamefont{M.}~\bibnamefont{Franz}},
  \bibinfo{journal}{Phys. Rev. B} \textbf{\bibinfo{volume}{84}},
  \bibinfo{pages}{201105} (\bibinfo{year}{2011}),
  \urlprefix\url{https://link.aps.org/doi/10.1103/PhysRevB.84.201105}.

\bibitem[{\citenamefont{Wang et~al.}(2015)\citenamefont{Wang, Zhou, Lian, and
  Zhang}}]{WangZhang15}
\bibinfo{author}{\bibfnamefont{J.}~\bibnamefont{Wang}},
  \bibinfo{author}{\bibfnamefont{Q.}~\bibnamefont{Zhou}},
  \bibinfo{author}{\bibfnamefont{B.}~\bibnamefont{Lian}}, \bibnamefont{and}
  \bibinfo{author}{\bibfnamefont{S.-C.} \bibnamefont{Zhang}},
  \bibinfo{journal}{Phys. Rev. B} \textbf{\bibinfo{volume}{92}},
  \bibinfo{pages}{064520} (\bibinfo{year}{2015}),
  \urlprefix\url{https://link.aps.org/doi/10.1103/PhysRevB.92.064520}.

\bibitem[{\citenamefont{Agterberg}(2003)}]{Agterberg03}
\bibinfo{author}{\bibfnamefont{D.~F.} \bibnamefont{Agterberg}},
  \bibinfo{journal}{Physica C: Superconductivity}
  \textbf{\bibinfo{volume}{387}}, \bibinfo{pages}{13} (\bibinfo{year}{2003}),
  ISSN \bibinfo{issn}{0921-4534},
  \urlprefix\url{https://www.sciencedirect.com/science/article/pii/S0921453403006348}.

\bibitem[{\citenamefont{Timm et~al.}(2017)\citenamefont{Timm, Schnyder,
  Agterberg, and Brydon}}]{TimmAgterberg17}
\bibinfo{author}{\bibfnamefont{C.}~\bibnamefont{Timm}},
  \bibinfo{author}{\bibfnamefont{A.~P.} \bibnamefont{Schnyder}},
  \bibinfo{author}{\bibfnamefont{D.~F.} \bibnamefont{Agterberg}},
  \bibnamefont{and} \bibinfo{author}{\bibfnamefont{P.~M.~R.}
  \bibnamefont{Brydon}}, \bibinfo{journal}{Physical Review B}
  \textbf{\bibinfo{volume}{96}}, \bibinfo{pages}{094526}
  (\bibinfo{year}{2017}), \bibinfo{note}{publisher: American Physical Society},
  \urlprefix\url{https://link.aps.org/doi/10.1103/PhysRevB.96.094526}.

\bibitem[{\citenamefont{Yuan and Fu}(2018)}]{YuanFu18}
\bibinfo{author}{\bibfnamefont{N.~F.~Q.} \bibnamefont{Yuan}} \bibnamefont{and}
  \bibinfo{author}{\bibfnamefont{L.}~\bibnamefont{Fu}},
  \bibinfo{journal}{Physical Review B} \textbf{\bibinfo{volume}{97}},
  \bibinfo{pages}{115139} (\bibinfo{year}{2018}), \bibinfo{note}{publisher:
  American Physical Society},
  \urlprefix\url{https://link.aps.org/doi/10.1103/PhysRevB.97.115139}.

\bibitem[{\citenamefont{Link et~al.}(2020)\citenamefont{Link, Boettcher, and
  Herbut}}]{LinkBoettcherHerbut20}
\bibinfo{author}{\bibfnamefont{J.~M.} \bibnamefont{Link}},
  \bibinfo{author}{\bibfnamefont{I.}~\bibnamefont{Boettcher}},
  \bibnamefont{and} \bibinfo{author}{\bibfnamefont{I.~F.}
  \bibnamefont{Herbut}}, \bibinfo{journal}{Physical Review B}
  \textbf{\bibinfo{volume}{101}}, \bibinfo{pages}{184503}
  (\bibinfo{year}{2020}), \bibinfo{note}{publisher: American Physical Society},
  \urlprefix\url{https://link.aps.org/doi/10.1103/PhysRevB.101.184503}.

\bibitem[{\citenamefont{Akbari and Thalmeier}(2022)}]{AkbariThalmeier22}
\bibinfo{author}{\bibfnamefont{A.}~\bibnamefont{Akbari}} \bibnamefont{and}
  \bibinfo{author}{\bibfnamefont{P.}~\bibnamefont{Thalmeier}},
  \bibinfo{journal}{Physical Review Research} \textbf{\bibinfo{volume}{4}},
  \bibinfo{pages}{023096} (\bibinfo{year}{2022}), \bibinfo{note}{publisher:
  American Physical Society},
  \urlprefix\url{https://link.aps.org/doi/10.1103/PhysRevResearch.4.023096}.

\bibitem[{\citenamefont{Banerjee et~al.}(2022)\citenamefont{Banerjee, Ikegaya,
  and Schnyder}}]{BanerjeeSchnyder22}
\bibinfo{author}{\bibfnamefont{S.}~\bibnamefont{Banerjee}},
  \bibinfo{author}{\bibfnamefont{S.}~\bibnamefont{Ikegaya}}, \bibnamefont{and}
  \bibinfo{author}{\bibfnamefont{A.~P.} \bibnamefont{Schnyder}},
  \bibinfo{journal}{Physical Review Research} \textbf{\bibinfo{volume}{4}},
  \bibinfo{pages}{L042049} (\bibinfo{year}{2022}), \bibinfo{note}{publisher:
  American Physical Society},
  \urlprefix\url{https://link.aps.org/doi/10.1103/PhysRevResearch.4.L042049}.

\bibitem[{\citenamefont{Babkin et~al.}(2024)\citenamefont{Babkin, Higginbotham,
  and Serbyn}}]{BabkinSerbyn24}
\bibinfo{author}{\bibfnamefont{S.~S.} \bibnamefont{Babkin}},
  \bibinfo{author}{\bibfnamefont{A.~P.} \bibnamefont{Higginbotham}},
  \bibnamefont{and} \bibinfo{author}{\bibfnamefont{M.}~\bibnamefont{Serbyn}},
  \bibinfo{journal}{SciPost Physics} \textbf{\bibinfo{volume}{16}},
  \bibinfo{pages}{115} (\bibinfo{year}{2024}), ISSN \bibinfo{issn}{2542-4653},
  \urlprefix\url{https://scipost.org/10.21468/SciPostPhys.16.5.115}.

\bibitem[{\citenamefont{Mateos et~al.}(2024)\citenamefont{Mateos, Tosi,
  Braggio, Taddei, and Arrachea}}]{Mateos24}
\bibinfo{author}{\bibfnamefont{J.~H.} \bibnamefont{Mateos}},
  \bibinfo{author}{\bibfnamefont{L.}~\bibnamefont{Tosi}},
  \bibinfo{author}{\bibfnamefont{A.}~\bibnamefont{Braggio}},
  \bibinfo{author}{\bibfnamefont{F.}~\bibnamefont{Taddei}}, \bibnamefont{and}
  \bibinfo{author}{\bibfnamefont{L.}~\bibnamefont{Arrachea}},
  \bibinfo{journal}{Physical Review B} \textbf{\bibinfo{volume}{110}},
  \bibinfo{pages}{075415} (\bibinfo{year}{2024}), \bibinfo{note}{publisher:
  American Physical Society},
  \urlprefix\url{https://link.aps.org/doi/10.1103/PhysRevB.110.075415}.

\bibitem[{\citenamefont{Sano et~al.}(2025)\citenamefont{Sano, Sato, Sasaki,
  Ikegaya, Kobayashi, and Asano}}]{Sano25}
\bibinfo{author}{\bibfnamefont{T.}~\bibnamefont{Sano}},
  \bibinfo{author}{\bibfnamefont{T.}~\bibnamefont{Sato}},
  \bibinfo{author}{\bibfnamefont{A.}~\bibnamefont{Sasaki}},
  \bibinfo{author}{\bibfnamefont{S.}~\bibnamefont{Ikegaya}},
  \bibinfo{author}{\bibfnamefont{S.}~\bibnamefont{Kobayashi}},
  \bibnamefont{and} \bibinfo{author}{\bibfnamefont{Y.}~\bibnamefont{Asano}},
  \bibinfo{journal}{Physical Review B} \textbf{\bibinfo{volume}{111}},
  \bibinfo{pages}{174503} (\bibinfo{year}{2025}), \bibinfo{note}{publisher:
  American Physical Society},
  \urlprefix\url{https://link.aps.org/doi/10.1103/PhysRevB.111.174503}.

\bibitem[{\citenamefont{Wei et~al.}(2025)\citenamefont{Wei, Xiang, Xu, Wang,
  and Wang}}]{Wei25}
\bibinfo{author}{\bibfnamefont{M.}~\bibnamefont{Wei}},
  \bibinfo{author}{\bibfnamefont{L.}~\bibnamefont{Xiang}},
  \bibinfo{author}{\bibfnamefont{F.}~\bibnamefont{Xu}},
  \bibinfo{author}{\bibfnamefont{B.}~\bibnamefont{Wang}}, \bibnamefont{and}
  \bibinfo{author}{\bibfnamefont{J.}~\bibnamefont{Wang}}
  (\bibinfo{year}{2025}), \bibinfo{note}{arXiv:2505.23278 [cond-mat]},
  \urlprefix\url{http://arxiv.org/abs/2505.23278}.

\bibitem[{\citenamefont{Devizorova et~al.}(2021)\citenamefont{Devizorova,
  Putilov, Chaykin, Mironov, and Buzdin}}]{DevizoravaBuzdin21}
\bibinfo{author}{\bibfnamefont{Z.}~\bibnamefont{Devizorova}},
  \bibinfo{author}{\bibfnamefont{A.~V.} \bibnamefont{Putilov}},
  \bibinfo{author}{\bibfnamefont{I.}~\bibnamefont{Chaykin}},
  \bibinfo{author}{\bibfnamefont{S.}~\bibnamefont{Mironov}}, \bibnamefont{and}
  \bibinfo{author}{\bibfnamefont{A.~I.} \bibnamefont{Buzdin}},
  \bibinfo{journal}{Phys. Rev. B} \textbf{\bibinfo{volume}{103}},
  \bibinfo{pages}{064504} (\bibinfo{year}{2021}),
  \urlprefix\url{https://link.aps.org/doi/10.1103/PhysRevB.103.064504}.

\bibitem[{\citenamefont{Ikeda et~al.}(2022)\citenamefont{Ikeda, Daido, and
  Yanase}}]{IkedaDaidoYanase22}
\bibinfo{author}{\bibfnamefont{Y.}~\bibnamefont{Ikeda}},
  \bibinfo{author}{\bibfnamefont{A.}~\bibnamefont{Daido}}, \bibnamefont{and}
  \bibinfo{author}{\bibfnamefont{Y.}~\bibnamefont{Yanase}},
  \bibinfo{journal}{arXiv preprint arXiv:2212.09211}  (\bibinfo{year}{2022}).

\bibitem[{\citenamefont{Hasan et~al.}(2025)\citenamefont{Hasan, Shaffer,
  Khodas, and Levchenko}}]{HasanShafferKhodasLevchenko25}
\bibinfo{author}{\bibfnamefont{J.}~\bibnamefont{Hasan}},
  \bibinfo{author}{\bibfnamefont{D.}~\bibnamefont{Shaffer}},
  \bibinfo{author}{\bibfnamefont{M.}~\bibnamefont{Khodas}}, \bibnamefont{and}
  \bibinfo{author}{\bibfnamefont{A.}~\bibnamefont{Levchenko}}
  (\bibinfo{year}{2025}), \bibinfo{note}{arXiv:2502.09421 [cond-mat]},
  \urlprefix\url{http://arxiv.org/abs/2502.09421}.

\bibitem[{\citenamefont{Aoyama}(2024)}]{Aoyama24}
\bibinfo{author}{\bibfnamefont{K.}~\bibnamefont{Aoyama}},
  \bibinfo{journal}{Phys. Rev. B} \textbf{\bibinfo{volume}{109}},
  \bibinfo{pages}{024516} (\bibinfo{year}{2024}),
  \urlprefix\url{https://link.aps.org/doi/10.1103/PhysRevB.109.024516}.

\bibitem[{\citenamefont{Zhuang et~al.}(2025)\citenamefont{Zhuang, Shaffer,
  Hasan, and Levchenko}}]{ZhuangShafferHasanLevchenko25}
\bibinfo{author}{\bibfnamefont{Z.}~\bibnamefont{Zhuang}},
  \bibinfo{author}{\bibfnamefont{D.}~\bibnamefont{Shaffer}},
  \bibinfo{author}{\bibfnamefont{J.}~\bibnamefont{Hasan}}, \bibnamefont{and}
  \bibinfo{author}{\bibfnamefont{A.}~\bibnamefont{Levchenko}}
  (\bibinfo{year}{2025}), \eprint{2510.18963},
  \urlprefix\url{https://arxiv.org/abs/2510.18963}.

\bibitem[{\citenamefont{Putilov et~al.}(2024)\citenamefont{Putilov, Mironov,
  and Buzdin}}]{PutilovBuzdin24}
\bibinfo{author}{\bibfnamefont{A.~V.} \bibnamefont{Putilov}},
  \bibinfo{author}{\bibfnamefont{S.~V.} \bibnamefont{Mironov}},
  \bibnamefont{and} \bibinfo{author}{\bibfnamefont{A.~I.}
  \bibnamefont{Buzdin}}, \bibinfo{journal}{Phys. Rev. B}
  \textbf{\bibinfo{volume}{109}}, \bibinfo{pages}{014510}
  (\bibinfo{year}{2024}),
  \urlprefix\url{https://link.aps.org/doi/10.1103/PhysRevB.109.014510}.

\bibitem[{\citenamefont{Plastovets and Buzdin}(2024)}]{PlastovetsBuzdin24}
\bibinfo{author}{\bibfnamefont{V.}~\bibnamefont{Plastovets}} \bibnamefont{and}
  \bibinfo{author}{\bibfnamefont{A.}~\bibnamefont{Buzdin}},
  \bibinfo{journal}{Phys. Rev. B} \textbf{\bibinfo{volume}{110}},
  \bibinfo{pages}{144521} (\bibinfo{year}{2024}),
  \urlprefix\url{https://link.aps.org/doi/10.1103/PhysRevB.110.144521}.

\bibitem[{\citenamefont{Mazanik et~al.}(2025)\citenamefont{Mazanik, Kokkeler,
  Tokatly, and Bergeret}}]{MazanikBergeret25}
\bibinfo{author}{\bibfnamefont{A.~A.} \bibnamefont{Mazanik}},
  \bibinfo{author}{\bibfnamefont{T.}~\bibnamefont{Kokkeler}},
  \bibinfo{author}{\bibfnamefont{I.~V.} \bibnamefont{Tokatly}},
  \bibnamefont{and} \bibinfo{author}{\bibfnamefont{F.~S.}
  \bibnamefont{Bergeret}}, \bibinfo{journal}{Phys. Rev. B}
  \textbf{\bibinfo{volume}{112}}, \bibinfo{pages}{024514}
  (\bibinfo{year}{2025}),
  \urlprefix\url{https://link.aps.org/doi/10.1103/dsxt-phyk}.

\bibitem[{\citenamefont{Park et~al.}(2011)\citenamefont{Park, Kim, Yu, Han, and
  Kim}}]{Park11}
\bibinfo{author}{\bibfnamefont{S.~R.} \bibnamefont{Park}},
  \bibinfo{author}{\bibfnamefont{C.~H.} \bibnamefont{Kim}},
  \bibinfo{author}{\bibfnamefont{J.}~\bibnamefont{Yu}},
  \bibinfo{author}{\bibfnamefont{J.~H.} \bibnamefont{Han}}, \bibnamefont{and}
  \bibinfo{author}{\bibfnamefont{C.}~\bibnamefont{Kim}},
  \bibinfo{journal}{Phys. Rev. Lett.} \textbf{\bibinfo{volume}{107}},
  \bibinfo{pages}{156803} (\bibinfo{year}{2011}),
  \urlprefix\url{https://link.aps.org/doi/10.1103/PhysRevLett.107.156803}.

\bibitem[{\citenamefont{Saunderson et~al.}(2025)\citenamefont{Saunderson,
  Gradhand, Go, Annett, Mercaldo, Cuoco, Kläui, Gayles, and
  Mokrousov}}]{Saunderson25}
\bibinfo{author}{\bibfnamefont{T.~G.} \bibnamefont{Saunderson}},
  \bibinfo{author}{\bibfnamefont{M.}~\bibnamefont{Gradhand}},
  \bibinfo{author}{\bibfnamefont{D.}~\bibnamefont{Go}},
  \bibinfo{author}{\bibfnamefont{J.~F.} \bibnamefont{Annett}},
  \bibinfo{author}{\bibfnamefont{M.~T.} \bibnamefont{Mercaldo}},
  \bibinfo{author}{\bibfnamefont{M.}~\bibnamefont{Cuoco}},
  \bibinfo{author}{\bibfnamefont{M.}~\bibnamefont{Kläui}},
  \bibinfo{author}{\bibfnamefont{J.}~\bibnamefont{Gayles}}, \bibnamefont{and}
  \bibinfo{author}{\bibfnamefont{Y.}~\bibnamefont{Mokrousov}}
  (\bibinfo{year}{2025}), \eprint{2504.01271},
  \urlprefix\url{https://arxiv.org/abs/2504.01271}.

\bibitem[{\citenamefont{de~Picoli et~al.}(2023)\citenamefont{de~Picoli, Blood,
  Lyanda-Geller, and V\"ayrynen}}]{dePicolliVayrynen23}
\bibinfo{author}{\bibfnamefont{T.}~\bibnamefont{de~Picoli}},
  \bibinfo{author}{\bibfnamefont{Z.}~\bibnamefont{Blood}},
  \bibinfo{author}{\bibfnamefont{Y.}~\bibnamefont{Lyanda-Geller}},
  \bibnamefont{and} \bibinfo{author}{\bibfnamefont{J.~I.}
  \bibnamefont{V\"ayrynen}}, \bibinfo{journal}{Phys. Rev. B}
  \textbf{\bibinfo{volume}{107}}, \bibinfo{pages}{224518}
  (\bibinfo{year}{2023}),
  \urlprefix\url{https://link.aps.org/doi/10.1103/PhysRevB.107.224518}.

\bibitem[{\citenamefont{Meyer and Houzet}(2024)}]{MeyerHouzet24}
\bibinfo{author}{\bibfnamefont{J.~S.} \bibnamefont{Meyer}} \bibnamefont{and}
  \bibinfo{author}{\bibfnamefont{M.}~\bibnamefont{Houzet}},
  \bibinfo{journal}{Applied Physics Letters} \textbf{\bibinfo{volume}{125}},
  \bibinfo{pages}{022603} (\bibinfo{year}{2024}), ISSN
  \bibinfo{issn}{0003-6951},
  \eprint{https://pubs.aip.org/aip/apl/article-pdf/doi/10.1063/5.0211491/20041558/022603_1_5.0211491.pdf},
  \urlprefix\url{https://doi.org/10.1063/5.0211491}.

\bibitem[{\citenamefont{Kokkeler et~al.}(2024)\citenamefont{Kokkeler, Tokatly,
  and Bergeret}}]{KokkelerTokatlyBergeret24}
\bibinfo{author}{\bibfnamefont{T.}~\bibnamefont{Kokkeler}},
  \bibinfo{author}{\bibfnamefont{I.}~\bibnamefont{Tokatly}}, \bibnamefont{and}
  \bibinfo{author}{\bibfnamefont{F.~S.} \bibnamefont{Bergeret}},
  \bibinfo{journal}{SciPost Physics} \textbf{\bibinfo{volume}{16}},
  \bibinfo{pages}{055} (\bibinfo{year}{2024}), ISSN \bibinfo{issn}{2542-4653},
  \urlprefix\url{https://scipost.org/10.21468/SciPostPhys.16.2.055}.

\bibitem[{\citenamefont{Ili\ifmmode~\acute{c}\else \'{c}\fi{}
  et~al.}(2024)\citenamefont{Ili\ifmmode~\acute{c}\else \'{c}\fi{}, Virtanen,
  Crawford, Heikkil\"a, and Bergeret}}]{IlicBergeret24}
\bibinfo{author}{\bibfnamefont{S.}~\bibnamefont{Ili\ifmmode~\acute{c}\else
  \'{c}\fi{}}}, \bibinfo{author}{\bibfnamefont{P.}~\bibnamefont{Virtanen}},
  \bibinfo{author}{\bibfnamefont{D.}~\bibnamefont{Crawford}},
  \bibinfo{author}{\bibfnamefont{T.~T.} \bibnamefont{Heikkil\"a}},
  \bibnamefont{and} \bibinfo{author}{\bibfnamefont{F.~S.}
  \bibnamefont{Bergeret}}, \bibinfo{journal}{Phys. Rev. B}
  \textbf{\bibinfo{volume}{110}}, \bibinfo{pages}{L140501}
  (\bibinfo{year}{2024}),
  \urlprefix\url{https://link.aps.org/doi/10.1103/PhysRevB.110.L140501}.

\bibitem[{\citenamefont{Nunchot and
  Yanase}(2025{\natexlab{b}})}]{NunchotYanase25_2}
\bibinfo{author}{\bibfnamefont{N.}~\bibnamefont{Nunchot}} \bibnamefont{and}
  \bibinfo{author}{\bibfnamefont{Y.}~\bibnamefont{Yanase}}
  (\bibinfo{year}{2025}{\natexlab{b}}), \eprint{2507.21897},
  \urlprefix\url{https://arxiv.org/abs/2507.21897}.

\bibitem[{\citenamefont{Oh and Nagaosa}(2024)}]{OhNagaosa24}
\bibinfo{author}{\bibfnamefont{T.}~\bibnamefont{Oh}} \bibnamefont{and}
  \bibinfo{author}{\bibfnamefont{N.}~\bibnamefont{Nagaosa}},
  \bibinfo{journal}{Phys. Rev. B} \textbf{\bibinfo{volume}{110}},
  \bibinfo{pages}{134507} (\bibinfo{year}{2024}),
  \urlprefix\url{https://link.aps.org/doi/10.1103/PhysRevB.110.134507}.

\bibitem[{\citenamefont{Banerjee and
  Scheurer}(2024{\natexlab{b}})}]{BanerjeeScheurer24alt}
\bibinfo{author}{\bibfnamefont{S.}~\bibnamefont{Banerjee}} \bibnamefont{and}
  \bibinfo{author}{\bibfnamefont{M.~S.} \bibnamefont{Scheurer}},
  \bibinfo{journal}{Phys. Rev. B} \textbf{\bibinfo{volume}{110}},
  \bibinfo{pages}{024503} (\bibinfo{year}{2024}{\natexlab{b}}),
  \urlprefix\url{https://link.aps.org/doi/10.1103/PhysRevB.110.024503}.

\bibitem[{\citenamefont{Kang et~al.}(2024)\citenamefont{Kang, Barth, Costa,
  Garcia-Ruiz, Mre\ifmmode \acute{n}\else
  \'{n}\fi{}ca-Kolasi\ifmmode~\acute{n}\else \'{n}\fi{}ska, Liu, and
  Kochan}}]{KangKochan24}
\bibinfo{author}{\bibfnamefont{W.-H.} \bibnamefont{Kang}},
  \bibinfo{author}{\bibfnamefont{M.}~\bibnamefont{Barth}},
  \bibinfo{author}{\bibfnamefont{A.}~\bibnamefont{Costa}},
  \bibinfo{author}{\bibfnamefont{A.}~\bibnamefont{Garcia-Ruiz}},
  \bibinfo{author}{\bibfnamefont{A.}~\bibnamefont{Mre\ifmmode \acute{n}\else
  \'{n}\fi{}ca-Kolasi\ifmmode~\acute{n}\else \'{n}\fi{}ska}},
  \bibinfo{author}{\bibfnamefont{M.-H.} \bibnamefont{Liu}}, \bibnamefont{and}
  \bibinfo{author}{\bibfnamefont{D.}~\bibnamefont{Kochan}},
  \bibinfo{journal}{Phys. Rev. Lett.} \textbf{\bibinfo{volume}{133}},
  \bibinfo{pages}{216201} (\bibinfo{year}{2024}),
  \urlprefix\url{https://link.aps.org/doi/10.1103/PhysRevLett.133.216201}.

\bibitem[{\citenamefont{Bankier et~al.}(2025)\citenamefont{Bankier, Attias,
  Levchenko, and Khodas}}]{BankierLevchenkoKhodas25}
\bibinfo{author}{\bibfnamefont{I.}~\bibnamefont{Bankier}},
  \bibinfo{author}{\bibfnamefont{L.}~\bibnamefont{Attias}},
  \bibinfo{author}{\bibfnamefont{A.}~\bibnamefont{Levchenko}},
  \bibnamefont{and} \bibinfo{author}{\bibfnamefont{M.}~\bibnamefont{Khodas}}
  (\bibinfo{year}{2025}), \bibinfo{note}{arXiv:2503.15115 [cond-mat]},
  \urlprefix\url{http://arxiv.org/abs/2503.15115}.

\bibitem[{\citenamefont{Kealhofer et~al.}(2023)\citenamefont{Kealhofer, Jeong,
  Rashidi, Balents, and Stemmer}}]{KealhofferBalents23}
\bibinfo{author}{\bibfnamefont{R.}~\bibnamefont{Kealhofer}},
  \bibinfo{author}{\bibfnamefont{H.}~\bibnamefont{Jeong}},
  \bibinfo{author}{\bibfnamefont{A.}~\bibnamefont{Rashidi}},
  \bibinfo{author}{\bibfnamefont{L.}~\bibnamefont{Balents}}, \bibnamefont{and}
  \bibinfo{author}{\bibfnamefont{S.}~\bibnamefont{Stemmer}},
  \bibinfo{journal}{Phys. Rev. B} \textbf{\bibinfo{volume}{107}},
  \bibinfo{pages}{L100504} (\bibinfo{year}{2023}),
  \urlprefix\url{https://link.aps.org/doi/10.1103/PhysRevB.107.L100504}.

\bibitem[{\citenamefont{Nunchot and Yanase}(2024)}]{NunchotYanase24}
\bibinfo{author}{\bibfnamefont{N.}~\bibnamefont{Nunchot}} \bibnamefont{and}
  \bibinfo{author}{\bibfnamefont{Y.}~\bibnamefont{Yanase}},
  \bibinfo{journal}{Physical Review B} \textbf{\bibinfo{volume}{109}},
  \bibinfo{pages}{054508} (\bibinfo{year}{2024}), \bibinfo{note}{publisher:
  American Physical Society},
  \urlprefix\url{https://link.aps.org/doi/10.1103/PhysRevB.109.054508}.

\bibitem[{\citenamefont{He et~al.}(2023)\citenamefont{He, Tanaka, and
  Nagaosa}}]{HeNagaosa23}
\bibinfo{author}{\bibfnamefont{J.~J.} \bibnamefont{He}},
  \bibinfo{author}{\bibfnamefont{Y.}~\bibnamefont{Tanaka}}, \bibnamefont{and}
  \bibinfo{author}{\bibfnamefont{N.}~\bibnamefont{Nagaosa}},
  \bibinfo{journal}{Nature Communications} \textbf{\bibinfo{volume}{14}},
  \bibinfo{pages}{3330} (\bibinfo{year}{2023}).

\bibitem[{\citenamefont{Li and He}(2025)}]{LiHe25}
\bibinfo{author}{\bibfnamefont{C.}~\bibnamefont{Li}} \bibnamefont{and}
  \bibinfo{author}{\bibfnamefont{J.~J.} \bibnamefont{He}},
  \bibinfo{journal}{Phys. Rev. B} \textbf{\bibinfo{volume}{111}},
  \bibinfo{pages}{224504} (\bibinfo{year}{2025}),
  \urlprefix\url{https://link.aps.org/doi/10.1103/PhysRevB.111.224504}.

\bibitem[{\citenamefont{Cuozzo and Léonard}(2025)}]{Cuozzo25}
\bibinfo{author}{\bibfnamefont{J.~J.} \bibnamefont{Cuozzo}} \bibnamefont{and}
  \bibinfo{author}{\bibfnamefont{F.}~\bibnamefont{Léonard}}
  (\bibinfo{year}{2025}), \eprint{2504.02948},
  \urlprefix\url{https://arxiv.org/abs/2504.02948}.

\bibitem[{\citenamefont{Buzdin}(2008)}]{Buzdin08}
\bibinfo{author}{\bibfnamefont{A.}~\bibnamefont{Buzdin}},
  \bibinfo{journal}{Physical Review Letters} \textbf{\bibinfo{volume}{101}},
  \bibinfo{pages}{107005} (\bibinfo{year}{2008}), \bibinfo{note}{publisher:
  American Physical Society},
  \urlprefix\url{https://link.aps.org/doi/10.1103/PhysRevLett.101.107005}.

\bibitem[{\citenamefont{Dolcini et~al.}(2015)\citenamefont{Dolcini, Houzet, and
  Meyer}}]{DolciniMeyerHouzet15}
\bibinfo{author}{\bibfnamefont{F.}~\bibnamefont{Dolcini}},
  \bibinfo{author}{\bibfnamefont{M.}~\bibnamefont{Houzet}}, \bibnamefont{and}
  \bibinfo{author}{\bibfnamefont{J.~S.} \bibnamefont{Meyer}},
  \bibinfo{journal}{Phys. Rev. B} \textbf{\bibinfo{volume}{92}},
  \bibinfo{pages}{035428} (\bibinfo{year}{2015}),
  \urlprefix\url{https://link.aps.org/doi/10.1103/PhysRevB.92.035428}.

\bibitem[{\citenamefont{Hasan et~al.}(2022)\citenamefont{Hasan, Nesterov, Li,
  Houzet, Meyer, and Levchenko}}]{HasanSongciLevchenko22}
\bibinfo{author}{\bibfnamefont{J.}~\bibnamefont{Hasan}},
  \bibinfo{author}{\bibfnamefont{K.~N.} \bibnamefont{Nesterov}},
  \bibinfo{author}{\bibfnamefont{S.}~\bibnamefont{Li}},
  \bibinfo{author}{\bibfnamefont{M.}~\bibnamefont{Houzet}},
  \bibinfo{author}{\bibfnamefont{J.~S.} \bibnamefont{Meyer}}, \bibnamefont{and}
  \bibinfo{author}{\bibfnamefont{A.}~\bibnamefont{Levchenko}},
  \bibinfo{journal}{Physical Review B} \textbf{\bibinfo{volume}{106}},
  \bibinfo{pages}{214518} (\bibinfo{year}{2022}), \bibinfo{note}{publisher:
  American Physical Society},
  \urlprefix\url{https://link.aps.org/doi/10.1103/PhysRevB.106.214518}.

\bibitem[{\citenamefont{Amundsen et~al.}(2024)\citenamefont{Amundsen, Linder,
  Robinson, \ifmmode \check{Z}\else \v{Z}\fi{}uti\ifmmode~\acute{c}\else
  \'{c}\fi{}, and Banerjee}}]{AmundsenZutic24}
\bibinfo{author}{\bibfnamefont{M.}~\bibnamefont{Amundsen}},
  \bibinfo{author}{\bibfnamefont{J.}~\bibnamefont{Linder}},
  \bibinfo{author}{\bibfnamefont{J.~W.~A.} \bibnamefont{Robinson}},
  \bibinfo{author}{\bibfnamefont{I.}~\bibnamefont{\ifmmode \check{Z}\else
  \v{Z}\fi{}uti\ifmmode~\acute{c}\else \'{c}\fi{}}}, \bibnamefont{and}
  \bibinfo{author}{\bibfnamefont{N.}~\bibnamefont{Banerjee}},
  \bibinfo{journal}{Rev. Mod. Phys.} \textbf{\bibinfo{volume}{96}},
  \bibinfo{pages}{021003} (\bibinfo{year}{2024}),
  \urlprefix\url{https://link.aps.org/doi/10.1103/RevModPhys.96.021003}.

\bibitem[{\citenamefont{Roig et~al.}(2024{\natexlab{a}})\citenamefont{Roig,
  Kotetes, and Andersen}}]{Roig24}
\bibinfo{author}{\bibfnamefont{M.}~\bibnamefont{Roig}},
  \bibinfo{author}{\bibfnamefont{P.}~\bibnamefont{Kotetes}}, \bibnamefont{and}
  \bibinfo{author}{\bibfnamefont{B.~M.} \bibnamefont{Andersen}},
  \bibinfo{journal}{Phys. Rev. B} \textbf{\bibinfo{volume}{109}},
  \bibinfo{pages}{144503} (\bibinfo{year}{2024}{\natexlab{a}}),
  \urlprefix\url{https://link.aps.org/doi/10.1103/PhysRevB.109.144503}.

\bibitem[{\citenamefont{Osin et~al.}(2024)\citenamefont{Osin, Levchenko, and
  Khodas}}]{OsinLevchenkoKhodas24}
\bibinfo{author}{\bibfnamefont{A.~S.} \bibnamefont{Osin}},
  \bibinfo{author}{\bibfnamefont{A.}~\bibnamefont{Levchenko}},
  \bibnamefont{and} \bibinfo{author}{\bibfnamefont{M.}~\bibnamefont{Khodas}},
  \bibinfo{journal}{Physical Review B} \textbf{\bibinfo{volume}{109}},
  \bibinfo{pages}{184512} (\bibinfo{year}{2024}), \bibinfo{note}{publisher:
  American Physical Society},
  \urlprefix\url{https://link.aps.org/doi/10.1103/PhysRevB.109.184512}.

\bibitem[{\citenamefont{Davydova et~al.}(2022)\citenamefont{Davydova, Prembabu,
  and Fu}}]{DavydovaFu22}
\bibinfo{author}{\bibfnamefont{M.}~\bibnamefont{Davydova}},
  \bibinfo{author}{\bibfnamefont{S.}~\bibnamefont{Prembabu}}, \bibnamefont{and}
  \bibinfo{author}{\bibfnamefont{L.}~\bibnamefont{Fu}},
  \bibinfo{journal}{Science Advances} \textbf{\bibinfo{volume}{8}},
  \bibinfo{pages}{eabo0309} (\bibinfo{year}{2022}), \bibinfo{note}{publisher:
  American Association for the Advancement of Science},
  \urlprefix\url{https://www.science.org/doi/10.1126/sciadv.abo0309}.

\bibitem[{\citenamefont{Zhang}(2024)}]{Zhang24}
\bibinfo{author}{\bibfnamefont{X.-P.} \bibnamefont{Zhang}},
  \bibinfo{journal}{Physical Review B} \textbf{\bibinfo{volume}{109}},
  \bibinfo{pages}{184513} (\bibinfo{year}{2024}), \bibinfo{note}{publisher:
  American Physical Society},
  \urlprefix\url{https://link.aps.org/doi/10.1103/PhysRevB.109.184513}.

\bibitem[{\citenamefont{Huang et~al.}(2024)\citenamefont{Huang, de~Picoli, and
  Väyrynen}}]{HuangVayrynen24}
\bibinfo{author}{\bibfnamefont{H.}~\bibnamefont{Huang}},
  \bibinfo{author}{\bibfnamefont{T.}~\bibnamefont{de~Picoli}},
  \bibnamefont{and} \bibinfo{author}{\bibfnamefont{J.~I.}
  \bibnamefont{Väyrynen}}, \bibinfo{journal}{Applied Physics Letters}
  \textbf{\bibinfo{volume}{125}}, \bibinfo{pages}{032602}
  (\bibinfo{year}{2024}), ISSN \bibinfo{issn}{0003-6951},
  \eprint{https://pubs.aip.org/aip/apl/article-pdf/doi/10.1063/5.0213137/20051756/032602_1_5.0213137.pdf},
  \urlprefix\url{https://doi.org/10.1063/5.0213137}.

\bibitem[{\citenamefont{Karabassov et~al.}(2024)\citenamefont{Karabassov,
  Bobkova, Stolyarov, Silkin, and Vasenko}}]{KarabassovBobkovaVasenko24}
\bibinfo{author}{\bibfnamefont{T.}~\bibnamefont{Karabassov}},
  \bibinfo{author}{\bibfnamefont{I.~V.} \bibnamefont{Bobkova}},
  \bibinfo{author}{\bibfnamefont{V.~S.} \bibnamefont{Stolyarov}},
  \bibinfo{author}{\bibfnamefont{V.~M.} \bibnamefont{Silkin}},
  \bibnamefont{and} \bibinfo{author}{\bibfnamefont{A.~S.}
  \bibnamefont{Vasenko}} (\bibinfo{year}{2024}),
  \bibinfo{note}{arXiv:2407.14852 [cond-mat]},
  \urlprefix\url{http://arxiv.org/abs/2407.14852}.

\bibitem[{\citenamefont{Zazunov et~al.}(2009)\citenamefont{Zazunov, Egger,
  Jonckheere, and Martin}}]{ZazunovEggerMartin09}
\bibinfo{author}{\bibfnamefont{A.}~\bibnamefont{Zazunov}},
  \bibinfo{author}{\bibfnamefont{R.}~\bibnamefont{Egger}},
  \bibinfo{author}{\bibfnamefont{T.}~\bibnamefont{Jonckheere}},
  \bibnamefont{and} \bibinfo{author}{\bibfnamefont{T.}~\bibnamefont{Martin}},
  \bibinfo{journal}{Physical Review Letters} \textbf{\bibinfo{volume}{103}},
  \bibinfo{pages}{147004} (\bibinfo{year}{2009}), \bibinfo{note}{publisher:
  American Physical Society},
  \urlprefix\url{https://link.aps.org/doi/10.1103/PhysRevLett.103.147004}.

\bibitem[{\citenamefont{Reynoso et~al.}(2012)\citenamefont{Reynoso, Usaj,
  Balseiro, Feinberg, and Avignon}}]{ReynosoAvignon12}
\bibinfo{author}{\bibfnamefont{A.~A.} \bibnamefont{Reynoso}},
  \bibinfo{author}{\bibfnamefont{G.}~\bibnamefont{Usaj}},
  \bibinfo{author}{\bibfnamefont{C.~A.} \bibnamefont{Balseiro}},
  \bibinfo{author}{\bibfnamefont{D.}~\bibnamefont{Feinberg}}, \bibnamefont{and}
  \bibinfo{author}{\bibfnamefont{M.}~\bibnamefont{Avignon}},
  \bibinfo{journal}{Physical Review B} \textbf{\bibinfo{volume}{86}},
  \bibinfo{pages}{214519} (\bibinfo{year}{2012}), \bibinfo{note}{publisher:
  American Physical Society},
  \urlprefix\url{https://link.aps.org/doi/10.1103/PhysRevB.86.214519}.

\bibitem[{\citenamefont{Yokoyama et~al.}(2013)\citenamefont{Yokoyama, Eto, and
  V.~Nazarov}}]{YokoyamaNazarov13}
\bibinfo{author}{\bibfnamefont{T.}~\bibnamefont{Yokoyama}},
  \bibinfo{author}{\bibfnamefont{M.}~\bibnamefont{Eto}}, \bibnamefont{and}
  \bibinfo{author}{\bibfnamefont{Y.}~\bibnamefont{V.~Nazarov}},
  \bibinfo{journal}{Journal of the Physical Society of Japan}
  \textbf{\bibinfo{volume}{82}}, \bibinfo{pages}{054703}
  (\bibinfo{year}{2013}), ISSN \bibinfo{issn}{0031-9015},
  \bibinfo{note}{publisher: The Physical Society of Japan},
  \urlprefix\url{https://journals-jps-jp.ezproxy.library.wisc.edu/doi/10.7566/JPSJ.82.054703}.

\bibitem[{\citenamefont{Yokoyama et~al.}(2014)\citenamefont{Yokoyama, Eto, and
  Nazarov}}]{YokoyamaNazarov14}
\bibinfo{author}{\bibfnamefont{T.}~\bibnamefont{Yokoyama}},
  \bibinfo{author}{\bibfnamefont{M.}~\bibnamefont{Eto}}, \bibnamefont{and}
  \bibinfo{author}{\bibfnamefont{Y.~V.} \bibnamefont{Nazarov}},
  \bibinfo{journal}{Physical Review B} \textbf{\bibinfo{volume}{89}},
  \bibinfo{pages}{195407} (\bibinfo{year}{2014}), \bibinfo{note}{publisher:
  American Physical Society},
  \urlprefix\url{https://link.aps.org/doi/10.1103/PhysRevB.89.195407}.

\bibitem[{\citenamefont{Nesterov et~al.}(2016)\citenamefont{Nesterov, Houzet,
  and Meyer}}]{NesterovHouzetMeyer16}
\bibinfo{author}{\bibfnamefont{K.~N.} \bibnamefont{Nesterov}},
  \bibinfo{author}{\bibfnamefont{M.}~\bibnamefont{Houzet}}, \bibnamefont{and}
  \bibinfo{author}{\bibfnamefont{J.~S.} \bibnamefont{Meyer}},
  \bibinfo{journal}{Physical Review B} \textbf{\bibinfo{volume}{93}},
  \bibinfo{pages}{174502} (\bibinfo{year}{2016}), \bibinfo{note}{publisher:
  American Physical Society},
  \urlprefix\url{https://link.aps.org/doi/10.1103/PhysRevB.93.174502}.

\bibitem[{\citenamefont{Pekerten et~al.}(2022)\citenamefont{Pekerten, Pakizer,
  Hawn, and Matos-Abiague}}]{PekertenMatosAbiague22}
\bibinfo{author}{\bibfnamefont{B.}~\bibnamefont{Pekerten}},
  \bibinfo{author}{\bibfnamefont{J.~D.} \bibnamefont{Pakizer}},
  \bibinfo{author}{\bibfnamefont{B.}~\bibnamefont{Hawn}}, \bibnamefont{and}
  \bibinfo{author}{\bibfnamefont{A.}~\bibnamefont{Matos-Abiague}},
  \bibinfo{journal}{Physical Review B} \textbf{\bibinfo{volume}{105}},
  \bibinfo{pages}{054504} (\bibinfo{year}{2022}), \bibinfo{note}{publisher:
  American Physical Society},
  \urlprefix\url{https://link.aps.org/doi/10.1103/PhysRevB.105.054504}.

\bibitem[{\citenamefont{Wang et~al.}(2022)\citenamefont{Wang, Wang, and
  Wu}}]{WangWangWu22}
\bibinfo{author}{\bibfnamefont{D.}~\bibnamefont{Wang}},
  \bibinfo{author}{\bibfnamefont{Q.-H.} \bibnamefont{Wang}}, \bibnamefont{and}
  \bibinfo{author}{\bibfnamefont{C.}~\bibnamefont{Wu}} (\bibinfo{year}{2022}),
  \bibinfo{note}{arXiv:2209.12646 [cond-mat]},
  \urlprefix\url{http://arxiv.org/abs/2209.12646}.

\bibitem[{\citenamefont{Zhang et~al.}(2022)\citenamefont{Zhang, Gu, Li, Hu, and
  Jiang}}]{ZhangJiang22}
\bibinfo{author}{\bibfnamefont{Y.}~\bibnamefont{Zhang}},
  \bibinfo{author}{\bibfnamefont{Y.}~\bibnamefont{Gu}},
  \bibinfo{author}{\bibfnamefont{P.}~\bibnamefont{Li}},
  \bibinfo{author}{\bibfnamefont{J.}~\bibnamefont{Hu}}, \bibnamefont{and}
  \bibinfo{author}{\bibfnamefont{K.}~\bibnamefont{Jiang}},
  \bibinfo{journal}{Phys. Rev. X} \textbf{\bibinfo{volume}{12}},
  \bibinfo{pages}{041013} (\bibinfo{year}{2022}),
  \urlprefix\url{https://link.aps.org/doi/10.1103/PhysRevX.12.041013}.

\bibitem[{\citenamefont{Costa et~al.}(2023{\natexlab{a}})\citenamefont{Costa,
  Fabian, and Kochan}}]{CostaKochan23}
\bibinfo{author}{\bibfnamefont{A.}~\bibnamefont{Costa}},
  \bibinfo{author}{\bibfnamefont{J.}~\bibnamefont{Fabian}}, \bibnamefont{and}
  \bibinfo{author}{\bibfnamefont{D.}~\bibnamefont{Kochan}},
  \bibinfo{journal}{Physical Review B} \textbf{\bibinfo{volume}{108}},
  \bibinfo{pages}{054522} (\bibinfo{year}{2023}{\natexlab{a}}),
  \bibinfo{note}{publisher: American Physical Society},
  \urlprefix\url{https://link.aps.org/doi/10.1103/PhysRevB.108.054522}.

\bibitem[{\citenamefont{Costa et~al.}(2023{\natexlab{b}})\citenamefont{Costa,
  Baumgartner, Reinhardt, Berger, Gronin, Gardner, Lindemann, Manfra, Fabian,
  Kochan et~al.}}]{CostaManfraKochanParadisoStrunk23}
\bibinfo{author}{\bibfnamefont{A.}~\bibnamefont{Costa}},
  \bibinfo{author}{\bibfnamefont{C.}~\bibnamefont{Baumgartner}},
  \bibinfo{author}{\bibfnamefont{S.}~\bibnamefont{Reinhardt}},
  \bibinfo{author}{\bibfnamefont{J.}~\bibnamefont{Berger}},
  \bibinfo{author}{\bibfnamefont{S.}~\bibnamefont{Gronin}},
  \bibinfo{author}{\bibfnamefont{G.~C.} \bibnamefont{Gardner}},
  \bibinfo{author}{\bibfnamefont{T.}~\bibnamefont{Lindemann}},
  \bibinfo{author}{\bibfnamefont{M.~J.} \bibnamefont{Manfra}},
  \bibinfo{author}{\bibfnamefont{J.}~\bibnamefont{Fabian}},
  \bibinfo{author}{\bibfnamefont{D.}~\bibnamefont{Kochan}},
  \bibnamefont{et~al.}, \bibinfo{journal}{Nature Nanotechnology}
  \textbf{\bibinfo{volume}{18}}, \bibinfo{pages}{1266}
  (\bibinfo{year}{2023}{\natexlab{b}}), ISSN \bibinfo{issn}{1748-3395},
  \bibinfo{note}{publisher: Nature Publishing Group},
  \urlprefix\url{https://www-nature-com.ezproxy.library.wisc.edu/articles/s41565-023-01451-x}.

\bibitem[{\citenamefont{Huang}(2023)}]{Huang23}
\bibinfo{author}{\bibfnamefont{C.-S.} \bibnamefont{Huang}}
  (\bibinfo{year}{2023}), \bibinfo{note}{arXiv:2312.02692 [cond-mat]},
  \urlprefix\url{http://arxiv.org/abs/2312.02692}.

\bibitem[{\citenamefont{Liu et~al.}(2024{\natexlab{a}})\citenamefont{Liu,
  Smith, Andreev, and Spivak}}]{LiuAndreevSpivak24}
\bibinfo{author}{\bibfnamefont{T.}~\bibnamefont{Liu}},
  \bibinfo{author}{\bibfnamefont{M.}~\bibnamefont{Smith}},
  \bibinfo{author}{\bibfnamefont{A.~V.} \bibnamefont{Andreev}},
  \bibnamefont{and} \bibinfo{author}{\bibfnamefont{B.~Z.}
  \bibnamefont{Spivak}}, \bibinfo{journal}{Physical Review B}
  \textbf{\bibinfo{volume}{109}}, \bibinfo{pages}{L020501}
  (\bibinfo{year}{2024}{\natexlab{a}}), \bibinfo{note}{publisher: American
  Physical Society},
  \urlprefix\url{https://link.aps.org/doi/10.1103/PhysRevB.109.L020501}.

\bibitem[{\citenamefont{Pekerten et~al.}(2024)\citenamefont{Pekerten, Brandão,
  Bussiere, Monroe, Zhou, Han, Shabani, Matos-Abiague, and
  Žutić}}]{PekertenMatosAbiagueZutic24}
\bibinfo{author}{\bibfnamefont{B.}~\bibnamefont{Pekerten}},
  \bibinfo{author}{\bibfnamefont{D.~S.} \bibnamefont{Brandão}},
  \bibinfo{author}{\bibfnamefont{B.}~\bibnamefont{Bussiere}},
  \bibinfo{author}{\bibfnamefont{D.}~\bibnamefont{Monroe}},
  \bibinfo{author}{\bibfnamefont{T.}~\bibnamefont{Zhou}},
  \bibinfo{author}{\bibfnamefont{J.~E.} \bibnamefont{Han}},
  \bibinfo{author}{\bibfnamefont{J.}~\bibnamefont{Shabani}},
  \bibinfo{author}{\bibfnamefont{A.}~\bibnamefont{Matos-Abiague}},
  \bibnamefont{and} \bibinfo{author}{\bibfnamefont{I.}~\bibnamefont{Žutić}},
  \bibinfo{journal}{Applied Physics Letters} \textbf{\bibinfo{volume}{124}},
  \bibinfo{pages}{252602} (\bibinfo{year}{2024}), ISSN
  \bibinfo{issn}{0003-6951},
  \eprint{https://pubs.aip.org/aip/apl/article-pdf/doi/10.1063/5.0214920/20072919/252602_1_5.0214920.pdf},
  \urlprefix\url{https://doi.org/10.1063/5.0214920}.

\bibitem[{\citenamefont{Wang et~al.}(2024{\natexlab{b}})\citenamefont{Wang,
  Miao, and Chen}}]{WangChen24}
\bibinfo{author}{\bibfnamefont{G.}~\bibnamefont{Wang}},
  \bibinfo{author}{\bibfnamefont{J.}~\bibnamefont{Miao}}, \bibnamefont{and}
  \bibinfo{author}{\bibfnamefont{W.-Q.} \bibnamefont{Chen}}
  (\bibinfo{year}{2024}{\natexlab{b}}), \bibinfo{note}{arXiv:2408.15661
  [cond-mat]}, \urlprefix\url{http://arxiv.org/abs/2408.15661}.

\bibitem[{\citenamefont{Soori}(2024)}]{Soori24}
\bibinfo{author}{\bibfnamefont{A.}~\bibnamefont{Soori}},
  \bibinfo{journal}{Journal of Physics: Condensed Matter}
  \textbf{\bibinfo{volume}{36}}, \bibinfo{pages}{335303}
  (\bibinfo{year}{2024}), ISSN \bibinfo{issn}{0953-8984},
  \bibinfo{note}{publisher: IOP Publishing},
  \urlprefix\url{https://dx.doi.org/10.1088/1361-648X/ad4aad}.

\bibitem[{\citenamefont{Ruiz et~al.}(2025)\citenamefont{Ruiz, Mateos, Tosi,
  Strunk, Balseiro, and Arrachea}}]{RuizStrunk25}
\bibinfo{author}{\bibfnamefont{G.~F.~R.} \bibnamefont{Ruiz}},
  \bibinfo{author}{\bibfnamefont{J.~H.} \bibnamefont{Mateos}},
  \bibinfo{author}{\bibfnamefont{L.}~\bibnamefont{Tosi}},
  \bibinfo{author}{\bibfnamefont{C.}~\bibnamefont{Strunk}},
  \bibinfo{author}{\bibfnamefont{C.}~\bibnamefont{Balseiro}}, \bibnamefont{and}
  \bibinfo{author}{\bibfnamefont{L.}~\bibnamefont{Arrachea}},
  \emph{\bibinfo{title}{Antichiral edge states and bogoliubov fermi surfaces in
  a two-dimensional proximity-induced superconductor}} (\bibinfo{year}{2025}),
  \eprint{2506.11256}, \urlprefix\url{https://arxiv.org/abs/2506.11256}.

\bibitem[{\citenamefont{Krekels et~al.}(2025)\citenamefont{Krekels, Levajac,
  Moors, Simion, and Sor\'ee}}]{Krekels25}
\bibinfo{author}{\bibfnamefont{S.}~\bibnamefont{Krekels}},
  \bibinfo{author}{\bibfnamefont{V.}~\bibnamefont{Levajac}},
  \bibinfo{author}{\bibfnamefont{K.}~\bibnamefont{Moors}},
  \bibinfo{author}{\bibfnamefont{G.}~\bibnamefont{Simion}}, \bibnamefont{and}
  \bibinfo{author}{\bibfnamefont{B.}~\bibnamefont{Sor\'ee}},
  \bibinfo{journal}{Phys. Rev. B} \textbf{\bibinfo{volume}{112}},
  \bibinfo{pages}{144508} (\bibinfo{year}{2025}),
  \urlprefix\url{https://link.aps.org/doi/10.1103/1xbv-bvch}.

\bibitem[{\citenamefont{Kopasov et~al.}(2021)\citenamefont{Kopasov, Kutlin, and
  Mel'nikov}}]{KopasovMelnikov21}
\bibinfo{author}{\bibfnamefont{A.~A.} \bibnamefont{Kopasov}},
  \bibinfo{author}{\bibfnamefont{A.~G.} \bibnamefont{Kutlin}},
  \bibnamefont{and} \bibinfo{author}{\bibfnamefont{A.~S.}
  \bibnamefont{Mel'nikov}}, \bibinfo{journal}{Physical Review B}
  \textbf{\bibinfo{volume}{103}}, \bibinfo{pages}{144520}
  (\bibinfo{year}{2021}), \bibinfo{note}{publisher: American Physical Society},
  \urlprefix\url{https://link.aps.org/doi/10.1103/PhysRevB.103.144520}.

\bibitem[{\citenamefont{Cayao et~al.}(2024)\citenamefont{Cayao, Nagaosa, and
  Tanaka}}]{CayaoNagaosaTanaka24}
\bibinfo{author}{\bibfnamefont{J.}~\bibnamefont{Cayao}},
  \bibinfo{author}{\bibfnamefont{N.}~\bibnamefont{Nagaosa}}, \bibnamefont{and}
  \bibinfo{author}{\bibfnamefont{Y.}~\bibnamefont{Tanaka}},
  \bibinfo{journal}{Physical Review B} \textbf{\bibinfo{volume}{109}},
  \bibinfo{pages}{L081405} (\bibinfo{year}{2024}), \bibinfo{note}{publisher:
  American Physical Society},
  \urlprefix\url{https://link.aps.org/doi/10.1103/PhysRevB.109.L081405}.

\bibitem[{\citenamefont{Liu et~al.}(2024{\natexlab{b}})\citenamefont{Liu,
  Huang, and Wang}}]{LiuWang24}
\bibinfo{author}{\bibfnamefont{Z.}~\bibnamefont{Liu}},
  \bibinfo{author}{\bibfnamefont{L.}~\bibnamefont{Huang}}, \bibnamefont{and}
  \bibinfo{author}{\bibfnamefont{J.}~\bibnamefont{Wang}},
  \bibinfo{journal}{Physical Review B} \textbf{\bibinfo{volume}{110}},
  \bibinfo{pages}{014519} (\bibinfo{year}{2024}{\natexlab{b}}),
  \bibinfo{note}{publisher: American Physical Society},
  \urlprefix\url{https://link.aps.org/doi/10.1103/PhysRevB.110.014519}.

\bibitem[{\citenamefont{Mondal et~al.}(2025)\citenamefont{Mondal, Fu, and
  Cayao}}]{MondalCayao25}
\bibinfo{author}{\bibfnamefont{S.}~\bibnamefont{Mondal}},
  \bibinfo{author}{\bibfnamefont{P.-H.} \bibnamefont{Fu}}, \bibnamefont{and}
  \bibinfo{author}{\bibfnamefont{J.}~\bibnamefont{Cayao}},
  \bibinfo{journal}{Phys. Rev. B} \textbf{\bibinfo{volume}{112}},
  \bibinfo{pages}{144506} (\bibinfo{year}{2025}),
  \urlprefix\url{https://link.aps.org/doi/10.1103/79tj-c3y4}.

\bibitem[{\citenamefont{Guan and An}(2025)}]{GuanAn25}
\bibinfo{author}{\bibfnamefont{S.-T.} \bibnamefont{Guan}} \bibnamefont{and}
  \bibinfo{author}{\bibfnamefont{J.}~\bibnamefont{An}} (\bibinfo{year}{2025}),
  \eprint{2506.16959}, \urlprefix\url{https://arxiv.org/abs/2506.16959}.

\bibitem[{\citenamefont{Geng et~al.}(2023)\citenamefont{Geng, Hijano, Ili{\'c},
  Ilyn, Maasilta, Monfardini, Spies, Strambini, Virtanen, Calvo
  et~al.}}]{GengBergeretHeikkila23}
\bibinfo{author}{\bibfnamefont{Z.}~\bibnamefont{Geng}},
  \bibinfo{author}{\bibfnamefont{A.}~\bibnamefont{Hijano}},
  \bibinfo{author}{\bibfnamefont{S.}~\bibnamefont{Ili{\'c}}},
  \bibinfo{author}{\bibfnamefont{M.}~\bibnamefont{Ilyn}},
  \bibinfo{author}{\bibfnamefont{I.}~\bibnamefont{Maasilta}},
  \bibinfo{author}{\bibfnamefont{A.}~\bibnamefont{Monfardini}},
  \bibinfo{author}{\bibfnamefont{M.}~\bibnamefont{Spies}},
  \bibinfo{author}{\bibfnamefont{E.}~\bibnamefont{Strambini}},
  \bibinfo{author}{\bibfnamefont{P.}~\bibnamefont{Virtanen}},
  \bibinfo{author}{\bibfnamefont{M.}~\bibnamefont{Calvo}},
  \bibnamefont{et~al.}, \bibinfo{journal}{Superconductor Science and
  Technology} \textbf{\bibinfo{volume}{36}}, \bibinfo{pages}{123001}
  (\bibinfo{year}{2023}).

\bibitem[{\citenamefont{Mitrovic and Bocko}(2025)}]{Mitrovic25}
\bibinfo{author}{\bibfnamefont{A.}~\bibnamefont{Mitrovic}} \bibnamefont{and}
  \bibinfo{author}{\bibfnamefont{M.}~\bibnamefont{Bocko}},
  \bibinfo{journal}{Phys. Rev. Appl.} \textbf{\bibinfo{volume}{23}},
  \bibinfo{pages}{067001} (\bibinfo{year}{2025}),
  \urlprefix\url{https://link.aps.org/doi/10.1103/PhysRevApplied.23.067001}.

\bibitem[{\citenamefont{Minutillo et~al.}(2018)\citenamefont{Minutillo,
  Giuliano, Lucignano, Tagliacozzo, and Campagnano}}]{MinutilloCampagnano18}
\bibinfo{author}{\bibfnamefont{M.}~\bibnamefont{Minutillo}},
  \bibinfo{author}{\bibfnamefont{D.}~\bibnamefont{Giuliano}},
  \bibinfo{author}{\bibfnamefont{P.}~\bibnamefont{Lucignano}},
  \bibinfo{author}{\bibfnamefont{A.}~\bibnamefont{Tagliacozzo}},
  \bibnamefont{and}
  \bibinfo{author}{\bibfnamefont{G.}~\bibnamefont{Campagnano}},
  \bibinfo{journal}{Physical Review B} \textbf{\bibinfo{volume}{98}},
  \bibinfo{pages}{144510} (\bibinfo{year}{2018}), \bibinfo{note}{publisher:
  American Physical Society},
  \urlprefix\url{https://link.aps.org/doi/10.1103/PhysRevB.98.144510}.

\bibitem[{\citenamefont{Hess et~al.}(2023)\citenamefont{Hess, Legg, Loss, and
  Klinovaja}}]{HessLeggLossKlinovaja23}
\bibinfo{author}{\bibfnamefont{R.}~\bibnamefont{Hess}},
  \bibinfo{author}{\bibfnamefont{H.~F.} \bibnamefont{Legg}},
  \bibinfo{author}{\bibfnamefont{D.}~\bibnamefont{Loss}}, \bibnamefont{and}
  \bibinfo{author}{\bibfnamefont{J.}~\bibnamefont{Klinovaja}}
  (\bibinfo{year}{2023}), \bibinfo{note}{arXiv:2308.04817 [cond-mat]},
  \urlprefix\url{http://arxiv.org/abs/2308.04817}.

\bibitem[{\citenamefont{Hikino}(2025)}]{Hikino25}
\bibinfo{author}{\bibfnamefont{S.-i.} \bibnamefont{Hikino}},
  \bibinfo{journal}{Journal of the Physical Society of Japan}
  \textbf{\bibinfo{volume}{94}}, \bibinfo{pages}{094701}
  (\bibinfo{year}{2025}).

\bibitem[{\citenamefont{Grein et~al.}(2009)\citenamefont{Grein, Eschrig,
  Metalidis, and Sch\"on}}]{GreinSchon09}
\bibinfo{author}{\bibfnamefont{R.}~\bibnamefont{Grein}},
  \bibinfo{author}{\bibfnamefont{M.}~\bibnamefont{Eschrig}},
  \bibinfo{author}{\bibfnamefont{G.}~\bibnamefont{Metalidis}},
  \bibnamefont{and} \bibinfo{author}{\bibfnamefont{G.}~\bibnamefont{Sch\"on}},
  \bibinfo{journal}{Phys. Rev. Lett.} \textbf{\bibinfo{volume}{102}},
  \bibinfo{pages}{227005} (\bibinfo{year}{2009}),
  \urlprefix\url{https://link.aps.org/doi/10.1103/PhysRevLett.102.227005}.

\bibitem[{\citenamefont{Margaris et~al.}(2010)\citenamefont{Margaris,
  Paltoglou, and Flytzanis}}]{MargarisFlytzanis10}
\bibinfo{author}{\bibfnamefont{I.}~\bibnamefont{Margaris}},
  \bibinfo{author}{\bibfnamefont{V.}~\bibnamefont{Paltoglou}},
  \bibnamefont{and}
  \bibinfo{author}{\bibfnamefont{N.}~\bibnamefont{Flytzanis}},
  \bibinfo{journal}{Journal of Physics: Condensed Matter}
  \textbf{\bibinfo{volume}{22}}, \bibinfo{pages}{445701}
  (\bibinfo{year}{2010}), ISSN \bibinfo{issn}{0953-8984},
  \urlprefix\url{https://dx.doi.org/10.1088/0953-8984/22/44/445701}.

\bibitem[{\citenamefont{Cheng et~al.}(2024)\citenamefont{Cheng, Mao, and
  Sun}}]{ChengMaoSun24}
\bibinfo{author}{\bibfnamefont{Q.}~\bibnamefont{Cheng}},
  \bibinfo{author}{\bibfnamefont{Y.}~\bibnamefont{Mao}}, \bibnamefont{and}
  \bibinfo{author}{\bibfnamefont{Q.-F.} \bibnamefont{Sun}},
  \bibinfo{journal}{Physical Review B} \textbf{\bibinfo{volume}{110}},
  \bibinfo{pages}{014518} (\bibinfo{year}{2024}), \bibinfo{note}{publisher:
  American Physical Society},
  \urlprefix\url{https://link.aps.org/doi/10.1103/PhysRevB.110.014518}.

\bibitem[{\citenamefont{Jiang et~al.}(2025)\citenamefont{Jiang, Liu, and
  Wang}}]{Jiang25}
\bibinfo{author}{\bibfnamefont{Y.}~\bibnamefont{Jiang}},
  \bibinfo{author}{\bibfnamefont{H.-L.} \bibnamefont{Liu}}, \bibnamefont{and}
  \bibinfo{author}{\bibfnamefont{J.}~\bibnamefont{Wang}},
  \bibinfo{journal}{Chinese Physics B}  (\bibinfo{year}{2025}).

\bibitem[{\citenamefont{Wang et~al.}(2025{\natexlab{b}})\citenamefont{Wang,
  Miao, and Chen}}]{WangChen25}
\bibinfo{author}{\bibfnamefont{G.-Q.} \bibnamefont{Wang}},
  \bibinfo{author}{\bibfnamefont{J.-J.} \bibnamefont{Miao}}, \bibnamefont{and}
  \bibinfo{author}{\bibfnamefont{W.-Q.} \bibnamefont{Chen}},
  \bibinfo{journal}{Phys. Rev. B} \textbf{\bibinfo{volume}{112}},
  \bibinfo{pages}{014508} (\bibinfo{year}{2025}{\natexlab{b}}),
  \urlprefix\url{https://link.aps.org/doi/10.1103/kwz7-5stj}.

\bibitem[{\citenamefont{Patil et~al.}(2024)\citenamefont{Patil, Tang, and
  Belzig}}]{PatilBelzig24}
\bibinfo{author}{\bibfnamefont{S.}~\bibnamefont{Patil}},
  \bibinfo{author}{\bibfnamefont{G.}~\bibnamefont{Tang}}, \bibnamefont{and}
  \bibinfo{author}{\bibfnamefont{W.}~\bibnamefont{Belzig}}
  (\bibinfo{year}{2024}), \bibinfo{note}{arXiv:2411.04061 [cond-mat,
  physics:physics]}, \urlprefix\url{http://arxiv.org/abs/2411.04061}.

\bibitem[{\citenamefont{Boruah et~al.}(2025)\citenamefont{Boruah, Acharjee, and
  Saikia}}]{Boruah25}
\bibinfo{author}{\bibfnamefont{A.}~\bibnamefont{Boruah}},
  \bibinfo{author}{\bibfnamefont{S.}~\bibnamefont{Acharjee}}, \bibnamefont{and}
  \bibinfo{author}{\bibfnamefont{P.~K.} \bibnamefont{Saikia}},
  \bibinfo{journal}{Phys. Rev. B} \textbf{\bibinfo{volume}{112}},
  \bibinfo{pages}{054505} (\bibinfo{year}{2025}),
  \urlprefix\url{https://link.aps.org/doi/10.1103/qzq2-f9kv}.

\bibitem[{\citenamefont{Costa et~al.}(2024)\citenamefont{Costa, Kanehira,
  Matsueda, and Fabian}}]{CostaFabian24}
\bibinfo{author}{\bibfnamefont{A.}~\bibnamefont{Costa}},
  \bibinfo{author}{\bibfnamefont{O.}~\bibnamefont{Kanehira}},
  \bibinfo{author}{\bibfnamefont{H.}~\bibnamefont{Matsueda}}, \bibnamefont{and}
  \bibinfo{author}{\bibfnamefont{J.}~\bibnamefont{Fabian}}
  (\bibinfo{year}{2024}), \bibinfo{note}{arXiv:2411.11570 [cond-mat]},
  \urlprefix\url{http://arxiv.org/abs/2411.11570}.

\bibitem[{\citenamefont{Alidoust et~al.}(2021)\citenamefont{Alidoust, Shen, and
  \ifmmode \check{Z}\else \v{Z}\fi{}uti\ifmmode~\acute{c}\else
  \'{c}\fi{}}}]{AlidoustZutic21}
\bibinfo{author}{\bibfnamefont{M.}~\bibnamefont{Alidoust}},
  \bibinfo{author}{\bibfnamefont{C.}~\bibnamefont{Shen}}, \bibnamefont{and}
  \bibinfo{author}{\bibfnamefont{I.}~\bibnamefont{\ifmmode \check{Z}\else
  \v{Z}\fi{}uti\ifmmode~\acute{c}\else \'{c}\fi{}}}, \bibinfo{journal}{Phys.
  Rev. B} \textbf{\bibinfo{volume}{103}}, \bibinfo{pages}{L060503}
  (\bibinfo{year}{2021}),
  \urlprefix\url{https://link.aps.org/doi/10.1103/PhysRevB.103.L060503}.

\bibitem[{\citenamefont{Vakili et~al.}(2024)\citenamefont{Vakili, Ali, and
  Kovalev}}]{VakiliKovalev24}
\bibinfo{author}{\bibfnamefont{H.}~\bibnamefont{Vakili}},
  \bibinfo{author}{\bibfnamefont{M.}~\bibnamefont{Ali}}, \bibnamefont{and}
  \bibinfo{author}{\bibfnamefont{A.~A.} \bibnamefont{Kovalev}}
  (\bibinfo{year}{2024}), \bibinfo{note}{arXiv:2406.11127 [cond-mat]},
  \urlprefix\url{http://arxiv.org/abs/2406.11127}.

\bibitem[{\citenamefont{Sharma and Thakurathi}(2025)}]{SharmaThakurathi25}
\bibinfo{author}{\bibfnamefont{L.}~\bibnamefont{Sharma}} \bibnamefont{and}
  \bibinfo{author}{\bibfnamefont{M.}~\bibnamefont{Thakurathi}},
  \bibinfo{journal}{Phys. Rev. B} \textbf{\bibinfo{volume}{112}},
  \bibinfo{pages}{104506} (\bibinfo{year}{2025}),
  \urlprefix\url{https://link.aps.org/doi/10.1103/yqsg-xdg8}.

\bibitem[{\citenamefont{Maiellaro et~al.}(2024)\citenamefont{Maiellaro, Trama,
  Settino, Guarcello, Romeo, and Citro}}]{MaiellaroCitro24}
\bibinfo{author}{\bibfnamefont{A.}~\bibnamefont{Maiellaro}},
  \bibinfo{author}{\bibfnamefont{M.}~\bibnamefont{Trama}},
  \bibinfo{author}{\bibfnamefont{J.}~\bibnamefont{Settino}},
  \bibinfo{author}{\bibfnamefont{C.}~\bibnamefont{Guarcello}},
  \bibinfo{author}{\bibfnamefont{F.}~\bibnamefont{Romeo}}, \bibnamefont{and}
  \bibinfo{author}{\bibfnamefont{R.}~\bibnamefont{Citro}},
  \bibinfo{journal}{SciPost Physics} \textbf{\bibinfo{volume}{17}},
  \bibinfo{pages}{101} (\bibinfo{year}{2024}), ISSN \bibinfo{issn}{2542-4653},
  \urlprefix\url{https://scipost.org/10.21468/SciPostPhys.17.4.101}.

\bibitem[{\citenamefont{Levichev et~al.}(2023)\citenamefont{Levichev,
  Pashenkin, Gusev, and Vodolazov}}]{LevichevVodolazov23}
\bibinfo{author}{\bibfnamefont{M.~Y.} \bibnamefont{Levichev}},
  \bibinfo{author}{\bibfnamefont{I.~Y.} \bibnamefont{Pashenkin}},
  \bibinfo{author}{\bibfnamefont{N.~S.} \bibnamefont{Gusev}}, \bibnamefont{and}
  \bibinfo{author}{\bibfnamefont{D.~Y.} \bibnamefont{Vodolazov}},
  \bibinfo{journal}{Phys. Rev. B} \textbf{\bibinfo{volume}{108}},
  \bibinfo{pages}{094517} (\bibinfo{year}{2023}),
  \urlprefix\url{https://link.aps.org/doi/10.1103/PhysRevB.108.094517}.

\bibitem[{\citenamefont{Sim and Knolle}(2025)}]{SimKnolle25}
\bibinfo{author}{\bibfnamefont{G.}~\bibnamefont{Sim}} \bibnamefont{and}
  \bibinfo{author}{\bibfnamefont{J.}~\bibnamefont{Knolle}},
  \bibinfo{journal}{Phys. Rev. B} \textbf{\bibinfo{volume}{112}},
  \bibinfo{pages}{L020502} (\bibinfo{year}{2025}),
  \urlprefix\url{https://link.aps.org/doi/10.1103/b7rh-v7nq}.

\bibitem[{\citenamefont{Mukasa and Masaki}(2025)}]{MukasaMasaki25}
\bibinfo{author}{\bibfnamefont{K.}~\bibnamefont{Mukasa}} \bibnamefont{and}
  \bibinfo{author}{\bibfnamefont{Y.}~\bibnamefont{Masaki}},
  \bibinfo{journal}{Journal of the Physical Society of Japan}
  \textbf{\bibinfo{volume}{94}}, \bibinfo{pages}{064705}
  (\bibinfo{year}{2025}), \eprint{https://doi.org/10.7566/JPSJ.94.064705},
  \urlprefix\url{https://doi.org/10.7566/JPSJ.94.064705}.

\bibitem[{\citenamefont{Matsumoto et~al.}(2025)\citenamefont{Matsumoto, Yanase,
  and Daido}}]{MatsumotoYanase25}
\bibinfo{author}{\bibfnamefont{T.}~\bibnamefont{Matsumoto}},
  \bibinfo{author}{\bibfnamefont{Y.}~\bibnamefont{Yanase}}, \bibnamefont{and}
  \bibinfo{author}{\bibfnamefont{A.}~\bibnamefont{Daido}},
  \bibinfo{journal}{Phys. Rev. B} \textbf{\bibinfo{volume}{111}},
  \bibinfo{pages}{064501} (\bibinfo{year}{2025}),
  \urlprefix\url{https://link.aps.org/doi/10.1103/PhysRevB.111.064501}.

\bibitem[{\citenamefont{Xie et~al.}(2023)\citenamefont{Xie, Efetov, and
  Law}}]{XieLaw23}
\bibinfo{author}{\bibfnamefont{Y.-M.} \bibnamefont{Xie}},
  \bibinfo{author}{\bibfnamefont{D.~K.} \bibnamefont{Efetov}},
  \bibnamefont{and} \bibinfo{author}{\bibfnamefont{K.~T.} \bibnamefont{Law}},
  \bibinfo{journal}{Phys. Rev. Res.} \textbf{\bibinfo{volume}{5}},
  \bibinfo{pages}{023029} (\bibinfo{year}{2023}),
  \urlprefix\url{https://link.aps.org/doi/10.1103/PhysRevResearch.5.023029}.

\bibitem[{\citenamefont{Nakamura et~al.}(2024)\citenamefont{Nakamura, Daido,
  and Yanase}}]{NakamuraYanase24}
\bibinfo{author}{\bibfnamefont{K.}~\bibnamefont{Nakamura}},
  \bibinfo{author}{\bibfnamefont{A.}~\bibnamefont{Daido}}, \bibnamefont{and}
  \bibinfo{author}{\bibfnamefont{Y.}~\bibnamefont{Yanase}},
  \bibinfo{journal}{Physical Review B} \textbf{\bibinfo{volume}{109}},
  \bibinfo{pages}{094501} (\bibinfo{year}{2024}), \bibinfo{note}{publisher:
  American Physical Society},
  \urlprefix\url{https://link.aps.org/doi/10.1103/PhysRevB.109.094501}.

\bibitem[{\citenamefont{Agterberg and Sigrist}(1998)}]{AgterbergSigrist98}
\bibinfo{author}{\bibfnamefont{D.~F.} \bibnamefont{Agterberg}}
  \bibnamefont{and} \bibinfo{author}{\bibfnamefont{M.}~\bibnamefont{Sigrist}},
  \bibinfo{journal}{Phys. Rev. Lett.} \textbf{\bibinfo{volume}{80}},
  \bibinfo{pages}{2689} (\bibinfo{year}{1998}),
  \urlprefix\url{https://link.aps.org/doi/10.1103/PhysRevLett.80.2689}.

\bibitem[{\citenamefont{Leridon et~al.}(2007)\citenamefont{Leridon, Ng, and
  Varma}}]{LeridonVarma07}
\bibinfo{author}{\bibfnamefont{B.}~\bibnamefont{Leridon}},
  \bibinfo{author}{\bibfnamefont{T.-K.} \bibnamefont{Ng}}, \bibnamefont{and}
  \bibinfo{author}{\bibfnamefont{C.~M.} \bibnamefont{Varma}},
  \bibinfo{journal}{Phys. Rev. Lett.} \textbf{\bibinfo{volume}{99}},
  \bibinfo{pages}{027002} (\bibinfo{year}{2007}),
  \urlprefix\url{https://link.aps.org/doi/10.1103/PhysRevLett.99.027002}.

\bibitem[{\citenamefont{Yerin et~al.}(2024)\citenamefont{Yerin, Drechsler,
  Varlamov, Cuoco, and Giazotto}}]{YerinGiazotto24}
\bibinfo{author}{\bibfnamefont{Y.}~\bibnamefont{Yerin}},
  \bibinfo{author}{\bibfnamefont{S.-L.} \bibnamefont{Drechsler}},
  \bibinfo{author}{\bibfnamefont{A.~A.} \bibnamefont{Varlamov}},
  \bibinfo{author}{\bibfnamefont{M.}~\bibnamefont{Cuoco}}, \bibnamefont{and}
  \bibinfo{author}{\bibfnamefont{F.}~\bibnamefont{Giazotto}},
  \bibinfo{journal}{Phys. Rev. B} \textbf{\bibinfo{volume}{110}},
  \bibinfo{pages}{054501} (\bibinfo{year}{2024}),
  \urlprefix\url{https://link.aps.org/doi/10.1103/PhysRevB.110.054501}.

\bibitem[{\citenamefont{Kanasugi and Yanase}(2022)}]{KanasugiYanase22}
\bibinfo{author}{\bibfnamefont{S.}~\bibnamefont{Kanasugi}} \bibnamefont{and}
  \bibinfo{author}{\bibfnamefont{Y.}~\bibnamefont{Yanase}},
  \bibinfo{journal}{Communications Physics} \textbf{\bibinfo{volume}{5}},
  \bibinfo{pages}{39} (\bibinfo{year}{2022}).

\bibitem[{\citenamefont{de~Azambuja and Möckli}(2025)}]{AzambujMockli25}
\bibinfo{author}{\bibfnamefont{M.~K.} \bibnamefont{de~Azambuja}}
  \bibnamefont{and} \bibinfo{author}{\bibfnamefont{D.}~\bibnamefont{Möckli}}
  (\bibinfo{year}{2025}), \eprint{2506.05550},
  \urlprefix\url{https://arxiv.org/abs/2506.05550}.

\bibitem[{\citenamefont{Yuan}(2025)}]{Yuan25}
\bibinfo{author}{\bibfnamefont{N.~F.~Q.} \bibnamefont{Yuan}}
  (\bibinfo{year}{2025}), \eprint{2502.18075},
  \urlprefix\url{https://arxiv.org/abs/2502.18075}.

\bibitem[{\citenamefont{Liu}(2017)}]{Liu17}
\bibinfo{author}{\bibfnamefont{C.-X.} \bibnamefont{Liu}},
  \bibinfo{journal}{Phys. Rev. Lett.} \textbf{\bibinfo{volume}{118}},
  \bibinfo{pages}{087001} (\bibinfo{year}{2017}),
  \urlprefix\url{https://link.aps.org/doi/10.1103/PhysRevLett.118.087001}.

\bibitem[{\citenamefont{M\"ockli et~al.}(2018)\citenamefont{M\"ockli, Yanase,
  and Sigrist}}]{MockliYanaseSigrsit18}
\bibinfo{author}{\bibfnamefont{D.}~\bibnamefont{M\"ockli}},
  \bibinfo{author}{\bibfnamefont{Y.}~\bibnamefont{Yanase}}, \bibnamefont{and}
  \bibinfo{author}{\bibfnamefont{M.}~\bibnamefont{Sigrist}},
  \bibinfo{journal}{Phys. Rev. B} \textbf{\bibinfo{volume}{97}},
  \bibinfo{pages}{144508} (\bibinfo{year}{2018}),
  \urlprefix\url{https://link.aps.org/doi/10.1103/PhysRevB.97.144508}.

\bibitem[{\citenamefont{Xie and Law}(2023)}]{XieLaw23orb}
\bibinfo{author}{\bibfnamefont{Y.-M.} \bibnamefont{Xie}} \bibnamefont{and}
  \bibinfo{author}{\bibfnamefont{K.~T.} \bibnamefont{Law}},
  \bibinfo{journal}{Phys. Rev. Lett.} \textbf{\bibinfo{volume}{131}},
  \bibinfo{pages}{016001} (\bibinfo{year}{2023}),
  \urlprefix\url{https://link.aps.org/doi/10.1103/PhysRevLett.131.016001}.

\bibitem[{\citenamefont{Wan et~al.}(2023)\citenamefont{Wan, Zheliuk, Yuan,
  Peng, Zhang, Liang, Zeitler, Wiedmann, Hussey, Palstra et~al.}}]{Wan23}
\bibinfo{author}{\bibfnamefont{P.}~\bibnamefont{Wan}},
  \bibinfo{author}{\bibfnamefont{O.}~\bibnamefont{Zheliuk}},
  \bibinfo{author}{\bibfnamefont{N.~F.} \bibnamefont{Yuan}},
  \bibinfo{author}{\bibfnamefont{X.}~\bibnamefont{Peng}},
  \bibinfo{author}{\bibfnamefont{L.}~\bibnamefont{Zhang}},
  \bibinfo{author}{\bibfnamefont{M.}~\bibnamefont{Liang}},
  \bibinfo{author}{\bibfnamefont{U.}~\bibnamefont{Zeitler}},
  \bibinfo{author}{\bibfnamefont{S.}~\bibnamefont{Wiedmann}},
  \bibinfo{author}{\bibfnamefont{N.~E.} \bibnamefont{Hussey}},
  \bibinfo{author}{\bibfnamefont{T.~T.} \bibnamefont{Palstra}},
  \bibnamefont{et~al.}, \bibinfo{journal}{Nature}
  \textbf{\bibinfo{volume}{619}}, \bibinfo{pages}{46} (\bibinfo{year}{2023}).

\bibitem[{\citenamefont{Zhao et~al.}(2023{\natexlab{b}})\citenamefont{Zhao,
  Debbeler, K{\"u}hne, Fecher, Gross, and Smet}}]{Zhao23}
\bibinfo{author}{\bibfnamefont{D.}~\bibnamefont{Zhao}},
  \bibinfo{author}{\bibfnamefont{L.}~\bibnamefont{Debbeler}},
  \bibinfo{author}{\bibfnamefont{M.}~\bibnamefont{K{\"u}hne}},
  \bibinfo{author}{\bibfnamefont{S.}~\bibnamefont{Fecher}},
  \bibinfo{author}{\bibfnamefont{N.}~\bibnamefont{Gross}}, \bibnamefont{and}
  \bibinfo{author}{\bibfnamefont{J.}~\bibnamefont{Smet}},
  \bibinfo{journal}{Nature Physics} \textbf{\bibinfo{volume}{19}},
  \bibinfo{pages}{1599} (\bibinfo{year}{2023}{\natexlab{b}}).

\bibitem[{\citenamefont{Clepkens and Kee}(2024)}]{Clepkens24}
\bibinfo{author}{\bibfnamefont{J.}~\bibnamefont{Clepkens}} \bibnamefont{and}
  \bibinfo{author}{\bibfnamefont{H.-Y.} \bibnamefont{Kee}},
  \bibinfo{journal}{Phys. Rev. B} \textbf{\bibinfo{volume}{109}},
  \bibinfo{pages}{214512} (\bibinfo{year}{2024}),
  \urlprefix\url{https://link.aps.org/doi/10.1103/PhysRevB.109.214512}.

\bibitem[{\citenamefont{Nag et~al.}(2024)\citenamefont{Nag, Schirmer, Rossi,
  Liu, and Jain}}]{NagJain24}
\bibinfo{author}{\bibfnamefont{U.}~\bibnamefont{Nag}},
  \bibinfo{author}{\bibfnamefont{J.}~\bibnamefont{Schirmer}},
  \bibinfo{author}{\bibfnamefont{E.}~\bibnamefont{Rossi}},
  \bibinfo{author}{\bibfnamefont{C.~X.} \bibnamefont{Liu}}, \bibnamefont{and}
  \bibinfo{author}{\bibfnamefont{J.~K.} \bibnamefont{Jain}}
  (\bibinfo{year}{2024}), \eprint{2408.00689},
  \urlprefix\url{https://arxiv.org/abs/2408.00689}.

\bibitem[{\citenamefont{Cao et~al.}(2024)\citenamefont{Cao, Liao, Yan, Zhu,
  Zhang, Watanabe, Taniguchi, Morpurgo, Liu, Xue et~al.}}]{Cao24}
\bibinfo{author}{\bibfnamefont{Z.}~\bibnamefont{Cao}},
  \bibinfo{author}{\bibfnamefont{M.}~\bibnamefont{Liao}},
  \bibinfo{author}{\bibfnamefont{H.}~\bibnamefont{Yan}},
  \bibinfo{author}{\bibfnamefont{Y.}~\bibnamefont{Zhu}},
  \bibinfo{author}{\bibfnamefont{L.}~\bibnamefont{Zhang}},
  \bibinfo{author}{\bibfnamefont{K.}~\bibnamefont{Watanabe}},
  \bibinfo{author}{\bibfnamefont{T.}~\bibnamefont{Taniguchi}},
  \bibinfo{author}{\bibfnamefont{A.~F.} \bibnamefont{Morpurgo}},
  \bibinfo{author}{\bibfnamefont{H.}~\bibnamefont{Liu}},
  \bibinfo{author}{\bibfnamefont{Q.-K.} \bibnamefont{Xue}},
  \bibnamefont{et~al.} (\bibinfo{year}{2024}), \eprint{2409.00373},
  \urlprefix\url{https://arxiv.org/abs/2409.00373}.

\bibitem[{\citenamefont{Yan et~al.}(2024)\citenamefont{Yan, Liu, Liu, Zhang,
  and Xie}}]{Yan24}
\bibinfo{author}{\bibfnamefont{H.}~\bibnamefont{Yan}},
  \bibinfo{author}{\bibfnamefont{H.}~\bibnamefont{Liu}},
  \bibinfo{author}{\bibfnamefont{Y.}~\bibnamefont{Liu}},
  \bibinfo{author}{\bibfnamefont{D.}~\bibnamefont{Zhang}}, \bibnamefont{and}
  \bibinfo{author}{\bibfnamefont{X.~C.} \bibnamefont{Xie}}
  (\bibinfo{year}{2024}), \eprint{2409.20336},
  \urlprefix\url{https://arxiv.org/abs/2409.20336}.

\bibitem[{\citenamefont{Zhao et~al.}(2025)\citenamefont{Zhao, Guo, Yan, Yuan,
  Zhao, Guan, Lan, Li, Liu, and Wang}}]{Zhao25}
\bibinfo{author}{\bibfnamefont{X.}~\bibnamefont{Zhao}},
  \bibinfo{author}{\bibfnamefont{G.}~\bibnamefont{Guo}},
  \bibinfo{author}{\bibfnamefont{C.}~\bibnamefont{Yan}},
  \bibinfo{author}{\bibfnamefont{N.~F.~Q.} \bibnamefont{Yuan}},
  \bibinfo{author}{\bibfnamefont{C.}~\bibnamefont{Zhao}},
  \bibinfo{author}{\bibfnamefont{H.}~\bibnamefont{Guan}},
  \bibinfo{author}{\bibfnamefont{C.}~\bibnamefont{Lan}},
  \bibinfo{author}{\bibfnamefont{Y.}~\bibnamefont{Li}},
  \bibinfo{author}{\bibfnamefont{X.}~\bibnamefont{Liu}}, \bibnamefont{and}
  \bibinfo{author}{\bibfnamefont{S.}~\bibnamefont{Wang}}
  (\bibinfo{year}{2025}), \eprint{2411.08980},
  \urlprefix\url{https://arxiv.org/abs/2411.08980}.

\bibitem[{\citenamefont{Zhu et~al.}(2025)\citenamefont{Zhu, Chou, Huang, and
  Das~Sarma}}]{ZhuDasSarma25}
\bibinfo{author}{\bibfnamefont{J.}~\bibnamefont{Zhu}},
  \bibinfo{author}{\bibfnamefont{Y.-Z.} \bibnamefont{Chou}},
  \bibinfo{author}{\bibfnamefont{Y.}~\bibnamefont{Huang}}, \bibnamefont{and}
  \bibinfo{author}{\bibfnamefont{S.}~\bibnamefont{Das~Sarma}},
  \bibinfo{journal}{Phys. Rev. B} \textbf{\bibinfo{volume}{112}},
  \bibinfo{pages}{L020507} (\bibinfo{year}{2025}),
  \urlprefix\url{https://link.aps.org/doi/10.1103/bqns-1ld3}.

\bibitem[{\citenamefont{Chazono and Yanase}(2025)}]{ChazonoYanase25}
\bibinfo{author}{\bibfnamefont{M.}~\bibnamefont{Chazono}} \bibnamefont{and}
  \bibinfo{author}{\bibfnamefont{Y.}~\bibnamefont{Yanase}}
  (\bibinfo{year}{2025}), \eprint{2506.19706},
  \urlprefix\url{https://arxiv.org/abs/2506.19706}.

\bibitem[{\citenamefont{Yoshida et~al.}(2012)\citenamefont{Yoshida, Sigrist,
  and Yanase}}]{YoshidaSigristYanase12}
\bibinfo{author}{\bibfnamefont{T.}~\bibnamefont{Yoshida}},
  \bibinfo{author}{\bibfnamefont{M.}~\bibnamefont{Sigrist}}, \bibnamefont{and}
  \bibinfo{author}{\bibfnamefont{Y.}~\bibnamefont{Yanase}},
  \bibinfo{journal}{Phys. Rev. B} \textbf{\bibinfo{volume}{86}},
  \bibinfo{pages}{134514} (\bibinfo{year}{2012}),
  \urlprefix\url{https://link.aps.org/doi/10.1103/PhysRevB.86.134514}.

\bibitem[{\citenamefont{Yoshida et~al.}(2015)\citenamefont{Yoshida, Sigrist,
  and Yanase}}]{YoshidaSigristYanase15}
\bibinfo{author}{\bibfnamefont{T.}~\bibnamefont{Yoshida}},
  \bibinfo{author}{\bibfnamefont{M.}~\bibnamefont{Sigrist}}, \bibnamefont{and}
  \bibinfo{author}{\bibfnamefont{Y.}~\bibnamefont{Yanase}},
  \bibinfo{journal}{Phys. Rev. Lett.} \textbf{\bibinfo{volume}{115}},
  \bibinfo{pages}{027001} (\bibinfo{year}{2015}),
  \urlprefix\url{https://link.aps.org/doi/10.1103/PhysRevLett.115.027001}.

\bibitem[{\citenamefont{Nakamura and Yanase}(2017)}]{NakamuraYanase17}
\bibinfo{author}{\bibfnamefont{Y.}~\bibnamefont{Nakamura}} \bibnamefont{and}
  \bibinfo{author}{\bibfnamefont{Y.}~\bibnamefont{Yanase}},
  \bibinfo{journal}{Phys. Rev. B} \textbf{\bibinfo{volume}{96}},
  \bibinfo{pages}{054501} (\bibinfo{year}{2017}),
  \urlprefix\url{https://link.aps.org/doi/10.1103/PhysRevB.96.054501}.

\bibitem[{\citenamefont{Liu et~al.}(2024{\natexlab{c}})\citenamefont{Liu,
  Huang, Huang, Moritz, and Devereaux}}]{LiuDevereaux24}
\bibinfo{author}{\bibfnamefont{F.}~\bibnamefont{Liu}},
  \bibinfo{author}{\bibfnamefont{X.-X.} \bibnamefont{Huang}},
  \bibinfo{author}{\bibfnamefont{E.~W.} \bibnamefont{Huang}},
  \bibinfo{author}{\bibfnamefont{B.}~\bibnamefont{Moritz}}, \bibnamefont{and}
  \bibinfo{author}{\bibfnamefont{T.~P.} \bibnamefont{Devereaux}},
  \bibinfo{journal}{Phys. Rev. Lett.} \textbf{\bibinfo{volume}{133}},
  \bibinfo{pages}{156503} (\bibinfo{year}{2024}{\natexlab{c}}),
  \urlprefix\url{https://link.aps.org/doi/10.1103/PhysRevLett.133.156503}.

\bibitem[{\citenamefont{Burg et~al.}(2019)\citenamefont{Burg, Zhu, Taniguchi,
  Watanabe, MacDonald, and Tutuc}}]{BurgMacDonald19}
\bibinfo{author}{\bibfnamefont{G.~W.} \bibnamefont{Burg}},
  \bibinfo{author}{\bibfnamefont{J.}~\bibnamefont{Zhu}},
  \bibinfo{author}{\bibfnamefont{T.}~\bibnamefont{Taniguchi}},
  \bibinfo{author}{\bibfnamefont{K.}~\bibnamefont{Watanabe}},
  \bibinfo{author}{\bibfnamefont{A.~H.} \bibnamefont{MacDonald}},
  \bibnamefont{and} \bibinfo{author}{\bibfnamefont{E.}~\bibnamefont{Tutuc}},
  \bibinfo{journal}{Phys. Rev. Lett.} \textbf{\bibinfo{volume}{123}},
  \bibinfo{pages}{197702} (\bibinfo{year}{2019}),
  \urlprefix\url{https://link.aps.org/doi/10.1103/PhysRevLett.123.197702}.

\bibitem[{\citenamefont{Zondiner et~al.}(2020)\citenamefont{Zondiner, Rozen,
  Rodan-Legrain, Cao, Queiroz, Taniguchi, Watanabe, Oreg, von Oppen, Stern
  et~al.}}]{ZondinerVonOppen20}
\bibinfo{author}{\bibfnamefont{U.}~\bibnamefont{Zondiner}},
  \bibinfo{author}{\bibfnamefont{A.}~\bibnamefont{Rozen}},
  \bibinfo{author}{\bibfnamefont{D.}~\bibnamefont{Rodan-Legrain}},
  \bibinfo{author}{\bibfnamefont{Y.}~\bibnamefont{Cao}},
  \bibinfo{author}{\bibfnamefont{R.}~\bibnamefont{Queiroz}},
  \bibinfo{author}{\bibfnamefont{T.}~\bibnamefont{Taniguchi}},
  \bibinfo{author}{\bibfnamefont{K.}~\bibnamefont{Watanabe}},
  \bibinfo{author}{\bibfnamefont{Y.}~\bibnamefont{Oreg}},
  \bibinfo{author}{\bibfnamefont{F.}~\bibnamefont{von Oppen}},
  \bibinfo{author}{\bibfnamefont{A.}~\bibnamefont{Stern}},
  \bibnamefont{et~al.}, \bibinfo{journal}{Nature}
  \textbf{\bibinfo{volume}{582}}, \bibinfo{pages}{203} (\bibinfo{year}{2020}),
  \urlprefix\url{https://doi.org/10.1038/s41586-020-2373-y}.

\bibitem[{\citenamefont{Zhang et~al.}(2020)\citenamefont{Zhang, Hou, Zhao, Guo,
  Liu, Li, Ren, Sun, and He}}]{Zhang20}
\bibinfo{author}{\bibfnamefont{Y.}~\bibnamefont{Zhang}},
  \bibinfo{author}{\bibfnamefont{Z.}~\bibnamefont{Hou}},
  \bibinfo{author}{\bibfnamefont{Y.-X.} \bibnamefont{Zhao}},
  \bibinfo{author}{\bibfnamefont{Z.-H.} \bibnamefont{Guo}},
  \bibinfo{author}{\bibfnamefont{Y.-W.} \bibnamefont{Liu}},
  \bibinfo{author}{\bibfnamefont{S.-Y.} \bibnamefont{Li}},
  \bibinfo{author}{\bibfnamefont{Y.-N.} \bibnamefont{Ren}},
  \bibinfo{author}{\bibfnamefont{Q.-F.} \bibnamefont{Sun}}, \bibnamefont{and}
  \bibinfo{author}{\bibfnamefont{L.}~\bibnamefont{He}}, \bibinfo{journal}{Phys.
  Rev. B} \textbf{\bibinfo{volume}{102}}, \bibinfo{pages}{081403}
  (\bibinfo{year}{2020}),
  \urlprefix\url{https://link.aps.org/doi/10.1103/PhysRevB.102.081403}.

\bibitem[{\citenamefont{Li et~al.}(2021)\citenamefont{Li, Jiang, Shen, Zhang,
  Li, Tao, Devakul, Watanabe, Taniguchi, Fu et~al.}}]{LiFuMak21}
\bibinfo{author}{\bibfnamefont{T.}~\bibnamefont{Li}},
  \bibinfo{author}{\bibfnamefont{S.}~\bibnamefont{Jiang}},
  \bibinfo{author}{\bibfnamefont{B.}~\bibnamefont{Shen}},
  \bibinfo{author}{\bibfnamefont{Y.}~\bibnamefont{Zhang}},
  \bibinfo{author}{\bibfnamefont{L.}~\bibnamefont{Li}},
  \bibinfo{author}{\bibfnamefont{Z.}~\bibnamefont{Tao}},
  \bibinfo{author}{\bibfnamefont{T.}~\bibnamefont{Devakul}},
  \bibinfo{author}{\bibfnamefont{K.}~\bibnamefont{Watanabe}},
  \bibinfo{author}{\bibfnamefont{T.}~\bibnamefont{Taniguchi}},
  \bibinfo{author}{\bibfnamefont{L.}~\bibnamefont{Fu}}, \bibnamefont{et~al.},
  \bibinfo{journal}{Nature} \textbf{\bibinfo{volume}{600}},
  \bibinfo{pages}{641} (\bibinfo{year}{2021}),
  \urlprefix\url{https://doi.org/10.1038/s41586-021-04171-1}.

\bibitem[{\citenamefont{Saito et~al.}(2021)\citenamefont{Saito, Yang, Ge, Liu,
  Taniguchi, Watanabe, Li, Berg, and Young}}]{SaitoYoung21}
\bibinfo{author}{\bibfnamefont{Y.}~\bibnamefont{Saito}},
  \bibinfo{author}{\bibfnamefont{F.}~\bibnamefont{Yang}},
  \bibinfo{author}{\bibfnamefont{J.}~\bibnamefont{Ge}},
  \bibinfo{author}{\bibfnamefont{X.}~\bibnamefont{Liu}},
  \bibinfo{author}{\bibfnamefont{T.}~\bibnamefont{Taniguchi}},
  \bibinfo{author}{\bibfnamefont{K.}~\bibnamefont{Watanabe}},
  \bibinfo{author}{\bibfnamefont{J.~I.~A.} \bibnamefont{Li}},
  \bibinfo{author}{\bibfnamefont{E.}~\bibnamefont{Berg}}, \bibnamefont{and}
  \bibinfo{author}{\bibfnamefont{A.~F.} \bibnamefont{Young}},
  \bibinfo{journal}{Nature} \textbf{\bibinfo{volume}{592}},
  \bibinfo{pages}{220} (\bibinfo{year}{2021}),
  \urlprefix\url{https://doi.org/10.1038/s41586-021-03409-2}.

\bibitem[{\citenamefont{Yu et~al.}(2022)\citenamefont{Yu, Foutty, Han, Barber,
  Schattner, Watanabe, Taniguchi, Phillips, Shen, Kivelson
  et~al.}}]{YuFeldman22}
\bibinfo{author}{\bibfnamefont{J.}~\bibnamefont{Yu}},
  \bibinfo{author}{\bibfnamefont{B.~A.} \bibnamefont{Foutty}},
  \bibinfo{author}{\bibfnamefont{Z.}~\bibnamefont{Han}},
  \bibinfo{author}{\bibfnamefont{M.~E.} \bibnamefont{Barber}},
  \bibinfo{author}{\bibfnamefont{Y.}~\bibnamefont{Schattner}},
  \bibinfo{author}{\bibfnamefont{K.}~\bibnamefont{Watanabe}},
  \bibinfo{author}{\bibfnamefont{T.}~\bibnamefont{Taniguchi}},
  \bibinfo{author}{\bibfnamefont{P.}~\bibnamefont{Phillips}},
  \bibinfo{author}{\bibfnamefont{Z.-X.} \bibnamefont{Shen}},
  \bibinfo{author}{\bibfnamefont{S.~A.} \bibnamefont{Kivelson}},
  \bibnamefont{et~al.}, \bibinfo{journal}{Nature Physics}
  \textbf{\bibinfo{volume}{18}}, \bibinfo{pages}{825} (\bibinfo{year}{2022}),
  \urlprefix\url{https://doi.org/10.1038/s41567-022-01589-w}.

\bibitem[{\citenamefont{Xie et~al.}(2022)\citenamefont{Xie, Zhang, Hu, Mak, and
  Law}}]{XieMakLaw22}
\bibinfo{author}{\bibfnamefont{Y.-M.} \bibnamefont{Xie}},
  \bibinfo{author}{\bibfnamefont{C.-P.} \bibnamefont{Zhang}},
  \bibinfo{author}{\bibfnamefont{J.-X.} \bibnamefont{Hu}},
  \bibinfo{author}{\bibfnamefont{K.~F.} \bibnamefont{Mak}}, \bibnamefont{and}
  \bibinfo{author}{\bibfnamefont{K.~T.} \bibnamefont{Law}},
  \bibinfo{journal}{Phys. Rev. Lett.} \textbf{\bibinfo{volume}{128}},
  \bibinfo{pages}{026402} (\bibinfo{year}{2022}),
  \urlprefix\url{https://link.aps.org/doi/10.1103/PhysRevLett.128.026402}.

\bibitem[{\citenamefont{Zhou et~al.}(2021)\citenamefont{Zhou, Xie, Ghazaryan,
  Holder, Ehrets, Spanton, Taniguchi, Watanabe, Berg, Serbyn
  et~al.}}]{ZhouYoung21}
\bibinfo{author}{\bibfnamefont{H.}~\bibnamefont{Zhou}},
  \bibinfo{author}{\bibfnamefont{T.}~\bibnamefont{Xie}},
  \bibinfo{author}{\bibfnamefont{A.}~\bibnamefont{Ghazaryan}},
  \bibinfo{author}{\bibfnamefont{T.}~\bibnamefont{Holder}},
  \bibinfo{author}{\bibfnamefont{J.~R.} \bibnamefont{Ehrets}},
  \bibinfo{author}{\bibfnamefont{E.~M.} \bibnamefont{Spanton}},
  \bibinfo{author}{\bibfnamefont{T.}~\bibnamefont{Taniguchi}},
  \bibinfo{author}{\bibfnamefont{K.}~\bibnamefont{Watanabe}},
  \bibinfo{author}{\bibfnamefont{E.}~\bibnamefont{Berg}},
  \bibinfo{author}{\bibfnamefont{M.}~\bibnamefont{Serbyn}},
  \bibnamefont{et~al.}, \bibinfo{journal}{Nature}
  \textbf{\bibinfo{volume}{598}}, \bibinfo{pages}{429} (\bibinfo{year}{2021}),
  \urlprefix\url{https://doi.org/10.1038/s41586-021-03938-w}.

\bibitem[{\citenamefont{de~la Barrera et~al.}(2022)\citenamefont{de~la Barrera,
  Aronson, Zheng, Watanabe, Taniguchi, Ma, Jarillo-Herrero, and
  Ashoori}}]{delaBarrera22}
\bibinfo{author}{\bibfnamefont{S.~C.} \bibnamefont{de~la Barrera}},
  \bibinfo{author}{\bibfnamefont{S.}~\bibnamefont{Aronson}},
  \bibinfo{author}{\bibfnamefont{Z.}~\bibnamefont{Zheng}},
  \bibinfo{author}{\bibfnamefont{K.}~\bibnamefont{Watanabe}},
  \bibinfo{author}{\bibfnamefont{T.}~\bibnamefont{Taniguchi}},
  \bibinfo{author}{\bibfnamefont{Q.}~\bibnamefont{Ma}},
  \bibinfo{author}{\bibfnamefont{P.}~\bibnamefont{Jarillo-Herrero}},
  \bibnamefont{and} \bibinfo{author}{\bibfnamefont{R.}~\bibnamefont{Ashoori}},
  \bibinfo{journal}{Nature Physics} \textbf{\bibinfo{volume}{18}},
  \bibinfo{pages}{771} (\bibinfo{year}{2022}),
  \urlprefix\url{https://doi.org/10.1038/s41567-022-01616-w}.

\bibitem[{\citenamefont{Seiler et~al.}(2022)\citenamefont{Seiler, Geisenhof,
  Winterer, Watanabe, Taniguchi, Xu, Zhang, and Weitz}}]{Seiler22}
\bibinfo{author}{\bibfnamefont{A.~M.} \bibnamefont{Seiler}},
  \bibinfo{author}{\bibfnamefont{F.~R.} \bibnamefont{Geisenhof}},
  \bibinfo{author}{\bibfnamefont{F.}~\bibnamefont{Winterer}},
  \bibinfo{author}{\bibfnamefont{K.}~\bibnamefont{Watanabe}},
  \bibinfo{author}{\bibfnamefont{T.}~\bibnamefont{Taniguchi}},
  \bibinfo{author}{\bibfnamefont{T.}~\bibnamefont{Xu}},
  \bibinfo{author}{\bibfnamefont{F.}~\bibnamefont{Zhang}}, \bibnamefont{and}
  \bibinfo{author}{\bibfnamefont{R.~T.} \bibnamefont{Weitz}},
  \bibinfo{journal}{Nature} \textbf{\bibinfo{volume}{608}},
  \bibinfo{pages}{298} (\bibinfo{year}{2022}),
  \urlprefix\url{https://doi.org/10.1038/s41586-022-04937-1}.

\bibitem[{\citenamefont{Pantale{\'o}n et~al.}(2023)\citenamefont{Pantale{\'o}n,
  Jimeno-Pozo, Sainz-Cruz, Phong, Cea, and Guinea}}]{PantaleonGuinea23}
\bibinfo{author}{\bibfnamefont{P.~A.} \bibnamefont{Pantale{\'o}n}},
  \bibinfo{author}{\bibfnamefont{A.}~\bibnamefont{Jimeno-Pozo}},
  \bibinfo{author}{\bibfnamefont{H.}~\bibnamefont{Sainz-Cruz}},
  \bibinfo{author}{\bibfnamefont{V.}~\bibnamefont{Phong}},
  \bibinfo{author}{\bibfnamefont{T.}~\bibnamefont{Cea}}, \bibnamefont{and}
  \bibinfo{author}{\bibfnamefont{F.}~\bibnamefont{Guinea}},
  \bibinfo{journal}{Nature Reviews Physics} \textbf{\bibinfo{volume}{5}},
  \bibinfo{pages}{304} (\bibinfo{year}{2023}),
  \urlprefix\url{https://doi.org/10.1038/s42254-023-00575-2}.

\bibitem[{\citenamefont{Li et~al.}(2024{\natexlab{b}})\citenamefont{Li, Xu, Li,
  Li, Li, Watanabe, Taniguchi, Tong, Shen, Lu et~al.}}]{LiLi24}
\bibinfo{author}{\bibfnamefont{C.}~\bibnamefont{Li}},
  \bibinfo{author}{\bibfnamefont{F.}~\bibnamefont{Xu}},
  \bibinfo{author}{\bibfnamefont{B.}~\bibnamefont{Li}},
  \bibinfo{author}{\bibfnamefont{J.}~\bibnamefont{Li}},
  \bibinfo{author}{\bibfnamefont{G.}~\bibnamefont{Li}},
  \bibinfo{author}{\bibfnamefont{K.}~\bibnamefont{Watanabe}},
  \bibinfo{author}{\bibfnamefont{T.}~\bibnamefont{Taniguchi}},
  \bibinfo{author}{\bibfnamefont{B.}~\bibnamefont{Tong}},
  \bibinfo{author}{\bibfnamefont{J.}~\bibnamefont{Shen}},
  \bibinfo{author}{\bibfnamefont{L.}~\bibnamefont{Lu}}, \bibnamefont{et~al.},
  \bibinfo{journal}{Nature} \textbf{\bibinfo{volume}{631}},
  \bibinfo{pages}{300} (\bibinfo{year}{2024}{\natexlab{b}}),
  \urlprefix\url{https://doi.org/10.1038/s41586-024-07584-w}.

\bibitem[{\citenamefont{Holleis et~al.}(2025)\citenamefont{Holleis, Patterson,
  Zhang, Vituri, Yoo, Zhou, Taniguchi, Watanabe, Berg, Nadj-Perge
  et~al.}}]{HolleisYoung25}
\bibinfo{author}{\bibfnamefont{L.}~\bibnamefont{Holleis}},
  \bibinfo{author}{\bibfnamefont{C.~L.} \bibnamefont{Patterson}},
  \bibinfo{author}{\bibfnamefont{Y.}~\bibnamefont{Zhang}},
  \bibinfo{author}{\bibfnamefont{Y.}~\bibnamefont{Vituri}},
  \bibinfo{author}{\bibfnamefont{H.~M.} \bibnamefont{Yoo}},
  \bibinfo{author}{\bibfnamefont{H.}~\bibnamefont{Zhou}},
  \bibinfo{author}{\bibfnamefont{T.}~\bibnamefont{Taniguchi}},
  \bibinfo{author}{\bibfnamefont{K.}~\bibnamefont{Watanabe}},
  \bibinfo{author}{\bibfnamefont{E.}~\bibnamefont{Berg}},
  \bibinfo{author}{\bibfnamefont{S.}~\bibnamefont{Nadj-Perge}},
  \bibnamefont{et~al.}, \bibinfo{journal}{Nature Physics}
  \textbf{\bibinfo{volume}{21}}, \bibinfo{pages}{444} (\bibinfo{year}{2025}),
  \urlprefix\url{https://doi.org/10.1038/s41567-024-02776-7}.

\bibitem[{\citenamefont{Kumawat et~al.}(2025)\citenamefont{Kumawat,
  Vishwakarma, Zeeshan, Mal, Kumar, and Mani}}]{Kumawat25}
\bibinfo{author}{\bibfnamefont{S.}~\bibnamefont{Kumawat}},
  \bibinfo{author}{\bibfnamefont{C.~K.} \bibnamefont{Vishwakarma}},
  \bibinfo{author}{\bibfnamefont{M.}~\bibnamefont{Zeeshan}},
  \bibinfo{author}{\bibfnamefont{I.}~\bibnamefont{Mal}},
  \bibinfo{author}{\bibfnamefont{S.}~\bibnamefont{Kumar}}, \bibnamefont{and}
  \bibinfo{author}{\bibfnamefont{B.~K.} \bibnamefont{Mani}},
  \bibinfo{journal}{Phys. Rev. Mater.} \textbf{\bibinfo{volume}{9}},
  \bibinfo{pages}{054003} (\bibinfo{year}{2025}),
  \urlprefix\url{https://link.aps.org/doi/10.1103/PhysRevMaterials.9.054003}.

\bibitem[{\citenamefont{Po et~al.}(2018)\citenamefont{Po, Zou, Vishwanath, and
  Senthil}}]{PoSenthil18}
\bibinfo{author}{\bibfnamefont{H.~C.} \bibnamefont{Po}},
  \bibinfo{author}{\bibfnamefont{L.}~\bibnamefont{Zou}},
  \bibinfo{author}{\bibfnamefont{A.}~\bibnamefont{Vishwanath}},
  \bibnamefont{and} \bibinfo{author}{\bibfnamefont{T.}~\bibnamefont{Senthil}},
  \bibinfo{journal}{Phys. Rev. X} \textbf{\bibinfo{volume}{8}},
  \bibinfo{pages}{031089} (\bibinfo{year}{2018}),
  \urlprefix\url{https://link.aps.org/doi/10.1103/PhysRevX.8.031089}.

\bibitem[{\citenamefont{Liu et~al.}(2019)\citenamefont{Liu, Ma, Gao, and
  Dai}}]{LiuDai19}
\bibinfo{author}{\bibfnamefont{J.}~\bibnamefont{Liu}},
  \bibinfo{author}{\bibfnamefont{Z.}~\bibnamefont{Ma}},
  \bibinfo{author}{\bibfnamefont{J.}~\bibnamefont{Gao}}, \bibnamefont{and}
  \bibinfo{author}{\bibfnamefont{X.}~\bibnamefont{Dai}},
  \bibinfo{journal}{Phys. Rev. X} \textbf{\bibinfo{volume}{9}},
  \bibinfo{pages}{031021} (\bibinfo{year}{2019}),
  \urlprefix\url{https://link.aps.org/doi/10.1103/PhysRevX.9.031021}.

\bibitem[{\citenamefont{Chichinadze et~al.}(2020)\citenamefont{Chichinadze,
  Classen, and Chubukov}}]{ChichinadzeClassenChubukov20}
\bibinfo{author}{\bibfnamefont{D.~V.} \bibnamefont{Chichinadze}},
  \bibinfo{author}{\bibfnamefont{L.}~\bibnamefont{Classen}}, \bibnamefont{and}
  \bibinfo{author}{\bibfnamefont{A.~V.} \bibnamefont{Chubukov}},
  \bibinfo{journal}{Phys. Rev. B} \textbf{\bibinfo{volume}{102}},
  \bibinfo{pages}{125120} (\bibinfo{year}{2020}),
  \urlprefix\url{https://link.aps.org/doi/10.1103/PhysRevB.102.125120}.

\bibitem[{\citenamefont{Xu et~al.}(2020)\citenamefont{Xu, Wu, Jian, and
  Xu}}]{XuXu20}
\bibinfo{author}{\bibfnamefont{Y.}~\bibnamefont{Xu}},
  \bibinfo{author}{\bibfnamefont{X.-C.} \bibnamefont{Wu}},
  \bibinfo{author}{\bibfnamefont{C.-M.} \bibnamefont{Jian}}, \bibnamefont{and}
  \bibinfo{author}{\bibfnamefont{C.}~\bibnamefont{Xu}}, \bibinfo{journal}{Phys.
  Rev. B} \textbf{\bibinfo{volume}{101}}, \bibinfo{pages}{205426}
  (\bibinfo{year}{2020}),
  \urlprefix\url{https://link.aps.org/doi/10.1103/PhysRevB.101.205426}.

\bibitem[{\citenamefont{Hsu et~al.}(2020)\citenamefont{Hsu, Wu, and
  Das~Sarma}}]{HsuDasSarma20}
\bibinfo{author}{\bibfnamefont{Y.-T.} \bibnamefont{Hsu}},
  \bibinfo{author}{\bibfnamefont{F.}~\bibnamefont{Wu}}, \bibnamefont{and}
  \bibinfo{author}{\bibfnamefont{S.}~\bibnamefont{Das~Sarma}},
  \bibinfo{journal}{Phys. Rev. B} \textbf{\bibinfo{volume}{102}},
  \bibinfo{pages}{085103} (\bibinfo{year}{2020}),
  \urlprefix\url{https://link.aps.org/doi/10.1103/PhysRevB.102.085103}.

\bibitem[{\citenamefont{Lian et~al.}(2021)\citenamefont{Lian, Song, Regnault,
  Efetov, Yazdani, and Bernevig}}]{LianBernevig21}
\bibinfo{author}{\bibfnamefont{B.}~\bibnamefont{Lian}},
  \bibinfo{author}{\bibfnamefont{Z.-D.} \bibnamefont{Song}},
  \bibinfo{author}{\bibfnamefont{N.}~\bibnamefont{Regnault}},
  \bibinfo{author}{\bibfnamefont{D.~K.} \bibnamefont{Efetov}},
  \bibinfo{author}{\bibfnamefont{A.}~\bibnamefont{Yazdani}}, \bibnamefont{and}
  \bibinfo{author}{\bibfnamefont{B.~A.} \bibnamefont{Bernevig}},
  \bibinfo{journal}{Phys. Rev. B} \textbf{\bibinfo{volume}{103}},
  \bibinfo{pages}{205414} (\bibinfo{year}{2021}),
  \urlprefix\url{https://link.aps.org/doi/10.1103/PhysRevB.103.205414}.

\bibitem[{\citenamefont{Chichinadze et~al.}(2022)\citenamefont{Chichinadze,
  Classen, Wang, and Chubukov}}]{ChichinadzeClassenChubukov22}
\bibinfo{author}{\bibfnamefont{D.~V.} \bibnamefont{Chichinadze}},
  \bibinfo{author}{\bibfnamefont{L.}~\bibnamefont{Classen}},
  \bibinfo{author}{\bibfnamefont{Y.}~\bibnamefont{Wang}}, \bibnamefont{and}
  \bibinfo{author}{\bibfnamefont{A.~V.} \bibnamefont{Chubukov}},
  \bibinfo{journal}{npj Quantum Materials} \textbf{\bibinfo{volume}{7}},
  \bibinfo{pages}{114} (\bibinfo{year}{2022}).

\bibitem[{\citenamefont{Christos et~al.}(2022)\citenamefont{Christos, Sachdev,
  and Scheurer}}]{ChristosSachdev22}
\bibinfo{author}{\bibfnamefont{M.}~\bibnamefont{Christos}},
  \bibinfo{author}{\bibfnamefont{S.}~\bibnamefont{Sachdev}}, \bibnamefont{and}
  \bibinfo{author}{\bibfnamefont{M.~S.} \bibnamefont{Scheurer}},
  \bibinfo{journal}{Phys. Rev. X} \textbf{\bibinfo{volume}{12}},
  \bibinfo{pages}{021018} (\bibinfo{year}{2022}),
  \urlprefix\url{https://link.aps.org/doi/10.1103/PhysRevX.12.021018}.

\bibitem[{\citenamefont{Mandal and Fernandes}(2023)}]{MandalFernandes23}
\bibinfo{author}{\bibfnamefont{I.}~\bibnamefont{Mandal}} \bibnamefont{and}
  \bibinfo{author}{\bibfnamefont{R.~M.} \bibnamefont{Fernandes}},
  \bibinfo{journal}{Phys. Rev. B} \textbf{\bibinfo{volume}{107}},
  \bibinfo{pages}{125142} (\bibinfo{year}{2023}),
  \urlprefix\url{https://link.aps.org/doi/10.1103/PhysRevB.107.125142}.

\bibitem[{\citenamefont{Dong et~al.}(2023)\citenamefont{Dong, Levitov, and
  Chubukov}}]{DongLevitovChubukov23}
\bibinfo{author}{\bibfnamefont{Z.}~\bibnamefont{Dong}},
  \bibinfo{author}{\bibfnamefont{L.}~\bibnamefont{Levitov}}, \bibnamefont{and}
  \bibinfo{author}{\bibfnamefont{A.~V.} \bibnamefont{Chubukov}},
  \bibinfo{journal}{Phys. Rev. B} \textbf{\bibinfo{volume}{108}},
  \bibinfo{pages}{134503} (\bibinfo{year}{2023}),
  \urlprefix\url{https://link.aps.org/doi/10.1103/PhysRevB.108.134503}.

\bibitem[{\citenamefont{Wang et~al.}(2024{\natexlab{c}})\citenamefont{Wang,
  Vila, Zaletel, and Chatterjee}}]{WangZaletel24}
\bibinfo{author}{\bibfnamefont{T.}~\bibnamefont{Wang}},
  \bibinfo{author}{\bibfnamefont{M.}~\bibnamefont{Vila}},
  \bibinfo{author}{\bibfnamefont{M.~P.} \bibnamefont{Zaletel}},
  \bibnamefont{and}
  \bibinfo{author}{\bibfnamefont{S.}~\bibnamefont{Chatterjee}},
  \bibinfo{journal}{Phys. Rev. Lett.} \textbf{\bibinfo{volume}{132}},
  \bibinfo{pages}{116504} (\bibinfo{year}{2024}{\natexlab{c}}),
  \urlprefix\url{https://link.aps.org/doi/10.1103/PhysRevLett.132.116504}.

\bibitem[{\citenamefont{Lee et~al.}(2024)\citenamefont{Lee, Chichinadze, and
  Chubukov}}]{LeeChichinadzeChubukov24}
\bibinfo{author}{\bibfnamefont{Y.-C.} \bibnamefont{Lee}},
  \bibinfo{author}{\bibfnamefont{D.~V.} \bibnamefont{Chichinadze}},
  \bibnamefont{and} \bibinfo{author}{\bibfnamefont{A.~V.}
  \bibnamefont{Chubukov}}, \bibinfo{journal}{Phys. Rev. B}
  \textbf{\bibinfo{volume}{109}}, \bibinfo{pages}{155118}
  (\bibinfo{year}{2024}),
  \urlprefix\url{https://link.aps.org/doi/10.1103/PhysRevB.109.155118}.

\bibitem[{\citenamefont{Raines et~al.}(2024)\citenamefont{Raines, Glazman, and
  Chubukov}}]{RainesGlazmanChubukov24}
\bibinfo{author}{\bibfnamefont{Z.~M.} \bibnamefont{Raines}},
  \bibinfo{author}{\bibfnamefont{L.~I.} \bibnamefont{Glazman}},
  \bibnamefont{and} \bibinfo{author}{\bibfnamefont{A.~V.}
  \bibnamefont{Chubukov}}, \bibinfo{journal}{Phys. Rev. Lett.}
  \textbf{\bibinfo{volume}{133}}, \bibinfo{pages}{146501}
  (\bibinfo{year}{2024}),
  \urlprefix\url{https://link.aps.org/doi/10.1103/PhysRevLett.133.146501}.

\bibitem[{\citenamefont{Friedlan et~al.}(2025)\citenamefont{Friedlan, Li, and
  Kee}}]{Friedlan25}
\bibinfo{author}{\bibfnamefont{A.}~\bibnamefont{Friedlan}},
  \bibinfo{author}{\bibfnamefont{H.}~\bibnamefont{Li}}, \bibnamefont{and}
  \bibinfo{author}{\bibfnamefont{H.-Y.} \bibnamefont{Kee}},
  \bibinfo{journal}{Phys. Rev. B} \textbf{\bibinfo{volume}{111}},
  \bibinfo{pages}{024504} (\bibinfo{year}{2025}),
  \urlprefix\url{https://link.aps.org/doi/10.1103/PhysRevB.111.024504}.

\bibitem[{\citenamefont{Mayrhofer and Chubukov}(2025)}]{MayrhoferChubukov25}
\bibinfo{author}{\bibfnamefont{R.~D.} \bibnamefont{Mayrhofer}}
  \bibnamefont{and} \bibinfo{author}{\bibfnamefont{A.~V.}
  \bibnamefont{Chubukov}}, \bibinfo{journal}{Phys. Rev. B}
  \textbf{\bibinfo{volume}{111}}, \bibinfo{pages}{245114}
  (\bibinfo{year}{2025}),
  \urlprefix\url{https://link.aps.org/doi/10.1103/PhysRevB.111.245114}.

\bibitem[{\citenamefont{Han and Kivelson}(2022)}]{HanKivelson22}
\bibinfo{author}{\bibfnamefont{Z.}~\bibnamefont{Han}} \bibnamefont{and}
  \bibinfo{author}{\bibfnamefont{S.~A.} \bibnamefont{Kivelson}},
  \bibinfo{journal}{Phys. Rev. B} \textbf{\bibinfo{volume}{105}},
  \bibinfo{pages}{L100509} (\bibinfo{year}{2022}),
  \urlprefix\url{https://link.aps.org/doi/10.1103/PhysRevB.105.L100509}.

\bibitem[{\citenamefont{Wu et~al.}(2023)\citenamefont{Wu, Wu, and
  Wu}}]{WuWuWu23}
\bibinfo{author}{\bibfnamefont{Z.}~\bibnamefont{Wu}},
  \bibinfo{author}{\bibfnamefont{Y.-M.} \bibnamefont{Wu}}, \bibnamefont{and}
  \bibinfo{author}{\bibfnamefont{F.}~\bibnamefont{Wu}}, \bibinfo{journal}{Phys.
  Rev. B} \textbf{\bibinfo{volume}{107}}, \bibinfo{pages}{045122}
  (\bibinfo{year}{2023}),
  \urlprefix\url{https://link.aps.org/doi/10.1103/PhysRevB.107.045122}.

\bibitem[{\citenamefont{Castro et~al.}(2023)\citenamefont{Castro, Shaffer, Wu,
  and Santos}}]{CastroShafferWuSantos23}
\bibinfo{author}{\bibfnamefont{P.}~\bibnamefont{Castro}},
  \bibinfo{author}{\bibfnamefont{D.}~\bibnamefont{Shaffer}},
  \bibinfo{author}{\bibfnamefont{Y.-M.} \bibnamefont{Wu}}, \bibnamefont{and}
  \bibinfo{author}{\bibfnamefont{L.~H.} \bibnamefont{Santos}},
  \bibinfo{journal}{Phys. Rev. Lett.} \textbf{\bibinfo{volume}{131}},
  \bibinfo{pages}{026601} (\bibinfo{year}{2023}),
  \urlprefix\url{https://link.aps.org/doi/10.1103/PhysRevLett.131.026601}.

\bibitem[{\citenamefont{Gil and Berg}(2025)}]{GilBerg25}
\bibinfo{author}{\bibfnamefont{A.}~\bibnamefont{Gil}} \bibnamefont{and}
  \bibinfo{author}{\bibfnamefont{E.}~\bibnamefont{Berg}}
  (\bibinfo{year}{2025}), \eprint{2504.19321},
  \urlprefix\url{https://arxiv.org/abs/2504.19321}.

\bibitem[{\citenamefont{Parra-Mart\'{\i}nez
  et~al.}(2025)\citenamefont{Parra-Mart\'{\i}nez, Jimeno-Pozo, Phong,
  Sainz-Cruz, Kaplan, Emanuel, Oreg, Pantale\'on, Silva-Guill\'en, and
  Guinea}}]{ParraMartinezGuinea25}
\bibinfo{author}{\bibfnamefont{G.}~\bibnamefont{Parra-Mart\'{\i}nez}},
  \bibinfo{author}{\bibfnamefont{A.}~\bibnamefont{Jimeno-Pozo}},
  \bibinfo{author}{\bibfnamefont{V.~o.~T.} \bibnamefont{Phong}},
  \bibinfo{author}{\bibfnamefont{H.}~\bibnamefont{Sainz-Cruz}},
  \bibinfo{author}{\bibfnamefont{D.}~\bibnamefont{Kaplan}},
  \bibinfo{author}{\bibfnamefont{P.}~\bibnamefont{Emanuel}},
  \bibinfo{author}{\bibfnamefont{Y.}~\bibnamefont{Oreg}},
  \bibinfo{author}{\bibfnamefont{P.~A.} \bibnamefont{Pantale\'on}},
  \bibinfo{author}{\bibfnamefont{J.~A.} \bibnamefont{Silva-Guill\'en}},
  \bibnamefont{and} \bibinfo{author}{\bibfnamefont{F.}~\bibnamefont{Guinea}},
  \bibinfo{journal}{Phys. Rev. Lett.} \textbf{\bibinfo{volume}{135}},
  \bibinfo{pages}{136503} (\bibinfo{year}{2025}),
  \urlprefix\url{https://link.aps.org/doi/10.1103/zfmh-rjzc}.

\bibitem[{\citenamefont{Zhuang and Sun}(2025)}]{Zhuang25}
\bibinfo{author}{\bibfnamefont{Y.-C.} \bibnamefont{Zhuang}} \bibnamefont{and}
  \bibinfo{author}{\bibfnamefont{Q.-F.} \bibnamefont{Sun}}
  (\bibinfo{year}{2025}), \eprint{2501.00835},
  \urlprefix\url{https://arxiv.org/abs/2501.00835}.

\bibitem[{\citenamefont{Wei et~al.}(2022)\citenamefont{Wei, Liu, Wang, and
  Liu}}]{WeiLiu22}
\bibinfo{author}{\bibfnamefont{Y.-J.} \bibnamefont{Wei}},
  \bibinfo{author}{\bibfnamefont{H.-L.} \bibnamefont{Liu}},
  \bibinfo{author}{\bibfnamefont{J.}~\bibnamefont{Wang}}, \bibnamefont{and}
  \bibinfo{author}{\bibfnamefont{J.-F.} \bibnamefont{Liu}},
  \bibinfo{journal}{Physical Review B} \textbf{\bibinfo{volume}{106}},
  \bibinfo{pages}{165419} (\bibinfo{year}{2022}), \bibinfo{note}{publisher:
  American Physical Society},
  \urlprefix\url{https://link.aps.org/doi/10.1103/PhysRevB.106.165419}.

\bibitem[{\citenamefont{Hu et~al.}(2023)\citenamefont{Hu, Sun, Xie, and
  Law}}]{HuLaw23}
\bibinfo{author}{\bibfnamefont{J.-X.} \bibnamefont{Hu}},
  \bibinfo{author}{\bibfnamefont{Z.-T.} \bibnamefont{Sun}},
  \bibinfo{author}{\bibfnamefont{Y.-M.} \bibnamefont{Xie}}, \bibnamefont{and}
  \bibinfo{author}{\bibfnamefont{K.~T.} \bibnamefont{Law}},
  \bibinfo{journal}{Phys. Rev. Lett.} \textbf{\bibinfo{volume}{130}},
  \bibinfo{pages}{266003} (\bibinfo{year}{2023}),
  \urlprefix\url{https://link.aps.org/doi/10.1103/PhysRevLett.130.266003}.

\bibitem[{\citenamefont{Tavger and Zaitsev}(1956)}]{TavgerZaitsev56}
\bibinfo{author}{\bibfnamefont{B.}~\bibnamefont{Tavger}} \bibnamefont{and}
  \bibinfo{author}{\bibfnamefont{V.}~\bibnamefont{Zaitsev}},
  \bibinfo{journal}{Sov. Phys. JETP} \textbf{\bibinfo{volume}{3}},
  \bibinfo{pages}{430} (\bibinfo{year}{1956}).

\bibitem[{\citenamefont{Dzyaloshinskii}(1958)}]{Dzyaloshinskii58}
\bibinfo{author}{\bibfnamefont{I.~E.} \bibnamefont{Dzyaloshinskii}},
  \bibinfo{journal}{Sov. Phys. JETP} \textbf{\bibinfo{volume}{6}},
  \bibinfo{pages}{621} (\bibinfo{year}{1958}).

\bibitem[{\citenamefont{Dzyaloshinskii}(1960)}]{Dzyaloshinskii60}
\bibinfo{author}{\bibfnamefont{I.~E.} \bibnamefont{Dzyaloshinskii}},
  \bibinfo{journal}{Soviet Physics JETP} \textbf{\bibinfo{volume}{10}},
  \bibinfo{pages}{628} (\bibinfo{year}{1960}).

\bibitem[{\citenamefont{Eerenstein et~al.}(2006)\citenamefont{Eerenstein,
  Mathur, and Scott}}]{Eerenstein06}
\bibinfo{author}{\bibfnamefont{W.}~\bibnamefont{Eerenstein}},
  \bibinfo{author}{\bibfnamefont{N.~D.} \bibnamefont{Mathur}},
  \bibnamefont{and} \bibinfo{author}{\bibfnamefont{J.~F.} \bibnamefont{Scott}},
  \bibinfo{journal}{Nature} \textbf{\bibinfo{volume}{442}},
  \bibinfo{pages}{759} (\bibinfo{year}{2006}),
  \urlprefix\url{https://doi.org/10.1038/nature05023}.

\bibitem[{\citenamefont{Mineev}(2025)}]{Mineev25}
\bibinfo{author}{\bibfnamefont{V.~P.} \bibnamefont{Mineev}}
  (\bibinfo{year}{2025}), \eprint{2504.01686},
  \urlprefix\url{https://arxiv.org/abs/2504.01686}.

\bibitem[{\citenamefont{\ifmmode~\check{S}\else \v{S}\fi{}mejkal
  et~al.}(2022)\citenamefont{\ifmmode~\check{S}\else \v{S}\fi{}mejkal, Sinova,
  and Jungwirth}}]{SmejkalJungwirth22}
\bibinfo{author}{\bibfnamefont{L.}~\bibnamefont{\ifmmode~\check{S}\else
  \v{S}\fi{}mejkal}}, \bibinfo{author}{\bibfnamefont{J.}~\bibnamefont{Sinova}},
  \bibnamefont{and}
  \bibinfo{author}{\bibfnamefont{T.}~\bibnamefont{Jungwirth}},
  \bibinfo{journal}{Phys. Rev. X} \textbf{\bibinfo{volume}{12}},
  \bibinfo{pages}{040501} (\bibinfo{year}{2022}),
  \urlprefix\url{https://link.aps.org/doi/10.1103/PhysRevX.12.040501}.

\bibitem[{\citenamefont{Mazin}(2024)}]{Mazin24}
\bibinfo{author}{\bibfnamefont{I.}~\bibnamefont{Mazin}},
  \bibinfo{journal}{Physics} \textbf{\bibinfo{volume}{17}}, \bibinfo{pages}{4}
  (\bibinfo{year}{2024}), \bibinfo{note}{publisher: American Physical Society},
  \urlprefix\url{https://physics.aps.org/articles/v17/4}.

\bibitem[{\citenamefont{Roig et~al.}(2024{\natexlab{b}})\citenamefont{Roig,
  Kreisel, Yu, Andersen, and Agterberg}}]{RoigAgterberg24}
\bibinfo{author}{\bibfnamefont{M.}~\bibnamefont{Roig}},
  \bibinfo{author}{\bibfnamefont{A.}~\bibnamefont{Kreisel}},
  \bibinfo{author}{\bibfnamefont{Y.}~\bibnamefont{Yu}},
  \bibinfo{author}{\bibfnamefont{B.~M.} \bibnamefont{Andersen}},
  \bibnamefont{and} \bibinfo{author}{\bibfnamefont{D.~F.}
  \bibnamefont{Agterberg}}, \bibinfo{journal}{Phys. Rev. B}
  \textbf{\bibinfo{volume}{110}}, \bibinfo{pages}{144412}
  (\bibinfo{year}{2024}{\natexlab{b}}),
  \urlprefix\url{https://link.aps.org/doi/10.1103/PhysRevB.110.144412}.

\bibitem[{\citenamefont{Jiang et~al.}(2024)\citenamefont{Jiang, Song, Zhu,
  Fang, Weng, Liu, Yang, and Fang}}]{Jiang24}
\bibinfo{author}{\bibfnamefont{Y.}~\bibnamefont{Jiang}},
  \bibinfo{author}{\bibfnamefont{Z.}~\bibnamefont{Song}},
  \bibinfo{author}{\bibfnamefont{T.}~\bibnamefont{Zhu}},
  \bibinfo{author}{\bibfnamefont{Z.}~\bibnamefont{Fang}},
  \bibinfo{author}{\bibfnamefont{H.}~\bibnamefont{Weng}},
  \bibinfo{author}{\bibfnamefont{Z.-X.} \bibnamefont{Liu}},
  \bibinfo{author}{\bibfnamefont{J.}~\bibnamefont{Yang}}, \bibnamefont{and}
  \bibinfo{author}{\bibfnamefont{C.}~\bibnamefont{Fang}},
  \bibinfo{journal}{Phys. Rev. X} \textbf{\bibinfo{volume}{14}},
  \bibinfo{pages}{031039} (\bibinfo{year}{2024}),
  \urlprefix\url{https://link.aps.org/doi/10.1103/PhysRevX.14.031039}.

\bibitem[{\citenamefont{Mineev}(2024)}]{Mineev24}
\bibinfo{author}{\bibfnamefont{V.~P.} \bibnamefont{Mineev}},
  \emph{\bibinfo{title}{Altermagnetic and {Noncentrosymmetric} {Metals}}}
  (\bibinfo{year}{2024}), \bibinfo{note}{arXiv:2412.05875 [cond-mat]},
  \urlprefix\url{http://arxiv.org/abs/2412.05875}.

\bibitem[{\citenamefont{Maeda et~al.}(2025)\citenamefont{Maeda, Fukaya, Yada,
  Lu, Tanaka, and Cayao}}]{MaedaCayao25}
\bibinfo{author}{\bibfnamefont{K.}~\bibnamefont{Maeda}},
  \bibinfo{author}{\bibfnamefont{Y.}~\bibnamefont{Fukaya}},
  \bibinfo{author}{\bibfnamefont{K.}~\bibnamefont{Yada}},
  \bibinfo{author}{\bibfnamefont{B.}~\bibnamefont{Lu}},
  \bibinfo{author}{\bibfnamefont{Y.}~\bibnamefont{Tanaka}}, \bibnamefont{and}
  \bibinfo{author}{\bibfnamefont{J.}~\bibnamefont{Cayao}},
  \bibinfo{journal}{Physical Review B} \textbf{\bibinfo{volume}{111}},
  \bibinfo{pages}{144508} (\bibinfo{year}{2025}), \bibinfo{note}{publisher:
  American Physical Society},
  \urlprefix\url{https://link.aps.org/doi/10.1103/PhysRevB.111.144508}.

\bibitem[{\citenamefont{Parshukov et~al.}(2025)\citenamefont{Parshukov,
  Wiedmann, and Schnyder}}]{ParshukovSchnyder25}
\bibinfo{author}{\bibfnamefont{K.}~\bibnamefont{Parshukov}},
  \bibinfo{author}{\bibfnamefont{R.}~\bibnamefont{Wiedmann}}, \bibnamefont{and}
  \bibinfo{author}{\bibfnamefont{A.~P.} \bibnamefont{Schnyder}},
  \bibinfo{journal}{Physical Review B} \textbf{\bibinfo{volume}{111}},
  \bibinfo{pages}{224406} (\bibinfo{year}{2025}), \bibinfo{note}{publisher:
  American Physical Society},
  \urlprefix\url{https://link.aps.org/doi/10.1103/PhysRevB.111.224406}.

\bibitem[{\citenamefont{Jungwirth et~al.}(2025)\citenamefont{Jungwirth, Sinova,
  Fernandes, Liu, Watanabe, Murakami, Nakatsuji, and
  Smejkal}}]{JungwirthFernandes25}
\bibinfo{author}{\bibfnamefont{T.}~\bibnamefont{Jungwirth}},
  \bibinfo{author}{\bibfnamefont{J.}~\bibnamefont{Sinova}},
  \bibinfo{author}{\bibfnamefont{R.~M.} \bibnamefont{Fernandes}},
  \bibinfo{author}{\bibfnamefont{Q.}~\bibnamefont{Liu}},
  \bibinfo{author}{\bibfnamefont{H.}~\bibnamefont{Watanabe}},
  \bibinfo{author}{\bibfnamefont{S.}~\bibnamefont{Murakami}},
  \bibinfo{author}{\bibfnamefont{S.}~\bibnamefont{Nakatsuji}},
  \bibnamefont{and} \bibinfo{author}{\bibfnamefont{L.}~\bibnamefont{Smejkal}},
  \emph{\bibinfo{title}{Symmetry, microscopy and spectroscopy signatures of
  altermagnetism}} (\bibinfo{year}{2025}), \bibinfo{note}{arXiv:2506.22860
  [cond-mat]}, \urlprefix\url{http://arxiv.org/abs/2506.22860}.

\bibitem[{\citenamefont{Fukaya et~al.}(2025)\citenamefont{Fukaya, Lu, Yada,
  Tanaka, and Cayao}}]{FukayaCayao25}
\bibinfo{author}{\bibfnamefont{Y.}~\bibnamefont{Fukaya}},
  \bibinfo{author}{\bibfnamefont{B.}~\bibnamefont{Lu}},
  \bibinfo{author}{\bibfnamefont{K.}~\bibnamefont{Yada}},
  \bibinfo{author}{\bibfnamefont{Y.}~\bibnamefont{Tanaka}}, \bibnamefont{and}
  \bibinfo{author}{\bibfnamefont{J.}~\bibnamefont{Cayao}},
  \bibinfo{journal}{Journal of Physics: Condensed Matter}
  \textbf{\bibinfo{volume}{37}}, \bibinfo{pages}{313003}
  (\bibinfo{year}{2025}), ISSN \bibinfo{issn}{0953-8984},
  \bibinfo{note}{publisher: IOP Publishing},
  \urlprefix\url{https://dx.doi.org/10.1088/1361-648X/adf1cf}.

\bibitem[{\citenamefont{de~M.~Froldi and Freire}(2025)}]{Froldi25}
\bibinfo{author}{\bibfnamefont{I.}~\bibnamefont{de~M.~Froldi}}
  \bibnamefont{and} \bibinfo{author}{\bibfnamefont{H.}~\bibnamefont{Freire}}
  (\bibinfo{year}{2025}), \eprint{2510.07506},
  \urlprefix\url{https://arxiv.org/abs/2510.07506}.

\bibitem[{\citenamefont{Cheng and Sun}(2024)}]{ChengSun24}
\bibinfo{author}{\bibfnamefont{Q.}~\bibnamefont{Cheng}} \bibnamefont{and}
  \bibinfo{author}{\bibfnamefont{Q.-F.} \bibnamefont{Sun}},
  \bibinfo{journal}{Physical Review B} \textbf{\bibinfo{volume}{109}},
  \bibinfo{pages}{024517} (\bibinfo{year}{2024}), \bibinfo{note}{publisher:
  American Physical Society},
  \urlprefix\url{https://link.aps.org/doi/10.1103/PhysRevB.109.024517}.

\bibitem[{\citenamefont{Wei and Wang}(2024)}]{WeiWang24}
\bibinfo{author}{\bibfnamefont{Y.-J.} \bibnamefont{Wei}} \bibnamefont{and}
  \bibinfo{author}{\bibfnamefont{J.}~\bibnamefont{Wang}},
  \bibinfo{journal}{Europhysics Letters} \textbf{\bibinfo{volume}{148}},
  \bibinfo{pages}{56003} (\bibinfo{year}{2024}).

\bibitem[{\citenamefont{Kokkeler et~al.}(2025)\citenamefont{Kokkeler, Tokatly,
  and Bergeret}}]{KokkelerTokatlyBergeret25}
\bibinfo{author}{\bibfnamefont{T.}~\bibnamefont{Kokkeler}},
  \bibinfo{author}{\bibfnamefont{I.}~\bibnamefont{Tokatly}}, \bibnamefont{and}
  \bibinfo{author}{\bibfnamefont{F.~S.} \bibnamefont{Bergeret}},
  \bibinfo{journal}{SciPost Physics} \textbf{\bibinfo{volume}{18}},
  \bibinfo{pages}{178} (\bibinfo{year}{2025}).

\bibitem[{\citenamefont{Rabinovich et~al.}(2019)\citenamefont{Rabinovich,
  Bobkova, and Bobkov}}]{RabinovichBobkovs19}
\bibinfo{author}{\bibfnamefont{D.~S.} \bibnamefont{Rabinovich}},
  \bibinfo{author}{\bibfnamefont{I.~V.} \bibnamefont{Bobkova}},
  \bibnamefont{and} \bibinfo{author}{\bibfnamefont{A.~M.}
  \bibnamefont{Bobkov}}, \bibinfo{journal}{Phys. Rev. Res.}
  \textbf{\bibinfo{volume}{1}}, \bibinfo{pages}{033095} (\bibinfo{year}{2019}),
  \urlprefix\url{https://link.aps.org/doi/10.1103/PhysRevResearch.1.033095}.

\bibitem[{\citenamefont{Coey}(1987)}]{Coey87}
\bibinfo{author}{\bibfnamefont{J.~M.~D.} \bibnamefont{Coey}},
  \bibinfo{journal}{Canadian Journal of Physics} \textbf{\bibinfo{volume}{65}},
  \bibinfo{pages}{1210} (\bibinfo{year}{1987}), ISSN \bibinfo{issn}{0008-4204},
  \bibinfo{note}{publisher: NRC Research Press},
  \urlprefix\url{https://cdnsciencepub-com.ezproxy.library.wisc.edu/doi/abs/10.1139/p87-197}.

\bibitem[{\citenamefont{Cheong and Xu}(2022)}]{CheongXu22}
\bibinfo{author}{\bibfnamefont{S.-W.} \bibnamefont{Cheong}} \bibnamefont{and}
  \bibinfo{author}{\bibfnamefont{X.}~\bibnamefont{Xu}}, \bibinfo{journal}{npj
  Quantum Materials} \textbf{\bibinfo{volume}{7}}, \bibinfo{pages}{40}
  (\bibinfo{year}{2022}), ISSN \bibinfo{issn}{2397-4648},
  \bibinfo{note}{publisher: Nature Publishing Group},
  \urlprefix\url{https://www.nature.com/articles/s41535-022-00447-5}.

\bibitem[{\citenamefont{Cheong and Huang}(2024)}]{CheongHuang24}
\bibinfo{author}{\bibfnamefont{S.-W.} \bibnamefont{Cheong}} \bibnamefont{and}
  \bibinfo{author}{\bibfnamefont{F.-T.} \bibnamefont{Huang}},
  \bibinfo{journal}{npj Quantum Materials} \textbf{\bibinfo{volume}{9}},
  \bibinfo{pages}{1} (\bibinfo{year}{2024}), ISSN \bibinfo{issn}{2397-4648},
  \bibinfo{note}{number: 1 Publisher: Nature Publishing Group},
  \urlprefix\url{https://www.nature.com/articles/s41535-024-00626-6}.

\bibitem[{\citenamefont{Starykh}(2015)}]{Starykh15}
\bibinfo{author}{\bibfnamefont{O.~A.} \bibnamefont{Starykh}},
  \bibinfo{journal}{Reports on Progress in Physics}
  \textbf{\bibinfo{volume}{78}}, \bibinfo{pages}{052502}
  (\bibinfo{year}{2015}),
  \urlprefix\url{https://dx.doi.org/10.1088/0034-4885/78/5/052502}.

\bibitem[{\citenamefont{{Zel'Dovich}}(1958)}]{Zeldovich58}
\bibinfo{author}{\bibfnamefont{I.~B.} \bibnamefont{{Zel'Dovich}}},
  \bibinfo{journal}{Soviet Journal of Experimental and Theoretical Physics}
  \textbf{\bibinfo{volume}{6}}, \bibinfo{pages}{1184} (\bibinfo{year}{1958}).

\bibitem[{\citenamefont{Dubovik and Tugushev}(1990)}]{Dubovik90}
\bibinfo{author}{\bibfnamefont{V.}~\bibnamefont{Dubovik}} \bibnamefont{and}
  \bibinfo{author}{\bibfnamefont{V.}~\bibnamefont{Tugushev}},
  \bibinfo{journal}{Physics Reports} \textbf{\bibinfo{volume}{187}},
  \bibinfo{pages}{145} (\bibinfo{year}{1990}), ISSN \bibinfo{issn}{0370-1573},
  \urlprefix\url{https://www.sciencedirect.com/science/article/pii/037015739090042Z}.

\bibitem[{\citenamefont{Gorbatsevich and Kopaev}(1994)}]{GorbatsevichKopaev94}
\bibinfo{author}{\bibfnamefont{A.~A.} \bibnamefont{Gorbatsevich}}
  \bibnamefont{and} \bibinfo{author}{\bibfnamefont{Y.~V.}
  \bibnamefont{Kopaev}}, \bibinfo{journal}{Ferroelectrics}
  \textbf{\bibinfo{volume}{161}}, \bibinfo{pages}{321} (\bibinfo{year}{1994}),
  ISSN \bibinfo{issn}{0015-0193}, \bibinfo{note}{publisher: Taylor \& Francis
  \_eprint: https://doi.org/10.1080/00150199408213381},
  \urlprefix\url{https://doi.org/10.1080/00150199408213381}.

\bibitem[{\citenamefont{Spaldin et~al.}(2008)\citenamefont{Spaldin, Fiebig, and
  Mostovoy}}]{Spaldin08}
\bibinfo{author}{\bibfnamefont{N.~A.} \bibnamefont{Spaldin}},
  \bibinfo{author}{\bibfnamefont{M.}~\bibnamefont{Fiebig}}, \bibnamefont{and}
  \bibinfo{author}{\bibfnamefont{M.}~\bibnamefont{Mostovoy}},
  \bibinfo{journal}{Journal of Physics: Condensed Matter}
  \textbf{\bibinfo{volume}{20}}, \bibinfo{pages}{434203}
  (\bibinfo{year}{2008}),
  \urlprefix\url{https://doi.org/10.1088/0953-8984/20/43/434203}.

\bibitem[{\citenamefont{Kopaev}(2009)}]{Kopaev09}
\bibinfo{author}{\bibfnamefont{Y.~V.} \bibnamefont{Kopaev}},
  \bibinfo{journal}{Physics-Uspekhi} \textbf{\bibinfo{volume}{52}},
  \bibinfo{pages}{1111} (\bibinfo{year}{2009}), ISSN \bibinfo{issn}{1063-7869},
  \bibinfo{note}{publisher: IOP Publishing},
  \urlprefix\url{https://iopscience-iop-org.ezproxy.library.wisc.edu/article/10.3367/UFNe.0179.200911d.1175/meta}.

\bibitem[{\citenamefont{Hayami et~al.}(2014)\citenamefont{Hayami, Kusunose, and
  Motome}}]{Hayami14}
\bibinfo{author}{\bibfnamefont{S.}~\bibnamefont{Hayami}},
  \bibinfo{author}{\bibfnamefont{H.}~\bibnamefont{Kusunose}}, \bibnamefont{and}
  \bibinfo{author}{\bibfnamefont{Y.}~\bibnamefont{Motome}},
  \bibinfo{journal}{Physical Review B} \textbf{\bibinfo{volume}{90}},
  \bibinfo{pages}{024432} (\bibinfo{year}{2014}), \bibinfo{note}{publisher:
  American Physical Society},
  \urlprefix\url{https://link.aps.org/doi/10.1103/PhysRevB.90.024432}.

\bibitem[{\citenamefont{Pourovskii et~al.}(2025)\citenamefont{Pourovskii,
  Fiore~Mosca, Celiberti, Khmelevskyi, Paramekanti, and
  Franchini}}]{Pourovskii25}
\bibinfo{author}{\bibfnamefont{L.~V.} \bibnamefont{Pourovskii}},
  \bibinfo{author}{\bibfnamefont{D.}~\bibnamefont{Fiore~Mosca}},
  \bibinfo{author}{\bibfnamefont{L.}~\bibnamefont{Celiberti}},
  \bibinfo{author}{\bibfnamefont{S.}~\bibnamefont{Khmelevskyi}},
  \bibinfo{author}{\bibfnamefont{A.}~\bibnamefont{Paramekanti}},
  \bibnamefont{and}
  \bibinfo{author}{\bibfnamefont{C.}~\bibnamefont{Franchini}},
  \bibinfo{journal}{Nature Reviews Materials} \textbf{\bibinfo{volume}{10}},
  \bibinfo{pages}{674} (\bibinfo{year}{2025}),
  \urlprefix\url{https://doi.org/10.1038/s41578-025-00824-z}.

\bibitem[{\citenamefont{Sumita and Yanase}(2016)}]{SumitaYanase16}
\bibinfo{author}{\bibfnamefont{S.}~\bibnamefont{Sumita}} \bibnamefont{and}
  \bibinfo{author}{\bibfnamefont{Y.}~\bibnamefont{Yanase}},
  \bibinfo{journal}{Phys. Rev. B} \textbf{\bibinfo{volume}{93}},
  \bibinfo{pages}{224507} (\bibinfo{year}{2016}),
  \urlprefix\url{https://link.aps.org/doi/10.1103/PhysRevB.93.224507}.

\bibitem[{\citenamefont{Sumita et~al.}(2017)\citenamefont{Sumita, Nomoto, and
  Yanase}}]{SumitaYanase17}
\bibinfo{author}{\bibfnamefont{S.}~\bibnamefont{Sumita}},
  \bibinfo{author}{\bibfnamefont{T.}~\bibnamefont{Nomoto}}, \bibnamefont{and}
  \bibinfo{author}{\bibfnamefont{Y.}~\bibnamefont{Yanase}},
  \bibinfo{journal}{Phys. Rev. Lett.} \textbf{\bibinfo{volume}{119}},
  \bibinfo{pages}{027001} (\bibinfo{year}{2017}),
  \urlprefix\url{https://link.aps.org/doi/10.1103/PhysRevLett.119.027001}.

\bibitem[{\citenamefont{Sumita and Yanase}(2020)}]{SumitaYanase20}
\bibinfo{author}{\bibfnamefont{S.}~\bibnamefont{Sumita}} \bibnamefont{and}
  \bibinfo{author}{\bibfnamefont{Y.}~\bibnamefont{Yanase}},
  \bibinfo{journal}{Phys. Rev. Res.} \textbf{\bibinfo{volume}{2}},
  \bibinfo{pages}{033225} (\bibinfo{year}{2020}),
  \urlprefix\url{https://link.aps.org/doi/10.1103/PhysRevResearch.2.033225}.

\bibitem[{\citenamefont{Kamra and Fu}(2024)}]{KamraFu24}
\bibinfo{author}{\bibfnamefont{L.~J.} \bibnamefont{Kamra}} \bibnamefont{and}
  \bibinfo{author}{\bibfnamefont{L.}~\bibnamefont{Fu}},
  \bibinfo{journal}{arXiv.org}  (\bibinfo{year}{2024}),
  \urlprefix\url{https://arxiv.org/abs/2409.00223v1}.

\bibitem[{\citenamefont{Nikolić et~al.}(2025)\citenamefont{Nikolić, Schulz,
  Buzdin, and Eschrig}}]{NikolicBuzdin25}
\bibinfo{author}{\bibfnamefont{D.}~\bibnamefont{Nikolić}},
  \bibinfo{author}{\bibfnamefont{N.~L.} \bibnamefont{Schulz}},
  \bibinfo{author}{\bibfnamefont{A.~I.} \bibnamefont{Buzdin}},
  \bibnamefont{and} \bibinfo{author}{\bibfnamefont{M.}~\bibnamefont{Eschrig}},
  \emph{\bibinfo{title}{Spin-resolved josephson diode effect through strongly
  spin-polarized conical magnets}} (\bibinfo{year}{2025}), \eprint{2508.05868},
  \urlprefix\url{https://arxiv.org/abs/2508.05868}.

\bibitem[{\citenamefont{Frazier et~al.}(2025)\citenamefont{Frazier, Zhang, and
  Li}}]{FrazierLi25}
\bibinfo{author}{\bibfnamefont{G.~R.} \bibnamefont{Frazier}},
  \bibinfo{author}{\bibfnamefont{J.}~\bibnamefont{Zhang}}, \bibnamefont{and}
  \bibinfo{author}{\bibfnamefont{Y.}~\bibnamefont{Li}} (\bibinfo{year}{2025}),
  \eprint{2506.15661}, \urlprefix\url{https://arxiv.org/abs/2506.15661}.

\bibitem[{\citenamefont{Wu et~al.}(2024)\citenamefont{Wu, Amin, Yu, and
  Agterberg}}]{WuAgterberg24}
\bibinfo{author}{\bibfnamefont{H.}~\bibnamefont{Wu}},
  \bibinfo{author}{\bibfnamefont{A.}~\bibnamefont{Amin}},
  \bibinfo{author}{\bibfnamefont{Y.}~\bibnamefont{Yu}}, \bibnamefont{and}
  \bibinfo{author}{\bibfnamefont{D.~F.} \bibnamefont{Agterberg}},
  \bibinfo{journal}{Physical Review B} \textbf{\bibinfo{volume}{109}},
  \bibinfo{pages}{L220501} (\bibinfo{year}{2024}), \bibinfo{note}{publisher:
  American Physical Society},
  \urlprefix\url{https://link.aps.org/doi/10.1103/PhysRevB.109.L220501}.

\bibitem[{\citenamefont{Varma}(2014)}]{Varma14}
\bibinfo{author}{\bibfnamefont{C.~M.} \bibnamefont{Varma}},
  \bibinfo{journal}{Journal of Physics: Condensed Matter}
  \textbf{\bibinfo{volume}{26}}, \bibinfo{pages}{505701}
  (\bibinfo{year}{2014}),
  \urlprefix\url{https://doi.org/10.1088/0953-8984/26/50/505701}.

\bibitem[{\citenamefont{Bourges et~al.}(2021)\citenamefont{Bourges, Bounoua,
  and Sidis}}]{Bourges21}
\bibinfo{author}{\bibfnamefont{P.}~\bibnamefont{Bourges}},
  \bibinfo{author}{\bibfnamefont{D.}~\bibnamefont{Bounoua}}, \bibnamefont{and}
  \bibinfo{author}{\bibfnamefont{Y.}~\bibnamefont{Sidis}},
  \bibinfo{journal}{Comptes Rendus. Physique} \textbf{\bibinfo{volume}{22}},
  \bibinfo{pages}{1} (\bibinfo{year}{2021}).

\bibitem[{\citenamefont{Fernandes et~al.}(2025)\citenamefont{Fernandes, Birol,
  Ye, and Vanderbilt}}]{Fernandes25}
\bibinfo{author}{\bibfnamefont{R.~M.} \bibnamefont{Fernandes}},
  \bibinfo{author}{\bibfnamefont{T.}~\bibnamefont{Birol}},
  \bibinfo{author}{\bibfnamefont{M.}~\bibnamefont{Ye}}, \bibnamefont{and}
  \bibinfo{author}{\bibfnamefont{D.}~\bibnamefont{Vanderbilt}},
  \emph{\bibinfo{title}{Loop-current order through the kagome looking glass}}
  (\bibinfo{year}{2025}), \eprint{2502.16657},
  \urlprefix\url{https://arxiv.org/abs/2502.16657}.

\bibitem[{\citenamefont{Schultz et~al.}(2025)\citenamefont{Schultz, Palle, Kim,
  Fernandes, and Schmalian}}]{SchultzFernandesSchmalian25}
\bibinfo{author}{\bibfnamefont{D.~J.} \bibnamefont{Schultz}},
  \bibinfo{author}{\bibfnamefont{G.}~\bibnamefont{Palle}},
  \bibinfo{author}{\bibfnamefont{Y.~B.} \bibnamefont{Kim}},
  \bibinfo{author}{\bibfnamefont{R.~M.} \bibnamefont{Fernandes}},
  \bibnamefont{and}
  \bibinfo{author}{\bibfnamefont{J.}~\bibnamefont{Schmalian}},
  \emph{\bibinfo{title}{Superconductivity in kagome metals due to soft
  loop-current fluctuations}} (\bibinfo{year}{2025}), \eprint{2507.16892},
  \urlprefix\url{https://arxiv.org/abs/2507.16892}.

\bibitem[{\citenamefont{Varma}(2025)}]{Varma25}
\bibinfo{author}{\bibfnamefont{C.~M.} \bibnamefont{Varma}},
  \bibinfo{journal}{Phys. Rev. B} \textbf{\bibinfo{volume}{112}},
  \bibinfo{pages}{024513} (\bibinfo{year}{2025}),
  \urlprefix\url{https://link.aps.org/doi/10.1103/vmwz-c5qw}.

\bibitem[{\citenamefont{Shen and Zhang}(2025)}]{ShenZhang25}
\bibinfo{author}{\bibfnamefont{Q.-K.} \bibnamefont{Shen}} \bibnamefont{and}
  \bibinfo{author}{\bibfnamefont{Y.}~\bibnamefont{Zhang}},
  \bibinfo{journal}{Phys. Rev. B} \textbf{\bibinfo{volume}{111}},
  \bibinfo{pages}{174515} (\bibinfo{year}{2025}),
  \urlprefix\url{https://link.aps.org/doi/10.1103/PhysRevB.111.174515}.

\bibitem[{\citenamefont{Bhowmik and Saha}(2025)}]{BhowmikSaha25}
\bibinfo{author}{\bibfnamefont{S.}~\bibnamefont{Bhowmik}} \bibnamefont{and}
  \bibinfo{author}{\bibfnamefont{A.}~\bibnamefont{Saha}},
  \bibinfo{journal}{Phys. Rev. B} \textbf{\bibinfo{volume}{111}},
  \bibinfo{pages}{L161402} (\bibinfo{year}{2025}),
  \urlprefix\url{https://link.aps.org/doi/10.1103/PhysRevB.111.L161402}.

\bibitem[{\citenamefont{Bhowmik
  et~al.}(2025{\natexlab{b}})\citenamefont{Bhowmik, Samanta, Nandy, Saha, and
  Ghosh}}]{BhowmikGhosh25shiba2D}
\bibinfo{author}{\bibfnamefont{S.}~\bibnamefont{Bhowmik}},
  \bibinfo{author}{\bibfnamefont{D.}~\bibnamefont{Samanta}},
  \bibinfo{author}{\bibfnamefont{A.~K.} \bibnamefont{Nandy}},
  \bibinfo{author}{\bibfnamefont{A.}~\bibnamefont{Saha}}, \bibnamefont{and}
  \bibinfo{author}{\bibfnamefont{S.~K.} \bibnamefont{Ghosh}}
  (\bibinfo{year}{2025}{\natexlab{b}}), \eprint{2508.10832},
  \urlprefix\url{https://arxiv.org/abs/2508.10832}.

\bibitem[{\citenamefont{Sinner et~al.}(2024)\citenamefont{Sinner, Wang, Parkin,
  Ernst, Dugaev, and Chotorlishvili}}]{SinnerChotorlishvili24}
\bibinfo{author}{\bibfnamefont{A.}~\bibnamefont{Sinner}},
  \bibinfo{author}{\bibfnamefont{X.-G.} \bibnamefont{Wang}},
  \bibinfo{author}{\bibfnamefont{S.~S.~P.} \bibnamefont{Parkin}},
  \bibinfo{author}{\bibfnamefont{A.}~\bibnamefont{Ernst}},
  \bibinfo{author}{\bibfnamefont{V.}~\bibnamefont{Dugaev}}, \bibnamefont{and}
  \bibinfo{author}{\bibfnamefont{L.}~\bibnamefont{Chotorlishvili}},
  \bibinfo{journal}{Physical Review B} \textbf{\bibinfo{volume}{109}},
  \bibinfo{pages}{214510} (\bibinfo{year}{2024}), \bibinfo{note}{publisher:
  American Physical Society},
  \urlprefix\url{https://link.aps.org/doi/10.1103/PhysRevB.109.214510}.

\bibitem[{\citenamefont{Silaev et~al.}(2014)\citenamefont{Silaev, Aladyshkin,
  Silaeva, and Aladyshkina}}]{Silaev14}
\bibinfo{author}{\bibfnamefont{M.~A.} \bibnamefont{Silaev}},
  \bibinfo{author}{\bibfnamefont{A.~Y.} \bibnamefont{Aladyshkin}},
  \bibinfo{author}{\bibfnamefont{M.~V.} \bibnamefont{Silaeva}},
  \bibnamefont{and} \bibinfo{author}{\bibfnamefont{A.~S.}
  \bibnamefont{Aladyshkina}}, \bibinfo{journal}{Journal of Physics: Condensed
  Matter} \textbf{\bibinfo{volume}{26}}, \bibinfo{pages}{095702}
  (\bibinfo{year}{2014}),
  \urlprefix\url{https://doi.org/10.1088/0953-8984/26/9/095702}.

\bibitem[{\citenamefont{Pal and Benjamin}(2019)}]{PalBenjamin19}
\bibinfo{author}{\bibfnamefont{S.}~\bibnamefont{Pal}} \bibnamefont{and}
  \bibinfo{author}{\bibfnamefont{C.}~\bibnamefont{Benjamin}},
  \bibinfo{journal}{Europhysics Letters} \textbf{\bibinfo{volume}{126}},
  \bibinfo{pages}{57002} (\bibinfo{year}{2019}),
  \urlprefix\url{https://doi.org/10.1209/0295-5075/126/57002}.

\bibitem[{\citenamefont{Halterman et~al.}(2022)\citenamefont{Halterman,
  Alidoust, Smith, and Starr}}]{Halterman22}
\bibinfo{author}{\bibfnamefont{K.}~\bibnamefont{Halterman}},
  \bibinfo{author}{\bibfnamefont{M.}~\bibnamefont{Alidoust}},
  \bibinfo{author}{\bibfnamefont{R.}~\bibnamefont{Smith}}, \bibnamefont{and}
  \bibinfo{author}{\bibfnamefont{S.}~\bibnamefont{Starr}},
  \bibinfo{journal}{Phys. Rev. B} \textbf{\bibinfo{volume}{105}},
  \bibinfo{pages}{104508} (\bibinfo{year}{2022}),
  \urlprefix\url{https://link.aps.org/doi/10.1103/PhysRevB.105.104508}.

\bibitem[{\citenamefont{Schulz et~al.}(2025{\natexlab{a}})\citenamefont{Schulz,
  Nikoli\ifmmode~\acute{c}\else \'{c}\fi{}, and Eschrig}}]{Schulz25I}
\bibinfo{author}{\bibfnamefont{N.~L.} \bibnamefont{Schulz}},
  \bibinfo{author}{\bibfnamefont{D.}~\bibnamefont{Nikoli\ifmmode~\acute{c}\else
  \'{c}\fi{}}}, \bibnamefont{and}
  \bibinfo{author}{\bibfnamefont{M.}~\bibnamefont{Eschrig}},
  \bibinfo{journal}{Phys. Rev. B} \textbf{\bibinfo{volume}{112}},
  \bibinfo{pages}{104515} (\bibinfo{year}{2025}{\natexlab{a}}),
  \urlprefix\url{https://link.aps.org/doi/10.1103/4t18-yyx4}.

\bibitem[{\citenamefont{Schulz et~al.}(2025{\natexlab{b}})\citenamefont{Schulz,
  Nikoli\ifmmode~\acute{c}\else \'{c}\fi{}, and Eschrig}}]{Schulz25II}
\bibinfo{author}{\bibfnamefont{N.~L.} \bibnamefont{Schulz}},
  \bibinfo{author}{\bibfnamefont{D.}~\bibnamefont{Nikoli\ifmmode~\acute{c}\else
  \'{c}\fi{}}}, \bibnamefont{and}
  \bibinfo{author}{\bibfnamefont{M.}~\bibnamefont{Eschrig}},
  \bibinfo{journal}{Phys. Rev. B} \textbf{\bibinfo{volume}{112}},
  \bibinfo{pages}{104514} (\bibinfo{year}{2025}{\natexlab{b}}),
  \urlprefix\url{https://link.aps.org/doi/10.1103/nb38-v1jq}.

\bibitem[{\citenamefont{Mal'shukov}(2024)}]{Malshukov24}
\bibinfo{author}{\bibfnamefont{A.~G.} \bibnamefont{Mal'shukov}},
  \bibinfo{journal}{Physical Review B} \textbf{\bibinfo{volume}{110}},
  \bibinfo{pages}{144522} (\bibinfo{year}{2024}), \bibinfo{note}{publisher:
  American Physical Society},
  \urlprefix\url{https://link.aps.org/doi/10.1103/PhysRevB.110.144522}.

\bibitem[{\citenamefont{Chiu et~al.}(2016)\citenamefont{Chiu, Teo, Schnyder,
  and Ryu}}]{ChiuRyu16}
\bibinfo{author}{\bibfnamefont{C.-K.} \bibnamefont{Chiu}},
  \bibinfo{author}{\bibfnamefont{J.~C.~Y.} \bibnamefont{Teo}},
  \bibinfo{author}{\bibfnamefont{A.~P.} \bibnamefont{Schnyder}},
  \bibnamefont{and} \bibinfo{author}{\bibfnamefont{S.}~\bibnamefont{Ryu}},
  \bibinfo{journal}{Rev. Mod. Phys.} \textbf{\bibinfo{volume}{88}},
  \bibinfo{pages}{035005} (\bibinfo{year}{2016}),
  \urlprefix\url{https://link.aps.org/doi/10.1103/RevModPhys.88.035005}.

\bibitem[{\citenamefont{Sato and Ando}(2017)}]{SatoAndo17}
\bibinfo{author}{\bibfnamefont{M.}~\bibnamefont{Sato}} \bibnamefont{and}
  \bibinfo{author}{\bibfnamefont{Y.}~\bibnamefont{Ando}},
  \bibinfo{journal}{Reports on Progress in Physics}
  \textbf{\bibinfo{volume}{80}}, \bibinfo{pages}{076501}
  (\bibinfo{year}{2017}),
  \urlprefix\url{https://doi.org/10.1088/1361-6633/aa6ac7}.

\bibitem[{\citenamefont{Bernevig et~al.}(2022)\citenamefont{Bernevig, Felser,
  and Beidenkopf}}]{Bernevig22}
\bibinfo{author}{\bibfnamefont{B.~A.} \bibnamefont{Bernevig}},
  \bibinfo{author}{\bibfnamefont{C.}~\bibnamefont{Felser}}, \bibnamefont{and}
  \bibinfo{author}{\bibfnamefont{H.}~\bibnamefont{Beidenkopf}},
  \bibinfo{journal}{Nature} \textbf{\bibinfo{volume}{603}}, \bibinfo{pages}{41}
  (\bibinfo{year}{2022}),
  \urlprefix\url{https://doi.org/10.1038/s41586-021-04105-x}.

\bibitem[{\citenamefont{Haldane}(1988)}]{Haldane88}
\bibinfo{author}{\bibfnamefont{F.~D.~M.} \bibnamefont{Haldane}},
  \bibinfo{journal}{Phys. Rev. Lett.} \textbf{\bibinfo{volume}{61}},
  \bibinfo{pages}{2015} (\bibinfo{year}{1988}),
  \urlprefix\url{https://link.aps.org/doi/10.1103/PhysRevLett.61.2015}.

\bibitem[{\citenamefont{Read and Green}(2000)}]{ReadGreen00}
\bibinfo{author}{\bibfnamefont{N.}~\bibnamefont{Read}} \bibnamefont{and}
  \bibinfo{author}{\bibfnamefont{D.}~\bibnamefont{Green}},
  \bibinfo{journal}{Phys. Rev. B} \textbf{\bibinfo{volume}{61}},
  \bibinfo{pages}{10267} (\bibinfo{year}{2000}),
  \urlprefix\url{https://link.aps.org/doi/10.1103/PhysRevB.61.10267}.

\bibitem[{\citenamefont{Kane and Mele}(2005)}]{KaneMele05}
\bibinfo{author}{\bibfnamefont{C.~L.} \bibnamefont{Kane}} \bibnamefont{and}
  \bibinfo{author}{\bibfnamefont{E.~J.} \bibnamefont{Mele}},
  \bibinfo{journal}{Phys. Rev. Lett.} \textbf{\bibinfo{volume}{95}},
  \bibinfo{pages}{226801} (\bibinfo{year}{2005}),
  \urlprefix\url{https://link.aps.org/doi/10.1103/PhysRevLett.95.226801}.

\bibitem[{\citenamefont{Hasan and Moore}(2011)}]{HasanMoore11}
\bibinfo{author}{\bibfnamefont{M.~Z.} \bibnamefont{Hasan}} \bibnamefont{and}
  \bibinfo{author}{\bibfnamefont{J.~E.} \bibnamefont{Moore}},
  \bibinfo{journal}{Annual Review of Condensed Matter Physics}
  \textbf{\bibinfo{volume}{2}}, \bibinfo{pages}{55} (\bibinfo{year}{2011}),
  ISSN \bibinfo{issn}{1947-5462},
  \urlprefix\url{https://www.annualreviews.org/content/journals/10.1146/annurev-conmatphys-062910-140432}.

\bibitem[{\citenamefont{Nayak et~al.}(2008)\citenamefont{Nayak, Simon, Stern,
  Freedman, and Das~Sarma}}]{NayakSimonDasSarma08}
\bibinfo{author}{\bibfnamefont{C.}~\bibnamefont{Nayak}},
  \bibinfo{author}{\bibfnamefont{S.~H.} \bibnamefont{Simon}},
  \bibinfo{author}{\bibfnamefont{A.}~\bibnamefont{Stern}},
  \bibinfo{author}{\bibfnamefont{M.}~\bibnamefont{Freedman}}, \bibnamefont{and}
  \bibinfo{author}{\bibfnamefont{S.}~\bibnamefont{Das~Sarma}},
  \bibinfo{journal}{Rev. Mod. Phys.} \textbf{\bibinfo{volume}{80}},
  \bibinfo{pages}{1083} (\bibinfo{year}{2008}),
  \urlprefix\url{https://link.aps.org/doi/10.1103/RevModPhys.80.1083}.

\bibitem[{\citenamefont{Alicea}(2012)}]{Alicea12}
\bibinfo{author}{\bibfnamefont{J.}~\bibnamefont{Alicea}},
  \bibinfo{journal}{Reports on Progress in Physics}
  \textbf{\bibinfo{volume}{75}}, \bibinfo{pages}{076501}
  (\bibinfo{year}{2012}),
  \urlprefix\url{https://dx.doi.org/10.1088/0034-4885/75/7/076501}.

\bibitem[{\citenamefont{Beenakker}(2013)}]{Beenakker13}
\bibinfo{author}{\bibfnamefont{C.}~\bibnamefont{Beenakker}},
  \bibinfo{journal}{Annual Review of Condensed Matter Physics}
  \textbf{\bibinfo{volume}{4}}, \bibinfo{pages}{113} (\bibinfo{year}{2013}),
  ISSN \bibinfo{issn}{1947-5462},
  \urlprefix\url{https://www.annualreviews.org/content/journals/10.1146/annurev-conmatphys-030212-184337}.

\bibitem[{\citenamefont{Cai et~al.}(2023)\citenamefont{Cai, Žutić, and
  Han}}]{CaiZutic23}
\bibinfo{author}{\bibfnamefont{R.}~\bibnamefont{Cai}},
  \bibinfo{author}{\bibfnamefont{I.}~\bibnamefont{Žutić}}, \bibnamefont{and}
  \bibinfo{author}{\bibfnamefont{W.}~\bibnamefont{Han}},
  \bibinfo{journal}{Advanced Quantum Technologies}
  \textbf{\bibinfo{volume}{6}}, \bibinfo{pages}{2200080}
  (\bibinfo{year}{2023}),
  \eprint{https://advanced-onlinelibrary-wiley-com.ezproxy.library.wisc.edu/doi/pdfdirect/10.1002/qute.202200080},
  \urlprefix\url{https://advanced-onlinelibrary-wiley-com.ezproxy.library.wisc.edu/doi/abs/10.1002/qute.202200080}.

\bibitem[{\citenamefont{Schiela et~al.}(2024)\citenamefont{Schiela, Yu, and
  Shabani}}]{SchielaShabani24}
\bibinfo{author}{\bibfnamefont{W.~F.} \bibnamefont{Schiela}},
  \bibinfo{author}{\bibfnamefont{P.}~\bibnamefont{Yu}}, \bibnamefont{and}
  \bibinfo{author}{\bibfnamefont{J.}~\bibnamefont{Shabani}},
  \bibinfo{journal}{PRX Quantum} \textbf{\bibinfo{volume}{5}},
  \bibinfo{pages}{030102} (\bibinfo{year}{2024}),
  \urlprefix\url{https://link.aps.org/doi/10.1103/PRXQuantum.5.030102}.

\bibitem[{\citenamefont{Karabassov et~al.}(2022)\citenamefont{Karabassov,
  Bobkova, Golubov, and Vasenko}}]{KarabassovBobkovaVasenko22}
\bibinfo{author}{\bibfnamefont{T.}~\bibnamefont{Karabassov}},
  \bibinfo{author}{\bibfnamefont{I.~V.} \bibnamefont{Bobkova}},
  \bibinfo{author}{\bibfnamefont{A.~A.} \bibnamefont{Golubov}},
  \bibnamefont{and} \bibinfo{author}{\bibfnamefont{A.~S.}
  \bibnamefont{Vasenko}}, \bibinfo{journal}{Phys. Rev. B}
  \textbf{\bibinfo{volume}{106}}, \bibinfo{pages}{224509}
  (\bibinfo{year}{2022}),
  \urlprefix\url{https://link.aps.org/doi/10.1103/PhysRevB.106.224509}.

\bibitem[{\citenamefont{Kotetes et~al.}(2023)\citenamefont{Kotetes, Sura, and
  Andersen}}]{Kotetes23}
\bibinfo{author}{\bibfnamefont{P.}~\bibnamefont{Kotetes}},
  \bibinfo{author}{\bibfnamefont{H.~O.~M.} \bibnamefont{Sura}},
  \bibnamefont{and} \bibinfo{author}{\bibfnamefont{B.~M.}
  \bibnamefont{Andersen}}, \bibinfo{journal}{Phys. Rev. B}
  \textbf{\bibinfo{volume}{108}}, \bibinfo{pages}{155310}
  (\bibinfo{year}{2023}),
  \urlprefix\url{https://link.aps.org/doi/10.1103/PhysRevB.108.155310}.

\bibitem[{\citenamefont{Karabassov et~al.}(2023)\citenamefont{Karabassov,
  Bobkova, Silkin, Lvov, Golubov, and Vasenko}}]{KarabassovBobkovaVasenko23}
\bibinfo{author}{\bibfnamefont{T.}~\bibnamefont{Karabassov}},
  \bibinfo{author}{\bibfnamefont{I.~V.} \bibnamefont{Bobkova}},
  \bibinfo{author}{\bibfnamefont{V.~M.} \bibnamefont{Silkin}},
  \bibinfo{author}{\bibfnamefont{B.~G.} \bibnamefont{Lvov}},
  \bibinfo{author}{\bibfnamefont{A.~A.} \bibnamefont{Golubov}},
  \bibnamefont{and} \bibinfo{author}{\bibfnamefont{A.~S.}
  \bibnamefont{Vasenko}}, \bibinfo{journal}{Physica Scripta}
  \textbf{\bibinfo{volume}{99}}, \bibinfo{pages}{015010}
  (\bibinfo{year}{2023}),
  \urlprefix\url{https://dx.doi.org/10.1088/1402-4896/ad1376}.

\bibitem[{\citenamefont{Tanaka et~al.}(2022)\citenamefont{Tanaka, Lu, and
  Nagaosa}}]{TanakaLuNagaosa22}
\bibinfo{author}{\bibfnamefont{Y.}~\bibnamefont{Tanaka}},
  \bibinfo{author}{\bibfnamefont{B.}~\bibnamefont{Lu}}, \bibnamefont{and}
  \bibinfo{author}{\bibfnamefont{N.}~\bibnamefont{Nagaosa}},
  \bibinfo{journal}{Physical Review B} \textbf{\bibinfo{volume}{106}},
  \bibinfo{pages}{214524} (\bibinfo{year}{2022}), \bibinfo{note}{publisher:
  American Physical Society},
  \urlprefix\url{https://link.aps.org/doi/10.1103/PhysRevB.106.214524}.

\bibitem[{\citenamefont{Lu et~al.}(2023)\citenamefont{Lu, Ikegaya, Burset,
  Tanaka, and Nagaosa}}]{LuTanakaNagaosa23}
\bibinfo{author}{\bibfnamefont{B.}~\bibnamefont{Lu}},
  \bibinfo{author}{\bibfnamefont{S.}~\bibnamefont{Ikegaya}},
  \bibinfo{author}{\bibfnamefont{P.}~\bibnamefont{Burset}},
  \bibinfo{author}{\bibfnamefont{Y.}~\bibnamefont{Tanaka}}, \bibnamefont{and}
  \bibinfo{author}{\bibfnamefont{N.}~\bibnamefont{Nagaosa}},
  \bibinfo{journal}{Physical Review Letters} \textbf{\bibinfo{volume}{131}},
  \bibinfo{pages}{096001} (\bibinfo{year}{2023}), \bibinfo{note}{publisher:
  American Physical Society},
  \urlprefix\url{https://link.aps.org/doi/10.1103/PhysRevLett.131.096001}.

\bibitem[{\citenamefont{Chen et~al.}(2018)\citenamefont{Chen, He, Ali, Lee,
  Fong, and Law}}]{ChenLaw18}
\bibinfo{author}{\bibfnamefont{C.-Z.} \bibnamefont{Chen}},
  \bibinfo{author}{\bibfnamefont{J.~J.} \bibnamefont{He}},
  \bibinfo{author}{\bibfnamefont{M.~N.} \bibnamefont{Ali}},
  \bibinfo{author}{\bibfnamefont{G.-H.} \bibnamefont{Lee}},
  \bibinfo{author}{\bibfnamefont{K.~C.} \bibnamefont{Fong}}, \bibnamefont{and}
  \bibinfo{author}{\bibfnamefont{K.~T.} \bibnamefont{Law}},
  \bibinfo{journal}{Physical Review B} \textbf{\bibinfo{volume}{98}},
  \bibinfo{pages}{075430} (\bibinfo{year}{2018}), \bibinfo{note}{publisher:
  American Physical Society},
  \urlprefix\url{https://link.aps.org/doi/10.1103/PhysRevB.98.075430}.

\bibitem[{\citenamefont{Wei et~al.}(2023)\citenamefont{Wei, Wang, and
  Wang}}]{WeiWangWang23}
\bibinfo{author}{\bibfnamefont{Y.-J.} \bibnamefont{Wei}},
  \bibinfo{author}{\bibfnamefont{J.-J.} \bibnamefont{Wang}}, \bibnamefont{and}
  \bibinfo{author}{\bibfnamefont{J.}~\bibnamefont{Wang}},
  \bibinfo{journal}{Physical Review B} \textbf{\bibinfo{volume}{108}},
  \bibinfo{pages}{054521} (\bibinfo{year}{2023}), \bibinfo{note}{publisher:
  American Physical Society},
  \urlprefix\url{https://link.aps.org/doi/10.1103/PhysRevB.108.054521}.

\bibitem[{\citenamefont{Wang et~al.}(2024{\natexlab{d}})\citenamefont{Wang,
  Jiang, Wang, and Liu}}]{WangLiu24}
\bibinfo{author}{\bibfnamefont{J.}~\bibnamefont{Wang}},
  \bibinfo{author}{\bibfnamefont{Y.}~\bibnamefont{Jiang}},
  \bibinfo{author}{\bibfnamefont{J.~J.} \bibnamefont{Wang}}, \bibnamefont{and}
  \bibinfo{author}{\bibfnamefont{J.-F.} \bibnamefont{Liu}},
  \bibinfo{journal}{Physical Review B} \textbf{\bibinfo{volume}{109}},
  \bibinfo{pages}{075412} (\bibinfo{year}{2024}{\natexlab{d}}),
  \bibinfo{note}{publisher: American Physical Society},
  \urlprefix\url{https://link.aps.org/doi/10.1103/PhysRevB.109.075412}.

\bibitem[{\citenamefont{Fracassi et~al.}(2024)\citenamefont{Fracassi, Traverso,
  Traverso~Ziani, Carrega, Heun, and Sassetti}}]{FracassiSassetti24}
\bibinfo{author}{\bibfnamefont{S.}~\bibnamefont{Fracassi}},
  \bibinfo{author}{\bibfnamefont{S.}~\bibnamefont{Traverso}},
  \bibinfo{author}{\bibfnamefont{N.}~\bibnamefont{Traverso~Ziani}},
  \bibinfo{author}{\bibfnamefont{M.}~\bibnamefont{Carrega}},
  \bibinfo{author}{\bibfnamefont{S.}~\bibnamefont{Heun}}, \bibnamefont{and}
  \bibinfo{author}{\bibfnamefont{M.}~\bibnamefont{Sassetti}},
  \bibinfo{journal}{Applied Physics Letters} \textbf{\bibinfo{volume}{124}},
  \bibinfo{pages}{242601} (\bibinfo{year}{2024}), ISSN
  \bibinfo{issn}{0003-6951}, \urlprefix\url{https://doi.org/10.1063/5.0210660}.

\bibitem[{\citenamefont{Ding et~al.}(2024)\citenamefont{Ding, Wang, Li, Tao,
  and Wang}}]{DingWang24}
\bibinfo{author}{\bibfnamefont{Z.}~\bibnamefont{Ding}},
  \bibinfo{author}{\bibfnamefont{D.}~\bibnamefont{Wang}},
  \bibinfo{author}{\bibfnamefont{M.}~\bibnamefont{Li}},
  \bibinfo{author}{\bibfnamefont{Y.}~\bibnamefont{Tao}}, \bibnamefont{and}
  \bibinfo{author}{\bibfnamefont{J.}~\bibnamefont{Wang}},
  \bibinfo{journal}{Physical Review B} \textbf{\bibinfo{volume}{110}},
  \bibinfo{pages}{155405} (\bibinfo{year}{2024}), \bibinfo{note}{publisher:
  American Physical Society},
  \urlprefix\url{https://link.aps.org/doi/10.1103/PhysRevB.110.155405}.

\bibitem[{\citenamefont{Scharf et~al.}(2024)\citenamefont{Scharf, Kochan, and
  Matos-Abiague}}]{ScharfKochanMatosAbiague24}
\bibinfo{author}{\bibfnamefont{B.}~\bibnamefont{Scharf}},
  \bibinfo{author}{\bibfnamefont{D.}~\bibnamefont{Kochan}}, \bibnamefont{and}
  \bibinfo{author}{\bibfnamefont{A.}~\bibnamefont{Matos-Abiague}},
  \bibinfo{journal}{Physical Review B} \textbf{\bibinfo{volume}{110}},
  \bibinfo{pages}{134511} (\bibinfo{year}{2024}), \bibinfo{note}{publisher:
  American Physical Society},
  \urlprefix\url{https://link.aps.org/doi/10.1103/PhysRevB.110.134511}.

\bibitem[{\citenamefont{Guo et~al.}(2025)\citenamefont{Guo, Pan, Dong, and
  Liu}}]{Guo25}
\bibinfo{author}{\bibfnamefont{G.-L.} \bibnamefont{Guo}},
  \bibinfo{author}{\bibfnamefont{X.-H.} \bibnamefont{Pan}},
  \bibinfo{author}{\bibfnamefont{H.}~\bibnamefont{Dong}}, \bibnamefont{and}
  \bibinfo{author}{\bibfnamefont{X.}~\bibnamefont{Liu}} (\bibinfo{year}{2025}),
  \eprint{2508.21357}, \urlprefix\url{https://arxiv.org/abs/2508.21357}.

\bibitem[{\citenamefont{Fu and Kane}(2007)}]{FuKane07}
\bibinfo{author}{\bibfnamefont{L.}~\bibnamefont{Fu}} \bibnamefont{and}
  \bibinfo{author}{\bibfnamefont{C.~L.} \bibnamefont{Kane}},
  \bibinfo{journal}{Phys. Rev. B} \textbf{\bibinfo{volume}{76}},
  \bibinfo{pages}{045302} (\bibinfo{year}{2007}),
  \urlprefix\url{https://link.aps.org/doi/10.1103/PhysRevB.76.045302}.

\bibitem[{\citenamefont{Hughes et~al.}(2011)\citenamefont{Hughes, Prodan, and
  Bernevig}}]{HughesProdanBernevig11}
\bibinfo{author}{\bibfnamefont{T.~L.} \bibnamefont{Hughes}},
  \bibinfo{author}{\bibfnamefont{E.}~\bibnamefont{Prodan}}, \bibnamefont{and}
  \bibinfo{author}{\bibfnamefont{B.~A.} \bibnamefont{Bernevig}},
  \bibinfo{journal}{Phys. Rev. B} \textbf{\bibinfo{volume}{83}},
  \bibinfo{pages}{245132} (\bibinfo{year}{2011}),
  \urlprefix\url{https://link.aps.org/doi/10.1103/PhysRevB.83.245132}.

\bibitem[{\citenamefont{Legg et~al.}(2022)\citenamefont{Legg, Loss, and
  Klinovaja}}]{LeggLossKlinovaja22}
\bibinfo{author}{\bibfnamefont{H.~F.} \bibnamefont{Legg}},
  \bibinfo{author}{\bibfnamefont{D.}~\bibnamefont{Loss}}, \bibnamefont{and}
  \bibinfo{author}{\bibfnamefont{J.}~\bibnamefont{Klinovaja}},
  \bibinfo{journal}{Phys. Rev. B} \textbf{\bibinfo{volume}{106}},
  \bibinfo{pages}{104501} (\bibinfo{year}{2022}),
  \urlprefix\url{https://link.aps.org/doi/10.1103/PhysRevB.106.104501}.

\bibitem[{\citenamefont{Monroe et~al.}(2024)\citenamefont{Monroe, Shen,
  Tringali, Alidoust, Zhou, and {\v{Z}}uti{\'c}}}]{MonroeZutic24}
\bibinfo{author}{\bibfnamefont{D.}~\bibnamefont{Monroe}},
  \bibinfo{author}{\bibfnamefont{C.}~\bibnamefont{Shen}},
  \bibinfo{author}{\bibfnamefont{D.}~\bibnamefont{Tringali}},
  \bibinfo{author}{\bibfnamefont{M.}~\bibnamefont{Alidoust}},
  \bibinfo{author}{\bibfnamefont{T.}~\bibnamefont{Zhou}}, \bibnamefont{and}
  \bibinfo{author}{\bibfnamefont{I.}~\bibnamefont{{\v{Z}}uti{\'c}}},
  \bibinfo{journal}{Applied Physics Letters} \textbf{\bibinfo{volume}{125}}
  (\bibinfo{year}{2024}).

\bibitem[{\citenamefont{Soori}(2023{\natexlab{a}})}]{Soori23II}
\bibinfo{author}{\bibfnamefont{A.}~\bibnamefont{Soori}},
  \bibinfo{journal}{Physica E: Low-dimensional Systems and Nanostructures}
  \textbf{\bibinfo{volume}{146}}, \bibinfo{pages}{115545}
  (\bibinfo{year}{2023}{\natexlab{a}}), ISSN \bibinfo{issn}{1386-9477},
  \urlprefix\url{https://www.sciencedirect.com/science/article/pii/S138694772200368X}.

\bibitem[{\citenamefont{Wang et~al.}(2025{\natexlab{c}})\citenamefont{Wang, Li,
  Li, and Liu}}]{WangLiu25}
\bibinfo{author}{\bibfnamefont{B.-Z.} \bibnamefont{Wang}},
  \bibinfo{author}{\bibfnamefont{Z.-K.} \bibnamefont{Li}},
  \bibinfo{author}{\bibfnamefont{Z.-D.} \bibnamefont{Li}}, \bibnamefont{and}
  \bibinfo{author}{\bibfnamefont{X.-J.} \bibnamefont{Liu}},
  \emph{\bibinfo{title}{Giant and robust josephson diode effect in multiband
  topological nanowires}} (\bibinfo{year}{2025}{\natexlab{c}}),
  \eprint{2510.05772}, \urlprefix\url{https://arxiv.org/abs/2510.05772}.

\bibitem[{\citenamefont{Chen et~al.}(2024{\natexlab{b}})\citenamefont{Chen,
  Karki, and Hosur}}]{ChenHosur24}
\bibinfo{author}{\bibfnamefont{K.}~\bibnamefont{Chen}},
  \bibinfo{author}{\bibfnamefont{B.}~\bibnamefont{Karki}}, \bibnamefont{and}
  \bibinfo{author}{\bibfnamefont{P.}~\bibnamefont{Hosur}},
  \bibinfo{journal}{Phys. Rev. B} \textbf{\bibinfo{volume}{109}},
  \bibinfo{pages}{064511} (\bibinfo{year}{2024}{\natexlab{b}}),
  \urlprefix\url{https://link.aps.org/doi/10.1103/PhysRevB.109.064511}.

\bibitem[{\citenamefont{Liu et~al.}(2025)\citenamefont{Liu, Zhao, Cheng, and
  Sun}}]{LiuSun25}
\bibinfo{author}{\bibfnamefont{W.-T.} \bibnamefont{Liu}},
  \bibinfo{author}{\bibfnamefont{S.-C.} \bibnamefont{Zhao}},
  \bibinfo{author}{\bibfnamefont{Q.}~\bibnamefont{Cheng}}, \bibnamefont{and}
  \bibinfo{author}{\bibfnamefont{Q.-F.} \bibnamefont{Sun}},
  \bibinfo{journal}{Phys. Rev. B} \textbf{\bibinfo{volume}{112}},
  \bibinfo{pages}{054509} (\bibinfo{year}{2025}),
  \urlprefix\url{https://link.aps.org/doi/10.1103/tlqg-268j}.

\bibitem[{\citenamefont{Yu et~al.}(2025)\citenamefont{Yu, Bernevig, Queiroz,
  Rossi, T{\"o}rm{\"a}, and Yang}}]{YuBernevigTorma25}
\bibinfo{author}{\bibfnamefont{J.}~\bibnamefont{Yu}},
  \bibinfo{author}{\bibfnamefont{B.~A.} \bibnamefont{Bernevig}},
  \bibinfo{author}{\bibfnamefont{R.}~\bibnamefont{Queiroz}},
  \bibinfo{author}{\bibfnamefont{E.}~\bibnamefont{Rossi}},
  \bibinfo{author}{\bibfnamefont{P.}~\bibnamefont{T{\"o}rm{\"a}}},
  \bibnamefont{and} \bibinfo{author}{\bibfnamefont{B.-J.} \bibnamefont{Yang}},
  \bibinfo{journal}{npj Quantum Materials} \textbf{\bibinfo{volume}{10}},
  \bibinfo{pages}{101} (\bibinfo{year}{2025}),
  \urlprefix\url{https://doi.org/10.1038/s41535-025-00801-3}.

\bibitem[{\citenamefont{Peotta and T{\"o}rm{\"a}}(2015)}]{PeottaTorma15}
\bibinfo{author}{\bibfnamefont{S.}~\bibnamefont{Peotta}} \bibnamefont{and}
  \bibinfo{author}{\bibfnamefont{P.}~\bibnamefont{T{\"o}rm{\"a}}},
  \bibinfo{journal}{Nature Communications} \textbf{\bibinfo{volume}{6}},
  \bibinfo{pages}{8944} (\bibinfo{year}{2015}),
  \urlprefix\url{https://doi.org/10.1038/ncomms9944}.

\bibitem[{\citenamefont{Hu et~al.}(2024)\citenamefont{Hu, Chen, and
  Law}}]{HuLaw24}
\bibinfo{author}{\bibfnamefont{J.-X.} \bibnamefont{Hu}},
  \bibinfo{author}{\bibfnamefont{S.~A.} \bibnamefont{Chen}}, \bibnamefont{and}
  \bibinfo{author}{\bibfnamefont{K.~T.} \bibnamefont{Law}}
  (\bibinfo{year}{2024}), \bibinfo{note}{arXiv:2403.01080 [cond-mat]},
  \urlprefix\url{http://arxiv.org/abs/2403.01080}.

\bibitem[{\citenamefont{Dunbrack et~al.}(2025)\citenamefont{Dunbrack, Virtanen,
  and Heikkilä}}]{DunbrackHeikkila25}
\bibinfo{author}{\bibfnamefont{A.}~\bibnamefont{Dunbrack}},
  \bibinfo{author}{\bibfnamefont{P.}~\bibnamefont{Virtanen}}, \bibnamefont{and}
  \bibinfo{author}{\bibfnamefont{T.~T.} \bibnamefont{Heikkilä}}
  (\bibinfo{year}{2025}), \eprint{2503.14721},
  \urlprefix\url{https://arxiv.org/abs/2503.14721}.

\bibitem[{\citenamefont{Steiner et~al.}(2023)\citenamefont{Steiner, Melischek,
  Trahms, Franke, and von Oppen}}]{SteinerVonOppen23}
\bibinfo{author}{\bibfnamefont{J.~F.} \bibnamefont{Steiner}},
  \bibinfo{author}{\bibfnamefont{L.}~\bibnamefont{Melischek}},
  \bibinfo{author}{\bibfnamefont{M.}~\bibnamefont{Trahms}},
  \bibinfo{author}{\bibfnamefont{K.~J.} \bibnamefont{Franke}},
  \bibnamefont{and} \bibinfo{author}{\bibfnamefont{F.}~\bibnamefont{von
  Oppen}}, \bibinfo{journal}{Phys. Rev. Lett.} \textbf{\bibinfo{volume}{130}},
  \bibinfo{pages}{177002} (\bibinfo{year}{2023}),
  \urlprefix\url{https://link.aps.org/doi/10.1103/PhysRevLett.130.177002}.

\bibitem[{\citenamefont{Shaffer et~al.}(2025)\citenamefont{Shaffer, Li, Hasan,
  Titov, and Levchenko}}]{ShafferLiLevchenko25}
\bibinfo{author}{\bibfnamefont{D.}~\bibnamefont{Shaffer}},
  \bibinfo{author}{\bibfnamefont{S.}~\bibnamefont{Li}},
  \bibinfo{author}{\bibfnamefont{J.}~\bibnamefont{Hasan}},
  \bibinfo{author}{\bibfnamefont{M.}~\bibnamefont{Titov}}, \bibnamefont{and}
  \bibinfo{author}{\bibfnamefont{A.}~\bibnamefont{Levchenko}},
  \bibinfo{journal}{Phys. Rev. B} \textbf{\bibinfo{volume}{112}},
  \bibinfo{pages}{094509} (\bibinfo{year}{2025}),
  \urlprefix\url{https://link.aps.org/doi/10.1103/n9y8-31gc}.

\bibitem[{\citenamefont{Daido and Yanase}(2025)}]{DaidoYanase25}
\bibinfo{author}{\bibfnamefont{A.}~\bibnamefont{Daido}} \bibnamefont{and}
  \bibinfo{author}{\bibfnamefont{Y.}~\bibnamefont{Yanase}},
  \bibinfo{journal}{Physical Review B} \textbf{\bibinfo{volume}{111}},
  \bibinfo{pages}{L020508} (\bibinfo{year}{2025}), \bibinfo{note}{publisher:
  American Physical Society},
  \urlprefix\url{https://link.aps.org/doi/10.1103/PhysRevB.111.L020508}.

\bibitem[{\citenamefont{Matsyshyn and Song}(2025)}]{Matsyshyn25}
\bibinfo{author}{\bibfnamefont{O.}~\bibnamefont{Matsyshyn}} \bibnamefont{and}
  \bibinfo{author}{\bibfnamefont{J.~C.~W.} \bibnamefont{Song}}
  (\bibinfo{year}{2025}), \eprint{2507.06290},
  \urlprefix\url{https://arxiv.org/abs/2507.06290}.

\bibitem[{\citenamefont{Arora and Narang}(2025)}]{Arora25}
\bibinfo{author}{\bibfnamefont{A.}~\bibnamefont{Arora}} \bibnamefont{and}
  \bibinfo{author}{\bibfnamefont{P.}~\bibnamefont{Narang}}
  (\bibinfo{year}{2025}), \eprint{2501.17924},
  \urlprefix\url{https://arxiv.org/abs/2501.17924}.

\bibitem[{\citenamefont{Hu et~al.}(2007)\citenamefont{Hu, Wu, and
  Dai}}]{HuWuDai07}
\bibinfo{author}{\bibfnamefont{J.}~\bibnamefont{Hu}},
  \bibinfo{author}{\bibfnamefont{C.}~\bibnamefont{Wu}}, \bibnamefont{and}
  \bibinfo{author}{\bibfnamefont{X.}~\bibnamefont{Dai}},
  \bibinfo{journal}{Phys. Rev. Lett.} \textbf{\bibinfo{volume}{99}},
  \bibinfo{pages}{067004} (\bibinfo{year}{2007}),
  \urlprefix\url{https://link.aps.org/doi/10.1103/PhysRevLett.99.067004}.

\bibitem[{\citenamefont{Misaki and Nagaosa}(2021)}]{MisakiNagaosa21}
\bibinfo{author}{\bibfnamefont{K.}~\bibnamefont{Misaki}} \bibnamefont{and}
  \bibinfo{author}{\bibfnamefont{N.}~\bibnamefont{Nagaosa}},
  \bibinfo{journal}{Physical Review B} \textbf{\bibinfo{volume}{103}},
  \bibinfo{pages}{245302} (\bibinfo{year}{2021}), \bibinfo{note}{publisher:
  American Physical Society},
  \urlprefix\url{https://link.aps.org/doi/10.1103/PhysRevB.103.245302}.

\bibitem[{\citenamefont{Menditto et~al.}(2016)\citenamefont{Menditto,
  Sickinger, Weides, Kohlstedt, Koelle, Kleiner, and
  Goldobin}}]{MendittoGoldobin16}
\bibinfo{author}{\bibfnamefont{R.}~\bibnamefont{Menditto}},
  \bibinfo{author}{\bibfnamefont{H.}~\bibnamefont{Sickinger}},
  \bibinfo{author}{\bibfnamefont{M.}~\bibnamefont{Weides}},
  \bibinfo{author}{\bibfnamefont{H.}~\bibnamefont{Kohlstedt}},
  \bibinfo{author}{\bibfnamefont{D.}~\bibnamefont{Koelle}},
  \bibinfo{author}{\bibfnamefont{R.}~\bibnamefont{Kleiner}}, \bibnamefont{and}
  \bibinfo{author}{\bibfnamefont{E.}~\bibnamefont{Goldobin}},
  \bibinfo{journal}{Physical Review E} \textbf{\bibinfo{volume}{94}},
  \bibinfo{pages}{042202} (\bibinfo{year}{2016}), \bibinfo{note}{publisher:
  American Physical Society},
  \urlprefix\url{https://link.aps.org/doi/10.1103/PhysRevE.94.042202}.

\bibitem[{\citenamefont{Schmid et~al.}(2024)\citenamefont{Schmid, Jozani,
  Kleiner, Koelle, and Goldobin}}]{SchmidGoldobin24}
\bibinfo{author}{\bibfnamefont{C.}~\bibnamefont{Schmid}},
  \bibinfo{author}{\bibfnamefont{A.}~\bibnamefont{Jozani}},
  \bibinfo{author}{\bibfnamefont{R.}~\bibnamefont{Kleiner}},
  \bibinfo{author}{\bibfnamefont{D.}~\bibnamefont{Koelle}}, \bibnamefont{and}
  \bibinfo{author}{\bibfnamefont{E.}~\bibnamefont{Goldobin}}
  (\bibinfo{year}{2024}), \bibinfo{note}{arXiv:2408.01521 [cond-mat,
  physics:physics]}, \urlprefix\url{http://arxiv.org/abs/2408.01521}.

\bibitem[{\citenamefont{Su et~al.}(2024)\citenamefont{Su, Wang, Gao, Luo, Yan,
  Wu, Li, Shen, Lu, Pan et~al.}}]{Su24}
\bibinfo{author}{\bibfnamefont{H.}~\bibnamefont{Su}},
  \bibinfo{author}{\bibfnamefont{J.-Y.} \bibnamefont{Wang}},
  \bibinfo{author}{\bibfnamefont{H.}~\bibnamefont{Gao}},
  \bibinfo{author}{\bibfnamefont{Y.}~\bibnamefont{Luo}},
  \bibinfo{author}{\bibfnamefont{S.}~\bibnamefont{Yan}},
  \bibinfo{author}{\bibfnamefont{X.}~\bibnamefont{Wu}},
  \bibinfo{author}{\bibfnamefont{G.}~\bibnamefont{Li}},
  \bibinfo{author}{\bibfnamefont{J.}~\bibnamefont{Shen}},
  \bibinfo{author}{\bibfnamefont{L.}~\bibnamefont{Lu}},
  \bibinfo{author}{\bibfnamefont{D.}~\bibnamefont{Pan}}, \bibnamefont{et~al.},
  \bibinfo{journal}{Phys. Rev. Lett.} \textbf{\bibinfo{volume}{133}},
  \bibinfo{pages}{087001} (\bibinfo{year}{2024}),
  \urlprefix\url{https://link.aps.org/doi/10.1103/PhysRevLett.133.087001}.

\bibitem[{\citenamefont{Borgongino et~al.}(2025)\citenamefont{Borgongino,
  Souto, Paghi, Senesi, Skibinska, Sorba, Giazotto, and
  Strambini}}]{BorgonginoGiazotto25}
\bibinfo{author}{\bibfnamefont{L.}~\bibnamefont{Borgongino}},
  \bibinfo{author}{\bibfnamefont{R.~S.} \bibnamefont{Souto}},
  \bibinfo{author}{\bibfnamefont{A.}~\bibnamefont{Paghi}},
  \bibinfo{author}{\bibfnamefont{G.}~\bibnamefont{Senesi}},
  \bibinfo{author}{\bibfnamefont{K.}~\bibnamefont{Skibinska}},
  \bibinfo{author}{\bibfnamefont{L.}~\bibnamefont{Sorba}},
  \bibinfo{author}{\bibfnamefont{F.}~\bibnamefont{Giazotto}}, \bibnamefont{and}
  \bibinfo{author}{\bibfnamefont{E.}~\bibnamefont{Strambini}}
  (\bibinfo{year}{2025}), \bibinfo{note}{arXiv:2504.08691 [cond-mat]},
  \urlprefix\url{http://arxiv.org/abs/2504.08691}.

\bibitem[{\citenamefont{Wang et~al.}(2025{\natexlab{d}})\citenamefont{Wang,
  Zhu, Bai, Lyu, Yang, Zhao, Zhou, Gu, Xue, and Zhang}}]{Wang25}
\bibinfo{author}{\bibfnamefont{H.}~\bibnamefont{Wang}},
  \bibinfo{author}{\bibfnamefont{Y.}~\bibnamefont{Zhu}},
  \bibinfo{author}{\bibfnamefont{Z.}~\bibnamefont{Bai}},
  \bibinfo{author}{\bibfnamefont{Z.}~\bibnamefont{Lyu}},
  \bibinfo{author}{\bibfnamefont{J.}~\bibnamefont{Yang}},
  \bibinfo{author}{\bibfnamefont{L.}~\bibnamefont{Zhao}},
  \bibinfo{author}{\bibfnamefont{X.~J.} \bibnamefont{Zhou}},
  \bibinfo{author}{\bibfnamefont{G.}~\bibnamefont{Gu}},
  \bibinfo{author}{\bibfnamefont{Q.-K.} \bibnamefont{Xue}}, \bibnamefont{and}
  \bibinfo{author}{\bibfnamefont{D.}~\bibnamefont{Zhang}},
  \emph{\bibinfo{title}{Quantum superconducting diode effect with perfect
  efficiency above liquid-nitrogen temperature}}
  (\bibinfo{year}{2025}{\natexlab{d}}), \eprint{2509.24764},
  \urlprefix\url{https://arxiv.org/abs/2509.24764}.

\bibitem[{\citenamefont{Seoane~Souto et~al.}(2024)\citenamefont{Seoane~Souto,
  Leijnse, Schrade, Valentini, Katsaros, and Danon}}]{SeoaneSoutoDanon24}
\bibinfo{author}{\bibfnamefont{R.}~\bibnamefont{Seoane~Souto}},
  \bibinfo{author}{\bibfnamefont{M.}~\bibnamefont{Leijnse}},
  \bibinfo{author}{\bibfnamefont{C.}~\bibnamefont{Schrade}},
  \bibinfo{author}{\bibfnamefont{M.}~\bibnamefont{Valentini}},
  \bibinfo{author}{\bibfnamefont{G.}~\bibnamefont{Katsaros}}, \bibnamefont{and}
  \bibinfo{author}{\bibfnamefont{J.}~\bibnamefont{Danon}},
  \bibinfo{journal}{Physical Review Research} \textbf{\bibinfo{volume}{6}},
  \bibinfo{pages}{L022002} (\bibinfo{year}{2024}), \bibinfo{note}{publisher:
  American Physical Society},
  \urlprefix\url{https://link.aps.org/doi/10.1103/PhysRevResearch.6.L022002}.

\bibitem[{\citenamefont{Scheer et~al.}(2025)\citenamefont{Scheer, Seoane~Souto,
  Hassler, and Danon}}]{ScheerDanon25}
\bibinfo{author}{\bibfnamefont{D.}~\bibnamefont{Scheer}},
  \bibinfo{author}{\bibfnamefont{R.}~\bibnamefont{Seoane~Souto}},
  \bibinfo{author}{\bibfnamefont{F.}~\bibnamefont{Hassler}}, \bibnamefont{and}
  \bibinfo{author}{\bibfnamefont{J.}~\bibnamefont{Danon}},
  \bibinfo{journal}{New Journal of Physics} \textbf{\bibinfo{volume}{27}},
  \bibinfo{pages}{033013} (\bibinfo{year}{2025}), ISSN
  \bibinfo{issn}{1367-2630}, \bibinfo{note}{publisher: IOP Publishing},
  \urlprefix\url{https://dx.doi.org/10.1088/1367-2630/adba80}.

\bibitem[{\citenamefont{Soori}(2023{\natexlab{b}})}]{Soori23I}
\bibinfo{author}{\bibfnamefont{A.}~\bibnamefont{Soori}},
  \bibinfo{journal}{Physica Scripta} \textbf{\bibinfo{volume}{98}},
  \bibinfo{pages}{065917} (\bibinfo{year}{2023}{\natexlab{b}}), ISSN
  \bibinfo{issn}{1402-4896}, \bibinfo{note}{publisher: IOP Publishing},
  \urlprefix\url{https://dx.doi.org/10.1088/1402-4896/acd02f}.

\bibitem[{\citenamefont{Ortega-Taberner
  et~al.}(2023)\citenamefont{Ortega-Taberner, Jauho, and
  Paaske}}]{OrtegaTaberner23}
\bibinfo{author}{\bibfnamefont{C.}~\bibnamefont{Ortega-Taberner}},
  \bibinfo{author}{\bibfnamefont{A.-P.} \bibnamefont{Jauho}}, \bibnamefont{and}
  \bibinfo{author}{\bibfnamefont{J.}~\bibnamefont{Paaske}},
  \bibinfo{journal}{Phys. Rev. B} \textbf{\bibinfo{volume}{107}},
  \bibinfo{pages}{115165} (\bibinfo{year}{2023}),
  \urlprefix\url{https://link.aps.org/doi/10.1103/PhysRevB.107.115165}.

\bibitem[{\citenamefont{Tsarev and Fominov}(2025)}]{TsarevFominov25}
\bibinfo{author}{\bibfnamefont{P.~N.} \bibnamefont{Tsarev}} \bibnamefont{and}
  \bibinfo{author}{\bibfnamefont{Y.~V.} \bibnamefont{Fominov}},
  \emph{\bibinfo{title}{All fractional shapiro steps in the rsj model with two
  josephson harmonics}} (\bibinfo{year}{2025}), \eprint{2505.20502},
  \urlprefix\url{https://arxiv.org/abs/2505.20502}.

\bibitem[{\citenamefont{Qi et~al.}(2025{\natexlab{b}})\citenamefont{Qi, Lu,
  Liu, Chen, and Xie}}]{QiXie25}
\bibinfo{author}{\bibfnamefont{J.}~\bibnamefont{Qi}},
  \bibinfo{author}{\bibfnamefont{M.}~\bibnamefont{Lu}},
  \bibinfo{author}{\bibfnamefont{J.}~\bibnamefont{Liu}},
  \bibinfo{author}{\bibfnamefont{C.-Z.} \bibnamefont{Chen}}, \bibnamefont{and}
  \bibinfo{author}{\bibfnamefont{X.~C.} \bibnamefont{Xie}},
  \bibinfo{journal}{Phys. Rev. B} \textbf{\bibinfo{volume}{112}},
  \bibinfo{pages}{L060502} (\bibinfo{year}{2025}{\natexlab{b}}),
  \urlprefix\url{https://link.aps.org/doi/10.1103/n51c-17pn}.

\bibitem[{\citenamefont{Valentini et~al.}(2024)\citenamefont{Valentini, Sagi,
  Baghumyan, de~Gijsel, Jung, Calcaterra, Ballabio, Aguilera~Servin, Aggarwal,
  Janik et~al.}}]{ValentiniDanon24}
\bibinfo{author}{\bibfnamefont{M.}~\bibnamefont{Valentini}},
  \bibinfo{author}{\bibfnamefont{O.}~\bibnamefont{Sagi}},
  \bibinfo{author}{\bibfnamefont{L.}~\bibnamefont{Baghumyan}},
  \bibinfo{author}{\bibfnamefont{T.}~\bibnamefont{de~Gijsel}},
  \bibinfo{author}{\bibfnamefont{J.}~\bibnamefont{Jung}},
  \bibinfo{author}{\bibfnamefont{S.}~\bibnamefont{Calcaterra}},
  \bibinfo{author}{\bibfnamefont{A.}~\bibnamefont{Ballabio}},
  \bibinfo{author}{\bibfnamefont{J.}~\bibnamefont{Aguilera~Servin}},
  \bibinfo{author}{\bibfnamefont{K.}~\bibnamefont{Aggarwal}},
  \bibinfo{author}{\bibfnamefont{M.}~\bibnamefont{Janik}},
  \bibnamefont{et~al.}, \bibinfo{journal}{Nature Communications}
  \textbf{\bibinfo{volume}{15}}, \bibinfo{pages}{169} (\bibinfo{year}{2024}),
  ISSN \bibinfo{issn}{2041-1723}, \bibinfo{note}{publisher: Nature Publishing
  Group}, \urlprefix\url{https://www.nature.com/articles/s41467-023-44114-0}.

\bibitem[{\citenamefont{Mironov et~al.}(2024)\citenamefont{Mironov, Mel'nikov,
  and Buzdin}}]{MironovBuzdin24}
\bibinfo{author}{\bibfnamefont{S.~V.} \bibnamefont{Mironov}},
  \bibinfo{author}{\bibfnamefont{A.~S.} \bibnamefont{Mel'nikov}},
  \bibnamefont{and} \bibinfo{author}{\bibfnamefont{A.~I.}
  \bibnamefont{Buzdin}}, \bibinfo{journal}{Phys. Rev. B}
  \textbf{\bibinfo{volume}{109}}, \bibinfo{pages}{L220503}
  (\bibinfo{year}{2024}),
  \urlprefix\url{https://link.aps.org/doi/10.1103/PhysRevB.109.L220503}.

\bibitem[{\citenamefont{Volkov}(1995)}]{Volkov95}
\bibinfo{author}{\bibfnamefont{A.~F.} \bibnamefont{Volkov}},
  \bibinfo{journal}{Phys. Rev. Lett.} \textbf{\bibinfo{volume}{74}},
  \bibinfo{pages}{4730} (\bibinfo{year}{1995}),
  \urlprefix\url{https://link.aps.org/doi/10.1103/PhysRevLett.74.4730}.

\bibitem[{\citenamefont{Golubov et~al.}(1997)\citenamefont{Golubov, Wilhelm,
  and Zaikin}}]{GolubovWilhelmZaikin97}
\bibinfo{author}{\bibfnamefont{A.~A.} \bibnamefont{Golubov}},
  \bibinfo{author}{\bibfnamefont{F.~K.} \bibnamefont{Wilhelm}},
  \bibnamefont{and} \bibinfo{author}{\bibfnamefont{A.~D.}
  \bibnamefont{Zaikin}}, \bibinfo{journal}{Physical Review B}
  \textbf{\bibinfo{volume}{55}}, \bibinfo{pages}{1123} (\bibinfo{year}{1997}),
  \bibinfo{note}{publisher: American Physical Society},
  \urlprefix\url{https://link.aps.org/doi/10.1103/PhysRevB.55.1123}.

\bibitem[{\citenamefont{Wilhelm et~al.}(1998)\citenamefont{Wilhelm, Sch\"on,
  and Zaikin}}]{WilhelmZaikin98}
\bibinfo{author}{\bibfnamefont{F.~K.} \bibnamefont{Wilhelm}},
  \bibinfo{author}{\bibfnamefont{G.}~\bibnamefont{Sch\"on}}, \bibnamefont{and}
  \bibinfo{author}{\bibfnamefont{A.~D.} \bibnamefont{Zaikin}},
  \bibinfo{journal}{Phys. Rev. Lett.} \textbf{\bibinfo{volume}{81}},
  \bibinfo{pages}{1682} (\bibinfo{year}{1998}),
  \urlprefix\url{https://link.aps.org/doi/10.1103/PhysRevLett.81.1682}.

\bibitem[{\citenamefont{Virtanen and Heikkil\"a}(2004)}]{VirtanenHeikkila04}
\bibinfo{author}{\bibfnamefont{P.}~\bibnamefont{Virtanen}} \bibnamefont{and}
  \bibinfo{author}{\bibfnamefont{T.~T.} \bibnamefont{Heikkil\"a}},
  \bibinfo{journal}{Phys. Rev. Lett.} \textbf{\bibinfo{volume}{92}},
  \bibinfo{pages}{177004} (\bibinfo{year}{2004}),
  \urlprefix\url{https://link.aps.org/doi/10.1103/PhysRevLett.92.177004}.

\bibitem[{\citenamefont{Titov}(2008)}]{Titov08}
\bibinfo{author}{\bibfnamefont{M.}~\bibnamefont{Titov}},
  \bibinfo{journal}{Physical Review B} \textbf{\bibinfo{volume}{78}},
  \bibinfo{pages}{224521} (\bibinfo{year}{2008}), \bibinfo{note}{publisher:
  American Physical Society},
  \urlprefix\url{https://link.aps.org/doi/10.1103/PhysRevB.78.224521}.

\bibitem[{\citenamefont{Dolgirev et~al.}(2018)\citenamefont{Dolgirev, Kalenkov,
  and Zaikin}}]{DolgirevKalenkovZaikin18}
\bibinfo{author}{\bibfnamefont{P.~E.} \bibnamefont{Dolgirev}},
  \bibinfo{author}{\bibfnamefont{M.~S.} \bibnamefont{Kalenkov}},
  \bibnamefont{and} \bibinfo{author}{\bibfnamefont{A.~D.}
  \bibnamefont{Zaikin}}, \bibinfo{journal}{Phys. Rev. B}
  \textbf{\bibinfo{volume}{97}}, \bibinfo{pages}{054521}
  (\bibinfo{year}{2018}),
  \urlprefix\url{https://link.aps.org/doi/10.1103/PhysRevB.97.054521}.

\bibitem[{\citenamefont{Kalenkov and Zaikin}(2021)}]{KalenkovZaikin21}
\bibinfo{author}{\bibfnamefont{M.~S.} \bibnamefont{Kalenkov}} \bibnamefont{and}
  \bibinfo{author}{\bibfnamefont{A.~D.} \bibnamefont{Zaikin}},
  \bibinfo{journal}{JETP Letters} \textbf{\bibinfo{volume}{114}},
  \bibinfo{pages}{593} (\bibinfo{year}{2021}), ISSN \bibinfo{issn}{1090-6487},
  \urlprefix\url{https://doi.org/10.1134/S0021364021220021}.

\bibitem[{\citenamefont{Baselmans et~al.}(1999)\citenamefont{Baselmans,
  Morpurgo, van Wees, and Klapwijk}}]{Baselmans99}
\bibinfo{author}{\bibfnamefont{J.~J.~A.} \bibnamefont{Baselmans}},
  \bibinfo{author}{\bibfnamefont{A.~F.} \bibnamefont{Morpurgo}},
  \bibinfo{author}{\bibfnamefont{B.~J.} \bibnamefont{van Wees}},
  \bibnamefont{and} \bibinfo{author}{\bibfnamefont{T.~M.}
  \bibnamefont{Klapwijk}}, \bibinfo{journal}{Nature}
  \textbf{\bibinfo{volume}{397}}, \bibinfo{pages}{43} (\bibinfo{year}{1999}),
  ISSN \bibinfo{issn}{1476-4687}, \bibinfo{note}{publisher: Nature Publishing
  Group}, \urlprefix\url{https://www.nature.com/articles/16204}.

\bibitem[{\citenamefont{Shaikhaidarov et~al.}(2000)\citenamefont{Shaikhaidarov,
  Volkov, Takayanagi, Petrashov, and Delsing}}]{ShaikhaidarovVolkov00}
\bibinfo{author}{\bibfnamefont{R.}~\bibnamefont{Shaikhaidarov}},
  \bibinfo{author}{\bibfnamefont{A.~F.} \bibnamefont{Volkov}},
  \bibinfo{author}{\bibfnamefont{H.}~\bibnamefont{Takayanagi}},
  \bibinfo{author}{\bibfnamefont{V.~T.} \bibnamefont{Petrashov}},
  \bibnamefont{and} \bibinfo{author}{\bibfnamefont{P.}~\bibnamefont{Delsing}},
  \bibinfo{journal}{Phys. Rev. B} \textbf{\bibinfo{volume}{62}},
  \bibinfo{pages}{R14649} (\bibinfo{year}{2000}),
  \urlprefix\url{https://link.aps.org/doi/10.1103/PhysRevB.62.R14649}.

\bibitem[{\citenamefont{Bezuglyi et~al.}(2003)\citenamefont{Bezuglyi, Shumeiko,
  and Wendin}}]{Bezuglyi03}
\bibinfo{author}{\bibfnamefont{E.~V.} \bibnamefont{Bezuglyi}},
  \bibinfo{author}{\bibfnamefont{V.~S.} \bibnamefont{Shumeiko}},
  \bibnamefont{and} \bibinfo{author}{\bibfnamefont{G.}~\bibnamefont{Wendin}},
  \bibinfo{journal}{Physical Review B} \textbf{\bibinfo{volume}{68}},
  \bibinfo{pages}{134506} (\bibinfo{year}{2003}), \bibinfo{note}{publisher:
  American Physical Society},
  \urlprefix\url{https://link.aps.org/doi/10.1103/PhysRevB.68.134506}.

\bibitem[{\citenamefont{Sun and Linder}(2024)}]{SunLinder24}
\bibinfo{author}{\bibfnamefont{C.}~\bibnamefont{Sun}} \bibnamefont{and}
  \bibinfo{author}{\bibfnamefont{J.}~\bibnamefont{Linder}},
  \bibinfo{journal}{Physical Review B} \textbf{\bibinfo{volume}{110}},
  \bibinfo{pages}{224512} (\bibinfo{year}{2024}), \bibinfo{note}{publisher:
  American Physical Society},
  \urlprefix\url{https://link.aps.org/doi/10.1103/PhysRevB.110.224512}.

\bibitem[{\citenamefont{Hijano et~al.}(2021)\citenamefont{Hijano,
  Ili\ifmmode~\acute{c}\else \'{c}\fi{}, and Bergeret}}]{HijanoIlicBergeret21}
\bibinfo{author}{\bibfnamefont{A.}~\bibnamefont{Hijano}},
  \bibinfo{author}{\bibfnamefont{S.}~\bibnamefont{Ili\ifmmode~\acute{c}\else
  \'{c}\fi{}}}, \bibnamefont{and} \bibinfo{author}{\bibfnamefont{F.~S.}
  \bibnamefont{Bergeret}}, \bibinfo{journal}{Phys. Rev. B}
  \textbf{\bibinfo{volume}{104}}, \bibinfo{pages}{214515}
  (\bibinfo{year}{2021}),
  \urlprefix\url{https://link.aps.org/doi/10.1103/PhysRevB.104.214515}.

\bibitem[{\citenamefont{Margineda et~al.}(2023)\citenamefont{Margineda,
  Claydon, Qejvanaj, and Checkley}}]{MarginedaCheckley23}
\bibinfo{author}{\bibfnamefont{D.}~\bibnamefont{Margineda}},
  \bibinfo{author}{\bibfnamefont{J.~S.} \bibnamefont{Claydon}},
  \bibinfo{author}{\bibfnamefont{F.}~\bibnamefont{Qejvanaj}}, \bibnamefont{and}
  \bibinfo{author}{\bibfnamefont{C.}~\bibnamefont{Checkley}},
  \bibinfo{journal}{Physical Review B} \textbf{\bibinfo{volume}{107}},
  \bibinfo{pages}{L100502} (\bibinfo{year}{2023}), \bibinfo{note}{publisher:
  American Physical Society},
  \urlprefix\url{https://link.aps.org/doi/10.1103/PhysRevB.107.L100502}.

\bibitem[{\citenamefont{Tian and Zhang}(2025)}]{TianZhang25}
\bibinfo{author}{\bibfnamefont{W.}~\bibnamefont{Tian}} \bibnamefont{and}
  \bibinfo{author}{\bibfnamefont{H.}~\bibnamefont{Zhang}},
  \bibinfo{journal}{Applied Physics Letters} \textbf{\bibinfo{volume}{127}},
  \bibinfo{pages}{072604} (\bibinfo{year}{2025}), ISSN
  \bibinfo{issn}{0003-6951},
  \eprint{https://pubs.aip.org/aip/apl/article-pdf/doi/10.1063/5.0277751/20655713/072604_1_5.0277751.pdf},
  \urlprefix\url{https://doi.org/10.1063/5.0277751}.

\bibitem[{\citenamefont{Clarke and Wilhelm}(2008)}]{Clarke08}
\bibinfo{author}{\bibfnamefont{J.}~\bibnamefont{Clarke}} \bibnamefont{and}
  \bibinfo{author}{\bibfnamefont{F.~K.} \bibnamefont{Wilhelm}},
  \bibinfo{journal}{Nature} \textbf{\bibinfo{volume}{453}},
  \bibinfo{pages}{1031} (\bibinfo{year}{2008}).

\bibitem[{\citenamefont{Devoret and Schoelkopf}(2013)}]{Devoret13}
\bibinfo{author}{\bibfnamefont{M.~H.} \bibnamefont{Devoret}} \bibnamefont{and}
  \bibinfo{author}{\bibfnamefont{R.~J.} \bibnamefont{Schoelkopf}},
  \bibinfo{journal}{Science} \textbf{\bibinfo{volume}{339}},
  \bibinfo{pages}{1169} (\bibinfo{year}{2013}),
  \eprint{https://www.science.org/doi/pdf/10.1126/science.1231930},
  \urlprefix\url{https://www.science.org/doi/abs/10.1126/science.1231930}.

\bibitem[{\citenamefont{Huang et~al.}(2020)\citenamefont{Huang, Wu, Fan, and
  Zhu}}]{Huang20}
\bibinfo{author}{\bibfnamefont{H.-L.} \bibnamefont{Huang}},
  \bibinfo{author}{\bibfnamefont{D.}~\bibnamefont{Wu}},
  \bibinfo{author}{\bibfnamefont{D.}~\bibnamefont{Fan}}, \bibnamefont{and}
  \bibinfo{author}{\bibfnamefont{X.}~\bibnamefont{Zhu}},
  \bibinfo{journal}{Science China Information Sciences}
  \textbf{\bibinfo{volume}{63}}, \bibinfo{pages}{180501}
  (\bibinfo{year}{2020}).

\bibitem[{\citenamefont{AbuGhanem}(2025)}]{Abughanem25}
\bibinfo{author}{\bibfnamefont{M.}~\bibnamefont{AbuGhanem}},
  \bibinfo{journal}{EPJ Quantum Technology} \textbf{\bibinfo{volume}{12}},
  \bibinfo{pages}{102} (\bibinfo{year}{2025}).

\bibitem[{\citenamefont{Likharev and Semenov}(1991)}]{LikharevSemenov91}
\bibinfo{author}{\bibfnamefont{K.}~\bibnamefont{Likharev}} \bibnamefont{and}
  \bibinfo{author}{\bibfnamefont{V.}~\bibnamefont{Semenov}},
  \bibinfo{journal}{IEEE Transactions on Applied Superconductivity}
  \textbf{\bibinfo{volume}{1}}, \bibinfo{pages}{3} (\bibinfo{year}{1991}).

\bibitem[{\citenamefont{Likharev}(2012)}]{Likharev12}
\bibinfo{author}{\bibfnamefont{K.~K.} \bibnamefont{Likharev}},
  \bibinfo{journal}{Physica C: Superconductivity and its Applications}
  \textbf{\bibinfo{volume}{482}}, \bibinfo{pages}{6} (\bibinfo{year}{2012}),
  ISSN \bibinfo{issn}{0921-4534},
  \urlprefix\url{https://www.sciencedirect.com/science/article/pii/S0921453412002481}.

\bibitem[{\citenamefont{Rasmussen et~al.}(2021)\citenamefont{Rasmussen,
  Christensen, Pedersen, Kristensen, B\ae{}kkegaard, Loft, and
  Zinner}}]{Rasmussen21}
\bibinfo{author}{\bibfnamefont{S.}~\bibnamefont{Rasmussen}},
  \bibinfo{author}{\bibfnamefont{K.}~\bibnamefont{Christensen}},
  \bibinfo{author}{\bibfnamefont{S.}~\bibnamefont{Pedersen}},
  \bibinfo{author}{\bibfnamefont{L.}~\bibnamefont{Kristensen}},
  \bibinfo{author}{\bibfnamefont{T.}~\bibnamefont{B\ae{}kkegaard}},
  \bibinfo{author}{\bibfnamefont{N.}~\bibnamefont{Loft}}, \bibnamefont{and}
  \bibinfo{author}{\bibfnamefont{N.}~\bibnamefont{Zinner}},
  \bibinfo{journal}{PRX Quantum} \textbf{\bibinfo{volume}{2}},
  \bibinfo{pages}{040204} (\bibinfo{year}{2021}),
  \urlprefix\url{https://link.aps.org/doi/10.1103/PRXQuantum.2.040204}.

\bibitem[{\citenamefont{Kim et~al.}()\citenamefont{Kim, Jang, Jin, Shin, Shin,
  Luo, Siddiqi, Kim, Yoon, and Nguyen}}]{KimNguyen25}
\bibinfo{author}{\bibfnamefont{H.}~\bibnamefont{Kim}},
  \bibinfo{author}{\bibfnamefont{G.}~\bibnamefont{Jang}},
  \bibinfo{author}{\bibfnamefont{S.}~\bibnamefont{Jin}},
  \bibinfo{author}{\bibfnamefont{D.}~\bibnamefont{Shin}},
  \bibinfo{author}{\bibfnamefont{H.-J.} \bibnamefont{Shin}},
  \bibinfo{author}{\bibfnamefont{J.}~\bibnamefont{Luo}},
  \bibinfo{author}{\bibfnamefont{I.}~\bibnamefont{Siddiqi}},
  \bibinfo{author}{\bibfnamefont{Y.}~\bibnamefont{Kim}},
  \bibinfo{author}{\bibfnamefont{H.~H.} \bibnamefont{Yoon}}, \bibnamefont{and}
  \bibinfo{author}{\bibfnamefont{L.~B.} \bibnamefont{Nguyen}} (????),
  \eprint{2505.12724}, \urlprefix\url{https://arxiv.org/abs/2505.12724}.

\bibitem[{\citenamefont{Buck}(1956)}]{Buck56}
\bibinfo{author}{\bibfnamefont{D.~A.} \bibnamefont{Buck}},
  \bibinfo{journal}{Proceedings of the IRE} \textbf{\bibinfo{volume}{44}},
  \bibinfo{pages}{482} (\bibinfo{year}{1956}).

\bibitem[{\citenamefont{Rikken et~al.}(2001)\citenamefont{Rikken, F\"olling,
  and Wyder}}]{Rikken01}
\bibinfo{author}{\bibfnamefont{G.~L. J.~A.} \bibnamefont{Rikken}},
  \bibinfo{author}{\bibfnamefont{J.}~\bibnamefont{F\"olling}},
  \bibnamefont{and} \bibinfo{author}{\bibfnamefont{P.}~\bibnamefont{Wyder}},
  \bibinfo{journal}{Phys. Rev. Lett.} \textbf{\bibinfo{volume}{87}},
  \bibinfo{pages}{236602} (\bibinfo{year}{2001}),
  \urlprefix\url{https://link.aps.org/doi/10.1103/PhysRevLett.87.236602}.

\bibitem[{\citenamefont{Wakatsuki et~al.}(2017)\citenamefont{Wakatsuki, Saito,
  Hoshino, Itahashi, Ideue, Ezawa, Iwasa, and Nagaosa}}]{WakatsukiNagaosa17}
\bibinfo{author}{\bibfnamefont{R.}~\bibnamefont{Wakatsuki}},
  \bibinfo{author}{\bibfnamefont{Y.}~\bibnamefont{Saito}},
  \bibinfo{author}{\bibfnamefont{S.}~\bibnamefont{Hoshino}},
  \bibinfo{author}{\bibfnamefont{Y.~M.} \bibnamefont{Itahashi}},
  \bibinfo{author}{\bibfnamefont{T.}~\bibnamefont{Ideue}},
  \bibinfo{author}{\bibfnamefont{M.}~\bibnamefont{Ezawa}},
  \bibinfo{author}{\bibfnamefont{Y.}~\bibnamefont{Iwasa}}, \bibnamefont{and}
  \bibinfo{author}{\bibfnamefont{N.}~\bibnamefont{Nagaosa}},
  \bibinfo{journal}{Science Advances} \textbf{\bibinfo{volume}{3}},
  \bibinfo{pages}{e1602390} (\bibinfo{year}{2017}),
  \eprint{https://www.science.org/doi/pdf/10.1126/sciadv.1602390},
  \urlprefix\url{https://www.science.org/doi/abs/10.1126/sciadv.1602390}.

\bibitem[{\citenamefont{Wakatsuki and Nagaosa}(2018)}]{WakatsukiNagaosa18}
\bibinfo{author}{\bibfnamefont{R.}~\bibnamefont{Wakatsuki}} \bibnamefont{and}
  \bibinfo{author}{\bibfnamefont{N.}~\bibnamefont{Nagaosa}},
  \bibinfo{journal}{Phys. Rev. Lett.} \textbf{\bibinfo{volume}{121}},
  \bibinfo{pages}{026601} (\bibinfo{year}{2018}),
  \urlprefix\url{https://link.aps.org/doi/10.1103/PhysRevLett.121.026601}.

\bibitem[{\citenamefont{Daido and Yanase}(2024)}]{DaidoYanase24}
\bibinfo{author}{\bibfnamefont{A.}~\bibnamefont{Daido}} \bibnamefont{and}
  \bibinfo{author}{\bibfnamefont{Y.}~\bibnamefont{Yanase}},
  \bibinfo{journal}{Phys. Rev. Res.} \textbf{\bibinfo{volume}{6}},
  \bibinfo{pages}{L022009} (\bibinfo{year}{2024}),
  \urlprefix\url{https://link.aps.org/doi/10.1103/PhysRevResearch.6.L022009}.

\bibitem[{\citenamefont{Attias et~al.}(2024)\citenamefont{Attias, Michaeli, and
  Khodas}}]{AttiasKhodas24}
\bibinfo{author}{\bibfnamefont{L.}~\bibnamefont{Attias}},
  \bibinfo{author}{\bibfnamefont{K.}~\bibnamefont{Michaeli}}, \bibnamefont{and}
  \bibinfo{author}{\bibfnamefont{M.}~\bibnamefont{Khodas}},
  \bibinfo{journal}{Phys. Rev. B} \textbf{\bibinfo{volume}{110}},
  \bibinfo{pages}{014521} (\bibinfo{year}{2024}),
  \urlprefix\url{https://link.aps.org/doi/10.1103/PhysRevB.110.014521}.

\bibitem[{\citenamefont{Suemune et~al.}(2006)\citenamefont{Suemune, Akazaki,
  Tanaka, Jo, Uesugi, Endo, Kumano, Hanamura, Takayanagi, Yamanishi
  et~al.}}]{Suemune06}
\bibinfo{author}{\bibfnamefont{I.}~\bibnamefont{Suemune}},
  \bibinfo{author}{\bibfnamefont{T.}~\bibnamefont{Akazaki}},
  \bibinfo{author}{\bibfnamefont{K.}~\bibnamefont{Tanaka}},
  \bibinfo{author}{\bibfnamefont{M.}~\bibnamefont{Jo}},
  \bibinfo{author}{\bibfnamefont{K.}~\bibnamefont{Uesugi}},
  \bibinfo{author}{\bibfnamefont{M.}~\bibnamefont{Endo}},
  \bibinfo{author}{\bibfnamefont{H.}~\bibnamefont{Kumano}},
  \bibinfo{author}{\bibfnamefont{E.}~\bibnamefont{Hanamura}},
  \bibinfo{author}{\bibfnamefont{H.}~\bibnamefont{Takayanagi}},
  \bibinfo{author}{\bibfnamefont{M.}~\bibnamefont{Yamanishi}},
  \bibnamefont{et~al.}, \bibinfo{journal}{Japanese Journal of Applied Physics}
  \textbf{\bibinfo{volume}{45}}, \bibinfo{pages}{9264} (\bibinfo{year}{2006}),
  \urlprefix\url{https://doi.org/10.1143/JJAP.45.9264}.

\bibitem[{\citenamefont{Recher et~al.}(2010)\citenamefont{Recher, Nazarov, and
  Kouwenhoven}}]{RecherNazarovKouwenhoven10}
\bibinfo{author}{\bibfnamefont{P.}~\bibnamefont{Recher}},
  \bibinfo{author}{\bibfnamefont{Y.~V.} \bibnamefont{Nazarov}},
  \bibnamefont{and} \bibinfo{author}{\bibfnamefont{L.~P.}
  \bibnamefont{Kouwenhoven}}, \bibinfo{journal}{Physical Review Letters}
  \textbf{\bibinfo{volume}{104}}, \bibinfo{pages}{156802}
  (\bibinfo{year}{2010}), \bibinfo{note}{publisher: American Physical Society},
  \urlprefix\url{https://link.aps.org/doi/10.1103/PhysRevLett.104.156802}.

\bibitem[{\citenamefont{Zhang and Wang}(2023)}]{Zhang23}
\bibinfo{author}{\bibfnamefont{Y.}~\bibnamefont{Zhang}} \bibnamefont{and}
  \bibinfo{author}{\bibfnamefont{Z.}~\bibnamefont{Wang}},
  \bibinfo{journal}{Phys. Rev. B} \textbf{\bibinfo{volume}{107}},
  \bibinfo{pages}{224510} (\bibinfo{year}{2023}),
  \urlprefix\url{https://link.aps.org/doi/10.1103/PhysRevB.107.224510}.

\bibitem[{\citenamefont{Mao et~al.}(2024)\citenamefont{Mao, Yan, Zhuang, and
  Sun}}]{MaoSun24}
\bibinfo{author}{\bibfnamefont{Y.}~\bibnamefont{Mao}},
  \bibinfo{author}{\bibfnamefont{Q.}~\bibnamefont{Yan}},
  \bibinfo{author}{\bibfnamefont{Y.-C.} \bibnamefont{Zhuang}},
  \bibnamefont{and} \bibinfo{author}{\bibfnamefont{Q.-F.} \bibnamefont{Sun}},
  \bibinfo{journal}{Phys. Rev. Lett.} \textbf{\bibinfo{volume}{132}},
  \bibinfo{pages}{216001} (\bibinfo{year}{2024}),
  \urlprefix\url{https://link.aps.org/doi/10.1103/PhysRevLett.132.216001}.

\bibitem[{\citenamefont{Sun et~al.}(2025{\natexlab{b}})\citenamefont{Sun,
  Tjernshaugen, and Linder}}]{SunLinder25}
\bibinfo{author}{\bibfnamefont{C.}~\bibnamefont{Sun}},
  \bibinfo{author}{\bibfnamefont{J.~B.} \bibnamefont{Tjernshaugen}},
  \bibnamefont{and} \bibinfo{author}{\bibfnamefont{J.}~\bibnamefont{Linder}},
  \bibinfo{journal}{Phys. Rev. B} \textbf{\bibinfo{volume}{112}},
  \bibinfo{pages}{064504} (\bibinfo{year}{2025}{\natexlab{b}}),
  \urlprefix\url{https://link.aps.org/doi/10.1103/h2qg-qhf7}.

\bibitem[{\citenamefont{Martínez-Pérez and
  Giazotto}(2013)}]{MartinezPerezGiazotto13}
\bibinfo{author}{\bibfnamefont{M.~J.} \bibnamefont{Martínez-Pérez}}
  \bibnamefont{and} \bibinfo{author}{\bibfnamefont{F.}~\bibnamefont{Giazotto}},
  \bibinfo{journal}{Applied Physics Letters} \textbf{\bibinfo{volume}{102}},
  \bibinfo{pages}{182602} (\bibinfo{year}{2013}), ISSN
  \bibinfo{issn}{0003-6951},
  \eprint{https://pubs.aip.org/aip/apl/article-pdf/doi/10.1063/1.4804550/14272388/182602_1_online.pdf},
  \urlprefix\url{https://doi.org/10.1063/1.4804550}.

\bibitem[{\citenamefont{Trupiano et~al.}(2025)\citenamefont{Trupiano, Simoni,
  and Giazotto}}]{TrupianoGiazotto25}
\bibinfo{author}{\bibfnamefont{G.}~\bibnamefont{Trupiano}},
  \bibinfo{author}{\bibfnamefont{G.~D.} \bibnamefont{Simoni}},
  \bibnamefont{and} \bibinfo{author}{\bibfnamefont{F.}~\bibnamefont{Giazotto}},
  \emph{\bibinfo{title}{A thermally modulated sinis trasconductance amplifier}}
  (\bibinfo{year}{2025}), \eprint{2505.21341},
  \urlprefix\url{https://arxiv.org/abs/2505.21341}.

\bibitem[{\citenamefont{Debnath et~al.}(2025)\citenamefont{Debnath, Saha, and
  Dutta}}]{DebnathSaha25}
\bibinfo{author}{\bibfnamefont{D.}~\bibnamefont{Debnath}},
  \bibinfo{author}{\bibfnamefont{A.}~\bibnamefont{Saha}}, \bibnamefont{and}
  \bibinfo{author}{\bibfnamefont{P.}~\bibnamefont{Dutta}}
  (\bibinfo{year}{2025}), \eprint{2509.12198},
  \urlprefix\url{https://arxiv.org/abs/2509.12198}.

\bibitem[{\citenamefont{Zeng}(2025)}]{Zeng25}
\bibinfo{author}{\bibfnamefont{W.}~\bibnamefont{Zeng}}, \bibinfo{journal}{Phys.
  Rev. Lett.} \textbf{\bibinfo{volume}{134}}, \bibinfo{pages}{176002}
  (\bibinfo{year}{2025}),
  \urlprefix\url{https://link.aps.org/doi/10.1103/PhysRevLett.134.176002}.

\bibitem[{\citenamefont{Fu et~al.}(2025)\citenamefont{Fu, Xu, Liu, Lee, and
  Ang}}]{Fu25}
\bibinfo{author}{\bibfnamefont{P.-H.} \bibnamefont{Fu}},
  \bibinfo{author}{\bibfnamefont{Y.}~\bibnamefont{Xu}},
  \bibinfo{author}{\bibfnamefont{J.-F.} \bibnamefont{Liu}},
  \bibinfo{author}{\bibfnamefont{C.~H.} \bibnamefont{Lee}}, \bibnamefont{and}
  \bibinfo{author}{\bibfnamefont{Y.~S.} \bibnamefont{Ang}},
  \bibinfo{journal}{Phys. Rev. B} \textbf{\bibinfo{volume}{111}},
  \bibinfo{pages}{L020507} (\bibinfo{year}{2025}),
  \urlprefix\url{https://link.aps.org/doi/10.1103/PhysRevB.111.L020507}.

\bibitem[{\citenamefont{Parafilo et~al.}(2025)\citenamefont{Parafilo, Kovalev,
  and Savenko}}]{Parafilo25}
\bibinfo{author}{\bibfnamefont{A.~V.} \bibnamefont{Parafilo}},
  \bibinfo{author}{\bibfnamefont{V.~M.} \bibnamefont{Kovalev}},
  \bibnamefont{and} \bibinfo{author}{\bibfnamefont{I.~G.}
  \bibnamefont{Savenko}}, \bibinfo{journal}{Phys. Rev. B}
  \textbf{\bibinfo{volume}{112}}, \bibinfo{pages}{104501}
  (\bibinfo{year}{2025}),
  \urlprefix\url{https://link.aps.org/doi/10.1103/qj13-lw4p}.

\bibitem[{\citenamefont{Davydova et~al.}(2024)\citenamefont{Davydova, Geier,
  and Fu}}]{DavydovaFu24}
\bibinfo{author}{\bibfnamefont{M.}~\bibnamefont{Davydova}},
  \bibinfo{author}{\bibfnamefont{M.}~\bibnamefont{Geier}}, \bibnamefont{and}
  \bibinfo{author}{\bibfnamefont{L.}~\bibnamefont{Fu}},
  \bibinfo{journal}{Science Advances} \textbf{\bibinfo{volume}{10}},
  \bibinfo{pages}{eadr4817} (\bibinfo{year}{2024}), \bibinfo{note}{publisher:
  American Association for the Advancement of Science},
  \urlprefix\url{https://www.science.org/doi/10.1126/sciadv.adr4817}.

\bibitem[{\citenamefont{Hosur}(2024)}]{Hosur24}
\bibinfo{author}{\bibfnamefont{P.}~\bibnamefont{Hosur}} (\bibinfo{year}{2024}),
  \eprint{2410.23352}, \urlprefix\url{https://arxiv.org/abs/2410.23352}.

\bibitem[{\citenamefont{Upadhyay et~al.}(2024)\citenamefont{Upadhyay, Golubev,
  Chang, Thomas, Guthrie, Peltonen, and Pekola}}]{UpadhyayGolubev24}
\bibinfo{author}{\bibfnamefont{R.}~\bibnamefont{Upadhyay}},
  \bibinfo{author}{\bibfnamefont{D.~S.} \bibnamefont{Golubev}},
  \bibinfo{author}{\bibfnamefont{Y.-C.} \bibnamefont{Chang}},
  \bibinfo{author}{\bibfnamefont{G.}~\bibnamefont{Thomas}},
  \bibinfo{author}{\bibfnamefont{A.}~\bibnamefont{Guthrie}},
  \bibinfo{author}{\bibfnamefont{J.~T.} \bibnamefont{Peltonen}},
  \bibnamefont{and} \bibinfo{author}{\bibfnamefont{J.~P.}
  \bibnamefont{Pekola}}, \bibinfo{journal}{Nature Communications}
  \textbf{\bibinfo{volume}{15}}, \bibinfo{pages}{630} (\bibinfo{year}{2024}),
  \urlprefix\url{https://doi.org/10.1038/s41467-024-44908-w}.

\bibitem[{\citenamefont{Ingla-Ayn{\'e}s
  et~al.}(2025)\citenamefont{Ingla-Ayn{\'e}s, Hou, Wang, Chu, Mukhanov, Wei,
  and Moodera}}]{InglaAynesMoodera25}
\bibinfo{author}{\bibfnamefont{J.}~\bibnamefont{Ingla-Ayn{\'e}s}},
  \bibinfo{author}{\bibfnamefont{Y.}~\bibnamefont{Hou}},
  \bibinfo{author}{\bibfnamefont{S.}~\bibnamefont{Wang}},
  \bibinfo{author}{\bibfnamefont{E.-D.} \bibnamefont{Chu}},
  \bibinfo{author}{\bibfnamefont{O.~A.} \bibnamefont{Mukhanov}},
  \bibinfo{author}{\bibfnamefont{P.}~\bibnamefont{Wei}}, \bibnamefont{and}
  \bibinfo{author}{\bibfnamefont{J.~S.} \bibnamefont{Moodera}},
  \bibinfo{journal}{Nature Electronics} \textbf{\bibinfo{volume}{8}},
  \bibinfo{pages}{411} (\bibinfo{year}{2025}),
  \urlprefix\url{https://doi.org/10.1038/s41928-025-01375-5}.

\bibitem[{\citenamefont{Castellani et~al.}(2025)\citenamefont{Castellani,
  Medeiros, Buzzi, Foster, Colangelo, and Berggren}}]{Castellani25}
\bibinfo{author}{\bibfnamefont{M.}~\bibnamefont{Castellani}},
  \bibinfo{author}{\bibfnamefont{O.}~\bibnamefont{Medeiros}},
  \bibinfo{author}{\bibfnamefont{A.}~\bibnamefont{Buzzi}},
  \bibinfo{author}{\bibfnamefont{R.~A.} \bibnamefont{Foster}},
  \bibinfo{author}{\bibfnamefont{M.}~\bibnamefont{Colangelo}},
  \bibnamefont{and} \bibinfo{author}{\bibfnamefont{K.~K.}
  \bibnamefont{Berggren}} (\bibinfo{year}{2025}), \eprint{2406.12175},
  \urlprefix\url{https://arxiv.org/abs/2406.12175}.

\end{thebibliography}

\end{document}